\documentclass[aps,reprint,superscriptaddress,prb,longbibliography]{revtex4-2}
\usepackage{amsmath}
\usepackage{amssymb}
\usepackage{graphicx}
\usepackage[english]{babel}
\usepackage{bm}
\usepackage{color}
\usepackage{hyperref}
\hypersetup{
    colorlinks=true,
    linkcolor=blue,
    citecolor=blue,      
    urlcolor=blue,
}

\newcommand{\epsvac}{\varepsilon_0} 
\newcommand{\dee}{\mathrm{d}} 

\newcommand{\eref}[1]{Eq.~(\ref{#1})}  
\newcommand{\erefs}[2]{Eqs.~(\ref{#1})-(\ref{#2})}  
\newcommand{\figref}[1]{Fig.~\ref{#1}}  
\newcommand{\figrefs}[2]{Figs.~\ref{#1}-\ref{#2}}  
\newcommand{\secref}[1]{Sec.~\ref{#1}}  
\newcommand{\secrefs}[2]{Secs.~\ref{#1}-\ref{#2}}  
\newcommand{\tabref}[1]{Table \ref{#1}}  
\newcommand{\appref}[1]{Appendix \ref{#1}}  

\begin{document}

\title{First-principles theory of nonlinear long-range electron-phonon
interaction}

\author{Matthew Houtput}
\affiliation{Theory of Quantum Systems and Complex Systems, Universiteit Antwerpen, B-2000 Antwerpen, Belgium}
\affiliation{Faculty of Physics, Computational Materials Physics, University of Vienna,
Kolingasse 14-16, Vienna A-1090, Austria}

\author{Luigi Ranalli}
\affiliation{Faculty of Physics, Computational Materials Physics, University of Vienna,
Kolingasse 14-16, Vienna A-1090, Austria}

\author{Carla Verdi}
\affiliation{School of Mathematics and Physics, University of Queensland, 4072 Brisbane, Queensland, Australia}

\author{Serghei Klimin}
\affiliation{Theory of Quantum Systems and Complex Systems, Universiteit Antwerpen, B-2000 Antwerpen, Belgium}

\author{Stefano Ragni}
\affiliation{Faculty of Physics, Computational Materials Physics, University of Vienna,
Kolingasse 14-16, Vienna A-1090, Austria}

\author{Cesare Franchini}
\affiliation{Faculty of Physics, Computational Materials Physics, University of Vienna,
Kolingasse 14-16, Vienna A-1090, Austria}
\affiliation{Department of Physics and Astronomy, Alma Mater Studiorum - Universit\`a di Bologna, Bologna, 40127 Italy}

\author{Jacques Tempere}
\affiliation{Theory of Quantum Systems and Complex Systems, Universiteit Antwerpen, B-2000 Antwerpen, Belgium}

\begin{abstract}
Describing electron-phonon interactions in a solid requires knowledge of the electron-phonon matrix elements in the Hamiltonian. State-of-the-art first-principles calculations for the electron-phonon interaction are limited to the 1-electron-1-phonon matrix element, which is suitable for harmonic materials. However, there is no first-principles theory for 1-electron-2-phonon interactions, which occur in anharmonic materials with significant electron-phonon interaction such as halide perovskites and quantum paraelectrics. Here, we derive an analytical expression for the long-range part of the 1-electron-2-phonon matrix element, written in terms of microscopic quantities that can be calculated from first principles. We show that the long-range 1-electron-2-phonon interaction is described by the derivative of the phonon dynamical matrix with respect to an external electric field. We calculate the quasiparticle energy of a large polaron including 1-electron-2-phonon interaction, and show that it can be written in terms of a 1-electron-2-phonon spectral function $\mathcal{T}_{\alpha \beta}(\omega)$. We demonstrate how to calculate this spectral function and its temperature dependence for the benchmark materials LiF and KTaO$_3$, where it turns out that the effect is very small. The first-principles framework developed in this article is general, paving the way for future calculations of 1-electron-2-phonon interactions in materials where the effect may be larger.
\end{abstract}

\maketitle

\section{Introduction} \label{sec:Introduction}
Electrons in a crystalline solid can interact with the lattice vibrations of the nuclei. This electron-phonon interaction is responsible for a wide range of interesting properties of materials, including formation of polarons \cite{landau1933, landau1948, frohlich1954, holstein1959, holstein1959a, devreese2009, sio2019, sio2019a, lafuente-bartolome2022, lafuente-bartolome2022a}, band gap renormalization \cite{feynman1955, allen1976, giustino2017}, conventional superconductivity \cite{eliashberg1960, mcmillan1968, allen1975, marsiglio2008}, and contributions to the electrical conductivity \cite{feynman1962, kadanoff1963, tempere2001, ponce2020}. In general, the Hamiltonian describing a coupled electron-phonon system consists of three contributions: the Hamiltonian of the electron system $\hat{H}_{\text{el}}$, the Hamiltonian of the phonon system $\hat{H}_{\text{ph}}$, and a term $\hat{H}_{\text{el-ph}}$ that describes how electrons and phonons interact with each other.

Many treatments of a coupled electron-phonon system rely on the harmonic approximation, where the Hamiltonian is expanded up to lowest order in the ionic displacements. This means the phonon Hamiltonian $\hat{H}_{\text{ph}}$ is approximated up to second order in the ionic displacements. Additionally, the electron-phonon interaction $\hat{H}_{\text{el-ph}}$ is approximated up to first order: this is known as linear electron-phonon interaction. In the harmonic approximation, the Hamiltonian is entirely described by three inputs: $\hat{H}_{\text{el}}$ depends only on the electron bands $\epsilon_{\mathbf{k},n}$, $\hat{H}_{\text{ph}}$ depends only on the phonon bands $\omega_{\mathbf{q},\nu}$, and $\hat{H}_{\text{el-ph}}$ only depends on the electron-phonon matrix element $g_{mn\nu}(\mathbf{k},\mathbf{q})$ (see also \secref{sec:MatrixElements}, \erefs{HamEl}{Ham1e1ph}. This formulation can be applied in several different contexts. For example, one could choose analytic models for the electron bands, phonon bands, and electron-phonon matrix elements: in particular, the Fr\"ohlich \cite{frohlich1954} and Holstein Hamiltonian \cite{holstein1959, holstein1959a} are both special cases of this formulation. On the other hand, keeping $\epsilon_{\mathbf{k},n}$, $\omega_{\mathbf{q},\nu}$, and $g_{mn\nu}(\mathbf{k},\mathbf{q})$ as general functions allows one to interface the Hamiltonian with first-principles calculations. Indeed, they can represent the electron bands, phonon bands, and electron-phonon matrix elements of a specific material, which can be calculated with a first-principles code such as VASP \cite{kresse1993, kresse1996, kresse1996a}, Quantum ESPRESSO \cite{giannozzi2009, giannozzi2017}, ABINIT \cite{gonze2020, romero2020}, EPW \cite{lee2023}, or Perturbo \cite{zhou2021}.

Physically, the electron-phonon matrix element $g_{mn\nu}(\mathbf{k},\mathbf{q})$ can be split into a long-range and a short-range contribution. In polar materials, $g_{mn\nu}(\mathbf{k},\mathbf{q})$ has a $|\mathbf{q}|^{-1}$ divergence around $\mathbf{q} \approx \mathbf{0}$, due to the appearance of nonzero Born effective charges: this allows long-wavelength phonons to generate macroscopic electric fields, which couple to the electrons. The $|\mathbf{q}|^{-1}$ divergence is the long-range contribution. It is the basis for the Fr\"ohlich Hamiltonian \cite{frohlich1954}, and in general, it has a semi-analytic expression which was first calculated in \cite{verdi2015}. In some contexts, the long-range electron-phonon interaction dominates the short-range interaction, which can then be neglected: most notably, this happens in the large polaron problem, where the electron wavefunction spans numerous unit cells. Then, the electron only interacts with the $\mathbf{q} \approx \mathbf{0}$ phonon modes, which are the modes whose momentum $\hbar\mathbf{q}$ is nonzero but very small with respect to the edge of the Brillouin zone \cite{frohlich1954, sio2019}.

For anharmonic materials, the ionic displacements can be large, and higher order terms in the ionic displacements should be added to both $\hat{H}_{\text{ph}}$ and $\hat{H}_{\text{el-ph}}$. The higher order terms in $\hat{H}_{\text{ph}}$ are due to phonon anharmonicity, and manifest themselves as 3-phonon interaction, 4-phonon interactions, etc. The higher order terms in $\hat{H}_{\text{el-ph}}$ are known as nonlinear electron-phonon interactions: they are expected to be important for materials with strong electron-phonon interaction and strong anharmonicity. Examples of such materials include the quantum paraelectrics SrTiO$_3$ and KTaO$_3$ \cite{collignon2019, gastiasoro2020, gupta2022, ranalli2023, verdi2023}, halide perovskites \cite{saidi2016, jena2019, schilcher2021, yamada2022}, and high-pressure hydrides \cite{errea2014, drozdov2015, errea2015, somayazulu2019, errea2020, hirsch2021, shipley2021, troyan2021, zhang2022}. The treatment of phonon anharmonicity is well-known in the literature \cite{deinzer2003, ward2009} and can be tackled on the first-principles level with codes such as Phono3Py \cite{togo2023, togo2023a} or SSCHA \cite{monacelli2021}. However, the nonlinear electron-phonon interaction has been much less studied in the literature. The main obstacle for a general treatment is the increased complexity of the nonlinear electron-phonon matrix elements. Even for the 1-electron-2-phonon interaction, which is the lowest order nonlinear term, there is no general theory or code available for a practical calculation of the matrix element $g_{mn\nu_1\nu_2}(\mathbf{k},\mathbf{q}_1,\mathbf{q}_2)$. Therefore, almost all literature treatments of nonlinear electron-phonon interaction that the authors are aware of start from a simplified model: either from a Holstein-type lattice Hamiltonian \cite{riseborough1984, adolphs2013, adolphs2014, li2015, li2015a, kennes2017, sentef2017, dee2020, grandi2021, sous2021, prokofev2022, ragni2023, zhang2023, han2024, kovac2024, klimin2024}, or from a model Hamiltonian for 1-electron-2-phonon interaction in SrTiO$_3$ \cite{ngai1974, epifanov1981, vandermarel2019, gastiasoro2020, kiselov2021, kumar2021, nazaryan2021, klimin2024}. Both of these Hamiltonians include phenomenological parameters, which makes them incompatible with a full first-principles approach. Notable exceptions are the first-principles calculation of renormalized electronic quantities based on adiabatic supercell approaches \cite{zacharias2016, monserrat2018, zacharias2020, kundu2021}, the first-principles treatment of conventional superconductivity in MgB$_2$ based on a second-order expansion of the deformation potential \cite{yildirim2001, liu2001}, and the first-principles theory of nonlinear electron-phonon interaction for conventional superconductivity in anharmonic metals such as PdH \cite{bianco2023}. Of these three treatments, only the last one provides a way to obtain the 1-electron-2-phonon matrix element. However, it is limited to metals, where only the short-range part of the electron-phonon interaction is considered because the long-range part is essentially suppressed due to very large screening. A theory for the long-range electron-phonon interaction in semiconductors, due to the generation of a macroscopic electric field, is still lacking.

In this article, we provide a general theory of long-range 1-electron-2-phonon interaction that can be combined with first-principles calculations. We propose a derivation for the long-range part of the nonlinear electron-phonon interaction, which we use to find the semi-analytic expression for the long-range part of the 1-electron-2-phonon interaction matrix element in arbitrary materials. This derivation contains the result of \cite{verdi2015} for the first-order long-range interaction as a special case, and can be straightforwardly generalized to arbitrary order. The only anharmonic quantity that appears is the derivative of the phonon dynamical matrix with respect to an external electric field, which plays a similar role as the Born effective charge tensor in the 1-electron-1-phonon interaction \cite{verdi2015}. As a demonstration for the theory presented in this article, we then use this matrix element to calculate the quasiparticle energy shifts $\Delta \varepsilon_{\mathbf{k},n}$ of a large polaron. We show that the result only depends on well-known harmonic quantities and a new 1-electron-2-phonon spectral function $\bm{\mathcal{T}}(\omega)$, which represents the strength of the long-range nonlinear interaction in a material. We explicitly show how to calculate $\bm{\mathcal{T}}(\omega)$ from first principles for the benchmark material LiF and KTaO$_3$ at finite temperatures.

This paper is structured as follows. In \secref{sec:MatrixElements}, we analytically derive an expression for the long-range part of the 1-electron-2-phonon interaction in terms of quantities that can be calculated from first principles. In \secref{sec:EnergyMass}, we use this matrix element to calculate the ground state energy and effective mass of a polaron in a material that only has long-range electron-phonon interaction, including the 1-electron-2-phonon interaction up to lowest order. In \secref{sec:FirstPrinciples}, we calculate the 1-electron-2-phonon matrix element from first principles for LiF and KTaO$_3$, and build a framework to analyse the strength of the 1-electron-2-phonon interaction and to reconstruct where its main contributions originate from. We also calculate the 1-electron-2-phonon spectral function $\bm{\mathcal{T}}(\omega)$ at finite temperatures and show that it grows larger as the temperature increases, in contrast with the equivalent 1-electron-1-phonon spectral function which is independent of temperature. We discuss the results and conclude in \secref{sec:Conclusion}.

\section{Analytical derivation of the long-range electron-phonon matrix elements} 
\label{sec:MatrixElements}

\subsection{Definition of the electron-phonon Hamiltonian} \label{subsec:ephDef}
Consider a crystalline material where the electrons interact with the phonons of the lattice. The Hamiltonian for such a system is given by:
\begin{align} \label{HamTot}
\hat{H} = \hat{H}_{\text{el}} + \hat{H}_{\text{ph}} + \hat{H}_{\text{el-ph}}.
\end{align}
If we limit ourselves to terms that are quadratic in the phonon coordinates, and electron-electron interactions are neglected, the three terms that appear in this Hamiltonian are given by \cite{giustino2017}:
\begin{align}
 \hat{H}_{\text{el}} & = \sum_{\mathbf{k},n} \epsilon_{\mathbf{k},n} \hat{c}^{\dagger}_{\mathbf{k},n} \hat{c}_{\mathbf{k},n}, \label{HamEl} \\
 \hat{H}_{\text{ph}} & = \sum_{\mathbf{q},\nu} \hbar \omega_{\mathbf{q},\nu} \left( \hat{a}^{\dagger}_{\mathbf{q},\nu} \hat{a}_{\mathbf{q},\nu} + \frac{1}{2} \right), \label{HamPh} \\
 \hat{H}_{\text{el-ph}} & = \sqrt{\frac{\Omega_0}{\Omega}}\sum_{\mathbf{k},\mathbf{q},n,m,\nu} g_{mn\nu}(\mathbf{k},\mathbf{q}) \hat{A}_{\mathbf{q},\nu} \hat{c}^{\dagger}_{\mathbf{k}+\mathbf{q},m} \hat{c}_{\mathbf{k},n} \label{Ham1e1ph} \\
& \hspace{-15pt} + \frac{\Omega_0}{\Omega} \sum_{\mathbf{k}, \mathbf{q}_1, \mathbf{q}_2} \sum_{m,n, \nu_1, \nu_2} g_{mn\nu_1\nu_2}(\mathbf{k},\mathbf{q}_1,\mathbf{q}_2) \hat{A}_{\mathbf{q}_1,\nu_1}\hat{A}_{\mathbf{q}_2,\nu_2} \nonumber \\
& \hspace{110pt} \times \hat{c}^{\dagger}_{\mathbf{k}+\mathbf{q}_1+\mathbf{q}_2,m} \hat{c}_{\mathbf{k},n}. \label{Ham1e2ph}
\end{align}
Here, $\hat{c}^{\dagger}_{\mathbf{k},n}$ and $\hat{c}_{\mathbf{k},n}$ are the fermionic electron operators which create and annihilate an electron in a Bloch state $|\psi_{\mathbf{k},n}\rangle$ with crystal momentum $\hbar \mathbf{k}$ in band $n$, and $\hat{A}_{\mathbf{q},\nu} := \hat{a}_{\mathbf{q},\nu}  + \hat{a}^{\dagger}_{-\mathbf{q},\nu}$, where $\hat{a}^{\dagger}_{\mathbf{q},\nu}$ and $\hat{a}_{\mathbf{q},\nu}$ are the bosonic phonon operators which create and annihilate a phonon with crystal momentum $\hbar \mathbf{q}$ in branch $\nu$. $\Omega_0$ and $\Omega$ represent the volume of the unit cell and the system volume, respectively. We will also introduce a Born-von K\'arm\'an supercell of volume $\Omega_{\text{sc}}$ for first-principles calculations. Throughout this article, it is always assumed that $\Omega \gg \Omega_{\text{sc}} \gg \Omega_0$. The assumption $\Omega_{\text{sc}} \gg \Omega_0$ is necessary for convergence of the first-principles calculations, and the assumption $\Omega \gg \Omega_{\text{sc}}$ is the thermodynamic limit, which implies that the sums over the momenta $\mathbf{k}, \mathbf{q}$ can be interpreted as integrals over the first Brillouin zone (1BZ). 
\begin{figure}
\centering
\includegraphics[width=8.6cm]{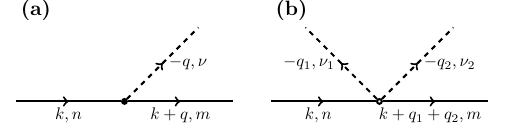}
\caption{Feynman diagrams of electron-phonon interactions that will be considered in this article. \textbf{(a)} Linear 1-electron-1-phonon interaction with vertex factor $g_{mn\nu}(\mathbf{k},\mathbf{q})$. \textbf{(b)} Nonlinear 1-electron-2-phonon interaction with vertex factor $g_{mn\nu_1\nu_2}(\mathbf{k},\mathbf{q}_1,\mathbf{q}_2)$. $k = (\mathbf{k},\omega)$ and $q = (\mathbf{q},\omega')$ represent four-momenta, $m$ and $n$ represent electron bands, and $
\nu$ represents a phonon branch. }
\label{fig:InteractionDiagrams}
\end{figure}
The Feynman diagrams for the 1-electron-1-phonon and 1-electron-2-phonon interactions are shown in \figref{fig:InteractionDiagrams}; the electron-phonon matrix elements $g_{mn\nu}(\mathbf{k},\mathbf{q})$ and $g_{mn\nu_1\nu_2}(\mathbf{k},\mathbf{q}_1,\mathbf{q}_2)$ respectively contain all the physics of these interactions. It is worth pointing out that the 1-electron-2-phonon matrix element is written as $g^{\text{DW}}_{mn\nu_1\nu_2}(\mathbf{k},\mathbf{q}_1, \mathbf{q}_2)$ in \cite{giustino2017}. The superscript DW can be traced back to the fact that the 1-electron-2-phonon interaction is often only used to treat the Debye-Waller term in the self-energy \cite{allen1976, giustino2010, giustino2017}. In the Debye-Waller term, $g_{mn\nu\nu}(\mathbf{k},\mathbf{q}, -\mathbf{q})$ can be approximated by an expression that contains two 1-electron-1-phonon matrix elements \cite{giustino2010}, but this is no longer true for the general 1-electron-2-phonon matrix element $g_{mn\nu_1\nu_2}(\mathbf{k},\mathbf{q}_1, \mathbf{q}_2)$. Since we apply the 1-electron-2-phonon interaction beyond only the Debye-Waller diagram in this article, we will also drop the superscript DW from the notation.

\subsection{Classical derivation of the long-range electron-ion Hamiltonian} \label{subsec:ClassicalDerivation}
The central goal of this article is to find expressions for the long-range parts of the electron-phonon matrix elements, which are due to the classical electrostatic interaction between the electric field $\bm{\mathcal{E}}$ of the electrons and the polarization field $\mathbf{P}$ of the ions in the material. For the 1-electron-1-phonon interaction, these expressions have been derived already \cite{frohlich1954, verdi2015}: here, a derivation is presented that can be generalized to higher orders in the phonon coordinates. To do this, we start from a rather general Hamiltonian of the ionic lattice. Let us write the lattice basis vectors as $\mathbf{a}_1$, $\mathbf{a}_2$, $\mathbf{a}_3$, such that the lattice vectors $\bm{\ell}$ and supercell lattice vectors $\mathbf{T}$ are defined as:
\begin{align}
\bm{\ell} & = m_1 \mathbf{a}_1 + m_2 \mathbf{a}_2 + m_3 \mathbf{a}_3,  \hspace{15pt} m_j \in \{0, 1, \ldots, N_j-1 \}, \\
\mathbf{T} & = n_1 N_1\mathbf{a}_1 + n_2 N_2\mathbf{a}_2 + n_3 N_3\mathbf{a}_3, \hspace{15pt} n_j \in \mathbb{Z}.
\end{align}
This splits the crystal into supercells of size $N_1 \times N_2 \times N_3$, each of which is labeled by a unique supercell vector $\mathbf{T}$. A unique unit cell can be labeled by $\bm{\ell}+\mathbf{T}$, one lattice vector and one supercell vector. This unit cell contains $N$ atoms, which are labelled with an index $\kappa \in \{1, \ldots, N\}$. The equilibrium positions of these atoms within the unit cell are denoted as $\bm{\tau}_{\kappa}$. As the ions move, they will be brought out of their equilibrium positions to a new position $\mathbf{T} + \bm{\ell} + \bm{\tau}_{\kappa} + \mathbf{u}_{\kappa}(\bm{\ell}+\mathbf{T})$, where $\mathbf{u}_{\kappa}(\bm{\ell}+\mathbf{T})$ is the instantaneous displacement of atom $\kappa$ in cell $\bm{\ell}+\mathbf{T}$. Then, the general Hamiltonian of this lattice can be written as:
\begin{equation} \label{HamExtremelyGeneral}
    H = \sum_{\kappa \alpha} \sum_{\bm{\ell}, \mathbf{T}} \frac{P_{\kappa \alpha}^2(\bm{\ell}+\mathbf{T})}{2 m_{\kappa}} + \tilde{E}_0,
\end{equation}
where $\alpha \in \{x, y, z\}$ denotes Cartesian directions, $m_{\kappa}$ is the mass of atom $\kappa$ and $P_{\kappa \alpha}(\bm{\ell}+\mathbf{T}) = m_{\kappa} \dot{u}_{\kappa \alpha}(\bm{\ell}+\mathbf{T})$ is its momentum. Furthermore, $\tilde{E}_0$ is an interaction energy that depends only on the ionic displacements $u_{\kappa \alpha}(\bm{\ell}+\mathbf{T})$ and the macroscopic electric displacement field $\mathbf{D}(\mathbf{r})$. The change in $\tilde{E}_0$ due to a change $\delta u$ in the ionic positions and a change $\delta \mathbf{D}$ in the electric displacement field satisfies (\cite{landau2013}, \S 10):
\begin{align} \label{IonicTerm}
\delta \tilde{E}_0 & = - \sum_{\kappa \alpha} \sum_{\bm{\ell}, \mathbf{T}}m_{\kappa} \ddot{u}_{\kappa \alpha}(\bm{\ell}+\mathbf{T}) \delta u_{\kappa \alpha}(\bm{\ell}+\mathbf{T}) \nonumber \\
& + \int_\Omega \bm{\mathcal{E}}_{\text{tot}}(\mathbf{r}) \cdot \delta \mathbf{D}(\mathbf{r}) d^3 \mathbf{r}.
\end{align}
where $\bm{\mathcal{E}}_{\text{tot}}(\mathbf{r})$ is the macroscopic electric field in the material.

\begin{figure}
    \centering
    \includegraphics[width=8.6cm]{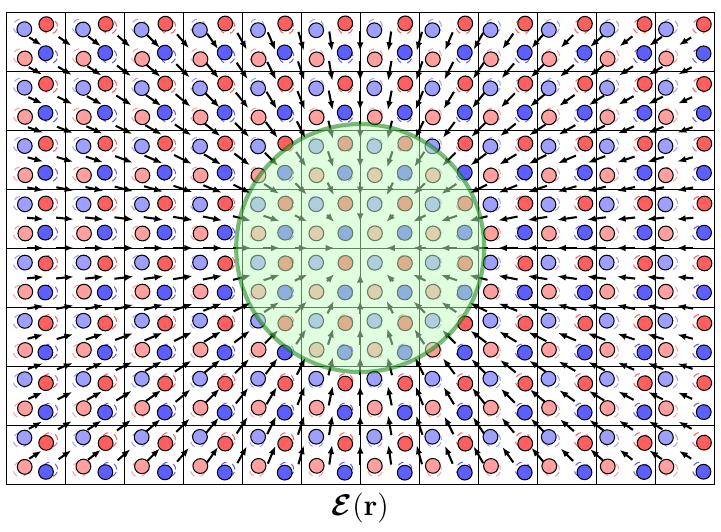}
    \caption{Schematic drawing of a large polaron in a crystalline solid. When the electron wavefunction is much larger than the size of the unit cell, the electric field due to the electron can be treated approximately as a constant external electric field in the unit cell. }
    \label{fig:LargePolaron}
\end{figure}
The general Hamiltonian in \eref{HamExtremelyGeneral} is too difficult to treat computationally for a system in the thermodynamic limit. To simplify this Hamiltonian, we make a key assumption, which is inspired by an observation in the large polaron problem in \figref{fig:LargePolaron}. For a large polaron, the electron wavefunction is very delocalized over the crystal, and the electric field generated by the electron will be approximately constant on the scale of a unit cell or even a supercell. This approximation becomes exact in the large polaron limit. In the more general context of electron-phonon interaction, the same assumption can be made: for the long-range interaction a constant external electric displacement field $\mathbf{D}$ can be assumed, and any spatial variations in this field should be included in the short-range interaction. With this assumption, the relevant Hamiltonian to study is the Hamiltonian of a single supercell, which then takes the following form:
\begin{align}
H_{\text{sc}} & = \sum_{\kappa \alpha} \sum_{\bm{\ell}} \frac{P_{\kappa \alpha}^2(\bm{\ell})}{2 m_{\kappa}} + E_0(\{ \mathbf{u}_{\kappa}(\bm{\ell}) \}, \mathbf{D}),   \label{HamClassicalED} \\
\delta E_0 & = - \sum_{\kappa \alpha} \sum_{\bm{\ell}, \mathbf{T}}m_{\kappa} \ddot{u}_{\kappa \alpha}(\bm{\ell}) \delta u_{\kappa \alpha}(\bm{\ell}) + \Omega_{\text{sc}} \bm{\mathcal{E}}_{\text{tot}} \cdot \delta \mathbf{D}. \label{deltaE0}
\end{align}
The total Hamiltonian is then approximately given by summing the Hamiltonians for all of the supercells $\mathbf{T}$, where we take the supercell average for the externally applied field $\mathbf{D}(\mathbf{T})$:
\begin{align}
H & \approx \sum_{\mathbf{T}} H_{\text{sc}}(\mathbf{T}), \label{HsumSC} \\
\mathbf{D}(\mathbf{T}) & := \frac{1}{\Omega_{\text{sc}}} \int_{\Omega_{\text{sc}}(\mathbf{T})}  \mathbf{D}(\mathbf{r}) \dee^3 \mathbf{r}. \label{DscDef}
\end{align}
This approximation is valid if the supercell is large enough, $\Omega_{\text{sc}} \gg \Omega_0$, such that the interactions between atoms in neighbouring supercells are negligible. By making the simplification in \eref{HsumSC}, all supercells in the system will decouple, and only a single supercell must be treated under the application of a constant electric field. This is computationally possible with first-principles theory.

So far, the $\mathbf{D}$-dependence of $E_0$ was kept general. However, in practice, the external electric field is small, and $E_0$ may be expanded in orders of $\mathbf{D}$. In this article, we will expand $E_0$ up to second order in $\mathbf{D}$, which gives the following expression for the supercell Hamiltonian in \eref{HamClassicalED}:
\begin{align}
H_{\text{sc}} & \approx \sum_{\kappa \alpha,\bm{\ell}}  \frac{P_{\kappa \alpha}^2(\bm{\ell})}{2 m_{\kappa}} + \left. E_0(\{\mathbf{u}_{\kappa}(\bm{\ell})\}) \right|_{\mathbf{D} = \mathbf{0}} \nonumber \\
& + \mathbf{D} \cdot \left. \frac{\partial E_0(\{\mathbf{u}_{\kappa}(\bm{\ell})\})}{\partial \mathbf{D}} \right|_{\mathbf{D} = \mathbf{0}} \nonumber \\
& + \mathbf{D} \cdot \left. \frac{\partial^2 E_0(\{\mathbf{u}_{\kappa}(\bm{\ell})\})}{\partial \mathbf{D} \partial \mathbf{D}} \right|_{\mathbf{D} = \mathbf{0}} \cdot \mathbf{D} + O(\mathbf{D}^3).
\end{align}
Each of the expansion coefficients is still a function of the ionic displacements $\mathbf{u}_{\kappa}(\bm{\ell})$. We will additionally neglect this ionic dependence for the second order term, and write $\left. \frac{\partial^2 E_0(\{\mathbf{u}_{\kappa}(\bm{\ell})\})}{\partial \mathbf{D} \partial \mathbf{D}} \right|_{\mathbf{D} = \mathbf{0}} \approx \left. \frac{\partial^2 E_0}{\partial \mathbf{D} \partial \mathbf{D}} \right|_{\mathbf{D} = \mathbf{u}_{\kappa}(\bm{\ell}) = \mathbf{0}}$. Later, the first order term in $\mathbf{D}$ will lead to all of the 1-electron-$n$-phonon interactions. Therefore, approximating the $O(\mathbf{D}^2)$ term and neglecting all terms of $O(\mathbf{D}^3)$ and higher will not affect the results for the 1-electron-$n$-phonon interactions, which are the primary focus of this article.

We can give names to the expansion coefficients in $\mathbf{D}$, such that they will correspond to physically relevant quantities. Let us define the electron energy $H_{\text{KS}}$, phonon potential energy $E_{\text{ph}}(\{\mathbf{u}_{\kappa}(\bm{\ell})\})$, ionic polarization $\mathbf{P}_{\text{ion}}(\{\mathbf{u}_{\kappa}(\bm{\ell})\})$, and dielectric tensor $\bm{\varepsilon}_{\infty}$ as follows:
\begin{align}
H_{\text{KS}} & := \left.E_0(\{\mathbf{0}\})\right|_{\mathbf{D}=\mathbf{0}}, \label{HksDef} \\
E_{\text{ph}}(\{\mathbf{u}_{\kappa}(\bm{\ell})\}) & := \left. E_0(\{\mathbf{u}_{\kappa}(\bm{\ell})\}) \right|_{\mathbf{D} = \mathbf{0}} - H_{\text{KS}}, \label{EphDef} \\
\mathbf{P}_{\text{ion}}(\{\mathbf{u}_{\kappa}(\bm{\ell})\}) & := -\frac{\epsvac}{\Omega_{\text{sc}}} \bm{\varepsilon}_{\infty} \cdot \left. \frac{\partial E_0(\{\mathbf{u}_{\kappa}(\bm{\ell})\})}{\partial \mathbf{D}} \right|_{\mathbf{D}=\mathbf{0}}, \label{PionDef} \\
\bm{\varepsilon}_{\infty}^{-1} & := \frac{\epsvac}{\Omega_{\text{sc}}} \left. \frac{\partial^2 E_0}{\partial \mathbf{D} \partial \mathbf{D}} \right|_{\mathbf{D} = \mathbf{u}_{\kappa}(\bm{\ell}) = \mathbf{0}}. \label{epsInfDef}
\end{align}
$H_{\text{KS}}$ represents the energy of the system without external electric field and with all ions in their equilibrium positions. After quantization, $H_{\text{KS}}$ will take the role of the Kohn-Sham Hamiltonian. With the definitions in \erefs{EphDef}{epsInfDef}, the supercell Hamiltonian \eref{HamClassicalED} can be written as:
\begin{align} \label{HamSCgen}
H_{\text{sc}} & \approx H_{\text{KS}} + \sum_{\kappa \alpha,\bm{\ell}}  \frac{P_{\kappa \alpha}^2(\bm{\ell})}{2 m_{\kappa}} + E_{\text{ph}}(\{\mathbf{u}_{\kappa}(\bm{\ell})\}) \nonumber \\
& - \frac{\Omega_{\text{sc}}}{\epsvac} \mathbf{D} \cdot \bm{\varepsilon}_{\infty}^{-1} \cdot \mathbf{P}_{\text{ion}}(\{\mathbf{u}_{\kappa}(\bm{\ell})\}) + \frac{\Omega_{\text{sc}}}{2 \epsvac} \mathbf{D} \cdot \bm{\varepsilon}_{\infty}^{-1} \cdot \mathbf{D},
\end{align}
So far, this derivation is written in terms of the ground state energy $E_0$, which is a function of the electric displacement field $\mathbf{D}$. In the first principles literature, it is far more common to define the electric enthalpy $E$, which is a function of the external electric field $\bm{\mathcal{E}}$ \cite{nunes2001, souza2002}:
\begin{equation} \label{ElectricEnthalpy}
E(\{ \mathbf{u}_{\kappa}(\bm{\ell}) \}, \bm{\mathcal{E}}) = E_0(\{ \mathbf{u}_{\kappa}(\bm{\ell})\}, \mathbf{D}) - \Omega_{\text{sc}} \bm{\mathcal{E}} \cdot \mathbf{P}_{\text{ext}},
\end{equation}
where $\mathbf{P}_{\text{ext}}  = \mathbf{D} - \epsvac \bm{\mathcal{E}}$ is the polarization in the case where all the ions are stationary. Many useful quantities can be calculated as derivatives of the electric enthalpy. For example, in \appref{app:ElectricEnthalpy}, it is proven that the quantities defined in \erefs{EphDef}{PionDef} can also be written directly in terms of (derivatives of) the electric enthalpy:
\begin{align}
H_{\text{KS}} & = \left.E(\{\mathbf{0}\})\right|_{\bm{\mathcal{E}}=\mathbf{0}}, \label{HksUseful} \\
E_{\text{ph}}(\{\mathbf{u}_{\kappa}(\bm{\ell})\}) & = \left.E(\{\mathbf{u}_{\kappa}(\bm{\ell})\})\right|_{\bm{\mathcal{E}}=\mathbf{0}} - H_{\text{KS}}, \label{EphUseful} \\
\mathbf{P}_{\text{ion}}(\{\mathbf{u}_{\kappa}(\bm{\ell})\}) & = -\frac{1}{\Omega_{\text{sc}}} \left. \frac{\partial E}{\partial \bm{\mathcal{E}}} \right|_{\bm{\mathcal{E}}=\mathbf{0}}, \label{PionUseful}
\end{align}
and that $\bm{\varepsilon}_{\infty}$, defined through \eref{epsInfDef}, defines a relation between $\mathbf{D}$ and $\bm{\mathcal{E}}$ in the usual way \cite{gonze1997}:
\begin{equation} \label{DElink}
\mathbf{D} = \epsvac \bm{\varepsilon}_{\infty} \cdot \bm{\mathcal{E}}
\end{equation}
To describe the 1-electron-1-phonon interaction one also needs the Born effective charge tensor, which is defined from the electric enthalpy as follows \cite{gonze1997}:
\begin{equation} \label{Zdef}
\mathbf{Z}_{\kappa} = -\frac{1}{e} \left.\frac{\partial^2 E}{\partial \bm{\mathcal{E}} \partial \mathbf{u}_{\kappa}(\bm{\ell})}\right|_{\mathbf{u}_{\kappa}(\bm{\ell}) = \bm{\mathcal{E}} = \mathbf{0}} = \frac{\Omega_{\text{sc}}}{e} \left.\frac{\partial \mathbf{P}_{\text{ion}}}{\partial \mathbf{u}_{\kappa}(\bm{\ell})}\right|_{\mathbf{u}_{\kappa}(\bm{\ell}) = \mathbf{0}}
\end{equation}
which is independent of $\bm{\ell}$ because of the translation invariance of the crystal. Finally, to describe the phonon properties, one requires the harmonic force constants $\Phi_{\kappa \alpha, \kappa' \beta}(\bm{\ell})$ and the dynamical matrix $\mathcal{D}_{\kappa \alpha, \kappa' \beta}(\mathbf{q})$, as well as their dependence on the constant external electric field $\bm{\mathcal{E}}$:
\begin{align}
\Phi_{\kappa \alpha, \kappa' \beta}(\bm{\ell}' - \bm{\ell}; \bm{\mathcal{E}}) & = \left. \frac{\partial ^2 E(\{ \mathbf{u}_{\kappa}(\bm{\ell}) \}, \bm{\mathcal{E}})}{\partial u_{\kappa \alpha}(\bm{\ell}) \partial u_{\kappa' \beta}(\bm{\ell}')} \right|_{\mathbf{u}_{\kappa}(\bm{\ell}) = \mathbf{0}}, \label{PhiDefE} \\ 
\mathcal{D}_{\kappa \alpha, \kappa' \beta}(\mathbf{q}; \bm{\mathcal{E}}) & = \frac{1}{\sqrt{m_{\kappa} m_{\kappa'}}} \sum_{\bm{\ell}} \Phi_{\kappa \alpha, \kappa' \beta}(\bm{\ell}; \bm{\mathcal{E}}) e^{i \mathbf{q}\cdot \bm{\ell}}. \label{DynDefE}
\end{align}
The dynamical matrix can be diagonalized to yield the phonon eigenvectors $e_{\kappa \alpha, \nu}(\mathbf{q};\bm{\mathcal{E}})$ and the squares of the phonon frequencies $\omega^2_{\nu}(\mathbf{q};\bm{\mathcal{E}})$:
\begin{equation} \label{FreqsVecsDef}
\sum_{\kappa' \beta} \mathcal{D}_{\kappa \alpha, \kappa' \beta}(\mathbf{q}; \bm{\mathcal{E}}) e_{\kappa' \beta, \nu}(\mathbf{q};\bm{\mathcal{E}}) = \omega^2_{\nu}(\mathbf{q};\bm{\mathcal{E}}) e_{\kappa \alpha, \nu}(\mathbf{q};\bm{\mathcal{E}}).
\end{equation}
Unless otherwise indicated, throughout this section we use the d-type convention for the phonon eigenvectors (see e.g. \cite{srivastava1990}, \S 2.3), and we assume the Born-Huang convention $e^*_{\kappa \alpha, \nu}(\mathbf{q};\bm{\mathcal{E}}) = e_{\kappa \alpha, \nu}(-\mathbf{q};\bm{\mathcal{E}})$ \cite{guster2022}. The dependence on the electric field is naturally included from the electric field dependence of the electric enthalpy. When $\bm{\mathcal{E}} = \mathbf{0}$, the definitions in \erefs{PhiDefE}{DynDefE} reduce to the literature definitions of the force constants $\Phi_{\kappa \alpha, \kappa' \beta}(\bm{\ell}' - \bm{\ell})$, the dynamical matrix $\mathcal{D}_{\kappa \alpha, \kappa' \beta}(\mathbf{q})$, the phonon frequencies $\omega_{\mathbf{q},\nu}$, and the phonon eigenvectors $e_{\kappa \alpha, \nu}(\mathbf{q})$ \cite{gonze1997, srivastava1990}.

$\mathbf{P}_{\text{ion}}(\{\mathbf{u}_{\kappa}(\bm{\ell})\})$ can arbitrarily be written as a volume average of dipoles associated with each atom $(\kappa, \bm{\ell})$ in the supercell:
\begin{equation} \label{PionToDipolesSC}
\mathbf{P}_{\text{ion}}(\mathbf{T}) = \frac{1}{\Omega_{\text{sc}}} \sum_{\kappa ,\bm{\ell}} \mathbf{p}_{\kappa}(\bm{\ell}+\mathbf{T}).
\end{equation} 
Doing this for every supercell $\mathbf{T}$ allows us to define a position-dependent polarization field $\mathbf{P}_{\text{ion}}(\mathbf{r})$, defined by placing the dipoles $\mathbf{p}_{\kappa}(\bm{\ell}+\mathbf{T})$ at the equilibrium positions of the atoms $(\kappa, \bm{\ell}+\mathbf{T})$:
\begin{align} \label{PionToDipoles}
\mathbf{P}_{\text{ion}}(\mathbf{r}) := \sum_{\kappa, \bm{\ell}, \mathbf{T}}\mathbf{p}_{\kappa}(\bm{\ell}+\mathbf{T}) \delta(\mathbf{r} - \mathbf{T} - \bm{\ell} - \bm{\tau}_{\kappa}).
\end{align}
This field is defined in such a way that taking the volume average over a supercell $\Omega_{\text{sc}}$ yields the correct ionic polarization for that supercell, analogously to the definition for $\mathbf{D}(\mathbf{T})$ in \eref{DscDef}:
\begin{equation} \label{PionscDef}
\mathbf{P}_{\text{ion}}(\mathbf{T}) = \frac{1}{\Omega_{\text{sc}}} \int_{\Omega_{\text{sc}}(\mathbf{T})}  \mathbf{P}_{\text{ion}}(\mathbf{r}) \dee^3 \mathbf{r}.
\end{equation}
The Fourier transform of $\mathbf{P}_{\text{ion}}(\mathbf{r})$ is given by:
\begin{equation} \label{PionFourierDef}
\mathbf{P}_{\text{ion}}(\mathbf{q}+\mathbf{G}) = \frac{1}{\Omega} \sum_{\kappa, \bm{\ell}, \mathbf{T}} \mathbf{p}_{\kappa}(\bm{\ell}+\mathbf{T}) e^{-i \mathbf{q}\cdot (\bm{\ell}+ \mathbf{T})} e^{-i (\mathbf{q}+\mathbf{G})\cdot \bm{\tau}_{\kappa}}.
\end{equation}
where $\mathbf{G}$ is a reciprocal lattice vector. In \secref{subsec:IonExpansion}, we will derive explicit expressions for the dipoles $\mathbf{p}_{\kappa}(\bm{\ell}+\mathbf{T})$ from the derivatives of the electric enthalpy. The choices of how to write $\mathbf{P}_{\text{ion}}$ as a sum of dipoles, and how to associate these dipoles with a position, are arbitrary. However, only the low-momentum components of $\mathbf{P}_{\text{ion}}(\mathbf{q}+\mathbf{G})$ will appear in the final results. In this limit with $\mathbf{q}+\mathbf{G} \rightarrow \mathbf{0}$, any choice of dipoles $\mathbf{p}_{\kappa}(\bm{\ell}+\mathbf{T})$ that satisfies \eref{PionToDipolesSC} will lead to the same result for \eref{PionFourierDef}: therefore, the arbitrariness has no physical relevance. We make the above choices because they are implicitly made in \cite{verdi2015} for the harmonic problem, where $\mathbf{p}_{\kappa}(\bm{\ell} + \mathbf{T}) = e \mathbf{Z}_{\kappa} \cdot \mathbf{u}_{\kappa}(\bm{\ell} + \mathbf{T})$. By making the same choices, we will eventually obtain the same result for the long-range 1-electron-1-phonon matrix element.

Next, the Hamiltonian $H$ on the full system can be recovered by summing up the supercell Hamiltonians $H_{\text{sc}}(\mathbf{T})$ over all supercell vectors $\mathbf{T}$. In \eref{HamSCgen} for the supercell Hamiltonian, both $\mathbf{D}(\mathbf{T})$ and $\mathbf{P}_{\text{ion}}(\mathbf{T})$ depend on the specific supercell, so:
\begin{align}
H & = H_{\text{KS}} + \sum_{\mathbf{T}} \left[\sum_{\kappa \alpha,\bm{\ell}}  \frac{P_{\kappa \alpha}^2(\bm{\ell}+\mathbf{T})}{2 m_{\kappa}} + E_{\text{ph}}(\{\mathbf{u}_{\kappa}(\bm{\ell}+\mathbf{T})\})\right] \nonumber \\
& - \sum_{\mathbf{T}} \frac{\Omega_{\text{sc}}}{\epsvac} \mathbf{D}(\mathbf{T}) \cdot \bm{\varepsilon}_{\infty}^{-1} \cdot \mathbf{P}_{\text{ion}}(\mathbf{T}) \nonumber \\
& + \sum_{\mathbf{T}} \frac{\Omega_{\text{sc}}}{2 \epsvac} \mathbf{D}(\mathbf{T}) \cdot \bm{\varepsilon}_{\infty}^{-1} \cdot \mathbf{D(\mathbf{T})}.
\end{align}
This can be written in terms of the continuous field $\mathbf{D}(\mathbf{r})$ by substituting the definitions for $\mathbf{D}(\mathbf{T})$ and $\mathbf{P}(\mathbf{T})$ in \eref{DscDef} and \eref{PionscDef}, and by using that $\mathbf{D}(\mathbf{r}) \approx \mathbf{D}(\mathbf{r}')$ when $\mathbf{r}$ and $\mathbf{r}'$ are in the same supercell. Then, a short calculation yields:
\begin{align}
H & = H_{\text{KS}} + \sum_{\mathbf{T}} \left[\sum_{\kappa \alpha,\bm{\ell}}  \frac{P_{\kappa \alpha}^2(\bm{\ell}+\mathbf{T})}{2 m_{\kappa}} + E_{\text{ph}}(\{\mathbf{u}_{\kappa}(\bm{\ell}+\mathbf{T})\})\right] \nonumber \\
& - \frac{1}{\epsvac} \int_{\Omega} \mathbf{D}(\mathbf{r}) \cdot \bm{\varepsilon}_{\infty}^{-1} \cdot \mathbf{P}_{\text{ion}}(\mathbf{r}) \dee^3 \mathbf{r} \nonumber \\
& + \frac{1}{2 \epsvac} \int_{\Omega} \mathbf{D}(\mathbf{r}) \cdot \bm{\varepsilon}_{\infty}^{-1} \cdot \mathbf{D(\mathbf{r})} \dee^3 \mathbf{r}. \label{HtotalContinuum}
\end{align}
This Hamiltonian describes an ionic lattice interacting with a macroscopic electric displacement field $\mathbf{D}(\mathbf{r})$. In the case of long-range electron-phonon interaction, this electric displacement field is due to $N_{\text{el}}$ electrons. If these electrons are moving slowly enough, their electric field must be longitudinal, which means the electric field can be described through the electrostatic potential:
\begin{equation}
\bm{\mathcal{E}} = -\nabla \phi(\mathbf{r}).
\end{equation}
The connection between $\mathbf{D}$ and $\bm{\mathcal{E}}$, described in \eref{DElink}, then tells us that the electric displacement field must be equal to:
\begin{equation}
\mathbf{D}(\mathbf{r}) = - \epsvac \bm{\varepsilon}_{\infty} \cdot \nabla \phi(\mathbf{r}).
\end{equation}
Combining this with Maxwell's first equation yields the anisotropic Poisson equation for the scalar potential:
\begin{equation} \label{PoissonEquation}
\nabla \cdot \left( \bm{\varepsilon}_{\infty} \nabla \phi(\mathbf{r}) \right) = \frac{e}{\epsvac} \sum_{j = 1}^{N_{\text{el}}} \delta(\mathbf{r} - \mathbf{r}_{\text{el},j}).
\end{equation}
This equation can be solved by transforming to Fourier space: the derivation can be found in \cite{verdi2015}. The final result for the electric displacement field of the electrons is:
\begin{equation} \label{DExpression}
\mathbf{D}(\mathbf{r}) = -\frac{ie}{\Omega} \sum_{\mathbf{Q} \neq \mathbf{0}} \frac{\bm{\varepsilon}_{\infty} \cdot \mathbf{Q}}{\mathbf{Q} \cdot \bm{\varepsilon}_{\infty} \cdot \mathbf{Q}} \rho_{\mathbf{Q}} e^{-i \mathbf{Q}\cdot \mathbf{r}},
\end{equation}
where the sum over $\mathbf{Q}$ is over the entire reciprocal space, and the dimensionless density operator of the electrons $\rho_{\mathbf{Q}}$ is defined in the usual way:
\begin{equation} \label{rhoDef}
\rho_{\mathbf{Q}} = \sum_{j = 1}^{N_{\text{el}}} e^{i \mathbf{Q} \cdot \mathbf{r}_{\text{el},j}}.
\end{equation}
\eref{DExpression} can be used back in \eref{HtotalContinuum} to find a fairly general expression for the classical electron-phonon Hamiltonian. The integrals are straightforward; one should only remember to remove the self-interaction of the electrons. The result is:
\begin{align}
H & = H_{\text{KS}} + H_{\text{ph}} + H_{\text{el-ph}} + H_{\text{el-el}}, \label{Hclassical0} \\
H_{\text{ph}} & = \sum_{\kappa \alpha, \bm{\ell}, \mathbf{T}} \frac{\mathcal{P}^2_{\kappa \alpha}(\bm{\ell}+\mathbf{T})}{2 m_{\kappa}} + E_{\text{ph}}(\{\mathbf{u}_{\kappa}(\bm{\ell}+\mathbf{T})\}), \label{Hclassical1} \\
H_{\text{el-ph}} & = \frac{ie}{\epsvac} \sum_{\kappa, \bm{\ell}, \mathbf{T}} \sum_{\mathbf{Q} \neq \mathbf{0}} \frac{\mathbf{Q} \cdot \mathbf{P}_{\text{ion}}(\mathbf{Q}) }{\mathbf{Q} \cdot \bm{\varepsilon}_{\infty} \cdot \mathbf{Q}} \rho_{\mathbf{Q}}, \label{Hclassical2} \\
H_{\text{el-el}} & = \frac{e^2}{2 \Omega \epsvac} \sum_{\mathbf{Q}\neq \mathbf{0}} \frac{\rho_{\mathbf{Q}} \rho_{-\mathbf{Q}} - \rho_{\mathbf{0}}}{\mathbf{Q} \cdot \bm{\varepsilon}_{\infty} \cdot \mathbf{Q}}, \label{Hclassical3}
\end{align}
where we recall that $\mathbf{P}_{\text{ion}}(\mathbf{Q})$ is defined in \eref{PionFourierDef}. \eref{Hclassical3} is the well-known Coulomb interaction between the electrons. The focus of this article is on the electron-phonon interaction, and we will only be interested in single-electron properties: therefore, for the remainder of the article, we will ignore the Coulomb interaction $H_{\text{el}-\text{el}}$. However, note that $H_{\text{el}-\text{el}}$ can always be added to the Hamiltonian if desired.

The above Hamiltonian is classical, but can be readily written in first quantization by promoting $H_{\text{KS}}$, $\rho_{\mathbf{Q}}$, $\mathbf{u}_{\kappa}(\bm{\ell}+\mathbf{T})$, and $\bm{\mathcal{P}}_{\kappa}(\bm{\ell}+\mathbf{T})$ to operators. It can also be written in second quantization, following the recipe provided in \cite{mahan2000}, \S 1.2. Let us introduce the fermionic operators $\hat{c}^{\dagger}_{\mathbf{k},n}$ and $\hat{c}_{\mathbf{k},n}$, which create and annihilate an eigenstate $|\psi_{\mathbf{k},n}\rangle$ of the Kohn-Sham Hamiltonian $H_{\text{KS}}$ with corresponding energies $\epsilon_{\mathbf{k},n}$. Then, the second quantization expressions for $\hat{H}_{\text{KS}}$ and $\hat{\rho}_{\mathbf{Q}}$ are \cite{mahan2000}:
\begin{align}
\hat{H}_{\text{KS}} & = \sum_{\mathbf{k},n} \epsilon_{\mathbf{k},n} \hat{c}^{\dagger}_{\mathbf{k},n} \hat{c}_{\mathbf{k},n}, \\
\hat{\rho}_{\mathbf{q}+\mathbf{G}} & = \sum_{\mathbf{k},mn} \langle \psi_{\mathbf{k}+\mathbf{q},m} | e^{i (\mathbf{q}+\mathbf{G})\cdot \mathbf{r}} | \psi_{\mathbf{k},n} \rangle \hat{c}^{\dagger}_{\mathbf{k}+\mathbf{q},m} \hat{c}_{\mathbf{k},n},
\end{align}
where $\mathbf{Q}= \mathbf{q}+\mathbf{G}$ is written as the sum of a reciprocal lattice vector $\mathbf{G}$, and a lattice wavevector $\hbar \mathbf{q}$ in the first Brillouin zone. With these expressions, the Hamiltonian becomes:
\begin{align}
\hat{H} & = \hat{H}_{\text{el}} + \hat{H}_{\text{ph}} + \hat{H}_{\text{el-ph}}, \label{Hsecond0} \\
\hat{H}_{\text{el}} & = \sum_{\mathbf{k},n} \epsilon_{\mathbf{k},n} \hat{c}^{\dagger}_{\mathbf{k},n} \hat{c}_{\mathbf{k},n} , \label{Hsecond1}  \\
\hat{H}_{\text{ph}} & = \sum_{\kappa \alpha, \bm{\ell}, \mathbf{T}} \frac{\mathcal{P}^2_{\kappa \alpha}(\bm{\ell}+\mathbf{T})}{2 m_{\kappa}} + E_{\text{ph}}(\{\mathbf{u}_{\kappa}(\bm{\ell}+\mathbf{T})\}), \label{Hsecond2} \\
\hat{H}_{\text{el-ph}} & = \frac{ie}{\epsvac} \sum_{\mathbf{k}, \mathbf{q}, mn} \sum_{\mathbf{G} \neq -\mathbf{q}} \frac{(\mathbf{q}+\mathbf{G}) \cdot \mathbf{P}_{\text{ion}}(\mathbf{q}+\mathbf{G}) }{(\mathbf{q}+\mathbf{G}) \cdot \bm{\varepsilon}_{\infty} \cdot (\mathbf{q}+\mathbf{G})} \times \\
& \hspace{25pt} \times \langle \psi_{\mathbf{k}+\mathbf{q},m} | e^{i (\mathbf{q}+\mathbf{G}) \cdot \mathbf{r}} | \psi_{\mathbf{k},n} \rangle \hat{c}^{\dagger}_{\mathbf{k}+\mathbf{q},m} \hat{c}_{\mathbf{k},n}. \label{Hsecond3}
\end{align}This Hamiltonian is the furthest one can go without assuming anything about the $\{\mathbf{u}_{\kappa}(\bm{\ell}+\mathbf{T})\}$-dependence of $E_{\text{ph}}$ or $\mathbf{P}_{\text{ion}}(\mathbf{q}+\mathbf{G})$. Respectively, these would be assumptions about the phonon anharmonicity, and about the electron-phonon anharmonicity. 

Each of the terms in the above Hamiltonian has a clear meaning. \eref{Hsecond1} is the electron Hamiltonian with all ions in the equilibrium positions, and in the absence of an external electric field. \eref{Hsecond2} is the phonon Hamiltonian in the absence of an external electric field, or equivalently, in the absence of electron-phonon coupling. It contains the entire energy landscape $E_{\text{ph}}(\{\mathbf{u}_{\kappa}(\bm{\ell}+\mathbf{T})\})$ and is therefore valid up to arbitrary orders of anharmonicity. Finally, \eref{Hsecond3} is the long-range electron-phonon interaction: it contains all the processes where one electron interacts with the phonons. This is again up to arbitrary orders of anharmonicity, since $\mathbf{P}_{\text{ion}}(\mathbf{q}+\mathbf{G})$ and $\mathbf{p}_{\kappa}(\bm{\ell}+\mathbf{T})$ are ultimately defined from the general ionic polarization $\mathbf{P}_{\text{ion}}(\{\mathbf{u}_{\kappa}(\bm{\ell}+\mathbf{T})\})$.

Note that the derivation requires that $\mathbf{P}_{\text{ion}}$ is approximately constant over the size of a supercell. Therefore, for self-consistency, we must assume that $\mathbf{P}_{\text{ion}}(\mathbf{q}+\mathbf{G})$ only consists of low-momentum components. In particular, the resulting electron-phonon Hamiltonian \eref{Hsecond3} is only valid in the limit $\mathbf{q}+\mathbf{G} \rightarrow \mathbf{0}$. This is how one can define the long-range limit for the electron-phonon interaction in general: only the momentum appearing in $\rho_{\mathbf{q}+\mathbf{G}}$ must be small, but all other momenta can be arbitrarily large. Throughout this article, we will frequently take the long-range limit in several expressions for self-consistency.

\begin{widetext}
\subsection{Expansion in the ion coordinates} \label{subsec:IonExpansion}
\eref{Hclassical2} for the electron-phonon interaction is not useful until one has a useful expression for the dipoles $\mathbf{p}_{\kappa}(\bm{\ell}+\mathbf{T})$. Such an expression can be obtained by expanding $\mathbf{P}_{\text{ion}}(\{\mathbf{u}_{\kappa}(\bm{\ell})\})$ in terms of the ionic displacements $\mathbf{u}_{\kappa}(\bm{\ell})$. In this article we focus on the 1-electron-2-phonon interaction and ignore phonon anharmonicity, so $E_{\text{ph}}(\{\mathbf{u}_{\kappa}(\bm{\ell}+\mathbf{T})\})$ and $\mathbf{P}_{\text{ion}}(\{\mathbf{u}_{\kappa}(\bm{\ell})\})$ are both expanded up to second order. The first order derivatives of $E_{\text{ph}}(\{\mathbf{u}_{\kappa}(\bm{\ell}+\mathbf{T})\})$ are zero since the derivatives are evaluated at the equilibrium positions of the ions, and its second order derivatives are the harmonic force constants defined in \eref{PhiDefE}. Therefore, the phonon Hamiltonian $H_{\text{ph}}$ from \eref{Hclassical1} can be written in the following form:
\begin{equation} \label{HamPhononHarmonicClassical}
H_{\text{ph}} = \sum_{\kappa \alpha, \bm{\ell}, \mathbf{T}} \frac{\mathcal{P}^2_{\kappa \alpha}(\bm{\ell}+\mathbf{T})}{2 m_{\kappa}} + \frac{1}{2} \sum_{\kappa \alpha, \kappa' \beta, \bm{\ell}, \bm{\ell}', \mathbf{T}} u_{\kappa \alpha}(\bm{\ell}+\mathbf{T})  \Phi_{\kappa \alpha, \kappa' \beta}(\bm{\ell} - \bm{\ell}') u_{\kappa' \beta}(\bm{\ell}'+\mathbf{T}),
\end{equation}
where it is assumed that the supercell is large enough, so that the force constants are zero when the two ions do not belong to the same supercell. This is the usual harmonic lattice Hamiltonian: its further treatment can be found in the literature  \cite{maradudin1968, giustino2017}. Here, the main steps are outlined. The Hamiltonian can be written in second quantization in terms of phonon ladder operators $\hat{a}_{\mathbf{q},\nu}, \hat{a}^{\dagger}_{\mathbf{q},\nu}$, if the ionic displacements and momenta are written as follows \cite{giustino2017}:
\begin{align}
u_{\kappa \alpha}(\bm{\ell}+\mathbf{T}) & = \sqrt{\frac{\Omega_0}{\Omega}} \sum_{\mathbf{q}, \nu} \sqrt{\frac{\hbar}{2 m_{\kappa} \omega_{\mathbf{q},\nu}}} \left( \hat{a}_{\mathbf{q},\nu} + \hat{a}^{\dagger}_{-\mathbf{q},\nu} \right) e_{\kappa \alpha,\nu}(\mathbf{q}) e^{i \mathbf{q} \cdot (\bm{\ell}+\mathbf{T})}, \label{uToLadder} \\
\mathcal{P}_{\kappa \alpha}(\bm{\ell}+\mathbf{T}) & = \frac{1}{i}\sqrt{\frac{\Omega_0}{\Omega}} \sum_{\mathbf{q}, \nu} \sqrt{\frac{\hbar  m_{\kappa} \omega_{\mathbf{q},\nu}}{2}} \left( \hat{a}_{\mathbf{q},\nu} - \hat{a}^{\dagger}_{-\mathbf{q},\nu} \right) e_{\kappa \alpha,\nu}(\mathbf{q}) e^{i \mathbf{q} \cdot (\bm{\ell}+\mathbf{T})}, \label{pToLadder}
\end{align}
and the bosonic commutation relations $[\hat{a}_{\mathbf{q},\nu}, \hat{a}^{\dagger}_{\mathbf{q}',\nu'}] = \delta_{\mathbf{q},\mathbf{q}'} \delta_{\nu, \nu'}$ are imposed. Substituting \erefs{uToLadder}{pToLadder} into \eref{HamPhononHarmonicClassical} and using the definitions in \erefs{DynDefE}{FreqsVecsDef} yields the familiar expression for the harmonic phonon Hamiltonian in second quantization:
\begin{equation}
\hat{H}_{\text{ph}} = \sum_{\mathbf{q},\nu} \hbar \omega_{\mathbf{q},\nu} \left( \hat{a}^{\dagger}_{\mathbf{q},\nu} \hat{a}_{\mathbf{q},\nu} + \frac{1}{2} \right).
\end{equation}
This is precisely \eref{HamPh}, where the phonon frequencies are obtained in the usual way from the diagonalization of the dynamical matrix.

To expand $\mathbf{P}_{\text{ion}}$, recall that it is defined as an electric field derivative of the energy according to \eref{PionUseful}. Therefore, the expansion coefficients can be written as derivatives of the energy, where it is again assumed that the supercell is large enough so that the derivatives are zero when the ions do not belong to the same supercell:
\begin{align}
\mathbf{P}_{\text{ion}}(\{\mathbf{u}_{\kappa}(\bm{\ell}+\mathbf{T})\}) & \approx -\frac{1}{\Omega_{\text{sc}}} \sum_{\kappa \alpha, \bm{\ell}, \mathbf{T}} u_{\kappa \alpha}(\bm{\ell}+\mathbf{T}) \left. \frac{\partial^2 E}{\partial \bm{\mathcal{E}} \partial u_{\kappa \alpha}(\bm{\ell}+\mathbf{T})} \right|_{\mathbf{u}_{\kappa}(\bm{\ell})=\bm{\mathcal{E}}=\mathbf{0}} \nonumber \\
& -\frac{1}{2\Omega_{\text{sc}}} \sum_{\kappa \alpha, \kappa' \beta, \bm{\ell}, \bm{\ell}', \mathbf{T}} u_{\kappa \alpha}(\bm{\ell}+\mathbf{T}) u_{\kappa' \beta}(\bm{\ell}'+\mathbf{T}) \left. \frac{\partial^3 E}{\partial \bm{\mathcal{E}} \partial u_{\kappa \alpha}(\bm{\ell}+\mathbf{T}) u_{\kappa' \beta}(\bm{\ell}'+\mathbf{T})} \right|_{\mathbf{u}_{\kappa}(\bm{\ell}+\mathbf{T})=\bm{\mathcal{E}}=\mathbf{0}}.
\end{align}
These derivatives can be written in terms of the Born effective charge tensor and the harmonic force constants using \erefs{Zdef}{PhiDefE}:
\begin{align}
\mathbf{P}_{\text{ion}}(\{\mathbf{u}_{\kappa}(\bm{\ell}+\mathbf{T})\})  & \approx \frac{1}{\Omega_{\text{sc}}} \sum_{\kappa, \bm{\ell}, \mathbf{T}} e \mathbf{Z}_{\kappa} \cdot \mathbf{u}_{\kappa}(\bm{\ell}+\mathbf{T}) -\frac{1}{2\Omega_{\text{sc}}} \sum_{\kappa \alpha, \kappa' \beta, \bm{\ell}, \bm{\ell}', \mathbf{T}} u_{\kappa \alpha}(\bm{\ell}+\mathbf{T}) \left. \frac{\partial \Phi_{\kappa \alpha, \kappa' \beta}(\bm{\ell}'-\bm{\ell};\bm{\mathcal{E}})}{\partial \bm{\mathcal{E}}} \right|_{\bm{\mathcal{E}}=\mathbf{0}} u_{\kappa' \beta}(\bm{\ell}'+\mathbf{T}). \label{PionAnharmonic}
\end{align}
This also provides a natural definition for the ionic dipoles $\mathbf{p}_{\kappa}(\bm{\ell})$ in \eref{PionToDipolesSC}, such that $\mathbf{P}_{\text{ion}}$ is the volume average of these dipoles in the supercell:
\begin{equation} \label{DipolesAnharmonic}
\mathbf{p}_{\kappa}(\bm{\ell}+\mathbf{T}) \approx e \mathbf{Z}_{\kappa} \cdot \mathbf{u}_{\kappa}(\bm{\ell}+\mathbf{T}) -\frac{1}{2} \sum_{\alpha, \kappa' \beta, \bm{\ell}'} u_{\kappa \alpha}(\bm{\ell}+\mathbf{T}) \left. \frac{\partial \Phi_{\kappa \alpha, \kappa' \beta}(\bm{\ell}'-\bm{\ell};\bm{\mathcal{E}})}{\partial \bm{\mathcal{E}}} \right|_{\bm{\mathcal{E}}=\mathbf{0}} u_{\kappa' \beta}(\bm{\ell}'+\mathbf{T}).
\end{equation}
Note here that the first term is a standard expression for the ionic dipole moment, which is used in the harmonic derivation in \cite{verdi2015}. It will lead to the 1-electron-1-phonon interaction, and the central quantity that controls it is the Born effective charge tensor $\mathbf{Z}_{\kappa}$. The second term is the next-higher order, which will lead to the 1-electron-2-phonon interaction. This derivation immediately clarifies that $\left. \frac{\partial \Phi_{\kappa \alpha, \kappa' \beta}(\bm{\ell}'-\bm{\ell};\bm{\mathcal{E}})}{\partial \bm{\mathcal{E}}} \right|_{\bm{\mathcal{E}}=\mathbf{0}}$ will be the central quantity that will control the 1-electron-2-phonon interaction. Although we will be satisfied with the second-order expansion in this article, this idea is easily extended to higher orders by including higher-order derivatives of the ionic coordinates in \eqref{PionAnharmonic}. It can straightforwardly be seen that the central quantity that controls the 1-electron-$n$-phonon interaction would be the electric field derivative of the $n$-phonon matrix elements, as defined in e.g. \cite{deinzer2003}.

To find an expression for the long-range electron-phonon Hamiltonian, \eref{DipolesAnharmonic} and \eref{uToLadder} should be substituted into \eref{Hclassical2}. The calculation is straightforward; only the main steps are presented here. First, $\mathbf{P}_{\text{ion}}(\mathbf{q}+\mathbf{G})$ is calculated with \eref{PionFourierDef}, \eref{uToLadder} and \eref{DipolesAnharmonic}, which gives:
\begin{align}
\mathbf{P}_{\text{ion}}(\mathbf{q}+\mathbf{G})
& = \frac{e}{\sqrt{\Omega \Omega_0}} \sum_{\kappa \nu} \sqrt{\frac{\hbar}{2 m_{\kappa} \omega_{\mathbf{q},\nu}}} \mathbf{Z}_{\kappa} \cdot \mathbf{e}_{\kappa, \nu}(\mathbf{q}) e^{-i (\mathbf{q}+\mathbf{G})\cdot \bm{\tau}_{\kappa}} \left( \hat{a}_{\mathbf{q},\nu} + \hat{a}^{\dagger}_{-\mathbf{q},\nu} \right) \nonumber \\
& - \frac{ie}{2 \Omega} \sum_{\mathbf{q}',\nu,\nu'} \mathbf{Y}_{\nu_1 \nu_2}(\mathbf{q}', \mathbf{q}+\mathbf{G}) \left( \hat{a}_{\mathbf{q}'-\mathbf{q},\nu} + \hat{a}^{\dagger}_{\mathbf{q}-\mathbf{q}',\nu} \right)\left( \hat{a}_{\mathbf{q}',\nu} + \hat{a}^{\dagger}_{-\mathbf{q}',\nu} \right), \label{PionExpanded}
\end{align}
where the auxiliary vector $ \mathbf{Y}_{\nu_1 \nu_2}(\mathbf{q}', \mathbf{q}+\mathbf{G})$ is defined as:
\begin{equation}  \label{YdefTooGeneral}
\mathbf{Y}_{\nu_1\nu_2}(\mathbf{q}', \mathbf{q}+\mathbf{G}) := \frac{1}{ie} \sqrt{\frac{\hbar}{2\omega_{\mathbf{q}-\mathbf{q}',\nu_1}} \frac{\hbar}{2\omega_{\mathbf{q}',\nu_2}} } \sum_{\kappa \alpha, \kappa' \beta} e^{-i (\mathbf{q}+\mathbf{G}) \cdot \bm{\tau}_{\kappa}} e_{\kappa \alpha,\nu_1}(\mathbf{q}-\mathbf{q}') \frac{\partial \mathcal{D}_{\kappa \alpha,\kappa' \beta}(\mathbf{q}')}{\partial \bm{\mathcal{E}}}  e_{\kappa' \beta,\nu_2}(\mathbf{q}').
\end{equation}
Here, the shorthand notation $\frac{\partial \mathcal{D}_{\kappa \alpha,\kappa' \beta}(\mathbf{q})}{\partial \bm{\mathcal{E}}} := \left. \frac{\partial \mathcal{D}_{\kappa \alpha,\kappa' \beta}(\mathbf{q};\bm{\mathcal{E}})}{\partial \bm{\mathcal{E}}} \right|_{\bm{\mathcal{E}}=\mathbf{0}}$ was introduced for the derivative at zero electric field: we will use this shorthand notation throughout the article. Now, \eref{PionExpanded} can be substituted back into \eref{Hsecond3} for the electron-phonon interaction. This yields a Hamiltonian of the form \erefs{Ham1e1ph}{Ham1e2ph}, with the following expressions for the electron-phonon matrix elements:
\begin{align}
g^{\text{(long)}}_{mn\nu}(\mathbf{k},\mathbf{q}) & = \frac{i e^2}{\epsvac \Omega_0} \sum_{\mathbf{G} \neq -\mathbf{q}} \sum_{\kappa} \sqrt{\frac{\hbar}{2 m_{\kappa} \omega_{\mathbf{q},\nu}}}  \frac{(\mathbf{q}+\mathbf{G}) \cdot \mathbf{Z}_{\kappa} \cdot  \mathbf{e}_{\nu,\kappa}(\mathbf{q}) }{(\mathbf{q}+\mathbf{G}) \cdot \bm{\varepsilon}_{\infty} \cdot (\mathbf{q}+\mathbf{G})} \left\langle \psi_{\mathbf{k}+\mathbf{q},m} \left| e^{i (\mathbf{q}+\mathbf{G}) \cdot (\mathbf{r}-\bm{\tau}_{\kappa})} \right|\psi_{\mathbf{k},n} \right\rangle, \label{g1LongAlmost} \\
 g^{\text{(long)}}_{mn\nu_1\nu_2}(\mathbf{k},\mathbf{q}_1,\mathbf{q}_2) & = \frac{e^2}{2 \epsvac \Omega_0} \sum_{\mathbf{G} \neq -\mathbf{q}_1 - \mathbf{q}_2} \frac{(\mathbf{q}_1+\mathbf{q}_2+\mathbf{G}) \cdot \mathbf{Y}_{\nu_1\nu_2}(\mathbf{q}_2, \mathbf{q}_1 + \mathbf{q}_2 + \mathbf{G}) }{(\mathbf{q}_1+\mathbf{q}_2+\mathbf{G}) \cdot \bm{\varepsilon}_{\infty} \cdot (\mathbf{q}_1+\mathbf{q}_2+\mathbf{G})} \langle \psi_{\mathbf{k}+\mathbf{q}_1+\mathbf{q}_2,m} | e^{i (\mathbf{q}_1+\mathbf{q}_2+\mathbf{G})\cdot \mathbf{r}} | \psi_{\mathbf{k},n} \label{g2LongAlmost} \rangle.
\end{align}
\eref{g1LongAlmost} is the literature result for the long-range part of the 1-electron-1-phonon matrix element \cite{verdi2015, giustino2017}. This confirms that the derivation presented in this section is a generalization of current literature theory, which can be applied to arbitrary orders of anharmonicity. 

We explicitly write the superscript ``(long)'' in \erefs{g1LongAlmost}{g2LongAlmost} to indicate that these are not the full electron-phonon matrix elements, but only the long-range parts. Recall from the discussion at the end of \secref{subsec:ClassicalDerivation} that the long-range part can be identified as the $\mathbf{Q} \rightarrow \mathbf{0}$ limit of whichever wavevector $\mathbf{Q}$ is in the density operator, or equivalently, in the electron matrix element in \erefs{g1LongAlmost}{g2LongAlmost}. For the 1-electron-1-phonon matrix element this is $\mathbf{q}+\mathbf{G}$, and for the 1-electron-2-phonon matrix element it is $\mathbf{q}_1 + \mathbf{q}_2 + \mathbf{G}$, so:
\begin{align}
\lim_{\mathbf{q}+\mathbf{G} \rightarrow \mathbf{0}} g_{mn\nu}(\mathbf{k},\mathbf{q}) & = g^{\text{(long)}}_{mn\nu}(\mathbf{k},\mathbf{q}), \\
\lim_{\mathbf{q}_1+\mathbf{q}_2+\mathbf{G} \rightarrow \mathbf{0}} g_{mn\nu_1\nu_2}(\mathbf{k},\mathbf{q}_1,\mathbf{q}_2) & =  g^{\text{(long)}}_{mn\nu_1\nu_2}(\mathbf{k},\mathbf{q}_1,\mathbf{q}_2).
\end{align}
The limit $\mathbf{G} \rightarrow \mathbf{0}$ in these expressions should be understood as follows: in the sums over $\mathbf{G}$ that appear in \erefs{g1LongAlmost}{g2LongAlmost}, one may always neglect any term that has $\mathbf{G}\neq\mathbf{0}$ without changing the physical results. In practice, the sum over $\mathbf{G}$ is only necessary when interpolating the entire electron-phonon matrix element \cite{verdi2015}, which is outside the scope of this article.

By definition, \eref{g1LongAlmost} is only valid in the $\mathbf{q}+\mathbf{G} \rightarrow \mathbf{0}$ limit, and \eref{g2LongAlmost} is only valid in the $\mathbf{q}_1+\mathbf{q}_2+\mathbf{G} \rightarrow \mathbf{0}$ limit. Therefore, for self-consistency, these limits should be taken in \erefs{g1LongAlmost}{g2LongAlmost} as well. For the 1-electron-2-phonon interaction, this is easily done by replacing $\mathbf{Y}_{\nu_1 \nu_2}(\mathbf{q}_2, \mathbf{q}_1 + \mathbf{q}_2 + \mathbf{G})$ with $\mathbf{Y}_{\nu_1 \nu_2}(\mathbf{q}_2, \mathbf{0}) := \mathbf{Y}_{\nu_1 \nu_2}(\mathbf{q}_2)$, which is given by:
\begin{equation}  \label{Ydef}
\mathbf{Y}_{\nu_1\nu_2}(\mathbf{q}) := \frac{1}{ie} \sqrt{\frac{\hbar}{2\omega_{\mathbf{q},\nu_1}} \frac{\hbar}{2\omega_{\mathbf{q},\nu_2}} } \sum_{\kappa \alpha, \kappa' \beta}  e^*_{\kappa \alpha,\nu_1}(\mathbf{q}) \frac{\partial \mathcal{D}_{\kappa \alpha,\kappa' \beta}(\mathbf{q})}{\partial \bm{\mathcal{E}}}  e_{\kappa' \beta,\nu_2}(\mathbf{q}).
\end{equation}
For the 1-electron-1-phonon interaction, it is convenient to introduce the mode polarities $\mathbf{p}_{\nu}(\hat{\mathbf{q}})$, which are used in the generalized Fr\"ohlich model for the 1-electron-1-phonon interaction \cite{miglio2020}. They are defined as:
\begin{equation} \label{ModePolaritiesDef}
\mathbf{p}_{\nu}(\hat{\mathbf{q}}) := \sum_{\kappa} \frac{\mathbf{Z}_{\kappa} \cdot \mathbf{e}_{\kappa,\nu}(\hat{\mathbf{q}})}{\sqrt{m_{\kappa}}}.
\end{equation}
In general, they may depend on the direction $\hat{\mathbf{q}}$ along which we take the $\mathbf{q} \rightarrow \mathbf{0}$ limit. Similarly, the phonon frequencies in the $\mathbf{q} \rightarrow \mathbf{0}$ limit can still have a non-analytic dependence on $\hat{\mathbf{q}}$ \cite{miglio2020}, so we write the phonon frequencies around $\Gamma$ as $\omega_{\hat{\mathbf{q}},\nu}$.
 
Taking the abovementioned limits in \erefs{g1LongAlmost}{g2LongAlmost} yields find the final expressions for the long-range electron-phonon matrix elements:
\begin{align}
g^{\text{(long)}}_{mn\nu}(\mathbf{k},\mathbf{q}) & = \frac{i e^2}{\epsvac \Omega_0}  \sqrt{\frac{\hbar}{2 \omega_{\hat{\mathbf{q}},\nu}}} \sum_{\mathbf{G} \neq -\mathbf{q}}  \frac{(\mathbf{q}+\mathbf{G}) \cdot \mathbf{p}_{\nu}(\hat{\mathbf{q}}) }{(\mathbf{q}+\mathbf{G}) \cdot \bm{\varepsilon}_{\infty} \cdot (\mathbf{q}+\mathbf{G})} \left\langle \psi_{\mathbf{k}+\mathbf{q},m} \left| e^{i (\mathbf{q}+\mathbf{G}) \cdot \mathbf{r}} \right|\psi_{\mathbf{k},n} \right\rangle, \label{g1Long} \\
 g^{\text{(long)}}_{mn\nu_1\nu_2}(\mathbf{k},\mathbf{q}_1,\mathbf{q}_2) & = \frac{e^2}{2 \epsvac \Omega_0} \sum_{\mathbf{G} \neq -\mathbf{q}_1 - \mathbf{q}_2} \frac{(\mathbf{q}_1+\mathbf{q}_2+\mathbf{G}) \cdot \mathbf{Y}_{\nu_1 \nu_2}(\mathbf{q}_2) }{(\mathbf{q}_1+\mathbf{q}_2+\mathbf{G}) \cdot \bm{\varepsilon}_{\infty} \cdot (\mathbf{q}_1+\mathbf{q}_2+\mathbf{G})} \langle \psi_{\mathbf{k}+\mathbf{q}_1+\mathbf{q}_2,m} | e^{i (\mathbf{q}_1+\mathbf{q}_2+\mathbf{G})\cdot \mathbf{r}} | \psi_{\mathbf{k},n} \label{g2Long} \rangle.
\end{align}
\end{widetext}
Note that \erefs{g1Long}{g2Long} look very similar in this form, and that the role of $\mathbf{Y}_{\nu_1\nu_2}(\mathbf{q})$ in the 1-electron-2-phonon interaction is analogous to the role of the mode polarities $\mathbf{p}_{\nu}(\hat{\mathbf{q}})$ in the 1-electron-1-phonon interaction.

We end this section with several remarks surrounding the long-range limit. Most notably, for the 1-electron-1-phonon interaction, the long-range limit is identical to the continuum limit, which is also obtained by letting $\mathbf{q} \rightarrow \mathbf{0}$ for all phonon properties. However, this is no longer the case for the 1-electron-2-phonon interaction: we only require that $\mathbf{q}_1 \approx -\mathbf{q}_2$, which leaves the phonon wavevector $\mathbf{q}_2$ completely free. This is reflected in \eref{g2Long}, where $\mathbf{Y}_{\nu_1 \nu_2}(\mathbf{q}_2)$ still depends on one phonon momentum which is free to explore the entire Brillouin zone. In contrast, $ \mathbf{p}_{\nu}(\hat{\mathbf{q}})$ is only evaluated very close to the $\Gamma$-point at $\mathbf{q}=\mathbf{0}$. One must therefore be careful to not confuse the continuum limit $\mathbf{q}_1, \mathbf{q}_2 \rightarrow \mathbf{0}$ with the long-range limit $\mathbf{q}_1 + \mathbf{q}_2 + \mathbf{G} \rightarrow \mathbf{0}$ when working with the 1-electron-2-phonon interaction. For example, the model of \cite{houtput2021} is derived in the continuum limit instead of the long-range limit, which is an additional hidden assumption of the model.

The long-range limit can be seen in terms of a series expansion of the full electron-phonon matrix elements $g_{mn\nu}(\mathbf{k},\mathbf{q})$ and $g_{mn\nu_1\nu_2}(\mathbf{k},\mathbf{q}_1,\mathbf{q}_2)$ as a function of the variables $\mathbf{Q} = \mathbf{q}+\mathbf{G}$ and $\mathbf{Q} = \mathbf{q}_1 + \mathbf{q}_2 + \mathbf{G}$, respectively. Then, the long-range part represents the lowest order term in the expansion, the dipole contribution, which is a $1/|\mathbf{Q}|$ divergence in both cases. The next order term would be a quadrupole contribution, a term of order $|\mathbf{Q}|^0$ but which may have a non-analytic angular dependence around $\mathbf{Q}=\mathbf{0}$. This term is known for the 1-electron-1-phonon interaction \cite{brunin2020, brunin2020a, park2020, jhalani2020}, but it will not be considered further in this article.

Finally, note that for cubic materials with a single non-degenerate conduction band that only interacts with a single dispersionless longitudinal optical (LO) phonon branch, \eref{g1Long} reduces to the well-known Fr\"ohlich interaction \cite{frohlich1954, verdi2015}. This interaction is often used on its own to describe large polarons, where it is assumed that the long-range interaction dominates over the short-range interaction, so that $g_{mn\nu}(\mathbf{k},\mathbf{q}) \approx g^{\text{(long)}}_{mn\nu}(\mathbf{k},\mathbf{q})$. For the remainder of this article, we make the same assumption for the 1-electron-2-phonon interaction, and assume that $g_{mn\nu_1\nu_2}(\mathbf{k},\mathbf{q}_1,\mathbf{q}_2) \approx g^{\text{(long)}}_{mn\nu_1\nu_2}(\mathbf{k},\mathbf{q}_1,\mathbf{q}_2)$ for a large polaron. In the next section, we will calculate the effect of the 1-electron-2-phonon interaction on the ground state energy of a large polaron. While the theory appears to be general, it is essential to apply it only to materials that host large polarons, where contributions to the ground state energy from short-range interactions are negligible. In contrast, for other types of materials, such as metals or those hosting small polarons, short-range effects become important and must be properly taken into account.

\section{Weak coupling treatment of the nonlinear electron-phonon Hamiltonian}
\label{sec:EnergyMass}
\subsection{Weak coupling electron self energy} \label{sec:SelfEnergyExpansion}
\begin{figure}
\centering
\includegraphics[width=8.6cm]{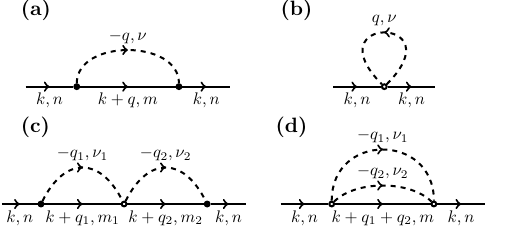}
\caption{Lowest order self-energy diagrams that will be considered in this article. Diagrams \textbf{(a)} and \textbf{(b)} are the Fan-Migdal and Debye-Waller diagrams. Only diagrams \textbf{(a)} and \textbf{(d)} give significant contributions when the short-range 1-electron-2-phonon interaction is neglected. $k = (\mathbf{k},\omega)$ and $q = (\mathbf{q},\omega')$ represent four-momenta, $m$ and $n$ represent electron bands, and $\nu$ represents a phonon branch.}
\label{fig:SelfEnergyDiagrams}
\end{figure}

With the Hamiltonian \erefs{HamTot}{Ham1e2ph} and with \eref{g1Long} and \eref{g2Long} for the electron-phonon matrix elements ready, we can proceed to investigate the effect of the long-range 1-electron-2-phonon interaction on a physically interesting quantity. One of the most fundamental properties is the renormalization of the electron energies to quasiparticle energies $\tilde{\epsilon}_{\mathbf{k},n}$. They are found as the poles of the electron Green's function $G_{mn}(\mathbf{k},\omega)$, which satisfies the following Dyson equation \cite{giustino2017, mahan2000}:
\begin{equation}
G^{-1}_{mn}(\mathbf{k},\omega) = {G^{(0)}_{mn}}^{-1}(\mathbf{k},\omega) + \Sigma^{-1}_{mn}(\mathbf{k},\omega).
\end{equation}
Here, $G^{(0)}_{mn}(\mathbf{k},\omega)$ is the free electron Green's function, given in terms of the Fermi-Dirac distribution $n_F(E)$ as follows \cite{mahan2000}:
\begin{equation} \label{ElectronGreensFunction}
G^{(0)}_{mn}(\mathbf{k},\omega) = \delta_{mn} \left( \frac{1 - n_F(\epsilon_{\mathbf{k},n})}{\omega-\frac{\epsilon_{\mathbf{k},n}}{\hbar}  + i \delta} + \frac{n_F(\epsilon_{\mathbf{k},n})}{\omega-\frac{\epsilon_{\mathbf{k},n}}{\hbar} - i \delta} \right),
\end{equation} 
and $\Sigma_{m n}(\mathbf{k},\omega)$ is the self energy of the electron. Often, the electron-phonon interaction is small enough that it is sufficient to obtain the quasiparticle energies up to lowest order in the self-energy. In this case, which is equivalent to Rayleigh-Schr\"odinger perturbation theory, the quasiparticle energy renormalization $\Delta \epsilon_{\mathbf{k},n} := \tilde{\epsilon}_{\mathbf{k},n} - \epsilon_{\mathbf{k},n}$ is given by \cite{giustino2017}:
\begin{equation} \label{QuasiparticleEnergies}
\Delta \epsilon_{\mathbf{k},n} \approx \hbar \Sigma_{nn}\left(\mathbf{k}, \frac{\epsilon_{\mathbf{k},n}}{\hbar} \right) + O(\Sigma^2).
\end{equation}
The self-energy itself can be obtained from a diagrammatic expansion (\cite{mahan2000}, \S 2-\S 3). Here, the perturbation is a sum of the 1-electron-1-phonon and 1-electron-2-phonon interactions, so one must build diagrams using both vertices in \figref{fig:InteractionDiagrams}. In this article, we will consider all diagrams that have at most two 1-electron-1-phonon vertices, and at most two 1-electron-2-phonon vertices. This leaves the four unique self-energy diagrams which have been depicted in \figref{fig:SelfEnergyDiagrams}. The diagram in \figref{fig:SelfEnergyDiagrams}a is the Fan-Migdal diagram, which is the only diagram that would remain if no 1-electron-2-phonon interaction was present. The diagram in \figref{fig:SelfEnergyDiagrams}b is the Debye-Waller diagram. Both of these diagrams are well-known in the literature \cite{giustino2017}, and are the usual self-energy diagrams that are kept when treating the 1-electron-1-phonon interaction: in that context, the Debye-Waller diagram is rewritten in terms of $g_{mn\nu}(\mathbf{k},\mathbf{0})$ using the rigid ion approximation \cite{giustino2010, giustino2017}. The diagrams in \figref{fig:SelfEnergyDiagrams}c and \figref{fig:SelfEnergyDiagrams}d are new: they both contain the 1-electron-2-phonon vertex $g_{mn\nu_1\nu_2}(\mathbf{k},\mathbf{q}_1,\mathbf{q}_2)$, for which the expression in \eref{g2Long} can be used.

We note that the choice of which diagrams to include in the expansion is somewhat arbitrary, because it is more difficult to define what ``lowest order in the electron-phonon interaction'' means when more than one interaction is present. For example, one might define the order of the diagram as the number of its interaction vertices, which indicates how many times the macroscopic electric field of the electron interacts with the phonons. By this metric, the diagrams in \figref{fig:SelfEnergyDiagrams}a and \figref{fig:SelfEnergyDiagrams}d have the same order. However, one might also define the order of the diagram by their number of phonon lines, or equivalently the order of the ionic displacements. By this metric, the diagram in \figref{fig:SelfEnergyDiagrams}d is of higher order than the diagram in \figref{fig:SelfEnergyDiagrams}a, and a self-consistent expansion that includes this diagram should also include all diagrams with four 1-electron-1-phonon interactions. In the theory of this section, we choose to limit ourselves to the diagrams in \figref{fig:SelfEnergyDiagrams} for the sake of simplicity and novelty, since results for the higher-order 1-electron-1-phonon processes already exist in the literature \cite{smondyrev1986, alexandrov2010, lee2020}. In practice, the decision of which diagrams to include should be based on the strength of the interaction vertices, which can change significantly from material to material. As a concrete example, the 1-electron-1-phonon interaction strength is strong in LiF and the higher-order 1-electron-1-phonon processes may not be neglected. Therefore, when we calculate the polaron energy for LiF in \secref{sec:FirstPrinciples}, we will use the literature result from \cite{smondyrev1986} that includes diagrams with up to six 1-electron-1-phonon interactions.

The diagrams in \figref{fig:SelfEnergyDiagrams} can be written as integrals over Green's functions using the Feynman rules (\cite{mahan2000}, \S 2.8). We follow the method and conventions of \cite{goldberg1985} to determine the prefactors for each diagram. Every 1-electron-1-phonon vertex corresponds to a factor $\sqrt{\frac{\Omega_0}{\Omega}} \frac{1}{\hbar} g_{mn\nu}(\mathbf{k},\mathbf{q})$, and every 1-electron-2-phonon vertex corresponds to a factor $\frac{\Omega_0}{\Omega} \frac{2}{\hbar} g_{mn\nu_1\nu_2}(\mathbf{k},\mathbf{q}_1,\mathbf{q}_2)$. The additional factor $2$ comes from the fact that the conventional factor $\frac{1}{2!}$ from \cite{goldberg1985} is missing from the Hamiltonian in \eref{Ham1e2ph}. Every electron line corresponds to the usual electron Green's function from \eref{ElectronGreensFunction}, and every phonon line corresponds to the usual phonon Green's function $D^{(0)}_{\nu \nu'}(\mathbf{q},\omega)$ \cite{mahan2000}:
 \begin{align} \label{D0Definition}
D^{(0)}_{\nu \nu'}(\mathbf{q},\omega) & = (1 + n_B(\omega_{\mathbf{q},\nu})) \frac{2\omega_{\mathbf{q},\nu} \delta_{\nu \nu'} }{\omega^2 - (\omega_{\mathbf{q},\nu} - i\delta)^2} \nonumber \\
& - n_B(\omega_{\mathbf{q},\nu}) \frac{2\omega_{\mathbf{q},\nu} \delta_{\nu \nu'} }{\omega^2 - (\omega_{\mathbf{q},\nu} + i\delta)^2},
\end{align} 
where $n_B(\omega)$ is the Bose-Einstein distribution. Finally, each diagram must be divided by a combinatorial factor, equal to the number of symmetries that leaves the diagram invariant according to the rules in \cite{goldberg1985}. This combinatorial factor is $1$ for the diagrams in \figref{fig:SelfEnergyDiagrams}a-c, and $2$ for the diagrams in \figref{fig:SelfEnergyDiagrams}b-d. Following these rules, the contribution of each of the diagrams in \figref{fig:SelfEnergyDiagrams} can be written as follows:
\begin{widetext}
\begin{align}
& \Sigma^{\text{(a)}}_{nn}(\mathbf{k},\omega) = i\frac{\Omega_0}{\Omega} \frac{1}{\hbar^2} \int_{-\infty}^{\infty} \frac{\dee \omega_1}{2\pi} \sum_{\mathbf{q}, \nu, m} |g_{mn\nu}(\mathbf{k},\mathbf{q})|^2 G^{(0)}_{mm}(\mathbf{k}+\mathbf{q}, \omega+\omega_1) D^{(0)}_{\nu \nu}(\mathbf{q}, -\omega_1), \label{sigmaAdef} \\
& \Sigma^{\text{(b)}}_{nn}(\mathbf{k},\omega) = i\frac{\Omega_0}{\Omega} \frac{1}{\hbar} \int_{-\infty}^{\infty} \frac{\dee \omega_1}{2\pi} \sum_{\mathbf{q}, \nu} g_{nn\nu\nu}(\mathbf{k},-\mathbf{q},\mathbf{q}) D^{(0)}_{\nu \nu}(-\mathbf{q},\omega_1), \label{sigmaBdef} \\
& \Sigma^{\text{(c)}}_{nn}(\mathbf{k},\omega) = -2\frac{\Omega_0^2}{\Omega^2}\frac{1}{\hbar^3} \sum_{m_1, m_2} \sum_{\nu_1, \nu_2} \sum_{\mathbf{q}_1, \mathbf{q}_2} \int_{-\infty}^{+\infty} \frac{\dee \omega_1}{2\pi} \int_{-\infty}^{+\infty}  \frac{\dee \omega_2}{2\pi} G^{(0)}_{m_1 m_1}(\mathbf{k}+\mathbf{q}_1, \omega+\omega_1) G^{(0)}_{m_2 m_2}(\mathbf{k}+\mathbf{q}_2, \omega+\omega_2) \nonumber \\
& \hspace{15pt} \times  D^{(0)}_{\nu_1 \nu_1}(\mathbf{q}_1, -\omega_1) D^{(0)}_{\nu_2 \nu_2}(\mathbf{q}_2, -\omega_2) g_{m_1 n \nu_1}(\mathbf{k},\mathbf{q}_1) g_{m_2 m_1 \nu_1 \nu_2}(\mathbf{k}+\mathbf{q}_1, -\mathbf{q}_1, \mathbf{q}_2) g^*_{m_2 n \nu_2}(\mathbf{k}, \mathbf{q}_2), \label{sigmaCdef} \\
& \Sigma^{\text{(d)}}_{nn}(\mathbf{k},\omega) = -2\frac{\Omega^2_0}{\Omega^2} \frac{1}{\hbar^2} \int_{-\infty}^{\infty} \frac{\dee \omega_1}{2\pi} \int_{-\infty}^{\infty} \frac{\dee \omega_2}{2\pi} \sum_{\mathbf{q}_1, \mathbf{q}_2, \nu_1, \nu_2, m} |g_{mn\nu_1 \nu_2}(\mathbf{k},\mathbf{q}_1,\mathbf{q}_2)|^2 \nonumber \\
& \hspace{15pt} \times G^{(0)}_{mm}(\mathbf{k}+\mathbf{q}_1+\mathbf{q}_2, \omega+\omega_1) D^{(0)}_{\nu_1 \nu_1}(-\mathbf{q}_1, -\omega_1 + \omega_2) D^{(0)}_{\nu_2 \nu_2}(-\mathbf{q}_2, -\omega_2).  \label{sigmaDdef}
\end{align}
\end{widetext}
We are interested in the case where there is only long-range electron-phonon interaction, e.g. in the case of a large polaron. In \appref{app:NegligibleDiagrams}, it is shown that in this case, the diagrams in \erefs{sigmaBdef}{sigmaCdef} have negligible contributions compared to the other two diagrams. Therefore, from here on, we will only consider the ``Fan-Migdal-like'' diagrams of \figref{fig:SelfEnergyDiagrams}a and \figref{fig:SelfEnergyDiagrams}d for the self-energy:
\begin{equation} \label{SelfEnergyContributions}
\Sigma_{nn}(\mathbf{k},\omega) = \Sigma^{\text{(a)}}_{nn}(\mathbf{k},\omega) + \Sigma^{\text{(d)}}_{nn}(\mathbf{k},\omega).
\end{equation}
In \eref{sigmaAdef} and \eref{sigmaDdef} for these two terms, the integrals over the frequencies $\omega_1$ and $\omega_2$ can be performed exactly. Firstly, note that the free phonon Green's function can be written as $D^{(0)}_{\nu \nu'}(\mathbf{q},\omega) = \mathcal{D}_0(\omega_{\mathbf{q},\nu},\omega) \delta_{\nu \nu'}$, where:
 \begin{align}
 \mathcal{D}_0(\omega',\omega) & = (1 + n_B(\omega')) \frac{2\omega'}{\omega^2 - (\omega' - i\delta)^2} \nonumber \\
 & - n_B(\omega') \frac{2\omega'}{\omega^2 - (\omega' + i\delta)^2}.
\end{align} 
Then, the following identity can be used in \eref{sigmaDdef}, which can be proven using complex contour integration on the upper half plane:
\begin{align}
& i \int_{-\infty}^{+\infty} \frac{\dee \omega'}{2\pi} \mathcal{D}_0(\omega_{\mathbf{q}_1,\nu_1}, \omega-\omega')\mathcal{D}_0(\omega_{\mathbf{q}_2,\nu_2}, \omega') \nonumber \\
& = [1+n_B(\omega_{\mathbf{q}_1,\nu_1})+n_B(\omega_{\mathbf{q}_2,\nu_2})] \mathcal{D}_0(\omega_{\mathbf{q}_1,\nu_1}+\omega_{\mathbf{q}_2,\nu_2}, \omega) \nonumber \\
& + |n_B(\omega_{\mathbf{q}_2,\nu_2})-n_B(\omega_{\mathbf{q}_1,\nu_1})| \mathcal{D}_0(|\omega_{\mathbf{q}_1,\nu_1}-\omega_{\mathbf{q}_2,\nu_2}|, \omega). \label{D0D0Integral}
\end{align}
The remaining integrals over $\omega_1$ are both of the following form:
\begin{align}
& i \int_{-\infty}^{+\infty} \frac{\dee \omega_1}{2\pi}  G^{(0)}_{nn}(\mathbf{k}, \omega+\omega_1) \mathcal{D}_0(\omega',-\omega_1) \nonumber \\
& = \frac{1 - n_F(\epsilon_{\mathbf{k},n}) + n_B(\omega')}{\omega - \omega' - \frac{\epsilon_{\mathbf{k},n}}{\hbar} + i\delta} + \frac{n_B(\omega') + n_F(\epsilon_{\mathbf{k},n})}{\omega + \omega' - \frac{\epsilon_{\mathbf{k},n}}{\hbar} + i\delta}. \label{G0D0Integral}
\end{align}
This identity can also be proven by complex contour integration on the upper half plane. It only holds after moving all the poles to the lower half plane after performing the integration, which means that the resulting self-energy will be the retarded one. This is a standard result at finite temperatures \cite{giustino2017, mahan2000}; this distinction will not be relevant in the remainder of this article.

By using the above two integral identities in \eref{sigmaAdef} and \eref{sigmaDdef}, the lowest order self-energy from \eref{SelfEnergyContributions} can be written in the following form:
\begin{align}
& \Sigma_{nn}\left(\mathbf{k},\omega\right) = \int_{-\infty}^{+\infty} \int_{-\infty}^{+\infty} \alpha^2 F_{\mathbf{k},n}(\varepsilon, \omega') \nonumber \\
& \times \left[ \frac{1 - n_F(\varepsilon) + n_B(\omega')}{\hbar \omega - \hbar \omega' - \varepsilon + i\delta} + \frac{n_F(\varepsilon) + n_B(\omega')}{\hbar \omega + \hbar \omega' - \varepsilon + i\delta} \right] \dee \varepsilon \dee \omega'.  \label{FMfromEliashberg}
\end{align}
In the above expression, the two-argument Eliashberg function $\alpha^2 F_{\mathbf{k},n}(\varepsilon, \omega)$ is defined as follows:
\begin{widetext}
\begin{align}
\alpha^2 F_{\mathbf{k},n}(\varepsilon, \omega) & := \frac{1}{\hbar} \sum_{\nu, m} \frac{1}{\Omega_{\text{1BZ}}} \int_{\text{1BZ}} |g_{mn\nu}(\mathbf{k},\mathbf{q})|^2 \delta(\varepsilon - \epsilon_{\mathbf{k}+\mathbf{q},m}) \delta(\omega - \omega_{\mathbf{q},\nu}) \dee^3 \mathbf{q} \nonumber \\
& + \frac{2}{\hbar} \sum_{\nu_1, \nu_2, m} \frac{1}{\Omega^2_{\text{1BZ}}} \int_{\text{1BZ}} \int_{\text{1BZ}} | g_{mn\nu_1 \nu_2}(\mathbf{k},\mathbf{q}_1,\mathbf{q}_2) |^2 \delta(\varepsilon - \epsilon_{\mathbf{k}+\mathbf{q}_1+\mathbf{q}_2,m}) \nonumber \\
& \hspace{15pt} \times \left( \begin{array}{l}
[1+n_B(\omega_{\mathbf{q}_1,\nu_1})+n_B(\omega_{\mathbf{q}_2,\nu_2})] \delta(\omega - \omega_{\mathbf{q}_1,\nu_1} - \omega_{\mathbf{q}_2,\nu_2}) \\
+|n_B(\omega_{\mathbf{q}_2,\nu_2})-n_B(\omega_{\mathbf{q}_1,\nu_1})| \delta(\omega - |\omega_{\mathbf{q}_1,\nu_1}-\omega_{\mathbf{q}_2,\nu_2}|)
\end{array}  \right)  \dee^3 \mathbf{q}_1 \dee^3 \mathbf{q}_2 \label{EliashbergDefinition}
\end{align}
\end{widetext}
Here, the sums over $\mathbf{q}$ have been replaced with integrals over the first Brillouin zone using the usual substitution $\sum_{\mathbf{q}} \rightarrow \frac{\Omega}{(2\pi)^3} \int_{\text{1BZ}} \dee^3 \mathbf{q}$ \cite{kittel2018}, and $\Omega_{\text{1BZ}} = \frac{(2\pi)^3}{\Omega_0}$ is the volume of the first Brillouin zone \cite{kittel2018}.

\eref{FMfromEliashberg} is the literature expression for the Fan-Migdal self energy \cite{giustino2017}. The 1-electron-2-phonon contribution from the diagram in \figref{fig:SelfEnergyDiagrams}d only enters the self energy through an additional term in the Eliashberg function. Indeed, the first line of \eref{EliashbergDefinition} is the familiar definition of the Eliashberg function for the 1-electron-1-phonon Hamiltonian \cite{giustino2017}. The other term is of a similar form, but is directly due to the 1-electron-2-phonon interaction. Curiously, this additional term is temperature dependent through the Bose-Einstein distributions: this temperature dependence comes directly from the integral identity in \eref{D0D0Integral}.

\subsection{The 1-electron-2-phonon spectral function}
Let us now consider the case where the long-range interaction is dominant over the short-range interaction, such as in the case of a large polaron. In this case, we may approximate the Eliashberg function by using the long-range electron-phonon matrix elements from \erefs{g1Long}{g2Long} in its definition from \eref{EliashbergDefinition}. Furthermore, because the 1-electron-1-phonon matrix element is dominated by processes where $\mathbf{q}+\mathbf{G} \approx \mathbf{0}$, we may replace the phonon frequencies $\omega_{\mathbf{q},\nu}$ in the first term with $\omega_{\hat{\mathbf{q}},\nu}$, the phonon frequencies around the $\Gamma$-point. Analogously, the 1-electron-2-phonon matrix element is dominated by processes where $\mathbf{q}_1 + \mathbf{q}_2 + \mathbf{G} \approx \mathbf{0}$, so we may set $\omega_{\mathbf{q}_1,\nu_1} \approx \omega_{-\mathbf{q}_2,\nu_1} = \omega_{\mathbf{q}_2,\nu_1}$ in the second term. This yields:
\begin{widetext}
\begin{align}
\alpha^2 F_{\mathbf{k},n}(\varepsilon, \omega) & \approx \frac{1}{\hbar} \sum_{\nu, m} \frac{1}{\Omega_{\text{1BZ}}} \int_{\text{1BZ}} |g^{\text{(long)}}_{mn\nu}(\mathbf{k},\mathbf{q})|^2 \delta(\varepsilon - \epsilon_{\mathbf{k}+\mathbf{q},m}) \delta(\omega - \omega_{\hat{\mathbf{q}},\nu}) \dee^3 \mathbf{q} \nonumber \\
& + \frac{2}{ \hbar} \sum_{\nu_1, \nu_2, m} \frac{1}{\Omega^2_{\text{1BZ}}} \int_{\text{1BZ}} \int_{\text{1BZ}} | g^{\text{(long)}}_{mn\nu_1 \nu_2}(\mathbf{k},\mathbf{q}_1,\mathbf{q}_2) |^2 \delta(\varepsilon - \epsilon_{\mathbf{k}+\mathbf{q}_1+\mathbf{q}_2,m}) \nonumber \\
& \hspace{15pt} \times \left( \begin{array}{l}
[1+n_B(\omega_{\mathbf{q}_2,\nu_1})+n_B(\omega_{\mathbf{q}_2,\nu_2})] \delta(\omega - \omega_{\mathbf{q}_2,\nu_1} - \omega_{\mathbf{q}_2,\nu_2}) \\
+|n_B(\omega_{\mathbf{q}_2,\nu_2})-n_B(\omega_{\mathbf{q}_2,\nu_1})| \delta(\omega - |\omega_{\mathbf{q}_2,\nu_1}-\omega_{\mathbf{q}_2,\nu_2}|)
\end{array}  \right) \dee^3 \mathbf{q}_1 \dee^3 \mathbf{q}_2.
\end{align}
Substituting \erefs{g1Long}{g2Long} into this expression yields, after a short calculation:
\begin{align}
\alpha^2 F_{\mathbf{k},n}(\varepsilon, \omega) & \approx \frac{e^2}{\epsvac \Omega_0} \frac{1}{\Omega_{\text{1BZ}}} \int_{\text{1BZ}} \sum_{\mathbf{G}_1, \mathbf{G}_2 \neq -\mathbf{q}} \frac{(\mathbf{q}+\mathbf{G}_1) \cdot (\mathcal{R}(\omega, \hat{\mathbf{q}}) \mathbf{I} +  \bm{\mathcal{T}}(\omega) ) \cdot (\mathbf{q}+\mathbf{G}_2)}{[(\mathbf{q}+\mathbf{G}_1) \cdot \bm{\varepsilon}_{\infty} \cdot (\mathbf{q}+\mathbf{G}_1)][(\mathbf{q}+\mathbf{G}_2) \cdot \bm{\varepsilon}_{\infty} \cdot (\mathbf{q}+\mathbf{G}_2)]} \nonumber \\
& \hspace{15pt} \times \left( \sum_{m} \delta(\varepsilon - \epsilon_{\mathbf{k}+\mathbf{q},m}) \langle \psi_{\mathbf{k},n} | e^{-i (\mathbf{q}+\mathbf{G}_2)\cdot \mathbf{r}} | \psi_{\mathbf{k}+\mathbf{q},m} \rangle \langle \psi_{\mathbf{k}+\mathbf{q},m} | e^{i (\mathbf{q}+\mathbf{G}_1)\cdot \mathbf{r}} | \psi_{\mathbf{k},n} \rangle \right) \dee^3 \mathbf{q}, \label{EliashbergIntermediate}
\end{align}
where we have introduced the following two dimensionless spectral functions, which represent the contributions from the 1-electron-1-phonon and 1-electron-2-phonon interactions, respectively:
\begin{align}
\mathcal{R}(\omega, \hat{\mathbf{q}}) & := \frac{e^2}{2 \epsvac \Omega_0 \omega} \sum_{\nu} |\hat{\mathbf{q}} \cdot \mathbf{p}_{\nu}(\hat{\mathbf{q}})|^2 \delta(\omega-\omega_{\hat{\mathbf{q}},\nu})  \label{RomegaDef} \\
\mathcal{T}_{\alpha \beta}(\omega) & := \frac{e^2}{2\hbar \epsvac \Omega_0} \sum_{\nu_1, \nu_2} \frac{1}{\Omega_{\text{1BZ}}} \int_{\text{1BZ}}   Y_{\nu_1\nu_2,\alpha}(\mathbf{q}) Y^*_{\nu_1\nu_2,\beta}(\mathbf{q})  \left( \begin{array}{l}
[1+n_B(\omega_{\mathbf{q},\nu_1})+n_B(\omega_{\mathbf{q},\nu_2})] \ \delta(\omega - \omega_{\mathbf{q},\nu_1} - \omega_{\mathbf{q},\nu_2}) \\
+|n_B(\omega_{\mathbf{q},\nu_2})-n_B(\omega_{\mathbf{q},\nu_1})| \ \delta(\omega - |\omega_{\mathbf{q},\nu_1}-\omega_{\mathbf{q},\nu_2}|)
\end{array}  \right) \dee^3\mathbf{q}. \label{TomegaDef}
\end{align}
In the common case where $\omega_{\hat{\mathbf{q}},\nu} = \omega_{\nu}$ does not depend on the direction $\hat{\mathbf{q}}$ (e.g. in cubic materials), the function $\mathcal{R}(\omega, \hat{\mathbf{q}})$ is simply a sum of delta peaks $\delta(\omega - \omega_{\nu})$. This is due to the long-range approximation. In practice, it means that $\mathcal{R}(\omega, \hat{\mathbf{q}})$ is merely a function to simplify notation: it cannot be calculated or plotted numerically.

By contrast, the 1-electron-2-phonon spectral function $\mathcal{T}_{\alpha \beta}(\omega)$ is an important quantity that can be calculated from first principles. Indeed, recalling the definition of $Y_{\nu_1\nu_2,\alpha}(\mathbf{q})$ in \eref{Ydef}, one only requires the phonon frequencies, phonon eigenvectors, and the $\bm{\mathcal{E}}$-derivative of the dynamical matrix to calculate $\mathcal{T}_{\alpha \beta}(\omega)$. It is the only new quantity in \eref{EliashbergIntermediate}: all other quantities are known, harmonic literature quantities. The intuitive meaning for $\mathcal{T}_{\alpha \beta}(\omega)$ is more clear in the zero temperature limit, where $n_B(\omega) = 0$:
\begin{equation} \label{TomegaDefTzero}
\mathcal{T}_{\alpha \beta}(\omega) = \frac{e^2}{2\hbar \epsvac \Omega_0} \sum_{\nu_1, \nu_2} \frac{1}{\Omega_{\text{1BZ}}} \int_{\text{1BZ}}   Y_{\nu_1\nu_2,\alpha}(\mathbf{q}) Y^*_{\nu_1\nu_2,\beta}(\mathbf{q})
 \delta(\omega - \omega_{\mathbf{q},\nu_1} - \omega_{\mathbf{q},\nu_2}) \dee^3\mathbf{q}.
\end{equation}
\end{widetext}
$\mathcal{T}_{\alpha \beta}(\omega)$ can be interpreted as the total strength of all the 1-electron-2-phonon processes where $\mathbf{q}_1 \approx -\mathbf{q}_2$ and the total phonon energy is $\hbar \omega$. Because $\mathcal{T}_{\alpha \beta}(\omega)$ is a dimensionless quantity, its value for a material will roughly indicate how strong we expect the 1-electron-2-phonon interaction to be for that material.

$\mathcal{T}_{\alpha \beta}(\omega)$ is a real and symmetric $3 \times 3$ tensor, which is a function of a single variable $\omega$. Furthermore, $\mathcal{T}_{\alpha \beta}(\omega) = 0$ unless $\omega \in [0, 2 \omega_{\text{max}}]$, where $\omega_{\text{max}}$ is the maximum phonon frequency of the material. This makes it possible to tabulate the 1-electron-2-phonon spectral function for a given material. Additionally, $\mathcal{T}_{\alpha \beta}(\omega)$ follows all the symmetries of the material. Expressed mathematically, if $\mathcal{G}$ is the point group of the material under consideration, then for every symmetry $\mathbf{R} \in \mathcal{G}$ of the material, the matrix $\bm{\mathcal{T}}(\omega)$ must satisfy:
\begin{equation} \label{Tsymmetry}
\bm{\mathcal{T}}(\omega) = \mathbf{R}^{-1} \cdot \bm{\mathcal{T}}(\omega)  \cdot \mathbf{R}.
\end{equation}
In particular, for any orthorhombic material this implies that $\mathcal{T}_{\alpha \beta}(\omega)$ is diagonal, and for any cubic material the 1-electron-2-phonon spectral function is a scalar:
\begin{equation} \label{TomegaCubic}
\mathcal{T}_{\alpha \beta}(\omega) = \mathcal{T}(\omega) \delta_{\alpha \beta}. \hspace{20pt} \text{(cubic materials)}
\end{equation}
For the remainder of this section, it is assumed that $\mathcal{T}_{\alpha \beta}(\omega)$ is an input parameter that has been calculated elsewhere. In \secref{sec:FirstPrinciples}, it is shown how to calculate $\mathcal{T}(\omega)$ in practice from first principles for the cubic materials LiF and KTaO$_3$. 

\subsection{Quasiparticle energy shifts}
With the 1-electron-2-phonon spectral function defined, we now return to \eref{EliashbergIntermediate} for the Eliashberg function. It can be simplified considerably by making appropriate approximations. Firstly, recall that due to the long-range approximation, it is allowed to neglect any term with a nonzero reciprocal lattice vector. Here, we keep only the terms where $\mathbf{G}_1 = \mathbf{G}_2$ in \eref{EliashbergIntermediate}. Then, in \appref{app:BlochSum}, we show that the sum over $m$ simplifies drastically, and the result no longer depends on the electronic states:
\begin{widetext}
\begin{equation}
\sum_{m} \delta(\varepsilon - \epsilon_{\mathbf{k}+\mathbf{q},m}) \langle \psi_{\mathbf{k},n} | e^{-i (\mathbf{q}+\mathbf{G})\cdot \mathbf{r}} | \psi_{\mathbf{k}+\mathbf{q},m} \rangle \langle \psi_{\mathbf{k}+\mathbf{q},m} | e^{i (\mathbf{q}+\mathbf{G})\cdot \mathbf{r}} | \psi_{\mathbf{k},n} \rangle = \delta(\varepsilon - \epsilon_{\mathbf{k}+\mathbf{q},n}).
\end{equation}
This gives the following expression for the Eliashberg function:
\begin{equation} \label{EliashbergGeneralAlmost}
\alpha^2 F_{\mathbf{k},n}(\varepsilon, \omega) \approx \frac{e^2}{\epsvac \Omega_0}  \sum_{\mathbf{G} \neq -\mathbf{q}} \frac{1}{\Omega_{\text{1BZ}}} \int_{\text{1BZ}} \frac{(\mathbf{q}+\mathbf{G}) \cdot \left[  \mathcal{R}(\omega, \hat{\mathbf{q}})\ \mathbf{I} + \bm{\mathcal{T}}(\omega) \right] \cdot (\mathbf{q}+\mathbf{G})}{[(\mathbf{q}+\mathbf{G}) \cdot \bm{\varepsilon}_{\infty} \cdot (\mathbf{q}+\mathbf{G})]^2} \delta(\varepsilon - \epsilon_{\mathbf{k}+\mathbf{q},n})  \dee^3 \mathbf{q}.
\end{equation}
The combination of integral over the first Brillouin zone $\Omega_{\text{1BZ}}$ and the sum over reciprocal lattice vectors $\mathbf{G}$ can be written as an integral over the entire reciprocal space, with an integration variable $\mathbf{q}+\mathbf{G} = \mathbf{Q}$. Using $\Omega_0 \Omega_{\text{1BZ}} = (2\pi)^3$, this yields:
\begin{equation} \label{EliashbergRT}
\alpha^2 F_{\mathbf{k},n}(\varepsilon, \omega) \approx \frac{e^2}{\epsvac (2\pi)^3} \int \frac{\mathbf{Q} \cdot \left[  \mathcal{R}(\omega, \hat{\mathbf{Q}})\ \mathbf{I} + \bm{\mathcal{T}}(\omega) \right] \cdot \mathbf{Q}}{(\mathbf{Q} \cdot \bm{\varepsilon}_{\infty} \cdot \mathbf{Q})^2} \delta(\varepsilon - \epsilon_{\mathbf{k}+\mathbf{Q},n})  \dee^3 \mathbf{Q}.
\end{equation}
This expression for the Eliashberg function can be combined with \eref{QuasiparticleEnergies} and \eref{FMfromEliashberg} to find an expression for the temperature-dependent quasiparticle energy shifts $\Delta \epsilon_{\mathbf{k},n}$:
\begin{equation}
\Delta \epsilon_{\mathbf{k},n} = \frac{\hbar e^2}{\epsvac (2\pi)^3} \int \int_{0}^{+\infty} \frac{\mathbf{Q} \cdot \left(  \mathcal{R}(\omega, \hat{\mathbf{Q}})\ \mathbf{I} + \bm{\mathcal{T}}(\omega) \right) \cdot \mathbf{Q}}{[\mathbf{Q} \cdot \bm{\varepsilon}_{\infty} \cdot \mathbf{Q}]^2} \left[ \frac{1 - n_F(\epsilon_{\mathbf{k}+\mathbf{Q},n}) + n_B(\omega)}{\epsilon_{\mathbf{k},n} - \epsilon_{\mathbf{k}+\mathbf{Q},n} - \hbar \omega + i\delta} + \frac{n_F(\epsilon_{\mathbf{k}+\mathbf{Q},n}) + n_B(\omega)}{\epsilon_{\mathbf{k},n} - \epsilon_{\mathbf{k}+\mathbf{Q},n} + \hbar \omega + i\delta} \right] \dee^3 \mathbf{Q} \dee \omega . \label{EnergiesEliashberg}
\end{equation}
\end{widetext}
where the lower bound for the $\omega$-integral was changed from $-\infty$ to $0$ because both $\mathcal{R}(\omega, \hat{\mathbf{Q}})$ and $\bm{\mathcal{T}}(\omega)$ are zero when $\omega < 0$. This expression can be evaluated once the electronic bands $\epsilon_{\mathbf{k},n}$ are known, for example after a first-principles calculation. In the remainder of this section, we will instead evaluate \eref{EnergiesEliashberg} by choosing appropriate model expressions for the electronic bands in the context of a large polaron.

\subsection{Ground state energy in the generalized Fr\"ohlich model} \label{sec:GeneralizedFrohlich}
The generalized Fr\"ohlich Hamiltonian was introduced in \cite{miglio2020}. Its only assumptions are that the electron-phonon interaction is dominated by the long-range interaction as in \erefs{g1Long}{g2Long}, that the conduction band minimum is located at $\Gamma$, and that all quantities are evaluated near $\Gamma$. In particular, it is assumed that only the electron bands around the conduction band minimum are important, and these bands are replaced by a parabolic approximation:
\begin{equation} \label{ParabolicBands}
\epsilon_{\mathbf{k},n} \approx \epsilon_c + \frac{\hbar^2 |\mathbf{k}|^2}{2 m^*_n(\hat{\mathbf{k}})}.
\end{equation}
Here, $\epsilon_c$ is the minimum energy of the conduction band at $\Gamma$. We are neglecting any band with $\epsilon_{\mathbf{0},n} \neq \epsilon_c$; in other words, we are only keeping those bands which are degenerate at the conduction band minimum. The degree of degeneracy at the conduction band minimum is denoted by $n_{\text{deg}}$.

With \eref{ParabolicBands}, it is possible to calculate the ground state energy $E_0$ of a large polaron, which is the energy shift of the conduction band bottom due to the electron-phonon interaction. It is equivalent to the conduction band zero-point renormalization in bandgap renormalization theory \cite{miglio2020}. For simplicity, we evaluate the ground state energy at temperature zero, such that $n_B(\omega) = 0$. Furthermore, we assume that the material is either undoped, or that the doping concentration of the material is very low, so that $E_F \lesssim \epsilon_c$ and $n_F(\epsilon_{\mathbf{k},n}) \approx 0$. Under those approximations, the ground state energy can be obtained from \eref{EnergiesEliashberg} as follows:
\begin{widetext}
\begin{equation}
E_0 = - \frac{1}{n_{\text{deg}}} \sum_{n=1}^{n_{\text{deg}}}\frac{\hbar e^2}{\epsvac (2\pi)^3} \int \int_{0}^{+\infty} \frac{\mathbf{Q} \cdot \left[  \mathcal{R}(\omega, \hat{\mathbf{Q}})\ \mathbf{I} + \bm{\mathcal{T}}(\omega) \right] \cdot \mathbf{Q}}{(\mathbf{Q} \cdot \bm{\varepsilon}_{\infty} \cdot \mathbf{Q})^2} \frac{1}{\frac{\hbar^2 |\mathbf{Q}|^2}{2 m^*_n(\hat{\mathbf{Q}})} + \hbar \omega} \dee^3 \mathbf{Q} \dee \omega .
\end{equation}
The integral over $\mathbf{Q}$ can be split into a radial integral over $Q^2 \dee Q$ and an angular integral $\dee^2 \hat{\mathbf{q}}$ over the unit sphere $S^2$:
\begin{equation}
E_0 = - \frac{1}{n_{\text{deg}}} \sum_{n=1}^{n_{\text{deg}}}\frac{\hbar e^2}{\epsvac (2\pi)^3} \int_{S^2} \int_{0}^{+\infty} \frac{\hat{\mathbf{q}} \cdot \left[  \mathcal{R}(\omega, \hat{\mathbf{q}})\ \mathbf{I} + \bm{\mathcal{T}}(\omega) \right] \cdot \hat{\mathbf{q}}}{(\hat{\mathbf{q}} \cdot \bm{\varepsilon}_{\infty} \cdot \hat{\mathbf{q}})^2} \left[\int_0^{+\infty}\frac{ \dee Q}{\frac{\hbar^2 Q^2}{2 m^*_n(\hat{\mathbf{q}})} + \hbar \omega} \right] \dee^2 \hat{\mathbf{q}} \dee \omega .
\end{equation}
The integral over $Q$ can now be evaluated analytically, which yields:
\begin{equation}
E_0 = -\frac{1}{n_{\text{deg}}} \sum_{n=1}^{n_{\text{deg}}} \frac{e^2}{(4\pi)^2 \epsvac} \sqrt{\frac{2}{\hbar}} \int_{0}^{+\infty} \int_{S^2} \frac{ \mathcal{R}(\omega, \hat{\mathbf{q}}) + \hat{\mathbf{q}} \cdot \bm{\mathcal{T}}(\omega) \cdot \hat{\mathbf{q}}}{\sqrt{\omega} (\hat{\mathbf{q}} \cdot \bm{\varepsilon}_{\infty} \cdot \hat{\mathbf{q}})^2} \sqrt{m^*_n(\hat{\mathbf{q}})} \dee \omega \dee^2 \hat{\mathbf{q}}.  
\end{equation}
The 1-electron-1-phonon contribution can still be simplified, because the $\omega$-integral over \eref{RomegaDef} for $\mathcal{R}(\omega, \hat{\mathbf{q}})$ can be performed analytically:
\begin{equation}
\int_{-\infty}^{+\infty} \frac{ \mathcal{R}(\omega, \hat{\mathbf{q}})}{\sqrt{\omega}} \dee \omega = \frac{e^2}{2 \epsvac \Omega_0} \sum_{\nu} \frac{ |\hat{\mathbf{q}}\cdot \mathbf{p}_{\nu}(\hat{\mathbf{q}})|^2 }{\omega_{\hat{\mathbf{q}},\nu}^{\frac{3}{2}}}.
\end{equation}
Similarly, in the 1-electron-2-phonon contribution, nothing depends on $\omega$ other than $\bm{\mathcal{T}}(\omega)$, so the 1-electron-2-phonon spectral function only enters the ground state energy through the following moment:
\begin{equation} \label{Tm12Def}
\bm{\mathcal{T}}_{-\frac{1}{2}} := \int_{0}^{+\infty} \frac{\bm{\mathcal{T}}(\omega)}{\sqrt{\omega}} \dee \omega.
\end{equation}
With these two integrals, the ground state energy can be written in its final form as a sum of the 1-electron-1-phonon and 1-electron-2-phonon contributions:
\begin{align}
E_0 & = E^{(\text{1e1ph})}_0 + E^{(\text{1e2ph})}_0, \label{ZPRGeneralFrohlich} \\
E^{(\text{1e1ph})}_0 & = - \sum_{\nu} \sum_{n=1}^{n_{\text{deg}}} \left(\frac{e^2}{4\pi \epsvac}\right)^2 \frac{\hbar}{\sqrt{2} \Omega_0 n_{\text{deg}}} \int_{S^2} \frac{\sqrt{m^*_n(\hat{\mathbf{q}})}}{(\hbar \omega_{\hat{\mathbf{q}},\nu})^{\frac{3}{2}}} \frac{ |\hat{\mathbf{q}}\cdot \mathbf{p}_{\nu}(\hat{\mathbf{q}})|^2 }{(\hat{\mathbf{q}} \cdot \bm{\varepsilon}_{\infty} \cdot \hat{\mathbf{q}})^2} \dee^2 \hat{\mathbf{q}}, \label{ZPRGeneralFrohlich1} \\
E^{(\text{1e2ph})}_0 & = - \sum_{n=1}^{n_{\text{deg}}} \frac{e^2}{4\pi \epsvac n_{\text{deg}}} \sqrt{\frac{2}{\hbar}} \frac{1}{4\pi} \int_{S^2} \frac{ \hat{\mathbf{q}} \cdot \bm{\mathcal{T}}_{-\frac{1}{2}} \cdot \hat{\mathbf{q}}}{(\hat{\mathbf{q}} \cdot \bm{\varepsilon}_{\infty} \cdot \hat{\mathbf{q}})^2} \sqrt{m^*_n(\hat{\mathbf{q}})} \dee^2 \hat{\mathbf{q}}. \label{ZPRGeneralFrohlich2}
\end{align}
\end{widetext}
\eref{ZPRGeneralFrohlich1} is the literature expression for the weak-coupling ground state energy in the generalized Fr\"ohlich model \cite{miglio2020}. We have shown here that the 1-electron-2-phonon interaction gives a contribution that can similarly be written as an angular average over $\dee^2 \hat{\mathbf{q}}$. Perhaps most interestingly, it shows that the $-\frac{1}{2}$ moment of the 1-electron-2-phonon spectral function has a physical meaning: it controls the strength of the contribution to the ground state energy. Combining \eref{Tm12Def} and \eref{TomegaDef}, we can find another useful expression for this moment at temperature zero:
\begin{align}
& \left(\bm{\mathcal{T}}_{-\frac{1}{2}}\right)_{\alpha \beta} \\
& \hspace{10pt} = \frac{e^2}{2\hbar \epsvac \Omega_0} \sum_{\nu_1, \nu_2} \frac{1}{\Omega_{\text{1BZ}}} \int_{\text{1BZ}} \frac{Y_{\nu_1\nu_2,\alpha}(\mathbf{q}) Y^*_{\nu_1\nu_2,\beta}(\mathbf{q})}{ \sqrt{ \omega_{\mathbf{q},\nu_1} + \omega_{\mathbf{q},\nu_2}} } \dee^3 \mathbf{q}. \nonumber 
\end{align}
This moment is a $3 \times 3$ matrix; for cubic materials it is even a simple scalar. Therefore, $\bm{\mathcal{T}}_{-\frac{1}{2}}$ is a useful quantity to tabulate for many materials to quantify the strength of the 1-electron-2-phonon interaction in that material. A downside is that $\bm{\mathcal{T}}_{-\frac{1}{2}}$ is no longer dimensionless, but has the same units as $\sqrt{\omega}$. If desired, one could calculate and tabulate $\bm{\mathcal{T}}_{-\frac{1}{2}}/\sqrt{\omega_{\text{ref}}}$ for some reference frequency $\omega_{\text{ref}}$, such as the maximum phonon frequency in the system.

\subsection{Polaron energy and effective mass in a Fr\"ohlich-type material}
Finally, we investigate the 1-electron-2-phonon interaction in the case of a material that satisfies all the assumptions of the Fr\"ohlich Hamiltonian. In particular, we assume that the material has cubic symmetry, and that it only has two atoms in the unit cell so that there is only a single longitudinal optical phonon branch. Additionally, we assume that the conduction band is not degenerate at its minimum, so that the electron band can be written as:
\begin{equation} \label{ParabolicBandFrohlich}
\epsilon_{\mathbf{k},c} \approx \epsilon_c + \frac{\hbar^2 |\mathbf{k}|^2}{2 m^*}.
\end{equation}
Since this model describes a cubic material, the 1-electron-2-phonon spectral function $\mathcal{T}(\omega)$ is a scalar function, as discussed in \eref{TomegaCubic}. The dielectric tensor $\bm{\varepsilon}_{\infty} = \varepsilon_{\infty} \mathbf{I}$ also becomes a scalar in this case. The 1-electron-1-phonon function $\mathcal{R}(\omega, \hat{\mathbf{q}})$ also simplifies considerably. Indeed, in this model, the mode polarities $\hat{\mathbf{q}} \cdot \mathbf{p}_{\nu}(\hat{\mathbf{q}})$ are zero for all phonon branches except for the LO branch, whose mode polarity we write as $p_{\text{LO}}$ \cite{guster2021}:
\begin{equation}
\hat{\mathbf{q}} \cdot \mathbf{p}_{\nu}(\hat{\mathbf{q}}) = p_{\text{LO}} \delta_{\nu, \text{LO}}.
\end{equation}
This simplifies $\mathcal{R}(\omega, \hat{\mathbf{q}})$ to a single delta peak:
\begin{equation} \label{RomegaFrohlich}
\mathcal{R}(\omega, \hat{\mathbf{q}}) =  \alpha \frac{4 \pi \epsvac \varepsilon_{\infty}^2}{e^2} \frac{(\hbar \omega_{\text{LO}})^2}{\sqrt{2 \hbar \omega_{\text{LO}} m^* }} \delta(\omega-\omega_{\text{LO}}),
\end{equation}
where the Fr\"ohlich coupling constant $\alpha$ is defined as \cite{guster2021}:
\begin{equation}
\alpha := \frac{4\pi}{\Omega_0} \sqrt{\frac{\hbar m^*}{2\omega_{\text{LO}}}} \left(\frac{e^2}{4\pi \epsvac \varepsilon_{\infty} \hbar \omega_{\text{LO}}} \right)^2 |p_{\text{LO}}|^2.
\end{equation}
The polaron energy shift $\Delta \epsilon_{\mathbf{k},c}$ in the conduction band can now be calculated from \eref{EnergiesEliashberg} in terms of $\alpha$ and $\mathcal{T}(\omega)$. Just like in \secref{sec:GeneralizedFrohlich}, we assume temperature zero and a very low doping concentration, such that we may write $n_B(\omega) = n_F(\varepsilon_{\mathbf{k},c}) = 0$. Using this, \eref{TomegaCubic}, and \eref{RomegaFrohlich} in \eref{EnergiesEliashberg} gives the following expression for the polaron energy shift:
\begin{align}
& \Delta \epsilon_{\mathbf{k},c} = -\frac{\hbar e^2}{\epsvac \varepsilon_{\infty}^2 (2\pi)^3} \nonumber \\
& \hspace{10pt} \times \int_{0}^{+\infty} \left[ \alpha \frac{4 \pi \epsvac \varepsilon_{\infty}^2}{e^2} \frac{(\hbar \omega_{\text{LO}})^2}{\sqrt{2 \hbar \omega_{\text{LO}} m^* }} \delta(\omega-\omega_{\text{LO}}) + \mathcal{T}(\omega) \right] \nonumber \\
& \hspace{10pt} \times \left(\int \frac{1}{\frac{\hbar^2 |\mathbf{k}+\mathbf{Q}|^2}{2 m^*} - \frac{\hbar^2 |\mathbf{k}|^2}{2 m^*} + \hbar \omega - i\delta} \frac{\dee^3 \mathbf{Q}}{|\mathbf{Q}|^2} \right) \dee \omega .
\end{align}
The remaining integral over $\mathbf{Q}$ also appears in the treatment of the Fr\"ohlich Hamiltonian. It can be performed analytically, and yields (see equations (7.10), (7.15), and (7.23) in \cite{mahan2000}):
\begin{align}
& \int \frac{1}{\frac{\hbar^2 |\mathbf{k}+\mathbf{Q}|^2}{2 m^*} - \frac{\hbar^2 |\mathbf{k}|^2}{2 m^*} + \hbar \omega - i\delta} \frac{\dee^3 \mathbf{Q}}{|\mathbf{Q}|^2} \nonumber \\
& \hspace{30pt} = 2\pi^2 \frac{2 m^*}{\hbar^2 |\mathbf{k}|} \arcsin\left(\sqrt{ \frac{\hbar}{2 m^* \omega}}  |\mathbf{k}| \right).
\end{align}
With the above integral, the quasiparticle energy of the conduction band can be written as:
\begin{align}
& \epsilon_{\mathbf{k},c} -  \epsilon_c = \frac{\hbar^2 |\mathbf{k}|^2}{2 m^*} -\hbar \omega_{\text{LO}} \alpha \frac{\arcsin\left( \sqrt{\frac{\hbar}{2 m^* \omega_{\text{LO}}}} |\mathbf{k}| \right)}{\sqrt{\frac{\hbar}{2 m^* \omega_{\text{LO}}}} |\mathbf{k}|} \nonumber \\
& \hspace{10pt} -\frac{e^2}{4\pi \epsvac \varepsilon^2_{\infty}} \frac{2 m^*}{\hbar |\mathbf{k}|} \int_{0}^{+\infty} \mathcal{T}(\omega) \arcsin\left(\sqrt{ \frac{\hbar}{2 m^* \omega}}  |\mathbf{k}| \right) \dee \omega.
\end{align}
For a slow-moving polaron, this energy band can be expanded up to order $|\mathbf{k}|^2$, which yields:
\begin{equation}
\epsilon_{\mathbf{k},c} \approx \epsilon_c + E_0 + \frac{\hbar^2 |\mathbf{k}|^2}{2 m_{\text{pol}}} + O(|\mathbf{k}|^4),
\end{equation}
where the ground state energy $E_0$ and polaron effective mass $m_{\text{pol}}$ are given by:
\begin{align}
E_0 & = -\alpha \hbar \omega_{\text{LO}} - \frac{e^2}{4\pi \epsvac \varepsilon^2_{\infty}} \sqrt{\frac{2 m^*}{\hbar}} \int_{0}^{+\infty} \frac{\mathcal{T}(\omega)}{\sqrt{\omega}} \dee \omega, \label{Epolaron} \\
\frac{m^*}{m_{\text{pol}}} & = 1-\frac{\alpha}{6} - \frac{1}{6\hbar} \frac{e^2}{4\pi \epsvac \varepsilon^2_{\infty}} \sqrt{\frac{2 m^*}{\hbar}} \int_{0}^{+\infty} \frac{\mathcal{T}(\omega)}{\omega^{\frac{3}{2}}} \dee \omega. \label{Mpolaron}
\end{align}
When there is no 1-electron-2-phonon interaction, $\mathcal{T}(\omega) = 0$ and these formulas reduce to the well-known weak-coupling results for the Fr\"ohlich Hamiltonian (Eqs. (2.20)-(2.22) in \cite{alexandrov2010} or Eq. (7.28) in \cite{mahan2000}). The 1-electron-2-phonon interaction simply adds contributions to the polaron ground state energy and inverse effective mass, and these contributions can be written in terms of $\mathcal{T}(\omega)$. As in the previous section, the contribution to the polaron energy only depends on the moment $\mathcal{T}_{-\frac{1}{2}}$ of the 1-electron-2-phonon spectral function. \eref{Mpolaron} shows that the effective polaron mass depends on a different moment:
\begin{equation}
\mathcal{T}_{-\frac{3}{2}} = \int_{0}^{+\infty} \frac{\mathcal{T}(\omega)}{\omega^{\frac{3}{2}}} \dee \omega
\end{equation}
This moment represents the strength of the 1-electron-2-phonon interaction in the context of the polaron mass, so it may also be useful to tabulate alongside $\mathcal{T}_{-\frac{1}{2}}$.

Note that \erefs{Epolaron}{Mpolaron} have been derived in the weak coupling approximation, since we only kept the lowest order diagrams in our diagrammatic expansion in \figref{fig:SelfEnergyDiagrams}. As discussed in \secref{sec:SelfEnergyExpansion}, this approximation is only valid when the 1-electron-1-phonon interaction is sufficiently small. For the case with only 1-electron-1-phonon interaction, more accurate results are known. Besides the all-coupling result obtained with the path integral method \cite{feynman1955}, the perturbation series in $\alpha$ for the polaron energy and effective mass are known up to higher orders in $\alpha$ \cite{smondyrev1986, alexandrov2010}:
\begin{align}
\frac{E_0}{\hbar \omega_{\text{LO}}} & = -\alpha - 0.015920 \alpha^2 - 0.000806 \alpha^3 - O(\alpha^4) \label{EpolaronFrohlich} \\
\frac{m^*}{m_{\text{pol}}} & = 1-\frac{\alpha}{6} - 0.00415 \alpha^2 - O(\alpha^3) \label{MpolaronFrohlich}
\end{align}
Therefore, one should only use \erefs{Epolaron}{Mpolaron} when $\alpha \lesssim 1$, and when $\mathcal{T}(\omega)$ is similarly appropriately small.

\section{First-principles application to lithium fluoride and potassium tantalate} 
\label{sec:FirstPrinciples}

\subsection{The dynamical matrix derivative} \label{sec:DynMatProperties}
In \secref{sec:EnergyMass} it was shown how to calculate the lowest-order effect of the long-range 1-electron-2-phonon interaction on the electron quasiparticle energies. The result was written in terms of a 1-electron-2-phonon spectral function $\mathcal{T}(\omega)$ which depends on the material. In this section, we show how to calculate $\mathcal{T}(\omega)$ from first principles, using the finite-displacement method to calculate phonon properties $\mathcal{D}_{\kappa \alpha,\kappa' \beta}(\mathbf{q})$ and $\frac{\partial \mathcal{D}_{\kappa \alpha,\kappa' \beta}(\mathbf{q})}{\partial \bm{\mathcal{E}}}$. As a benchmark example, this method is applied to two polar semiconductors: LiF, a cubic material with the rock salt structure and with significant long-range 1-electron-1-phonon interaction, and KTaO$_3$, a cubic quantum paraelectric with the perovskite structure and nonnegligible anharmonicity \cite{ranalli2024}. The electron polaron in LiF is a large polaron \cite{sio2019}, and LiF has one atom of Li and one atom of F in its primitive unit cell, so the 1-electron-1-phonon Hamiltonian reduces to the Fr\"ohlich Hamiltonian. This will make it easy to quantitatively compare our numerical results for the 1-electron-2-phonon interaction with those for the 1-electron-1-phonon interaction. For KTaO$_3$, the situation is slightly different. Because it has more than two atoms in its primitive unit cell, and because the electron band structure is degenerate around the conduction band minimum, the 1-electron-1-phonon Hamiltonian reduces to the generalized Fr\"ohlich Hamiltonian instead. However, because KTaO$_3$ has cubic symmetry, it can still be shown that the results from lowest-order perturbation theory in \erefs{ZPRGeneralFrohlich}{ZPRGeneralFrohlich2} reduce to:
\begin{align}
E_0 & = E_0^{\text{(1e1ph)}} + E_0^{\text{(1e2ph)}} \label{KTaO3_energy} \\
E_0^{\text{(1e1ph)}} & = - \frac{2\pi}{\Omega_0} \left( \frac{e^2}{4 \pi \epsvac \varepsilon_{\infty}} \right)^2 \sqrt{\frac{2m^*}{\hbar}} \sum_{\nu} \frac{|p_{\nu}|^2}{\omega_{\nu}^{\frac{3}{2}}} \label{KTaO3_energy1} \\
E_0^{\text{(1e2ph)}} & = - \frac{e^2}{4\pi \epsvac \varepsilon^2_{\infty}} \sqrt{\frac{2 m^*}{\hbar}} \int_{0}^{+\infty} \frac{\mathcal{T}(\omega)}{\sqrt{\omega}} \dee \omega  \label{KTaO3_energy2}
\end{align}
where $p_{\nu} := \hat{\mathbf{q}} \cdot \mathbf{p}_{\nu}(\hat{\mathbf{q}})$, and the single effective band mass $m^*$ is given by a square root average over all angles and all degenerate bands at the conduction band minimum \cite{guster2021}:
\begin{equation} \label{EffectiveBandMass}
\sqrt{m^*} = \frac{1}{4\pi n_{\text{deg}}} \sum_{n=1}^{n_{\text{deg}}} \int_{S^2} \sqrt{m_n^*(\hat{\mathbf{q}})} d^2 \hat{\mathbf{q}}
\end{equation}
With this definition of a single effective band mass, the contribution in \eref{KTaO3_energy2} due to the 1-electron-2-phonon interaction is precisely the same as in \eref{Epolaron}. Therefore, we can set up the same computational framework for both materials: in this framework, the 1-electron-1-phonon interaction is treated from the generalized Fr\"ohlich hamiltonian (which reduces to the usual Fr\"ohlich Hamiltonian for LiF), and the 1-electron-2-phonon spectral function $\mathcal{T}(\omega)$ can be calculated as a scalar.

We use the finite displacement method as implemented in PhonoPy \cite{togo2023, togo2023a} to calculate the required dynamical matrices. PhonoPy uses the c-type convention \cite{srivastava1990} for the dynamical matrix; therefore, expressions in this section will be written in the c-type convention as well. Note that \eref{Ydef} for $\mathbf{Y}_{\nu_1 \nu_2}(\mathbf{q})$ is invariant if we change between the c-type and d-type convention with the method described in \cite{srivastava1990}, \S 2.3; therefore, this change of convention has no influence on the 1-electron-2-phonon theory derived in the previous sections.

In the finite displacement method, first-principles calculations are performed on an $N \times N \times N$ supercell with one atom slightly displaced, in order to find an approximation for the force constants $\Phi_{\kappa \alpha, \kappa' \beta}(\bm{\ell})$; an approximation for the dynamical matrix $\mathcal{D}_{\kappa \alpha,\kappa' \beta}(\mathbf{q})$ is then obtained through \eref{DynDefE}. If the first-principles calculation is performed with an electric field $\bm{\mathcal{E}}$, then $\mathcal{D}_{\kappa \alpha,\kappa' \beta}(\mathbf{q}; \bm{\mathcal{E}})$ is automatically calculated as long as the first-principles calculation includes the effect of the electric field in the Hellman-Feynman forces. We note that first-principles calculations with an electric field are possible through the modern theory of polarization \cite{king-smith1993, nunes2001}, which circumvents the issue that the electric field introduces a non-periodic potential. The electric field derivative can then be calculated via the central difference approximation:
\begin{align}
& \frac{\partial \mathcal{D}_{\kappa \alpha,\kappa' \beta}(\mathbf{q})}{\partial \mathcal{E}_z} \\
& \hspace{10pt} \approx \frac{\mathcal{D}_{\kappa \alpha,\kappa' \beta}(\mathbf{q}; \mathcal{E} \mathbf{e}_z) - \mathcal{D}_{\kappa \alpha,\kappa' \beta}(\mathbf{q}; -\mathcal{E} \mathbf{e}_z)}{2 \mathcal{E}} + O(\mathcal{E}^3). \nonumber
\end{align}
Here, the value of $\mathcal{E}$ should be chosen appropriately: small enough such that the central difference approximation holds, and large enough such that the effect of the electric field is larger than the numerical precision of the calculations. Only the derivative in the $z$-direction is needed, since LiF and KTaO$_3$ are cubic materials.

Both LiF and KTaO$_3$ possess an inversion center which takes all atoms to an equivalent position. In this case, the c-type dynamical matrix $\mathcal{D}_{\kappa \alpha,\kappa' \beta}(\mathbf{q})$ is real, and its derivative $\frac{\partial \mathcal{D}_{\kappa \alpha,\kappa' \beta}(\mathbf{q})}{\partial \bm{\mathcal{E}}}$ is purely imaginary. This property is proven in \appref{app:InversionCenter}. Therefore, in the particular cases of LiF and KTaO$_3$, the following expression for the derivative of the dynamical matrix is numerically more stable:
\begin{align} \label{DynDerExpression}
& \frac{\partial \mathcal{D}_{\kappa \alpha,\kappa' \beta}(\mathbf{q})}{\partial \mathcal{E}_z} \\
& \hspace{10pt} \approx \text{Im}\left[\frac{\mathcal{D}_{\kappa \alpha,\kappa' \beta}(\mathbf{q}; \mathcal{E} \mathbf{e}_z) - \mathcal{D}_{\kappa \alpha,\kappa' \beta}(\mathbf{q}; -\mathcal{E} \mathbf{e}_z)}{2 \mathcal{E}}\right] + O(\mathcal{E}^3). \nonumber
\end{align}

The finite displacement method provides an approximation to the dynamical matrix which improves as the supercell size $N$ increases. In particular, the method is exact for all commensurate points $\mathbf{q}_c$, which are the wavevectors in the Brillouin zone which satisfy $e^{i \mathbf{q}_c\cdot \mathbf{T}}=1$ for all supercell lattice vectors $\mathbf{T}$ \cite{togo2023}. For all other $\mathbf{q}$-points, the standard technique is to use Fourier interpolation, which is implemented by default in PhonoPy \cite{togo2023, togo2023a}. The dynamical matrix is interpolated from the values at the commensurate points $\mathbf{q}_c$ with the following expression \cite{togo2023}:
\begin{widetext}
\begin{equation} \label{DynFourierOneEq}
\mathcal{D}_{\kappa \alpha, \kappa' \beta}(\mathbf{q}) \approx \sum_{\mathbf{q}_c} \mathcal{D}_{\kappa \alpha, \kappa' \beta}(\mathbf{q}_c) \left( \frac{1}{N_c} \sum_{\bm{\ell} \in \Omega_{\text{sc}}} \frac{1}{\#\{\mathbf{T}_{\kappa \kappa', \bm{\ell}} \}} \sum_{\mathbf{T} \in \{\mathbf{T}_{\kappa \kappa', \bm{\ell}} \} } e^{i (\mathbf{q}-\mathbf{q}_c)\cdot (\mathbf{T} + \bm{\ell} + \bm{\tau}_{\kappa'}-\bm{\tau}_{\kappa})} \right).
\end{equation}
Here, $N_c = N^3$ is the number of commensurate points in the Brillouin zone, $\{\mathbf{T}_{\kappa \kappa', \bm{\ell}}\}$ is the set of supercell vectors that minimizes $|\mathbf{T} + \bm{\ell} + \bm{\tau}_{\kappa'}-\bm{\tau}_{\kappa}|$, and $\#\{\mathbf{T}_{\kappa \kappa', \bm{\ell}} \}$ is the number of supercell vectors in that set. The factor between brackets only depends on the geometry of the unit cell: it has no dependence on the electric field. Therefore, one can straightforwardly take the electric field derivative of \eref{DynFourierOneEq} to find the Fourier interpolation rule for the derivative of the dynamical matrix:
\begin{equation} \label{DynFourierDer}
\frac{\partial \mathcal{D}_{\kappa \alpha, \kappa' \beta}(\mathbf{q})}{\partial \mathcal{E}_z} \approx \sum_{\mathbf{q}_c} \frac{\partial \mathcal{D}_{\kappa \alpha, \kappa' \beta}(\mathbf{q}_c)}{\partial \mathcal{E}_z} \left( \frac{1}{N_c} \sum_{\bm{\ell} \in \Omega_{\text{sc}}} \frac{1}{\#\{\mathbf{T}_{\kappa \kappa', \bm{\ell}} \}} \sum_{\mathbf{T} \in \{\mathbf{T}_{\kappa \kappa', \bm{\ell}} \} } e^{i (\mathbf{q}-\mathbf{q}_c)\cdot (\mathbf{T} + \bm{\ell} + \bm{\tau}_{\kappa'}-\bm{\tau}_{\kappa})} \right).
\end{equation}
In practice, $\frac{\partial \mathcal{D}_{\kappa \alpha, \kappa' \beta}(\mathbf{q}_c)}{\partial \mathcal{E}_z}$ is calculated at the commensurate points using \eref{DynDerExpression}, and then interpolated to arbitrary $\mathbf{q}$-points using \eref{DynFourierDer}.

Since the $\Gamma$-point is always a commensurate point, \erefs{DynFourierOneEq}{DynFourierDer} only work in practice if the dynamical matrix is well-defined in $\Gamma$. This is not the case in polar materials, where the dynamical matrix has a non-analytic contribution around $\Gamma$ \cite{gonze1997, togo2023}:
\begin{align}
\mathcal{D}^{\text{NAC}}_{\kappa \alpha, \kappa' \beta}(\mathbf{q}) & = \frac{1}{\sqrt{m_{\kappa} m_{\kappa'}}} \frac{e^2}{\epsvac \Omega_0} \sum_{\mathbf{G}\neq -\mathbf{q}} \sum_{\gamma, \delta} \frac{Z_{\gamma, \kappa \alpha} (q_{\gamma} + G_{\gamma}) (q_{\delta} + G_{\delta}) Z_{\delta, \kappa' \beta}}{(\mathbf{q}+\mathbf{G}) \cdot \bm{\varepsilon}_{\infty} \cdot (\mathbf{q}+\mathbf{G})} e^{i \mathbf{G}\cdot(\bm{\tau}_{\kappa}-\bm{\tau}_{\kappa'})} \nonumber \\ 
& - \frac{\delta_{\kappa \kappa'}}{\sqrt{m_{\kappa} m_{\kappa'}}} \frac{e^2}{\epsvac \Omega_0} \sum_{\kappa''} \sum_{\mathbf{G}\neq \mathbf{0}} \sum_{\gamma, \delta} \frac{Z_{\gamma, \kappa \alpha} G_{\gamma} G_{\delta} Z_{\delta, \kappa'' \beta}}{\mathbf{G} \cdot \bm{\varepsilon}_{\infty} \cdot \mathbf{G}} e^{i \mathbf{G}\cdot(\bm{\tau}_{\kappa}-\bm{\tau}_{\kappa''})}. \label{DynNAC}
\end{align}
\end{widetext}
For the dynamical matrix, the treatment of the non-analytic contribution is implemented in PhonoPy \cite{togo2023, togo2023a}. For the derivative of the dynamical matrix, it would have to be implemented from scratch. The non-analytic contribution to the derivative of the dynamical matrix $\frac{\partial \mathcal{D}^{\text{NAC}}_{\kappa \alpha, \kappa' \beta}(\mathbf{q})}{\partial \bm{\varepsilon}}$ can be obtained from \eref{DynNAC} by taking a derivative with respect to the electric field $\bm{\mathcal{E}}$. This requires knowledge of the following two tensors:
\begin{align}
\frac{\partial Z_{\beta, \kappa \alpha}}{\partial \mathcal{E}_{\gamma}} & = - \frac{1}{e} \frac{\partial^3 E}{\partial u_{\kappa \alpha} \partial \mathcal{E}_{\beta} \partial \mathcal{E}_{\gamma}} = \frac{\Omega_0 \epsvac}{e} \frac{\partial \chi_{\beta \gamma}}{\partial u_{\kappa \alpha}}, \label{RamanDef} \\
\frac{\partial \varepsilon^{(\infty)}_{\alpha \beta}}{\partial \mathcal{E}_{\gamma}} & = - \frac{1}{\epsvac} \frac{\partial^3 E}{\partial \mathcal{E}_{\alpha} \partial \mathcal{E}_{\beta} \partial \mathcal{E}_{\gamma}} = 2 \chi^{(2)}_{\alpha \beta \gamma}. \label{chi2Def}
\end{align}
Respectively, $\frac{\partial \chi_{\beta \gamma}}{\partial u_{\kappa \alpha}}$ and $\chi^{(2)}_{\alpha \beta \gamma}$ are the Raman susceptibility tensor and the nonlinear optical susceptibility of the material \cite{veithen2005}. In \appref{app:InversionCenter}, it is shown that both of these tensors are zero in materials with an inversion center that takes every atom to an equivalent position. Therefore, in the specific case of LiF and KTaO$_3$, $\frac{\partial \mathcal{D}_{\kappa \alpha, \kappa' \beta}(\mathbf{q}_c)}{\partial \mathcal{E}_z}$ is analytic in $\Gamma$, so that \eref{DynFourierDer} can be used without any further corrections.

\subsection{First-principles setup} \label{sec:FirstPrinciplesSetup}

All the first-principles calculations in this article have been performed using the Vienna Ab-initio Simulation Package (VASP) \cite{kresse1993, kresse1996, kresse1996a}.
Calculations in VASP were performed using the Perdew-Burke-Ernzerhof functional for solids (PBEsol) \cite{perdew2008} with a plane wave basis cutoff of 800~eV. The default  projector augmented wave (PAW) pseudopotentials were used for all calculations: the valence electron configurations and energy cutoffs of these pseudopotentials can be found in \tabref{tab:Pseudopotentials}. Unit-cell calculations for LiF and KTaO$_3$ were performed using a $12 \times 12 \times 12$ $\Gamma$-centered $\mathbf{k}$-grid and an $8 \times 8 \times 8$ Monkhorst-Pack $\mathbf{k}$-grid, respectively. The energy convergence tolerance was set to a strict value of $10^{-10}$~eV, in order to accurately evaluate the finite difference derivative with respect to $\mathcal{E}$. The calculations with a finite electric field were performed at $\mathcal{E} = 0.01$~$\frac{V}{\text{\AA}}$ for LiF and at $\mathcal{E} = 0.005$~$\frac{V}{\text{\AA}}$ for KTaO$_3$. These calculations were performed in VASP v6.5.0, since earlier versions do not include the effect of the electric field in the Hellman-Feynman forces. 

The PhonoPy package \cite{togo2023, togo2023a} was used to generate the required supercells and displacements for VASP calculations, and for the postprocessing step of generating the dynamical matrices $\frac{\partial \mathcal{D}_{\kappa \alpha, \kappa' \beta}(\mathbf{q})}{\partial \bm{\varepsilon}}$ from the Hellman-Feynman forces output by VASP. Calculations are performed on different supercell sizes, from $2 \times 2 \times 2$ to $6 \times 6 \times 6$ for LiF and from $2 \times 2 \times 2$ to $4 \times 4 \times 4$ for KTaO$_3$, to ensure that the results for $\mathcal{T}(\omega)$ are converged with respect to the size of the supercell. For supercell calculations, the $\mathbf{k}$-grid was appropriately scaled to maintain a constant scaling density. Phonon frequencies are reported as cycle frequencies in THz; one should multiply by $2\pi$ to obtain the radial frequency in $10^{12}~\frac{\text{rad}}{\text{s}}$.

\begin{table}
\centering
\begin{tabular}{c|c|c}
Element & Valence configuration & Energy cutoff \\ \hline
Li & 1s$^2$ 2s$^2$ & 499.034~eV \\
F & 2s$^2$ 2p$^5$ & 400~eV \\
K & 3s$^2$ 3p$^6$ 4s$^1$ & 259.264~eV \\
Ta & 5p$^6$ 5d$^4$ 6s$^1$ & 223.667~eV \\
O & 2s$^2$ 2p$^4$ & 400~eV
\end{tabular}
    \caption{Valence electron configurations and plane wave energy cutoffs of the pseudopotentials used for the first principles calculations in this article.}
    \label{tab:Pseudopotentials}
\end{table}

KTaO$_3$ contains a heavy tantalum atom with significant spin-orbit coupling. The spin-orbit coupling splits off one of the three degenerate electron bands at the conduction band minimum \cite{king2012}, leaving only a twofold degeneracy.
Therefore, we include spin-orbit coupling in the unit cell calculations for the electron bands, from which we calculate the effective band mass $m^*$ via equation \eref{EffectiveBandMass} with $n_{\text{deg}}=2$. However, we do not include spin-orbit coupling in the supercell calculations, since the inclusion of spin-orbit coupling does not significantly change the phonon frequencies in KTaO$_3$ \cite{esswein2023}. Instead, we perform calculations on larger supercell sizes, which is necessary for an accurate prediction of $\mathcal{T}(\omega)$ as we shall show in \secref{sec:TomegaPlots}.

Note that PhonoPy automatically performs a symmetry analysis in order to reduce the number of required calculations, but this symmetry analysis is performed using only the input structure; it is unaware that the electric field breaks all symmetries along the $z$-direction. To include this effect in the symmetry analysis, all calculations were performed with one of the atoms moved by a very small amount $\delta = 10^{-9}$ in the $z$-direction. For example, the FCC lattice of the rock salt structure remains unchanged:
\begin{align}
\mathbf{a}_1 & = 0 \mathbf{e}_x + \frac{a}{2} \mathbf{e}_y + \frac{a}{2} \mathbf{e}_z, \label{UnitDistorted1} \\
\mathbf{a}_2 & = \frac{a}{2} \mathbf{e}_x + 0 \mathbf{e}_y + \frac{a}{2} \mathbf{e}_z, \label{UnitDistorted2} \\
\mathbf{a}_3 & = \frac{a}{2} \mathbf{e}_x + \frac{a}{2} \mathbf{e}_y + 0 \mathbf{e}_z, \label{UnitDistorted3}
\end{align}
but the atomic positions are given by:
\begin{align}
\bm{\tau}_1 & = 0 \mathbf{e}_x + 0 \mathbf{e}_y + 0 \mathbf{e}_z, \label{UnitDistorted4} \\
\bm{\tau}_2 & = \frac{a}{2} \mathbf{e}_x + \frac{a}{2} \mathbf{e}_y + \left(\frac{1}{2} + \delta \right) a \mathbf{e}_z . \label{UnitDistorted5}
\end{align}

\begin{table}
    \centering
    \begin{tabular}{c|c|c|c}
       \textbf{LiF} & Experiments & Sio et al. \cite{sio2019a} & This article  \\ \hline
       $a$ & 4.02 $\text{\AA}$ \cite{dressler1987} & 4.058 $\text{\AA}$ & 4.004 $\text{\AA}$ \\
       $m^*/m_{\text{el}}$ &  & 0.88 & 0.88 \\
       $\omega_{\text{LO}}$ & 19.60 THz \cite{dolling1968} & 18.62 THz & 19.25 THz \\
       $\varepsilon(0)$ & 9.036 \cite{andeen1970} & 10.62 & 9.051 \\
       $\varepsilon_{\infty}$ & 1.92 \cite{levin1960} & 2.04 & 2.086 \\
       $\alpha$ &  & 4.92 & 4.52
    \end{tabular}
    \begin{tabular}{c|c|c|c}
       \textbf{KTaO$_3$} & Experiments & This article  \\ \hline
       $a$ & 3.989 $\text{\AA}$ \cite{wemple1965} & 3.99 $\text{\AA}$ \\
       $m^*/m_{\text{el}}$ & 0.2-0.6 \cite{senhouse1965, mattheiss1972} & 0.40 \\
       $\omega_{\text{LO}1}$ & 5.59 THz \cite{perry1989} & 5.14 THz \\
       $\omega_{\text{LO}2}$ & 12.52 THz \cite{perry1989} & 11.93 THz \\
       $\omega_{\text{LO}3}$ & 24.90 THz \cite{perry1989} & 23.80 THz \\
       $\varepsilon_{\infty}$ & 4.592 \cite{fujii1976} & 5.514 \\
    \end{tabular}
    \caption{Comparison of harmonic material parameters with experimental values \cite{dressler1987, dolling1968, andeen1970, levin1960, wemple1965, senhouse1965, mattheiss1972, perry1989, fujii1976}, for LiF (left) and KTaO$_3$ (right). For LiF, we also compare with the first-principles calculations in \cite{sio2019a}. $a$ is the lattice parameter of the conventional unit cell, $m^*/m_{\text{el}}$ is the electron band mass expressed in units of the electron mass, $\omega_{\text{LO}}$ is the LO phonon frequency at $\Gamma$ (LiF has one LO phonon branch at $\Gamma$, KTaO$_3$ has three), $\varepsilon(0)$ and $\varepsilon_{\infty}$ are the static and high-frequency dielectric constants, and $\alpha$ is the Fr\"ohlich coupling constant.}
    \label{tab:LiFharmonicparameters}
\end{table}
\tabref{tab:LiFharmonicparameters} shows several material parameters of LiF and KTaO$_3$ which are relevant for the Fr\"ohlich 1-electron-1-phonon interaction, calculated with the first-principles setup described in this section. The values of these material parameters are sufficiently close to the experimental values \cite{dressler1987, dolling1968, andeen1970, levin1960} and to the values calculated in \cite{sio2019a}.

\subsection{The 1-electron-2-phonon strengths $|Y_{\nu_1 \nu_2, z}(\mathbf{q})|^2$}
\secrefs{sec:DynMatProperties}{sec:FirstPrinciplesSetup} describe the first-principles setup which allows us to calculate $\omega_{\mathbf{q},\nu}$, $\mathbf{e}_{\kappa, \nu}(\mathbf{q})$, and $\frac{\partial \mathcal{D}_{\kappa \alpha, \kappa' \beta}(\mathbf{q})}{\partial \mathcal{E}_z}$ at arbitrary $\mathbf{q}$-points. This allows us to straightforwardly calculate $Y_{\nu_1 \nu_2,z}(\mathbf{q})$ with equation \eqref{Ydef}. This quantity will roughly indicate the strength of the 1-electron-2-phonon interaction, where the two phonons belong to branches $\nu_1, \nu_2$ and have momenta $\mathbf{q}_1 \approx -\mathbf{q}$ and $\mathbf{q}_2 \approx \mathbf{q}$.

\begin{figure}
\centering
\includegraphics[width=8.6cm]{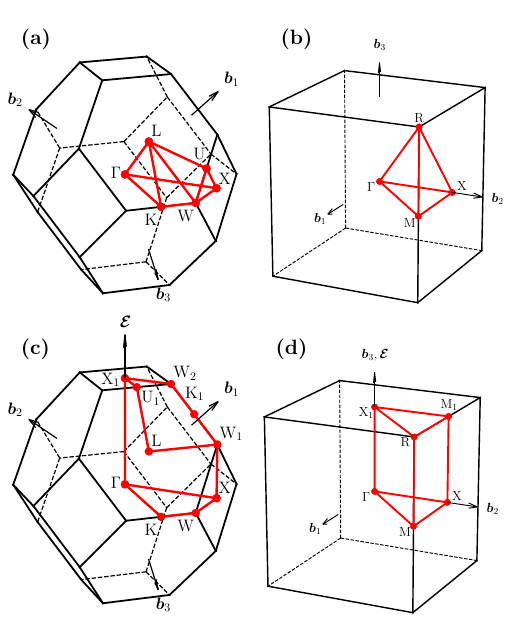}
\caption{The face-centered cubic and simple cubic Brillouin zones and two possible high-symmetry paths. \textbf{(a)-(b)} The standard paths from \cite{setyawan2010}, appropriate for a cubic cell. \textbf{(c)-(d)} More appropriate paths for the system with an applied electric field $\bm{\mathcal{E}} = \mathcal{E} \mathbf{e}_z$, based on the paths for tetragonal systems from \cite{setyawan2010}.}
\label{fig:FCCpaths}
\end{figure}

\begin{figure*}
\centering
\includegraphics[width=17.8cm]{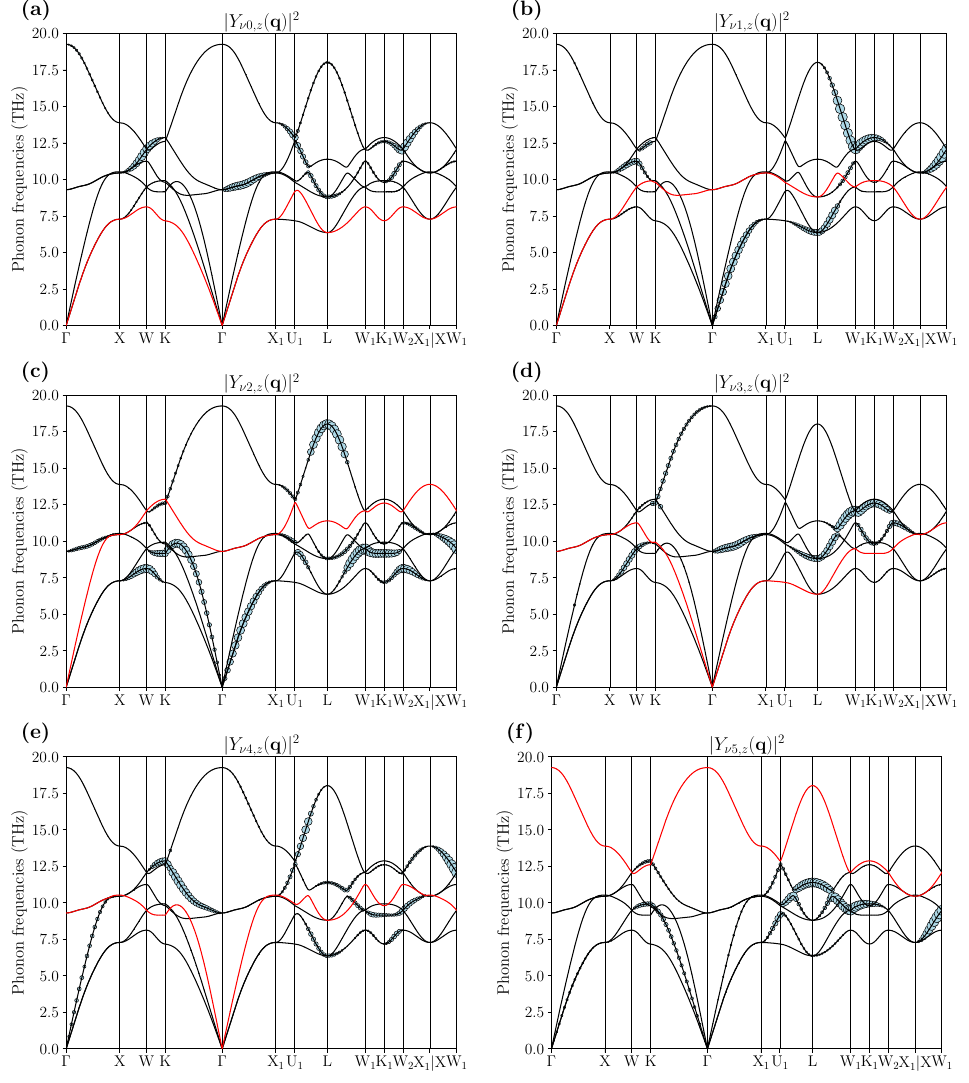}
\caption{Relative strength of the 1-electron-2-phonon interaction $|Y_{\nu_1 \nu_2,z}(\mathbf{q})|^2$ in LiF along the path shown in \figref{fig:FCCpaths}b, calculated on a $6 \times 6 \times 6$ supercell. The area of each circle is proportional to $|Y_{\nu_1 \nu_2,z}(\mathbf{q})|^2$, where $\nu_1$ is the branch the circle is on, and $\nu_2$ is a fixed branch which is highlighted in red. The largest circle in this plot corresponds to $|Y_{\nu_1 \nu_2,z}(\mathbf{q})|^2 = 7.79 \times 10^{-7} \text{\AA}^2$.}
\label{fig:YplotsBranches}
\end{figure*}

\begin{figure*}
\centering
\includegraphics[width=17.8cm]{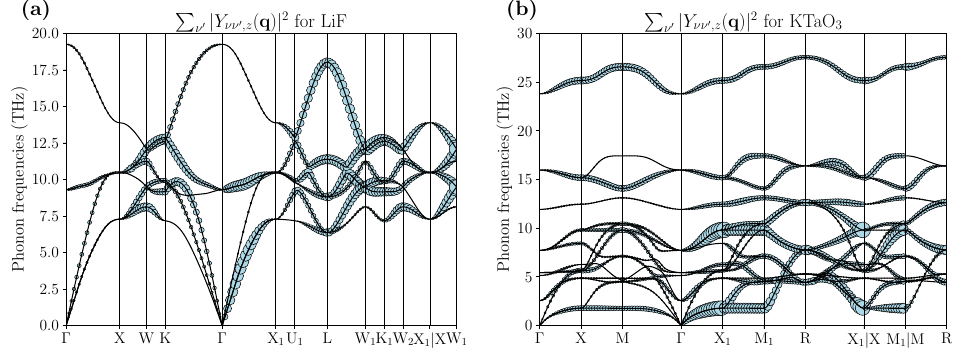}
\caption{Plot of the sum of contributions to $|Y_{\nu \nu',z}(\mathbf{q})|^2$, as defined in equation \eref{YplotSummedQuantity}. The 1-electron-2-phonon interaction has contributions from all over the Brillouin zone. Branches with very small circles do not participate in the 1-electron-2-phonon interaction at all. The largest circles correspond to a value of $\sum_{\nu'} |Y_{\nu \nu',z}(\mathbf{q})|^2 = 1.53 \times 10^{-6} \text{\AA}^2$ in \textbf{(a)}, and a value of $\sum_{\nu'} |Y_{\nu \nu',z}(\mathbf{q})|^2 = 1.11 \times 10^{-4} \text{\AA}^2$ in \textbf{(b)}.}
\label{fig:YplotsSummed}
\end{figure*}

In order to plot $Y_{\nu_1 \nu_2,z}(\mathbf{q})$, we should choose an appropriate high-symmetry path through the Brillouin zone. The phonon frequencies $\omega_{\mathbf{q},\nu}$ follow all the cubic symmetries, so they may be plotted on a high-symmetry path through the irreducible Brillouin zone. For example, in LiF, we may take the path suggested in \cite{setyawan2010}:
\begin{equation} \label{FCCpath}
\Gamma - X - W - K - \Gamma - L -U - W - L - K|U - X,
\end{equation}
which is shown in \figref{fig:FCCpaths}a. However, it is important to realize that $Y_{\nu_1 \nu_2,z}(\mathbf{q})$ does not follow the same symmetry as the phonon frequencies $\omega_{\mathbf{q},\nu}$: the electric field derivative breaks the symmetry along the $z$-direction. Therefore, it makes more sense to use a high-symmetry path that is appropriate for a tetragonal unit cell, rather than for a cubic unit cell. We choose the path shown in \figref{fig:FCCpaths}c:
\begin{equation} \label{HighSymmetryPath}
\Gamma - X - W - K - \Gamma - X_1 - U_1 - L - W_1 - W_2 - X_1 | X - W_1.
\end{equation}
This path is the suggested path for a tetragonal body-centered cell \cite{setyawan2010}, after transforming the coordinates to a tetragonal face-centered cell and then setting the dimensions of the unit cell equal to each other. In this path, we have introduced the labels $X_1, U_1, W_1, K_1,$ and $W_2$ for points that would be equivalent to the $X, U, W, $ and $K$ points in a cubic cell, but represent distinct points when the $z$-symmetry is broken. This means that the phonon frequencies are equal in e.g. $X$ and $X_1$, but $Y_{\nu_1 \nu_2,z}(\mathbf{q})$ in those two points may be different.

We find a similar path for KTaO$_3$ in an analogous way: we start from the suggested path for the simple tetragonal lattice from \cite{setyawan2010}, then set the dimensions of the unit cell equal to each other, and introduce new names for the labels based on those of the simple cubic lattice. This yields the following path:
\begin{equation} \label{HighSymmetryPath_CUB}
\Gamma - X - M - \Gamma - X_1 - M_1 - R - X_1 | X - M_1 | M - R.
\end{equation}
The new labels are $X_1$ and $M_1$, which represent points that have the same phonon frequencies as $X$ and $M$ but not necessarily the same value of $Y_{\nu_1 \nu_2,z}(\mathbf{q})$.

\figref{fig:YplotsBranches} shows the plot of $|Y_{\nu_1 \nu_2,z}(\mathbf{q})|^2$ for LiF along a high-symmetry path shown in \figref{fig:FCCpaths}b. There seem to be some pairs of branches that have much more interaction than other pairs of branches. For example, on the $\Gamma - X_1$ line, the transverse optical (TO) branches have significantly more interaction with the transverse acoustic (TA) branches than with the longitudinal bands. There also seems to be more interaction between the two highest-energy bands around the $L$-point; these are predominantly longitudinal optical (LO) and longitudinal acoustic (LA) bands, although as we will discuss in \secref{sec:LATOplots}, these labels are no longer exact. Finally, note that there is no 1-electron-2-phonon interaction when $\nu_1 = \nu_2$. This is visible in \figref{fig:YplotsBranches} as the fact that there are no circles on the red branches. This is due to the fact that LiF possesses an inversion center that takes all atoms to an equivalent position, which is proven in \appref{app:InversionCenter}.
\figref{fig:YplotsBranches} shows all the separate contributions to the 1-electron-2-phonon strengths, but requires a number of subplots equal to the number of phonon branches. This makes it impractical to make similar figures for materials with more atoms in the unit cell, such as KTaO$_3$ which would require 15 subfigures. To summarize the key info in a single plot, we overlap all the subplots in \figref{fig:YplotsBranches} and add up the areas of the overlapping circles. This corresponds to plotting the following quantity on top of the phonon branches $\omega_{\mathbf{q},\nu}$:
\begin{equation} \label{YplotSummedQuantity}
\sum_{\substack{\nu' \text{ degenerate}\\ \text{with } \nu}} \sum_{\nu''} |Y_{\nu' \nu'',z}(\mathbf{q})|^2.
\end{equation}
\figref{fig:YplotsSummed} shows a plot of this quantity for both LiF and KTaO$_3$. Whenever a branch in this figure has no circles on it, or the circles are very small, it does not participate in the long-range 1-electron-2-phonon interaction with any other branch. The most notable occurrences are all the branches on the $\Gamma - X$ line, or the longitudinal branches on the $\Gamma - X_1$ line.

\begin{figure*}
\centering
\includegraphics[width=17.8cm]{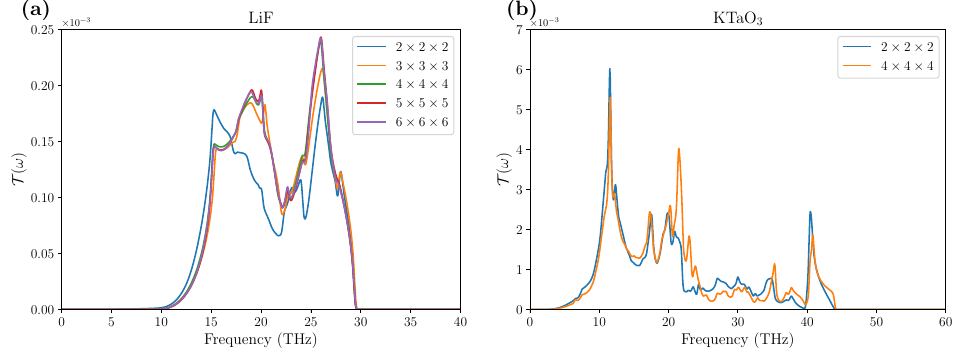}
\caption{1-electron-2-phonon spectral function $\mathcal{T}(\omega)$ at zero temperature of \textbf{(a)} LiF and \textbf{(b)} KTaO$_3$, calculated from first principles using \eref{TomegaSmeared} with $\sigma = 0.1~\text{THz}$ and a $128 \times 128 \times 128$ fine $\mathbf{q}$-grid. The spectral function converges rapidly with respect to the supercell size. The actual value of $\mathcal{T}(\omega)$ is very small, indicating that long-range 1-electron-2-phonon interaction is negligible in LiF and KTaO$_3$.}
\label{fig:LiF_Tomega}
\end{figure*}

The most striking feature of \figref{fig:YplotsSummed} is that the 1-electron-2-phonon interaction gets contributions from everywhere in the Brillouin zone. In fact, one can see that the contribution from $\Gamma$ itself is exactly zero for both LiF and KTaO$_3$. This is in stark contrast with the 1-electron-1-phonon long-range interaction, whose contributions from around $\Gamma$ dominate the other contributions. This is related to how the long-range approximation $\mathbf{q}_1 + \mathbf{q}_2 \approx \mathbf{0}$ is no longer equivalent to the continuum approximation $\mathbf{q}_1, \mathbf{q}_2 \approx \mathbf{0}$, as discussed at the end of \secref{sec:MatrixElements}. The reason that there is no contribution from $\Gamma$ is again the inversion center that takes all atoms to an equivalent position, which is present in both LiF and KTaO$_3$. In \appref{app:InversionCenter}, we present a simple symmetry argument which shows that $\frac{\partial \mathcal{D}_{\kappa \alpha, \kappa' \beta}(\mathbf{0})}{\partial \bm{\mathcal{E}}} = \mathbf{0}$ if such an inversion center is present, which immediately implies that $|Y_{\nu_1 \nu_2,z}(\mathbf{q})|^2$ is zero for all phonon branches.

\subsection{The 1-electron-2-phonon spectral function} \label{sec:TomegaPlots}
The 1-electron-2-phonon spectral function $\mathcal{T}(\omega)$ at zero temperature can be calculated from $Y_{\nu_1 \nu_2,z}(\mathbf{q})$ using \eref{TomegaDefTzero} and \eref{TomegaCubic}. The expression in \eref{TomegaDefTzero} features an integral over $\mathbf{q}$ where the integrand is a delta function, similar to the phonon density of states. We compute this integral numerically using the smearing method, where the delta function is replaced by a Gaussian peak with a finite width $\sigma$:
\begin{align} \label{TomegaSmeared}
\mathcal{T}(\omega) & \approx \frac{e^2}{2\hbar \epsvac \Omega_0} \sum_{\nu_1, \nu_2} \frac{1}{\Omega_{\text{1BZ}}} \int_{\text{1BZ}} |Y_{\nu_1\nu_2,z}(\mathbf{q})|^2 \nonumber \\
& \hspace{20pt} \times \frac{1}{\sqrt{2\pi} \sigma} e^{-\frac{(\omega - \omega_{\mathbf{q},\nu_1} - \omega_{\mathbf{q},\nu_2})^2}{2 \sigma^2}} \dee^3\mathbf{q}.
\end{align}
The integral over $\mathbf{q}$ is then computed on a fine grid using the composite trapezoid method. $\sigma$ should be chosen as small as possible, but large enough such that the variation of $\omega_{\mathbf{q},\nu_1} + \omega_{\mathbf{q},\nu_2}$ on the fine $\mathbf{q}$-grid is less than $\sigma$. The results in this section are reported for $\sigma = 0.1 \text{THz}$, which require a $128 \times 128 \times 128$ fine $\mathbf{q}$-grid for convergence. Once $\mathcal{T}(\omega)$ has been calculated, the moments $\mathcal{T}_n$ can also be calculated through:
\begin{equation} \label{Tmoments}
\mathcal{T}_n := \int_0^{+\infty} \mathcal{T}(\omega) \omega^n \dee \omega.
\end{equation}
\tabref{tab:LiFmoments} shows the moments $\mathcal{T}_{-\frac{1}{2}}$ and $\mathcal{T}_{-\frac{3}{2}}$, which are required in \erefs{Epolaron}{Mpolaron} for the polaron energy and effective mass. It also shows the moment $\mathcal{T}_{-1}$ because it is a naturally dimensionless quantity, and is therefore a useful summary of the entire spectrum.

\begin{table*}
    \centering
    \begin{tabular}{c|c|c|c|c|c|c}
    	& \multicolumn{3}{|c|}{\textbf{LiF}} & \multicolumn{3}{|c}{\textbf{KTaO$_3$}} \\ \hline
       Supercell & $\frac{1}{\sqrt{\omega_{\text{LO}}}}\mathcal{T}_{-\frac{1}{2}}$ & $\mathcal{T}_{-1}$ & $\sqrt{\omega_{\text{LO}}} \mathcal{T}_{-\frac{3}{2}}$ &  $\frac{1}{\sqrt{\omega_{\text{LO}3}}}\mathcal{T}_{-\frac{1}{2}}$ & $\mathcal{T}_{-1}$ & $\sqrt{\omega_{\text{LO}3}} \mathcal{T}_{-\frac{3}{2}}$ \\ \hline
       $2 \times 2 \times 2$ & 0.000097 & 0.000097 & 0.000097 & 0.001900 & 0.002372 & 0.003113 \\
       $3 \times 3 \times 3$ & 0.000108 & 0.000105 & 0.000104 & & & \\
       $4 \times 4 \times 4$ & 0.000111 & 0.000108 & 0.000107 & 0.001903 & 0.002313 & 0.002945 \\
       $5 \times 5 \times 5$ & 0.000111 & 0.000108 & 0.000106 & & & \\
       $6 \times 6 \times 6$ & 0.000111 & 0.000108 & 0.000107 & & &
    \end{tabular}
    \caption{Moments of $\mathcal{T}(\omega)$, calculated through equation \eref{Tmoments} with the data from \figref{fig:LiF_Tomega}. All moments are evaluated with a $128 \times 128 \times 128$ grid of interpolated $\mathbf{q}$-points and a smearing of $\sigma = 0.1$~THz. The moments were multiplied with appropriate powers of the frequencies in \tabref{tab:LiFharmonicparameters} in order to make them dimensionless.}
    \label{tab:LiFmoments}
\end{table*}

\figref{fig:LiF_Tomega} shows the calculated 1-electron-2-phonon spectrum $\mathcal{T}(\omega)$. The most important remark is that $\mathcal{T}(\omega)$ and its moments are on the order of $10^{-4}$ in LiF and on the order of $10^{-3}$ in KTaO$_3$. Since the 1-electron-1-phonon interaction is of order 1, as we shall see when calculating the polaron ground state energies, we can conclude that the long-range 1-electron-2-phonon interaction is essentially negligible in both LiF and KTaO$_3$. This means that the resulting polaron properties like $E_0$ or $m_{\text{pol}}$ will not be particularly interesting, since they will be essentially the same as the harmonic properties. Regardless, it is still interesting to focus on some of the more detailed aspects of $\mathcal{T}(\omega)$ and its calculation, which provides valuable insights for further calculations.

The calculations were performed on different supercells. For KTaO$_3$, the phonon spectrum contains imaginary modes on the $3 \times 3 \times 3$ supercell, so we only have data for the $2 \times 2 \times 2$ and $4 \times 4 \times 4$ supercells. On the contrary, we have much more data for LiF, which shows that $\mathcal{T}(\omega)$ converges rapidly with respect to the size of the supercell. The results are fully converged for a $6 \times 6 \times 6$ supercell, but even for a $4 \times 4 \times 4$ supercell, the results are sufficiently converged for most practical applications at a fraction of the computational cost. The $3 \times 3 \times 3$ supercell is overall also very good, but fails to predict the precise height of the second peak. The spectral function for the $2 \times 2 \times 2$ supercell is not quantitatively reliable, but it roughly matches the general two-peak structure. In KTaO$_3$, the result on the $2 \times 2 \times 2$ supercell is already quite reliable, although it misses the peak around $22~\text{THz}$. The moments $\mathcal{T}_n$ also converge rapidly, with a precision of three significant digits on the $4 \times 4 \times 4$ supercell, and a correct order-of-magnitude estimate on the $2 \times 2 \times 2$ supercell.

In LiF, $\mathcal{T}(\omega)$ is nonzero in a broad region between $10~\text{THz} < \omega < 30~\text{THz}$, with two peaks as the most prominent features. The lower bound of $10~\text{THz} \sim \omega_{\text{TO}}$ appears because \figref{fig:YplotsBranches} shows that there is very little interaction when $\nu_1$ and $\nu_2$ are both acoustic branches, so the interaction at the lowest frequency occurs when $\nu_1$ is acoustic and $\nu_2$ is a TO branch (or vice versa). The upper bound of $30~\text{THz}$ is lower than the theoretical maximum $2 \omega_{\text{max}} = 2 \omega_{\text{LO}} = 38.5~\text{THz}$. This is because LiF has no interaction when $\nu_1 = \nu_2$, so we must look at the highest two branches that still have any interaction. This happens at the L-point, where the sum of the two highest phonon frequencies is $18.0 + 11.4 = 29.4~\text{THz}$. At frequencies higher than this, we must therefore have $\mathcal{T}(\omega) = 0$.

In KTaO$_3$, the most noticable interaction in \figref{fig:YplotsSummed}b is on the $\Gamma - X_1 - M_1$ line, coming from the (mostly TA) lowest mode with frequency $\sim 2~\text{THz}$ and a (mostly TO) intermediate mode with frequency $\sim 10 ~\text{THz}$. This accounts for a significant contribution to the peak at $\sim 12 ~\text{THz}$ which is visible in \figref{fig:LiF_Tomega}. Other than this, the 1-electron-2-phonon spectral function in KTaO$_3$ has many more peaks at higher frequencies, which are hard to attribute to specific pairs of branches due to the many phonon bands all providing contributions from all over the Brillouin zone. $\mathcal{T}(\omega)$ becomes zero around $44~\text{THz}$, again well before the theoretical maximum value of $2 \omega_{\text{max}} = 55.1~\text{THz}$. The reason is the same as in LiF: there is no interaction when $\nu_1 = \nu_2$, and the highest phonon branch is well separated from the second highest one.

The moments $\mathcal{T}_{-\frac{1}{2}}$ and $\mathcal{T}_{-\frac{3}{2}}$ from \tabref{tab:LiFmoments} can be used in \erefs{Epolaron}{Mpolaron} to calculate the polaron ground state energy $E_0$ and effective mass $m_{\text{pol}}$. For the 1-electron-1-phonon interaction, we use \erefs{EpolaronFrohlich}{MpolaronFrohlich} instead because $\alpha = 4.52$ is too large to use lowest-order perturbation theory, as also discussed in \secref{sec:SelfEnergyExpansion}. This yields the following values:
\begin{align}
    E_0 & = -0.392279~\text{eV} - 0.000050~\text{eV} = -0.392~\text{eV}, \\
    \frac{m^*}{m_{\text{pol}}} & = 1 - 0.669274 - 0.000101, \hspace{10pt} \Leftrightarrow m_{\text{pol}} = 2.662 m_{\text{el}}.
\end{align}
In both cases, we have separately written the two terms in \erefs{Epolaron}{Mpolaron} that come from the 1-electron-1-phonon and 1-electron-2-phonon interactions. The 1-electron-2-phonon interaction contributes around $0.01 \%$ of the total contribution, so the rounded results are unchanged. For KTaO$_3$, we find a similar conclusion for the ground state energy. In order to use equation \eref{KTaO3_energy1} for the 1-electron-1-phonon contribution, we need values for the mode polarities $|p_{\nu}|^2$. We calculated these to be $|p_{\text{LO}1}|^2 = 0.00184~\text{(a.m.u.)}^{-1}$, $|p_{\text{LO}2}|^2 = 0.278~\text{(a.m.u.)}^{-1}$, and $|p_{\text{LO}3}|^2 = 3.24~\text{(a.m.u.)}^{-1}$ for the LO branches; all other mode polarities are zero. With the other values from \tabref{tab:LiFharmonicparameters} and the value of $\mathcal{T}_{-\frac{1}{2}}$ from \tabref{tab:LiFmoments}, the KTaO$_3$ polaron ground state energy can be calculated as:
\begin{align}
    E_0 & = -0.119432~\text{eV} - 0.000092~\text{eV} = -0.119~\text{eV}.
\end{align}
The relative contribution of the 1-electron-2-phonon interaction to the ground state energy is almost an order of magnitude larger in KTaO$_3$ than in LiF. However, this is still not large enough to be relevant compared to the 1-electron-1-phonon contribution.

It seems that in LiF and KTaO$_3$, at $T = 0~\text{K}$, the long-range 1-electron-2-phonon interaction is negligible with respect to the 1-electron-1-phonon interaction. This is a material-specific property: in other materials the 1-electron-2-phonon interaction may be larger. The results presented in this section serve as a proof of principle, showing that it is indeed possible to calculate $\mathcal{T}(\omega)$ for real materials from first principles.

\begin{figure*}
\centering
\includegraphics[width=16.0cm]{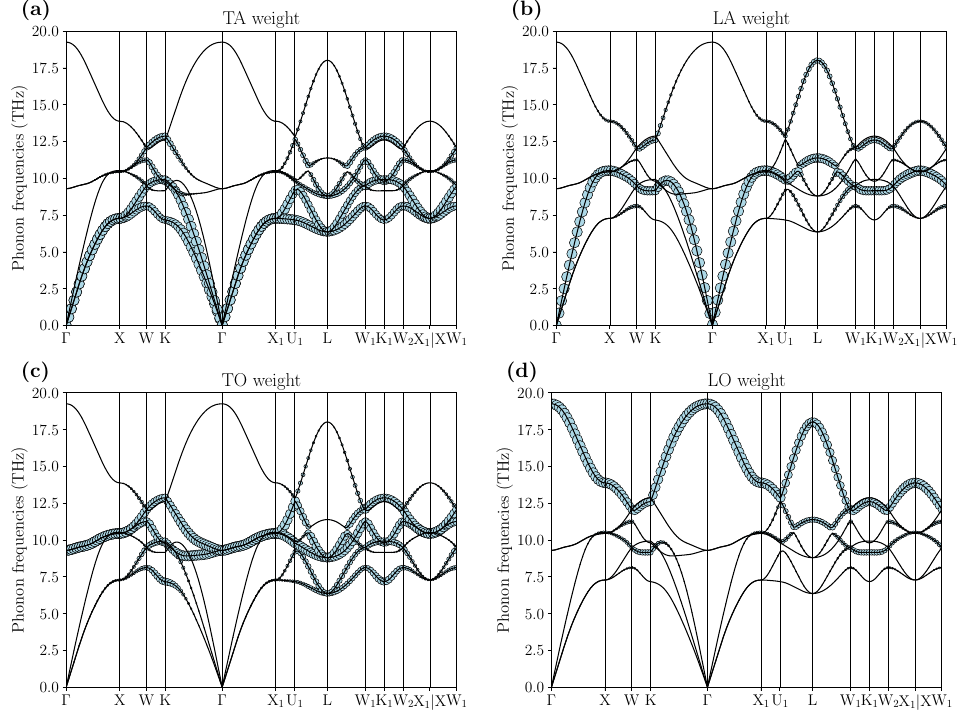}
\caption{\textbf{(a)} Transverse acoustic, \textbf{(b)} longitudinal acoustic, \textbf{(c)} transverse optical, and \textbf{(d)} longitudinal optical weights of the phonon bands in LiF, calculated on a $6 \times 6 \times 6$ supercell. For each frequency $\omega_{\mathbf{q},\nu}$, a circle is drawn with an area proportional to $P^{(\lambda)}_{\mathbf{q},\nu}$ as defined in \erefs{TAweight}{LOweight}. The phonon branches were plotted along the low-symmetry path of \figref{fig:FCCpaths}b, to facilitate comparison with \figref{fig:YplotsBranches}.}
\label{fig:LiF_LATO_weights}
\end{figure*}

\subsection{Longitudinal/transverse and acoustic/optical distinction} \label{sec:LATOplots}

In \figref{fig:YplotsBranches}, one can see that around $\Gamma$, there is significantly more interaction between the TA-TO branches than between any other pair of branches. It would be useful to track the separate contributions of the LA, LO, TA, and TO branches to $\mathcal{T}(\omega)$, but outside $\Gamma$, the phonon branches no longer uniquely fall into one of these four categories. Therefore, a mathematical measure is required that tracks how longitudinal/transverse and acoustic/optical a branch is. For every branch $(\mathbf{q},\nu)$, let us define real weights $P^{(\lambda)}_{\mathbf{q},\nu}$ with $\lambda \in \{\text{LA}, \text{TA}, \text{LO}, \text{TO}\}$, that tell us how longitudinal/transverse and acoustic/optical that branch is at that point:
\begin{figure*}
\centering
\includegraphics[width=17.8cm]{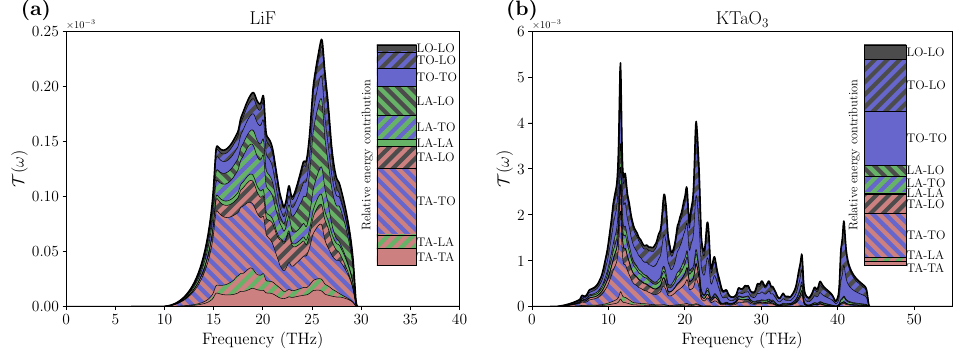}
\caption{The different longitudinal/transverse and acoustic/optical contributions of the 1-electron-2-phonon spectral function $\mathcal{T}(\omega)$, as defined by \eref{Tomega_resolved}: \textbf{(a)} in LiF on a $6 \times 6 \times 6$ supercell, and \textbf{(b)} in KTaO$_3$ on a $4 \times 4 \times 4$ supercell. The inset shows the relevant contributions to the moment $\mathcal{T}_{-\frac{1}{2}}$, as defined in \eref{TmomentsLATOdef}. Calculations were performed with a $128 \times 128 \times 128$ $\mathbf{q}$-grid and a smearing of $\sigma = 0.1~\text{THz}$.}
\label{fig:LiF_Tomega_LATO}
\end{figure*}
\begin{widetext}
\begin{align}
P^{(\text{TA})}_{\mathbf{q},\nu} & = \sum_{\kappa \alpha, \kappa' \beta} e^*_{\kappa \alpha, \nu}(\mathbf{q}) e_{\kappa' \beta, \nu}(\mathbf{q})  \frac{\sqrt{m_{\kappa} m_{\kappa'}}}{\sum_{\kappa''} m_{\kappa''}} (\delta_{\alpha \beta} - \hat{q}_{\alpha} \hat{q}_{\beta}), \label{TAweight} \\
P^{(\text{LA})}_{\mathbf{q},\nu} & = \sum_{\kappa \alpha, \kappa' \beta} e^*_{\kappa \alpha, \nu}(\mathbf{q}) e_{\kappa' \beta, \nu}(\mathbf{q})  \frac{\sqrt{m_{\kappa} m_{\kappa'}}}{\sum_{\kappa''} m_{\kappa''}} \hat{q}_{\alpha} \hat{q}_{\beta}, \label{LAweight} \\
P^{(\text{TO})}_{\mathbf{q},\nu} & = \sum_{\kappa \alpha, \kappa' \beta} e^*_{\kappa \alpha, \nu}(\mathbf{q}) e_{\kappa' \beta, \nu}(\mathbf{q}) \left(\delta_{\kappa \kappa'} - \frac{\sqrt{m_{\kappa} m_{\kappa'}}}{\sum_{\kappa''} m_{\kappa''}} \right) (\delta_{\alpha \beta} - \hat{q}_{\alpha} \hat{q}_{\beta}), \label{TOweight} \\
P^{(\text{LO})}_{\mathbf{q},\nu} & = \sum_{\kappa \alpha, \kappa' \beta} e^*_{\kappa \alpha, \nu}(\mathbf{q}) e_{\kappa' \beta, \nu}(\mathbf{q}) \left(\delta_{\kappa \kappa'} - \frac{\sqrt{m_{\kappa} m_{\kappa'}}}{\sum_{\kappa''} m_{\kappa''}} \right) \hat{q}_{\alpha} \hat{q}_{\beta}. \label{LOweight}
\end{align}
\end{widetext}
Because of the orthogonality of the phonon eigenvectors, we immediately find that these weights are normalized:
\begin{equation}
\sum_{\lambda} P^{(\lambda)}_{\mathbf{q},\nu} = 1.
\end{equation}
\figref{fig:LiF_LATO_weights} shows the weights $P^{(\lambda)}_{\mathbf{q},\nu}$ superimposed on the phonon spectrum $\omega_{\mathbf{q},\nu}$ for LiF. Around $\Gamma$ the characters TA, LA, TO, LO uniquely describe each band, but outside of $\Gamma$ most bands have a mix of at least two characters.

We can use the weights defined in \erefs{TAweight}{LOweight} to define separate contributions of $\mathcal{T}(\omega)$, that each tell us how much the longitudinal/transverse and the acoustic/optical branches are contributing. Because the weights sum up to one, these contributions can be defined as follows:
\begin{align}
\mathcal{T}(\omega) & = \sum_{\lambda_1 \lambda_2} \mathcal{T}^{(\lambda_1, \lambda_2)}(\omega), \\
\mathcal{T}^{(\lambda_1, \lambda_2)}(\omega) & = \frac{e^2}{2\hbar \epsvac \Omega_0} \sum_{\nu_1, \nu_2} \frac{1}{\Omega_{\text{1BZ}}} \int_{\text{1BZ}} |Y_{\nu_1\nu_2,z}(\mathbf{q})|^2 \nonumber \\
& \hspace{15pt} \times P^{(\lambda_1)}_{\mathbf{q},\nu_1} P^{(\lambda_2)}_{\mathbf{q},\nu_2}  \delta(\omega - \omega_{\mathbf{q},\nu_1} - \omega_{\mathbf{q},\nu_2}) \dee^3\mathbf{q}. \label{Tomega_resolved}
\end{align}
Similarly, we may define the separate contributions to the moments $\mathcal{T}_n$:
\begin{equation} \label{TmomentsLATOdef}
\mathcal{T}_n^{(\lambda_1, \lambda_2)} := \int_0^{+\infty} \mathcal{T}^{(\lambda_1, \lambda_2)}(\omega) \omega^n \dee \omega,
\end{equation}
Because $\mathcal{T}^{(\lambda_1, \lambda_2)}(\omega) = \mathcal{T}^{(\lambda_2, \lambda_1)}(\omega)$, this method distinguishes a total of 10 different contributions. \figref{fig:LiF_Tomega_LATO} shows these 10 different contributions and how they add up to form the full $\mathcal{T}(\omega)$ of \figref{fig:LiF_Tomega}. In LiF, there are two contributions that are larger than the others, which also roughly explains why $\mathcal{T}(\omega)$ has two main peaks. The first peak is largely due to TA-TO interactions, and the second peak is largely due to LA-LO and TA-TO interactions. It seems that in LiF, the 1-electron-2-phonon interaction is largest when one of the phonons is optical and the other is acoustic. However, note that even though the LA-LO and TA-TO contributions are the largest ones, together they only provide less than half of the total $\mathcal{T}(\omega)$, so all branches must be considered for a quantitative prediction. In KTaO$_3$, we may see that a significant contribution to the peak at $12 \text{THz}$ indeed comes from the interaction of the TA branch with a TO branch, as noted in the previous section. Overall, a large contribution seems to be coming from branches with TO weight, but this is likely due to the fact that 8 out of the 15 phonon branches in KTaO$_3$ are TO branches.

\subsection{Temperature dependence of the 1-electron-2-phonon spectral function} \label{sec:Tomega_temp}

\begin{figure*}
\centering
\includegraphics[width=17.8cm]{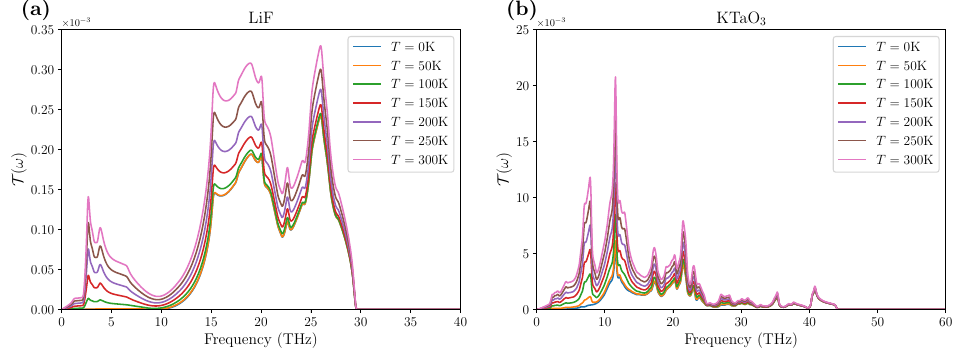}
\caption{Temperature dependence of the 1-electron-2-phonon spectral function $\mathcal{T}(\omega)$ at zero temperature of \textbf{(a)} LiF and \textbf{(b)} KTaO$_3$, calculated from first principles using \eref{TomegaSmearedTemp} with $\sigma = 0.1 ~\text{THz}$ and a $128 \times 128 \times 128$ fine $\mathbf{q}$-grid. The temperature dependence is larger in KTaO$_3$ because the phonon branches that participate in the 1-electron-2-phonon interaction have lower energy than in LiF.}
\label{fig:LiF_Tomega_temp}
\end{figure*}

So far, we have only calculated the 1-electron-2-phonon spectral function at temperature zero, for the sake of simplicity. We may also calculate it at finite temperatures, by applying the smearing method to \eref{TomegaDef}:
\begin{align}
& \mathcal{T}(\omega) := \frac{e^2}{2\hbar \epsvac \Omega_0} \sum_{\nu_1, \nu_2} \frac{1}{\Omega_{\text{1BZ}}} \int_{\text{1BZ}} |Y_{\nu_1\nu_2,z}(\mathbf{q})|^2 \frac{1}{\sqrt{2\pi} \sigma}  \nonumber \\
& \hspace{-3pt} \times \hspace{-4pt} \left( \begin{array}{l}
[1+n_B(\omega_{\mathbf{q},\nu_1})+n_B(\omega_{\mathbf{q},\nu_2})] e^{-\frac{(\omega - \omega_{\mathbf{q},\nu_1} - \omega_{\mathbf{q},\nu_2})^2}{2 \sigma^2}} \\
+|n_B(\omega_{\mathbf{q},\nu_2})-n_B(\omega_{\mathbf{q},\nu_1})| e^{-\frac{(\omega - |\omega_{\mathbf{q},\nu_1} - \omega_{\mathbf{q},\nu_2}|)^2}{2 \sigma^2}}
\end{array}  \right) \dee^3\mathbf{q}. \label{TomegaSmearedTemp}
\end{align}
Here $n_B(\omega)$ is simply the Bose-Einstein distribution function, which introduces a dependence on the temperature $T$:
\begin{equation}
n_B(\omega) = \frac{1}{\exp\left( \frac{\hbar \omega}{k_B T} \right) - 1 }
\end{equation}
Evaluating \eref{TomegaSmearedTemp} is computationally no more difficult than evaluating \eref{TomegaSmeared}, since $n_B(\omega)$ is analytically known. Because $n_B(\omega) > 0$ when $\omega > 0$, increasing the temperature will always make $\mathcal{T}(\omega)$ larger and therefore increase the effect of the 1-electron-2-phonon interaction. In particular, we expect the temperature effects to be most significant for low-frequency phonon branches, since this means $n_B(\omega_{\mathbf{q},\nu})$ will be larger in \eref{TomegaSmearedTemp}.

This is precisely what we see in \figref{fig:LiF_Tomega_temp}, which shows the 1-electron-2-phonon spectral function $\mathcal{T}(\omega)$ for both LiF and KTaO$_3$. In LiF, most of the long-range 1-electron-2-phonon interaction originates from high-frequency phonon branches, as we showed in \figref{fig:YplotsSummed}. Therefore, the 1-electron-2-phonon spectral function does not even visibly change in \figref{fig:LiF_Tomega_temp}a when increasing the temperature to $50~\text{K}$. At $T=300~\text{K}$, which corresponds to $\omega = \frac{k_B T}{\hbar} = 6.25~\text{THz}$, enough of the phonon branches are thermally activated to see a noticable increase in the 1-electron-2-phonon spectral function. As argued in \secref{sec:TomegaPlots}, the peak at $\sim 12 \text{THz}$ in KTaO$_3$ has a significant contribution from the lowest phonon mode with frequency $\omega \sim 2~\text{THz}$. This mode is thermally activated at much lower temperatures, which explains why the height of this peak increases with temperature much faster. On the contrary, the spectral function at higher frequencies is almost unaffected by temperature because none of the phonon modes are thermally activated in this region at $300~\text{K}$.

A noticable additional feature of the 1-electron-2-phonon spectral functions in \figref{fig:LiF_Tomega_temp} is the appearance of a peak at low frequencies, which is only present at finite temperatures. This peak is due to the second term in \eref{TomegaSmearedTemp}, which gives a contribution to the 1-electron-2-phonon spectral function at the difference $|\omega_{\mathbf{q},\nu_2}-\omega_{\mathbf{q},\nu_1}|$ of the two phonon frequencies, rather than at the sum. This contribution naturally shows up at lower frequencies. In KTaO$_3$, we can once again explain this with the contribution on the $\Gamma - X_1 - M_1$ line with two phonon branches of frequencies $\omega \sim 2 ~\text{THz}$ and $\omega \sim 10~\text{THz}$, which yields the finite temperature peak at $\omega \sim 8~\text{THz}$.

\begin{table*}
    \centering
    \begin{tabular}{c|c|c|c|c|c|c}
    	& \multicolumn{3}{|c|}{\textbf{LiF}} & \multicolumn{3}{|c}{\textbf{KTaO$_3$}} \\ \hline
       Temperature & $\frac{1}{\sqrt{\omega_{\text{LO}}}}\mathcal{T}_{-\frac{1}{2}}$ & $\mathcal{T}_{-1}$ & $\sqrt{\omega_{\text{LO}}} \mathcal{T}_{-\frac{3}{2}}$ &  $\frac{1}{\sqrt{\omega_{\text{LO}3}}}\mathcal{T}_{-\frac{1}{2}}$ & $\mathcal{T}_{-1}$ & $\sqrt{\omega_{\text{LO}3}} \mathcal{T}_{-\frac{3}{2}}$ \\ \hline
       0 K & 0.000111 & 0.000108 & 0.000107 & 0.001903 & 0.002313 & 0.002945 \\
       100 K & 0.000121 & 0.000126 & 0.000145 & 0.002938 & 0.004128 & 0.006592 \\
       200 K & 0.000169 & 0.000204 & 0.000311 & 0.005038 & 0.007656 & 0.013475 \\
       300 K & 0.000231 & 0.000298 & 0.000498 & 0.007315 & 0.011383 & 0.020539 \\
    \end{tabular}
    \caption{Moments of $\mathcal{T}(\omega)$, calculated through equation \eref{Tmoments} with the data from \figref{fig:LiF_Tomega}. All moments are evaluated with a $128 \times 128 \times 128$ grid of interpolated $\mathbf{q}$-points and a smearing of $\sigma = 0.1$~THz. The moments were multiplied with appropriate powers of the frequencies in \tabref{tab:LiFharmonicparameters} in order to make them dimensionless.}
    \label{tab:moments_temperature}
\end{table*}

\tabref{tab:moments_temperature} shows the moments of the 1-electron-2-phonon spectral function for LiF and KTaO$_3$, calculated from the data in \figref{fig:LiF_Tomega_temp}. It is clearly visible that the moments increase faster as a function of temperature in KTaO$_3$ than in LiF, due to the lower phonon frequencies. The low-frequency finite temperature peak of the 1-electron-2-phonon spectral function also causes a significant increase in the moments, since they are all calculated with a negative power and therefore favor contributions at lower frequencies. This is most clearly visible in LiF, where the moment $\mathcal{T}_{-\frac{3}{2}}$ is more than twice as large as the moment $\mathcal{T}_{-\frac{1}{2}}$ at $300~\text{K}$. 

Note that although higher temperatures favor the 1-electron-2-phonon interaction compared to the 1-electron-1-phonon interaction, the former is still negligible in LiF and KTaO$_3$ at room temperatures. The contribution of the 1-electron-2-phonon interaction to the polaron ground state energy was three to four orders of magnitude lower than that of the 1-electron-1-phonon interaction, and the temperature effects increase the 1-electron-2-phonon contribution by less than an order of magnitude. Regardless, it provides an important piece of information when looking for materials with significant 1-electron-2-phonon interaction: in this case, it is beneficial to look at materials with low phonon frequencies, since the temperature effects will be much larger in those materials. Indeed, $\mathcal{T}_{-\frac{3}{2}}$ increases by an order of magnitude in KTaO$_3$ when increasing the temperature from 0 K to 300 K, due to the stronger contributions from the low-frequency phonon modes.

\section{Conclusions and outlook} \label{sec:Conclusion}
The central result of this article is the analytical derivation of the long-range 1-electron-2-phonon matrix element $g^{\text{(long)}}_{mn\nu_1\nu_2}(\mathbf{k},\mathbf{q}_1,\mathbf{q}_2)$ in \eref{g2Long}, written in terms of quantities that are accessible via first-principles calculations. It was shown that the key quantity that controls the strength of the long-range 1-electron-2-phonon matrix element is $\frac{\partial \mathcal{D}_{\kappa \alpha, \kappa' \beta}(\mathbf{q})}{\partial \mathcal{E}}$, the derivative of the dynamical matrix with respect to an external electric field. This quantity is not implemented in standard first-principles codes, so we computed it using finite differences by calculating the dynamical matrix under a small positive and negative electric field. In principle, it should also be possible to calculate $\frac{\partial \mathcal{D}_{\kappa \alpha, \kappa' \beta}(\mathbf{q})}{\partial \mathcal{E}}$ using density functional perturbation theory with the use of the $2n+1$-theorem, similar to the methods described in \cite{veithen2005, refson2006}.

For a large polaron treated up to lowest order in the electron-phonon interactions, it was shown that the 1-electron-2-phonon interaction only enters the self-energy through a $3 \times 3$ spectral function $\mathcal{T}_{\alpha \beta}(\omega)$. This spectral function is a useful target quantity for first-principle calculations. It is expected that $\mathcal{T}_{\alpha \beta}(\omega)$ enters other electron-phonon quantities that only depend on the Fan-Migdal-like diagram in \figref{fig:SelfEnergyDiagrams}d: such quantities might include the superconducting Eliashberg function $\alpha^2 F(\omega)$ \cite{marsiglio2008} and the optical polaron conductivity \cite{tempere2001, houtput2022}.

As a benchmark, we calculated the 1-electron-2-phonon spectral function $\mathcal{T}(\omega)$ for LiF and KTaO$_3$. The framework developed in \secref{sec:FirstPrinciples} for extracting the 1-electron-2-phonon strength is applicable to any material. It can be used to determine the relative contributions from different phonon branches and from different parts of the Brillouin zone. This guides the construction of model Hamiltonians, where only a few phonon branches with dominant contributions are included in the Hamiltonian. The results for LiF show that $\mathcal{T}(\omega)$ has contributions from all over the Brillouin zone, which means that a simple model Hamiltonian that gives quantitative predictions for LiF cannot be constructed. It also means that it is not possible to make the continuum approximation for an accurate description of the long-range 1-electron-2-phonon interaction. 

The 1-electron-2-phonon interaction in LiF and KTaO$_3$ turns out to be small. Going forward, it will be more interesting to calculate $\mathcal{T}_{\alpha \beta}(\omega)$ for more anharmonic materials, since they naturally have larger phononic displacements. In such materials, it is common to describe the phonons with an effective anharmonic dynamical matrix $\mathcal{D}^{\text{(anh)}}_{\kappa \alpha, \kappa' \beta}(\mathbf{q})$, which is calculated from the entire ionic landscape with a method such as SSCHA \cite{monacelli2021}. Since the full electron-phonon interaction \eref{Hsecond3} originally comes from the electric field derivative of the entire ionic landscape through \eref{PionUseful}, one might capture the higher order electron-phonon effects within the same approximations by replacing $\frac{\partial \mathcal{D}_{\kappa \alpha, \kappa' \beta}(\mathbf{q})}{\partial \mathcal{E}}$ with $\frac{\partial \mathcal{D}^{\text{(anh)}}_{\kappa \alpha, \kappa' \beta}(\mathbf{q})}{\partial \mathcal{E}}$ in the theory of this article, e.g. in \eref{Ydef}.

Looking ahead, SrTiO$_3$ is a material where the calculation of $\mathcal{T}(\omega)$ might be interesting. There is already a model Hamiltonian for 1-electron-2-phonon coupling to the soft TO mode, which was first proposed in \cite{ngai1974, epifanov1981} using arguments similar to the ones proposed in \secref{sec:MatrixElements}.
This Hamiltonian was used to study the superconductivity \cite{kiselov2021, vandermarel2019} and resistivity \cite{kumar2021} of SrTiO$_3$.
 However, the Hamiltonian features a phenomenological parameter $g_2$, which cannot be calculated from first principles. Calculating $\mathcal{T}(\omega)$ for SrTiO$_3$ is challenging because of its strong phonon anharmonicity and electronic correlations that require a beyond-DFT treatment \cite{verdi2023}, but it would provide a fully ab-initio description of the long-range 1-electron-2-phonon interaction without phenomenological parameters. 
 
Halide perovskites such as CsPbI$_3$ or MAPbI$_3$ \cite{jena2019} are also promising, since they are expected to have much stronger long-range than short-range coupling \cite{yamada2022}. Additionally, they have relatively low phonon frequencies of less than $5~\text{THz}$ \cite{brivio2015, gu2021}: as motivated in \secref{sec:Tomega_temp}, this will make the 1-electron-2-phonon spectral function much larger at room temperature. It has already been shown that the higher-order electron-phonon interactions are necessary to find the correct temperature dependence for the band gap renormalization at ambient temperatures in these materials \cite{saidi2016}. A first-principles calculation of $\mathcal{T}(\omega)$ in CsPbI$_3$ or MAPbI$_3$ might reproduce this result and explain it in more detail.

\eref{g2Long} for the long-range 1-electron-2-phonon matrix element $g^{\text{(long)}}_{mn\nu_1\nu_2}(\mathbf{k},\mathbf{q}_1,\mathbf{q}_2)$ contains a $1/|\mathbf{q}_1+\mathbf{q}_2|$-divergence, similar to the $1/|\mathbf{q}|$-divergence in the 1-electron-1-phonon matrix element. The analytic expression of this divergence is used when interpolating the full 1-electron-1-phonon matrix element \cite{verdi2015, giustino2017}. Therefore, once it becomes computationally feasible to calculate the full 1-electron-2-phonon matrix element $g_{mn\nu_1\nu_2}(\mathbf{k},\mathbf{q}_1,\mathbf{q}_2)$ from first principles, knowledge of \eref{g2Long} for $g^{\text{(long)}}_{mn\nu_1\nu_2}(\mathbf{k},\mathbf{q}_1,\mathbf{q}_2)$ will also be necessary for robust  interpolation.

\section*{Data availability statement}
The first-principles data and postprocessing code used to generate \figrefs{fig:YplotsBranches}{fig:LiF_Tomega_temp} are publicly available online \cite{1e2ph-spectral}.

\acknowledgments

M.H. and J.T. acknowledge the Research Foundation Flanders (FWO), file numbers 1224724N, V472923N, G060820N, and G0AIY25N, for their funding of this research. C.F., J.T., S.R. and L.R. acknowledge support from the joint Austrian Science Fund (FWF) - FWO project I 4506. C.V. acknowledges support from the Australian Research Council (DE220101147). The computational results presented in this article were obtained using the Vienna Scientific Cluster (VSC).

\appendix

\section{The electric enthalpy and its derivatives} \label{app:ElectricEnthalpy}
The electric enthalpy is defined from the ground state energy through \eref{ElectricEnthalpy}, which we reproduce here for clarity:
\begin{equation}
E(\{ \mathbf{u}_{\kappa}(\bm{\ell}) \}, \bm{\mathcal{E}}) = E_0(\{ \mathbf{u}_{\kappa}(\bm{\ell})\}, \mathbf{D}) - \Omega_{\text{sc}} \bm{\mathcal{E}} \cdot \mathbf{P}_{\text{ext}},
\end{equation}
The main issue with defining derivatives of the electric enthalpy in the context of this article, is that this definition is not a Legendre transform. This is because $\bm{\mathcal{E}}$ represents only the external electric field, whereas the canonically conjugate variable $\frac{\partial E_0}{\partial \mathbf{D}} := \bm{\mathcal{E}}_{\text{tot}}$ represents the total electric field inside the material. In the context of this article, the polarization $\mathbf{P}_{\text{ion}}$ creates an internal electric field $\bm{\mathcal{E}}_{\text{int}}$ in the material when the ions are not in their equilibrium positions. This means that:
\begin{equation}
\bm{\mathcal{E}}_{\text{tot}} = \bm{\mathcal{E}}_{\text{int}} + \bm{\mathcal{E}}.
\end{equation}
Therefore, the definition for the electric enthalpy, as written in \eref{ElectricEnthalpy}, actually defines a function $E(\{ \mathbf{u}_{\kappa}(\bm{\ell}) \}, \bm{\mathcal{E}}, \mathbf{D})$ where both arguments $\bm{\mathcal{E}}$ and $\mathbf{D}$ are still present. In many literature definitions of the electric enthalpy \cite{nunes2001, souza2002}, it is implicitly assumed that $\bm{\mathcal{E}}_{\text{int}} = \mathbf{0}$. In this case, one has $\frac{\partial E_0}{\partial \mathbf{D}} = \bm{\mathcal{E}}$, and the definition \eref{ElectricEnthalpy} for the electric enthalpy is indeed a Legendre transform. Then, the electric enthalpy becomes a pure function of $\bm{\mathcal{E}}$, with no further $\mathbf{D}$-dependence.

The problem of the dependence on both $\bm{\mathcal{E}}$ and $\mathbf{D}$ can be circumvented by assuming a connection between the two. Indeed, they both represent an external field, so intuitively one might expect that they are related. This connection will be \eref{DElink}: let us motivate this connection here. In this article, we assume an optically linear material, where the response of the electric displacement field $\mathbf{D}$ to a homogeneous electric field $\bm{\mathcal{E}}_{\text{tot}}$ is defined by the dielectric function $\bm{\varepsilon}(\omega)$:
\begin{equation}
\mathbf{D}(\omega) = \epsvac \bm{\varepsilon}(\omega) \cdot \bm{\mathcal{E}}_{\text{tot}}(\omega).
\end{equation}
Suppose $\omega$ lies in the frequency range where it is low enough to not cause any transitions between the electronic levels, but fast enough so that the ions are too slow to follow the oscillation and are therefore almost stationary: $\mathbf{u}_{\kappa}(\bm{\ell}) = \mathbf{0}$. This limit is usually denoted as $\bm{\varepsilon}_{\infty}$,  and therefore $\bm{\varepsilon}_{\infty}$ can be formally defined as:
\begin{equation} \label{epsInfDefFormal}
\mathbf{D} = \epsvac \bm{\varepsilon}_{\infty} \cdot \bm{\mathcal{E}}_{\text{tot}}, \hspace{20pt} \text{ when } \mathbf{u}_{\kappa}(\bm{\ell})=\mathbf{0}.
\end{equation}
When all ions are in their equilibrium positions, $\bm{\mathcal{E}}_{\text{int}} = \mathbf{0}$ and we have that $\bm{\mathcal{E}}_{\text{tot}} = \bm{\mathcal{E}}$. Therefore, we may also write:
\begin{equation}
\mathbf{D} = \epsvac \bm{\varepsilon}_{\infty} \cdot \bm{\mathcal{E}}.
\end{equation}
This is \eref{DElink} from the main part of the article.

The above connection between $\mathbf{D}$ and $\bm{\mathcal{E}}$ can be used to eliminate the $\mathbf{D}$-dependence in the electric enthalpy from \eref{ElectricEnthalpy}. First, we note that $\mathbf{P}_{\text{ext}}$ in this expression is defined as the polarization when all ions are in their equilibrium positions. In this case, the only electric fields are $\mathbf{D}$ and $\bm{\mathcal{E}}$, and the standard definition of the polarization density $\mathbf{P}$ may be used to write:
\begin{equation} \label{PextDefinition2}
\mathbf{P}_{\text{ext}} = \mathbf{D} - \epsvac \bm{\mathcal{E}}.
\end{equation}
Now, if \eref{ElectricEnthalpy} is combined with the definition \eref{PextDefinition2} of $\mathbf{P}_{\text{ext}}$ and the expanded expression of the ground state energy $E_0$ from \eref{HamSCgen}, the electric enthalpy can be written as:
\begin{align}
E(\{\mathbf{u}_{\kappa}(\bm{\ell})\}, \bm{\mathcal{E}}; \mathbf{D}) & \approx H_{\text{KS}} + E_{\text{ph}}(\{\mathbf{u}_{\kappa}(\bm{\ell})\}) \nonumber \\
& - \frac{\Omega_{\text{sc}}}{\epsvac} \mathbf{D} \cdot \bm{\varepsilon}^{-1}_{\infty} \cdot \mathbf{P}_{\text{ion}}(\{\mathbf{u}_{\kappa}(\bm{\ell})\}) \nonumber \\
& + \frac{\Omega_{\text{sc}}}{2\epsvac} \mathbf{D} \cdot \bm{\varepsilon}^{-1}_{\infty} \cdot \mathbf{D} - \Omega_{\text{sc}} \bm{\mathcal{E}} \cdot (\mathbf{D} - \epsvac \bm{\mathcal{E}}) .
\end{align}
However,the $\mathbf{D}$-dependence can be eliminated in favor of $\bm{\mathcal{E}}$ by using \eref{DElink}. This gives the following expression for the electric enthalpy:
\begin{align} 
E(\{\mathbf{u}_{\kappa}(\bm{\ell})\}, \bm{\mathcal{E}}) & \approx H_{\text{KS}} + E_{\text{ph}}(\{\mathbf{u}_{\kappa}(\bm{\ell})\}) \nonumber \\
& - \Omega_{\text{sc}} \bm{\mathcal{E}} \cdot \mathbf{P}_{\text{ion}}(\{\mathbf{u}_{\kappa}(\bm{\ell})\}) \nonumber \\
& - \frac{\Omega_{\text{sc}} \epsvac}{2} \bm{\mathcal{E}} \cdot (\bm{\varepsilon}_{\infty} - 2 \mathbf{I}) \cdot \bm{\mathcal{E}} . \label{EnthalpyInt}
\end{align}
If the $\bm{\mathcal{E}}$-dependence is defined like this, the first order derivative with respect to the electric field can be read off immediately:
\begin{equation}
\mathbf{P}_{\text{ion}}(\{\mathbf{u}_{\kappa}(\bm{\ell})\}) = - \frac{1}{\Omega_{\text{sc}}} \left. \frac{\partial E}{\partial \bm{\mathcal{E}}} \right|_{\bm{\mathcal{E}} = \mathbf{0}}.
\end{equation}
This is the definition \eref{PionUseful} as presented in the main part of the article. Note that even for optically nonlinear materials, the above argument still holds: any higher order terms in $\bm{\mathcal{E}}$ that would need to be added to \eref{DElink} would only give contributions to \eref{EnthalpyInt} of at least second order in $\bm{\mathcal{E}}$. These contributions do not contribute to the above definition of $\mathbf{P}_{\text{ion}}$.

We must also prove that $\bm{\varepsilon}_{\infty}$ as defined from \eref{epsInfDefFormal} is equivalent to the definition in \eref{epsInfDef}. This can be proven by rederiving the relation between $\mathbf{D}$ and $\bm{\mathcal{E}}_{\text{tot}}$ from the ground state energy $E_0$. According to \eref{deltaE0}, we have that:
\begin{equation} \label{EtotFromE0}
\bm{\mathcal{E}}_{\text{tot}} = \frac{1}{\Omega_{\text{sc}}} \frac{\partial E_0}{\partial \mathbf{D}}.
\end{equation}
Now, note that we assume an optically linear material, which means that the ground state energy can always be written in the form of \eref{HamSCgen}:
\begin{align}
E_0(\{\mathbf{u}_{\kappa}(\bm{\ell})\}, \mathbf{D}) & \approx H_{\text{KS}} + E_{\text{ph}}(\{\mathbf{u}_{\kappa}(\bm{\ell})\}) \nonumber \\
& - \mathbf{D} \cdot \mathbf{A} \cdot \mathbf{P}_{\text{ion}}(\{\mathbf{u}_{\kappa}(\bm{\ell})\}) + \frac{1}{2} \mathbf{D} \cdot \mathbf{A} \cdot \mathbf{D},  \label{intE0linear}
\end{align}
where we have temporarily defined the shorthand matrix $\mathbf{A}$:
\begin{equation} \label{intAdef}
\mathbf{A} := \left. \frac{\partial^2 E_0}{\partial \mathbf{D} \partial \mathbf{D}} \right|_{u_{\kappa \alpha}(\bm{\ell}) = \mathbf{D} = \mathbf{0}},
\end{equation}
and $E_{\text{ph}}(\{\mathbf{u}_{\kappa}(\bm{\ell})\})$ and $\mathbf{P}_{\text{ion}}(\{\mathbf{u}_{\kappa}(\bm{\ell})\})$ are only functions of the ionic displacements $\mathbf{u}_{\kappa}(\bm{\ell})$, as defined through \eref{EphDef} and \eref{PionDef}. If we use \eref{EtotFromE0} to calculate the total electric field from \eref{intE0linear}, this gives:
\begin{align}
\bm{\mathcal{E}}_{\text{tot}} & = \frac{1}{\Omega_{\text{sc}}} \mathbf{A} \cdot \left( \mathbf{D} - \mathbf{P}_{\text{ion}}(\{\mathbf{u}_{\kappa}(\bm{\ell})\})\right), \\
\Leftrightarrow \mathbf{D} & = \Omega_{\text{sc}} \mathbf{A}^{-1} \cdot \bm{\mathcal{E}}_{\text{tot}} + \mathbf{P}_{\text{ion}}(\{\mathbf{u}_{\kappa}(\bm{\ell})\}). \label{intDEPlink}
\end{align}
If this expression is evaluated at $\mathbf{u}_{\kappa}(\bm{\ell}) = \mathbf{0}$, the second term is zero, and we can compare directly with \eref{epsInfDefFormal} of $\bm{\varepsilon}_{\infty}$ to find that:
\begin{align} \left. \frac{\partial^2 E_0}{\partial \mathbf{D} \partial \mathbf{D}} \right|_{u_{\kappa \alpha}(\bm{\ell}) = \mathbf{D} = \mathbf{0}} = \frac{\Omega_{\text{sc}}}{\epsvac} \bm{\varepsilon}_{\infty}^{-1},
\end{align}
and this is exactly the definition as presented in \eref{epsInfDef}.

Finally, \erefs{HksUseful}{EphUseful} for $H_{\text{KS}}$ and $E_{\text{ph}}$ are trivial, since they are defined at $\bm{\mathcal{E}} = \mathbf{0}$. In the case of no external electric field, the ground state energy $E_0$ and electric enthalpy $E$ are equal to each other, meaning that the definitions in \erefs{HksDef}{EphDef} carry over directly to the electric enthalpy.

\begin{widetext}
\section{Negligible diagrams} \label{app:NegligibleDiagrams}
In the diagrammatic expansion for the self-energy in \secref{sec:SelfEnergyExpansion}, we have neglected the following two contributions to the self-energy, which respectively come from the diagrams in \figref{fig:SelfEnergyDiagrams}b and \figref{fig:SelfEnergyDiagrams}c:
\begin{align}
& \Sigma^{\text{(b)}}_{nn}(\mathbf{k},\omega) = i\frac{\Omega_0}{\Omega} \frac{1}{\hbar} \int_{-\infty}^{\infty} \frac{\dee \omega_1}{2\pi} \sum_{\mathbf{q}, \nu} g_{nn\nu\nu}(\mathbf{k},-\mathbf{q},\mathbf{q}) D^{(0)}_{\nu \nu}(-\mathbf{q},\omega_1), \label{intSelfEnergyB} \\
& \Sigma^{\text{(c)}}_{nn}(\mathbf{k},\omega) = -2\frac{\Omega_0^2}{\Omega^2}\frac{1}{\hbar^3} \sum_{m_1, m_2} \sum_{\nu_1, \nu_2} \sum_{\mathbf{q}_1, \mathbf{q}_2} \int_{-\infty}^{+\infty} \frac{\dee \omega_1}{2\pi} \int_{-\infty}^{+\infty}  \frac{\dee \omega_2}{2\pi} \nonumber \\
& \hspace{15pt} \times G^{(0)}_{m_1 m_1}(\mathbf{k}+\mathbf{q}_1, \omega+\omega_1) G^{(0)}_{m_2 m_2}(\mathbf{k}+\mathbf{q}_2, \omega+\omega_2) D^{(0)}_{\nu_1 \nu_1}(\mathbf{q}_1, -\omega_1) D^{(0)}_{\nu_2 \nu_2}(\mathbf{q}_2, -\omega_2) \nonumber \\
& \hspace{15pt} \times  g_{m_1 n \nu_1}(\mathbf{k},\mathbf{q}_1) g_{m_2 m_1 \nu_1 \nu_2}(\mathbf{k}+\mathbf{q}_1, -\mathbf{q}_1, \mathbf{q}_2) g^*_{m_2 n \nu_2}(\mathbf{k}, \mathbf{q}_2). \label{intSelfEnergyC}
\end{align}
In this appendix, we show that these terms are either zero or very small when we use the long-range expressions from  \erefs{g1Long}{g2Long} for the matrix elements.

\subsection{The Debye-Waller diagram}
For the Debye-Waller self energy $\Sigma^{\text{(b)}}_{nn}(\mathbf{k},\omega)$, the argument is simple. One only needs the 1-electron-2-phonon matrix element, which is given by \eref{g2Long}. Because it is defined in the $\mathbf{q}_1 + \mathbf{q}_2 + \mathbf{G} \rightarrow \mathbf{0}$ limit, it is allowed to neglect all terms in the sum which have $\mathbf{G} \neq \mathbf{0}$. This gives the following approximate expression for the 1-electron-2-phonon matrix element:
\begin{equation}
 g^{\text{(long)}}_{mn\nu_1\nu_2}(\mathbf{k},\mathbf{q}_1,\mathbf{q}_2) = \left\{ \begin{array}{lcl}
\frac{e^2}{2 \epsvac \Omega_0} \frac{(\mathbf{q}_1+\mathbf{q}_2) \cdot \mathbf{Y}_{\nu_1 \nu_2}(\mathbf{q}_2) }{(\mathbf{q}_1+\mathbf{q}_2) \cdot \bm{\varepsilon}_{\infty} \cdot (\mathbf{q}_1+\mathbf{q}_2)} \langle \psi_{\mathbf{k}+\mathbf{q}_1+\mathbf{q}_2,m} | e^{i (\mathbf{q}_1+\mathbf{q}_2)\cdot \mathbf{r}} | \psi_{\mathbf{k},n} \rangle & \text{ if } \mathbf{q}_1 + \mathbf{q}_2 \neq \mathbf{0}, \\
0  & \text{ if } \mathbf{q}_1 + \mathbf{q}_2 = \mathbf{0}.
\end{array}
\right.
\end{equation}
The 1-electron-2-phonon matrix element only enters \eref{intSelfEnergyB} as $g^{\text{(long)}}_{nn\nu\nu}(\mathbf{k},-\mathbf{q},\mathbf{q})$. However, by the above definition, one immediately has that $g^{\text{(long)}}_{nn\nu\nu}(\mathbf{k},-\mathbf{q},\mathbf{q}) = 0$. This means the long-range interaction does not contribute to the Debye-Waller self-energy, as mentioned in the main text. Note that a similar argument can be used to show that $g^{\text{(long)}}_{m n \nu}(\mathbf{k},\mathbf{0}) = \mathbf{0}$, which implies that $\Sigma^{\text{(b)}}_{nn}(\mathbf{k},\omega) = 0$ is also true in the commonly used rigid-ion approximation for the Debye-Waller self-energy (see e.g. Eqs. 195-198 in \cite{giustino2017})

\subsection{The other diagram}
For the self-energy contribution in \eref{intSelfEnergyC}, the argument is less straightforward. The proof is divided into two big steps. In the first step, we study how the expression changes if we make the exchange of the labels $1 \leftrightarrow 2$, and show that this essentially comes down to taking the complex conjugate of all the matrix elements in this expression. In the second step, we prove that each of the matrix elements is approximately purely imaginary if we focus only on the long-range interaction. Combining those two properties will eventually yield the property $\Sigma_{nn}(\mathbf{k},\omega) \approx - \Sigma_{nn}(\mathbf{k},\omega)$, which is only possible when $\Sigma_{nn}(\mathbf{k},\omega) \approx 0$.

The first step is to study the exchange of the labels $1 \leftrightarrow 2$. If this substitution is performed in \eref{intSelfEnergyC}, the first two lines remain completely unchanged. Therefore, one should only check how the matrix elements change under this substitution. The 1-electron-1-phonon matrix elements change as:
\begin{equation}
g_{m_1 n \nu_1}(\mathbf{k},\mathbf{q}_1) g^*_{m_2 n \nu_2}(\mathbf{k}, \mathbf{q}_2) \leftrightarrow g_{m_2 n \nu_2}(\mathbf{k}, \mathbf{q}_2) g^*_{m_1 n \nu_1}(\mathbf{k},\mathbf{q}_1) ,
\end{equation}
which is the same as taking a complex conjugate. The 1-electron-2-phonon matrix element changes as:
\begin{equation}
g_{m_2 m_1 \nu_1 \nu_2}(\mathbf{k}+\mathbf{q}_1, -\mathbf{q}_1, \mathbf{q}_2) \leftrightarrow  g_{m_1 m_2 \nu_2 \nu_1}(\mathbf{k}+\mathbf{q}_2, -\mathbf{q}_2, \mathbf{q}_1).
\end{equation}
\eref{Ham1e2ph} for the Hamiltonian immediately implies the symmetry property $g_{mn\nu_1\nu_2}(\mathbf{k},\mathbf{q}_1,\mathbf{q}_2) = g_{mn\nu_2\nu_1}(\mathbf{k},\mathbf{q}_2,\mathbf{q}_1)$. With this symmetry property, the right-hand side can be rewritten as:
\begin{equation}
g_{m_2 m_1 \nu_1 \nu_2}(\mathbf{k}+\mathbf{q}_1, -\mathbf{q}_1, \mathbf{q}_2) \leftrightarrow  g_{m_1 m_2 \nu_1 \nu_2}(\mathbf{k}+\mathbf{q}_2, \mathbf{q}_1, -\mathbf{q}_2).
\end{equation}
Finally, we use that the 1-electron-2-phonon matrix element must satisfy the following condition such that the Hamiltonian in \eref{Ham1e2ph} is Hermitian:
\begin{equation}
g^{*}_{mn\nu_1\nu_2}(\mathbf{k},\mathbf{q}_1,\mathbf{q}_2) = g_{nm\nu_1\nu_2}(\mathbf{k}+\mathbf{q}_1+\mathbf{q}_2,-\mathbf{q}_1,-\mathbf{q}_2).
\end{equation}
Then, the right-hand side of the substitution changes as:
\begin{equation}
g_{m_2 m_1 \nu_1 \nu_2}(\mathbf{k}+\mathbf{q}_1, -\mathbf{q}_1, \mathbf{q}_2) \leftrightarrow  g^{*}_{m_2 m_1 \nu_1 \nu_2}(\mathbf{k}+\mathbf{q}_1, -\mathbf{q}_1, \mathbf{q}_2).
\end{equation}
Again, the exchange $1 \leftrightarrow 2$ simply gives a complex conjugate. Combining all of this, we find a second valid expression for $\Sigma_{nn}(\mathbf{k},\omega)$ through the substitution $1 \leftrightarrow 2$:
\begin{align}
\Sigma_{nn}(\mathbf{k},\omega) & = -\frac{2\Omega_0^2}{\hbar^3 (2\pi)^8}  \sum_{m_1 m_2} \sum_{\nu_1, \nu_2} \int_{\text{1BZ}} \dee^3 \mathbf{q}_1 \int_{\text{1BZ}} \dee \mathbf{q}_2 \int_{-\infty}^{+\infty} \dee \omega_1 \int_{-\infty}^{+\infty} \dee \omega_2 \nonumber \\
& \times G^{(0)}_{m_1 m_1}(\mathbf{k}+\mathbf{q}_1, \omega+\omega_1) G^{(0)}_{m_2 m_2}(\mathbf{k}+\mathbf{q}_2, \omega+\omega_2) \mathcal{D}_0(\omega_{\mathbf{q}_1,\nu_1}, -\omega_1)\mathcal{D}_0(\omega_{\mathbf{q}_2,\nu_2}, -\omega_2) \nonumber \\
& \times  (g_{m_1 n \nu_1}(\mathbf{k},\mathbf{q}_1) g_{m_2 m_1 \nu_1 \nu_2}(\mathbf{k}+\mathbf{q}_1, -\mathbf{q}_1, \mathbf{q}_2) g^*_{m_2 n \nu_2}(\mathbf{k}, \mathbf{q}_2) )^*. \label{intBdiagramExpr2}
\end{align}
This is the same expression as \eref{intSelfEnergyC}, but where all the electron-phonon matrix elements have been complex conjugated. In fact, \eref{intSelfEnergyC} and \eref{intBdiagramExpr2} can be added together to introduce a real part into the expression:
\begin{align}
\Sigma_{nn}(\mathbf{k},\omega) & = -\frac{2\Omega_0^2}{\hbar^3 (2\pi)^8}  \sum_{m_1 m_2} \sum_{\nu_1, \nu_2} \int_{\text{1BZ}} \dee^3 \mathbf{q}_1 \int_{\text{1BZ}} \dee \mathbf{q}_2 \int_{-\infty}^{+\infty} \dee \omega_1 \int_{-\infty}^{+\infty} \dee \omega_2 \nonumber \\
& \times G^{(0)}_{m_1 m_1}(\mathbf{k}+\mathbf{q}_1, \omega+\omega_1) G^{(0)}_{m_2 m_2}(\mathbf{k}+\mathbf{q}_2, \omega+\omega_2) \mathcal{D}_0(\omega_{\mathbf{q}_1,\nu_1}, -\omega_1)\mathcal{D}_0(\omega_{\mathbf{q}_2,\nu_2}, -\omega_2) \nonumber \\
& \times \text{Re}\left[ g_{m_1 n \nu_1}(\mathbf{k},\mathbf{q}_1) g_{m_2 m_1 \nu_1 \nu_2}(\mathbf{k}+\mathbf{q}_1, -\mathbf{q}_1, \mathbf{q}_2) g^*_{m_2 n \nu_2}(\mathbf{k}, \mathbf{q}_2) \right]. \label{intBdiagramExpr}
\end{align}
This expression essentially says that this diagram for the self-energy must be real with respect to complex conjugation of only the electron-phonon matrix elements.

The second step of the proof is to show that the relevant limits of the electron-phonon matrix elements are all purely imaginary. The long-range approximation for the 1-electron-1-phonon matrix elements in \figref{fig:SelfEnergyDiagrams}c implies that both phonon momenta must separately be very small: $\mathbf{q}_1 \approx \mathbf{q}_2 \approx \mathbf{0}$. In that case, the relevant limit of the electron-phonon matrix elements are:
\begin{align}
\lim_{|\mathbf{q}| \rightarrow \mathbf{0}} g_{mn\nu}(\mathbf{k},\mathbf{q}) & = \frac{i e^2}{\epsvac \Omega_0} \sqrt{\frac{\hbar}{2 \omega_{\hat{\mathbf{q}},\nu}}}  \frac{\mathbf{q}\cdot \mathbf{p}_{\nu}(\hat{\mathbf{q}})}{\mathbf{q} \cdot \bm{\varepsilon}_{\infty} \cdot \mathbf{q}} \left\langle \psi_{\mathbf{k},m} | \psi_{\mathbf{k},n} \right\rangle,  \label{int1e1ph} \\
\lim_{\mathbf{q}_1, \mathbf{q}_2 \rightarrow \mathbf{0}} g_{mn\nu_1\nu_2}(\mathbf{k},\mathbf{q}_1,\mathbf{q}_2)& = \frac{e^2}{2 \epsvac \Omega_0} \frac{(\mathbf{q}_1+\mathbf{q}_2) \cdot \mathbf{Y}_{\nu_1,\nu_2}(\hat{\mathbf{q}}_2) }{(\mathbf{q}_1+\mathbf{q}_2) \cdot \bm{\varepsilon}_{\infty} \cdot (\mathbf{q}_1+\mathbf{q}_2)} \langle \psi_{\mathbf{k},m} | \psi_{\mathbf{k},n} \rangle ,  \label{int1e2ph}
\end{align}
where $\mathbf{Y}_{\nu_1,\nu_2}(\hat{\mathbf{q}})$ is defined as the $\mathbf{q} \rightarrow \mathbf{0}$ limit of $\mathbf{Y}_{\nu_1,\nu_2}(\mathbf{q})$, which may still depend on the direction $\hat{\mathbf{q}}$. The Bloch states can be chosen such that they are orthonormal: therefore, the electron matrix elements in \erefs{int1e1ph}{int1e2ph} are real. Alternatively, a proof similar to the one in \appref{app:BlochSum} can be used to show that the sums over $m_1, m_2$ that appear in \eref{intBdiagramExpr} no longer contain the Bloch states. This means one only needs to investigate whether $\mathbf{p}_{\nu}(\hat{\mathbf{q}})$ and $\mathbf{Y}_{\nu_1,\nu_2}(\hat{\mathbf{q}}_2)$ are real or imaginary. This can be done by investigating the dynamical matrix $\mathcal{D}_{\kappa \alpha, \kappa' \beta}(\mathbf{q};\bm{\mathcal{E}})$ around $\Gamma$. The following property of the dynamical matrix follows immediately from its definition in \eref{DynDefE}:
\begin{equation}
\mathcal{D}^*_{\kappa \alpha, \kappa' \beta}(\mathbf{q};\bm{\mathcal{E}}) = \mathcal{D}_{\kappa \alpha, \kappa' \beta}(-\mathbf{q};\bm{\mathcal{E}}).
\end{equation}
This immediately implies that $\mathcal{D}_{\kappa \alpha, \kappa' \beta}(\hat{\mathbf{q}};\bm{\mathcal{E}})$ is real, since it is defined in the $|\mathbf{q}| \rightarrow \mathbf{0}$ limit and the non-analytic contribution from \eref{DynNAC} is real. This means that the usual dynamical matrix $\mathcal{D}_{\kappa \alpha, \kappa' \beta}(\mathbf{\hat{\mathbf{q}}})$, as well as its electric field derivative $\frac{\partial \mathcal{D}_{\kappa \alpha, \kappa' \beta}(\mathbf{\hat{\mathbf{q}}})}{\partial \bm{\mathcal{E}}}$, are real and symmetric. Therefore, we can choose the phonon eigenvectors to be real:
\begin{equation}
e_{\kappa \alpha, \nu}(\hat{\mathbf{q}}) = e^*_{\kappa \alpha, \nu}(\hat{\mathbf{q}}) = e_{\kappa \alpha, \nu}(-\hat{\mathbf{q}}).
\end{equation}
\eref{ModePolaritiesDef} then immediately implies that the mode polarities $\mathbf{p}_{\nu}(\hat{\mathbf{q}})$ are real. The same argument can be made for $\mathbf{Y}_{\nu_1,\nu_2}(\hat{\mathbf{q}})$:
\begin{align}
\mathbf{Y}_{\nu_1,\nu_2}(\hat{\mathbf{q}}) = \frac{1}{ie} \sqrt{\frac{\hbar}{2\omega_{\hat{\mathbf{q}},\nu_1}} \frac{\hbar}{2\omega_{\hat{\mathbf{q}},\nu_2}} } \sum_{\kappa \alpha} \sum_{\kappa' \beta} e^*_{\kappa \alpha,\nu_1}(\hat{\mathbf{q}}) \frac{\partial \mathcal{D}_{\kappa \alpha,\kappa' \beta}(\hat{\mathbf{q}})}{\partial \bm{\mathcal{E}}}  e_{\kappa' \beta,\nu_2}(\hat{\mathbf{q}}).
\end{align}
Here, everything is real except for the factor $i$ in front. This means $\mathbf{Y}_{\nu_1,\nu_2}(\hat{\mathbf{q}})$ is purely imaginary in general. Combining this with \erefs{int1e1ph}{int1e2ph} shows that each of the 1-electron-1-phonon and 1-electron-2-phonon matrix elements are purely imaginary, as mentioned earlier.

The proof can be completed as follows. \eref{intBdiagramExpr} features the real part of a product of three electron-phonon matrix elements. Due to the long-range limit, the dominant contribution to the integrals must come from the region with $\mathbf{q}_1 \approx \mathbf{q}_2 \approx \mathbf{0}$. In this region, each of the electron-phonon matrix elements is purely imaginary, so the real part of this product will be zero. If one uses \erefs{g1Long}{g2Long} instead of \erefs{int1e1ph}{int1e2ph} for the long-range electron-phonon matrix elements in \eref{intBdiagramExpr}, this argument only holds approximately. However, it still provides a reason to neglect this diagram compared to the Fan-Migdal-like diagram in \figref{fig:SelfEnergyDiagrams}d, where the argument of this appendix doesn't work because the long-range approximation in that diagram implies $\mathbf{q}_1 + \mathbf{q}_2 \approx \mathbf{0}$ instead of $\mathbf{q}_1 \approx \mathbf{q}_2 \approx \mathbf{0}$.

\section{Sum over Bloch states} \label{app:BlochSum}
In this appendix, it is shown that:
\begin{equation} \label{intBlochSum}
\lim_{\mathbf{q}+\mathbf{G} \rightarrow \mathbf{0}} \sum_{m} \delta(\varepsilon - \epsilon_{\mathbf{k}+\mathbf{q},m}) \langle \psi_{\mathbf{k},n} | e^{-i (\mathbf{q}+\mathbf{G})\cdot \mathbf{r}} | \psi_{\mathbf{k}+\mathbf{q},m} \rangle \langle \psi_{\mathbf{k}+\mathbf{q},m} | e^{i (\mathbf{q}+\mathbf{G})\cdot \mathbf{r}} | \psi_{\mathbf{k},n} \rangle = \delta(\varepsilon - \epsilon_{\mathbf{k}+\mathbf{q},n}).
\end{equation}
The argument is quite short. Firstly, because the Bloch states are eigenstates of a Hermitian Hamiltonian with the corresponding eigenvalues $\epsilon_{\mathbf{k},m}$, the matrix elements in \eref{intBlochSum} are zero unless $m$ and $n$ represent states with the same energy. Therefore, $\epsilon_{\mathbf{k}+\mathbf{q},m}$ can be replaced with $\epsilon_{\mathbf{k}+\mathbf{q},n}$ under the sum to obtain:
\begin{align}
& \lim_{\mathbf{q}+\mathbf{G} \rightarrow \mathbf{0}} \sum_{m} \delta(\varepsilon - \epsilon_{\mathbf{k}+\mathbf{q},m}) \langle \psi_{\mathbf{k},n} | e^{-i (\mathbf{q}+\mathbf{G})\cdot \mathbf{r}} | \psi_{\mathbf{k}+\mathbf{q},m} \rangle \langle \psi_{\mathbf{k}+\mathbf{q},m} | e^{i (\mathbf{q}+\mathbf{G})\cdot \mathbf{r}} | \psi_{\mathbf{k},n} \rangle \nonumber \\
& = \delta(\varepsilon - \epsilon_{\mathbf{k}+\mathbf{q},n})  \sum_{m}  \langle \psi_{\mathbf{k},n} | e^{-i (\mathbf{q}+\mathbf{G})\cdot \mathbf{r}} | \psi_{\mathbf{k}+\mathbf{q},m} \rangle \langle \psi_{\mathbf{k}+\mathbf{q},m} | e^{i (\mathbf{q}+\mathbf{G})\cdot \mathbf{r}} | \psi_{\mathbf{k},n} \rangle.
\end{align}
Furthermore, the Bloch states are complete and normalized:
\begin{align}
\sum_m | \psi_{\mathbf{k}+\mathbf{q},m} \rangle \langle \psi_{\mathbf{k}+\mathbf{q},m}| & = \hat{1}.
\end{align}
Then, one finds that:
\begin{align}
& \lim_{\mathbf{q}+\mathbf{G} \rightarrow \mathbf{0}} \sum_{m} \delta(\varepsilon - \epsilon_{\mathbf{k}+\mathbf{q},m}) \langle \psi_{\mathbf{k},n} | e^{-i (\mathbf{q}+\mathbf{G})\cdot \mathbf{r}} | \psi_{\mathbf{k}+\mathbf{q},m} \rangle \langle \psi_{\mathbf{k}+\mathbf{q},m} | e^{i (\mathbf{q}+\mathbf{G})\cdot \mathbf{r}} | \psi_{\mathbf{k},n} \rangle \nonumber \\
& = \delta(\varepsilon - \epsilon_{\mathbf{k}+\mathbf{q},n})  \langle \psi_{\mathbf{k},n} | e^{-i (\mathbf{q}+\mathbf{G})\cdot \mathbf{r}} e^{i (\mathbf{q}+\mathbf{G})\cdot \mathbf{r}} | \psi_{\mathbf{k},n} \rangle, \\
& = \delta(\varepsilon - \epsilon_{\mathbf{k}+\mathbf{q},n}) \langle \psi_{\mathbf{k},n} | \psi_{\mathbf{k},n} \rangle, \\
& = \delta(\varepsilon - \epsilon_{\mathbf{k}+\mathbf{q},n}),
\end{align}
where the final step used the fact that the Bloch states are normalized.
\end{widetext}

\section{Effect of an inversion center with $R(\kappa) = \kappa$} \label{app:InversionCenter}
The components of the dynamical matrix $\mathcal{D}_{\kappa \alpha, \kappa' \beta}(\mathbf{q}; \bm{\mathcal{E}})$ are not all independent of each other: they are linked to each other by symmetry constraints. Suppose $(\mathbf{R}, \bm{\tau})$ is an element of the space group of the material, consisting of a rotation $\mathbf{R}$ followed by a translation $\bm{\tau}$. Then, this symmetry imposes the following constraint on the dynamical matrix:
\begin{equation} \label{DynPropSymmetryCtype}
\bm{\mathcal{D}}_{\kappa \kappa'}(\mathbf{q},\bm{\mathcal{E}}) = \mathbf{R}^{-1} \cdot \bm{\mathcal{D}}_{R(\kappa)R(\kappa')}(\mathbf{R}\cdot \mathbf{q},\mathbf{R}\cdot \bm{\mathcal{E}}) \cdot \mathbf{R}.
\end{equation}
This expression is only valid in the c-type convention, which we assume throughout this appendix. Here, the index $R(\kappa)$ is the new index that the atom $\kappa$ is taken to after applying the symmetry $(\mathbf{R}, \bm{\tau})$, up to a lattice translation $\bm{\ell}$. Mathematically, $R(\kappa)$ is defined as:
\begin{equation}
\mathbf{R} \cdot \bm{\tau}_{\kappa} + \bm{\tau} = \bm{\tau}_{R(\kappa)} + \bm{\ell}.
\end{equation}
In LiF, the two atoms are distinguishable, so all symmetries $(\mathbf{R}, \bm{\tau})$ of the material must automatically satisfy $R(\kappa) = \kappa$; otherwise, $(\mathbf{R}, \bm{\tau})$ would not be a symmetry of the material. This is not true in general: for example, in the cubic perovskite structure ABX$_3$, the three X-ions may be scrambled by some of the rotations. Regardless, some of the symmetry elements may still satisfy $R(\kappa) = \kappa$. For example, the inversion center in the cubic perovskite structure still takes all atoms to an equivalent site up to a lattice translation, and therefore satisfies $R(\kappa) = \kappa$.

In this appendix, we study some properties of the dynamical matrix $\mathcal{D}_{\kappa \alpha, \kappa' \beta}(\mathbf{q}; \bm{\mathcal{E}})$ for a material that possesses an inversion center that satisfies $R(\kappa) = \kappa$. This includes both LiF and KTaO$_3$, the two materials studied in this article. The inversion center is represented by the space group operation $(-\mathbf{I}, \mathbf{0})$, so according to \eref{DynPropSymmetryCtype}, a material with such an inversion center necessarily satisfies:
\begin{equation} \label{DynPropSymmetryInversion}
\mathcal{D}_{\kappa \alpha, \kappa' \beta}(\mathbf{q},\bm{\mathcal{E}}) = \mathcal{D}_{\kappa \alpha, \kappa' \beta}(-\mathbf{q},-\bm{\mathcal{E}}).
\end{equation}
This property may now be used to prove several useful theorems about the material parameters that are present in this article.

\subsection{Imaginary dynamical matrix derivative}
\eref{DynPropSymmetryInversion} can be used to prove that $\frac{\partial \mathcal{D}_{\kappa \alpha, \kappa' \beta}(\mathbf{q})}{\partial \bm{\mathcal{E}}}$ is purely imaginary. From the definition \eref{DynDefE}, it immediately follows that the dynamical matrix satisfies the following property:
\begin{equation} \label{DynPropMinusQ}
\mathcal{D}_{\kappa \alpha, \kappa' \beta}(\mathbf{q},\bm{\mathcal{E}}) = \mathcal{D}^*_{\kappa \alpha, \kappa' \beta}(-\mathbf{q},\bm{\mathcal{E}}).
\end{equation}
Combining this with \eref{DynPropSymmetryInversion} shows that:
\begin{equation}
\mathcal{D}_{\kappa \alpha, \kappa' \beta}(\mathbf{q},\bm{\mathcal{E}}) = \mathcal{D}^*_{\kappa \alpha, \kappa' \beta}(\mathbf{q},-\bm{\mathcal{E}}),
\end{equation}
in all materials with an inversion center that satisfies $R(\kappa) = \kappa$. Evaluating this property at $\bm{\mathcal{E}} = \mathbf{0}$ immediately yields that the dynamical matrix is real. Taking the derivative with respect to $\bm{\mathcal{E}}$ on both sides yields that $\frac{\partial \mathcal{D}_{\kappa \alpha, \kappa' \beta}(\mathbf{q})}{\partial \bm{\mathcal{E}}}$ is purely imaginary. It should be emphasized that these properties are only true in the c-type convention.

Note that if we evaluate \eref{DynPropMinusQ} in $\mathbf{q} = \mathbf{0}$, it follows the dynamical matrix must be real in the $\Gamma$-point, regardless of the strength of the electric field. Together with the property that the dynamical matrix derivative must be purely imaginary, this implies that:
\begin{equation}
\frac{\partial \mathcal{D}_{\kappa \alpha, \kappa' \beta}(\mathbf{0})}{\partial \bm{\mathcal{E}}} = \mathbf{0}.
\end{equation}
In other words, there is no 1-electron-2-phonon interaction in $\Gamma$, which can be seen in \figrefs{fig:YplotsBranches}{fig:YplotsSummed}.

\subsection{No 1-electron-2-phonon interaction when $\omega_{\mathbf{q},\nu_1} = \omega_{\mathbf{q},\nu_2}$}
In \figref{fig:YplotsBranches}, it can be noticed that there is no 1-electron-2-phonon interaction in LiF when $\nu_1 = \nu_2$. Here, it is shown that this is true for any material with an inversion center that satisfies $R(\kappa) = \kappa$. Firstly, we need the following alternative expression for $\mathbf{Y}_{\nu_1 \nu_2}(\mathbf{q})$, which can be straightforwardly derived by taking the $\bm{\mathcal{E}}$-derivative of \eref{FreqsVecsDef} and inserting the result into \eref{Ydef}:
\begin{widetext}
\begin{equation} \label{YdefFreqsVecs}
\mathbf{Y}_{\nu_1\nu_2}(\mathbf{q}) = \frac{1}{ie} \sqrt{\frac{\hbar}{2\omega_{\mathbf{q},\nu_1}} \frac{\hbar}{2\omega_{\mathbf{q},\nu_2}} } \left[ \frac{\partial \omega^2_{\mathbf{q},\nu_1}}{\partial \bm{\mathcal{E}}} \delta_{\nu_1 \nu_2} + \left( \omega^2_{\mathbf{q},\nu_2} - \omega^2_{\mathbf{q},\nu_1} \right) \sum_{\kappa \alpha} e^*_{\kappa \alpha, \nu_1}(\mathbf{q}) \frac{\partial e_{\kappa \alpha, \nu_2}(\mathbf{q})}{\partial \bm{\mathcal{E}}} \right].
\end{equation}
This expression is written in terms of the electric field derivatives of the phonon frequencies and eigenvectors. \erefs{DynPropSymmetryInversion}{DynPropMinusQ} are properties for the dynamical matrix, but they also hold for its eigenvalues, such that:
\begin{equation}
\left . \begin{array}{lcl}
\omega_{\nu}(\mathbf{q},\bm{\mathcal{E}}) & = & \omega_{\nu}(-\mathbf{q},-\bm{\mathcal{E}}) \\
\omega_{\nu}(\mathbf{q},\bm{\mathcal{E}}) & = & \omega_{\nu}(-\mathbf{q},\bm{\mathcal{E}})
\end{array}
\right\} \Rightarrow \omega_{\nu}(\mathbf{q},\bm{\mathcal{E}}) = \omega_{\nu}(\mathbf{q},-\bm{\mathcal{E}}).
\end{equation}
From the final property it is immediately clear that $\frac{\partial \omega^2_{\mathbf{q},\nu_1}}{\partial \bm{\mathcal{E}}} = \mathbf{0}$. Therefore, for materials with an inversion center that satisfies $R(\kappa) = \kappa$, \eref{YdefFreqsVecs} reduces to:
\begin{equation}
\mathbf{Y}_{\nu_1\nu_2}(\mathbf{q}) = \frac{1}{ie} \sqrt{\frac{\hbar}{2\omega_{\mathbf{q},\nu_1}} \frac{\hbar}{2\omega_{\mathbf{q},\nu_2}} }  \left( \omega^2_{\mathbf{q},\nu_2} - \omega^2_{\mathbf{q},\nu_1} \right) \sum_{\kappa \alpha} e^*_{\kappa \alpha, \nu_1}(\mathbf{q}) \frac{\partial e_{\kappa \alpha, \nu_2}(\mathbf{q})}{\partial \bm{\mathcal{E}}}.
\end{equation}
In this expression, the factor $\left( \omega^2_{\mathbf{q},\nu_2} - \omega^2_{\mathbf{q},\nu_1} \right)$ makes it clear that $\mathbf{Y}_{\nu_1\nu_2}(\mathbf{q}) = \mathbf{0}$ when $\nu_1 = \nu_2$, or more generally for two degenerate branches.

\subsection{No non-analytic contribution for the dynamical matrix derivative}
In \secref{sec:DynMatProperties} of the main text, it was shown that the non-analytic contribution to the derivative of the dynamical matrix $\frac{\partial \mathcal{D}^{\text{NAC}}_{\kappa \alpha, \kappa' \beta}(\mathbf{q})}{\partial \bm{\mathcal{E}}}$ requires knowledge of the Raman susceptibility tensor $\frac{\partial \chi_{\beta \gamma}}{\partial u_{\kappa \alpha}}$ and the nonlinear optical susceptibility $\chi^{(2)}_{\alpha \beta \gamma}$, which were defined in \erefs{RamanDef}{chi2Def}. In general, if the material has a space group symmetry $(\mathbf{R}, \bm{\tau})$, these two tensors satisfy the following symmetry constraints:
\begin{align}
\frac{\partial \chi_{\beta \gamma}}{\partial u_{\kappa \alpha}} & = \sum_{\alpha' \beta' \gamma'} R_{\alpha' \alpha} R_{\beta' \beta} R_{\gamma' \gamma} \frac{\partial \chi_{\beta' \gamma'}}{\partial u_{R(\kappa) \alpha'}}, \\
\chi^{(2)}_{\alpha \beta \gamma} & = \sum_{\alpha' \beta' \gamma'} R_{\alpha' \alpha} R_{\beta' \beta} R_{\gamma' \gamma} \chi^{(2)}_{\alpha' \beta' \gamma'}.
\end{align}
Therefore, if the material has an inversion center $R_{\alpha \alpha'} = -\delta_{\alpha \alpha'}$ that satisfies $R(\kappa) = \kappa$, it imposes the following conditions:
\begin{align}
\frac{\partial \chi_{\beta \gamma}}{\partial u_{\kappa \alpha}} & = - \frac{\partial \chi_{\beta \gamma}}{\partial u_{\kappa \alpha}}, \\
\chi^{(2)}_{\alpha \beta \gamma} & = - \chi^{(2)}_{\alpha \beta \gamma},
\end{align}
which is only possible if both tensors are zero. Therefore, we find that $\frac{\partial \mathcal{D}^{\text{NAC}}_{\kappa \alpha, \kappa' \beta}(\mathbf{q})}{\partial \bm{\mathcal{E}}} = 0$ in such materials. This means the Fourier interpolation for the derivative of the dynamical matrix can be performed with \eref{DynFourierDer}, without subtracting and adding a non-analytic contribution.
\end{widetext}

\bibliography{References}

\begin{thebibliography}{125}%
\makeatletter
\providecommand \@ifxundefined [1]{%
 \@ifx{#1\undefined}
}%
\providecommand \@ifnum [1]{%
 \ifnum #1\expandafter \@firstoftwo
 \else \expandafter \@secondoftwo
 \fi
}%
\providecommand \@ifx [1]{%
 \ifx #1\expandafter \@firstoftwo
 \else \expandafter \@secondoftwo
 \fi
}%
\providecommand \natexlab [1]{#1}%
\providecommand \enquote  [1]{``#1''}%
\providecommand \bibnamefont  [1]{#1}%
\providecommand \bibfnamefont [1]{#1}%
\providecommand \citenamefont [1]{#1}%
\providecommand \href@noop [0]{\@secondoftwo}%
\providecommand \href [0]{\begingroup \@sanitize@url \@href}%
\providecommand \@href[1]{\@@startlink{#1}\@@href}%
\providecommand \@@href[1]{\endgroup#1\@@endlink}%
\providecommand \@sanitize@url [0]{\catcode `\\12\catcode `\$12\catcode
  `\&12\catcode `\#12\catcode `\^12\catcode `\_12\catcode `\%12\relax}%
\providecommand \@@startlink[1]{}%
\providecommand \@@endlink[0]{}%
\providecommand \url  [0]{\begingroup\@sanitize@url \@url }%
\providecommand \@url [1]{\endgroup\@href {#1}{\urlprefix }}%
\providecommand \urlprefix  [0]{URL }%
\providecommand \Eprint [0]{\href }%
\providecommand \doibase [0]{https://doi.org/}%
\providecommand \selectlanguage [0]{\@gobble}%
\providecommand \bibinfo  [0]{\@secondoftwo}%
\providecommand \bibfield  [0]{\@secondoftwo}%
\providecommand \translation [1]{[#1]}%
\providecommand \BibitemOpen [0]{}%
\providecommand \bibitemStop [0]{}%
\providecommand \bibitemNoStop [0]{.\EOS\space}%
\providecommand \EOS [0]{\spacefactor3000\relax}%
\providecommand \BibitemShut  [1]{\csname bibitem#1\endcsname}%
\let\auto@bib@innerbib\@empty
\bibitem [{\citenamefont {Landau}(1933)}]{landau1933}%
  \BibitemOpen
  \bibfield  {author} {\bibinfo {author} {\bibfnamefont {L.~D.}\ \bibnamefont
  {Landau}},\ }\bibfield  {title} {\bibinfo {title} {Electron motion in crystal
  lattices},\ }\href@noop {} {\bibfield  {journal} {\bibinfo  {journal} {Phys.
  Z. Sowjet.}\ }\textbf {\bibinfo {volume} {3}},\ \bibinfo {pages} {664}
  (\bibinfo {year} {1933})}\BibitemShut {NoStop}%
\bibitem [{\citenamefont {Landau}\ and\ \citenamefont
  {Pekar}(1948)}]{landau1948}%
  \BibitemOpen
  \bibfield  {author} {\bibinfo {author} {\bibfnamefont {{\relax
  LD}.}~\bibnamefont {Landau}}\ and\ \bibinfo {author} {\bibfnamefont {{\relax
  SI}.}~\bibnamefont {Pekar}},\ }\bibfield  {title} {\bibinfo {title}
  {Effective mass of a polaron},\ }\href@noop {} {\bibfield  {journal}
  {\bibinfo  {journal} {Zhurnal {\'E}ksperimentalnoi i Teoreticheskoi Fiziki}\
  }\textbf {\bibinfo {volume} {18}},\ \bibinfo {pages} {419} (\bibinfo {year}
  {1948})}\BibitemShut {NoStop}%
\bibitem [{\citenamefont {Fr{\"o}hlich}(1954)}]{frohlich1954}%
  \BibitemOpen
  \bibfield  {author} {\bibinfo {author} {\bibfnamefont {H.}~\bibnamefont
  {Fr{\"o}hlich}},\ }\bibfield  {title} {\bibinfo {title} {Electrons in lattice
  fields},\ }\href {https://doi.org/10.1080/00018735400101213} {\bibfield
  {journal} {\bibinfo  {journal} {Advances in Physics}\ }\textbf {\bibinfo
  {volume} {3}},\ \bibinfo {pages} {325} (\bibinfo {year} {1954})}\BibitemShut
  {NoStop}%
\bibitem [{\citenamefont {Holstein}(1959{\natexlab{a}})}]{holstein1959}%
  \BibitemOpen
  \bibfield  {author} {\bibinfo {author} {\bibfnamefont {T.}~\bibnamefont
  {Holstein}},\ }\bibfield  {title} {\bibinfo {title} {Studies of polaron
  motion: {{Part I}}. {{The}} molecular-crystal model},\ }\href
  {https://doi.org/10.1016/0003-4916(59)90002-8} {\bibfield  {journal}
  {\bibinfo  {journal} {Annals of Physics}\ }\textbf {\bibinfo {volume} {8}},\
  \bibinfo {pages} {325} (\bibinfo {year} {1959}{\natexlab{a}})}\BibitemShut
  {NoStop}%
\bibitem [{\citenamefont {Holstein}(1959{\natexlab{b}})}]{holstein1959a}%
  \BibitemOpen
  \bibfield  {author} {\bibinfo {author} {\bibfnamefont {T.}~\bibnamefont
  {Holstein}},\ }\bibfield  {title} {\bibinfo {title} {Studies of polaron
  motion: {{Part II}}. {{The}} ``small'' polaron},\ }\href
  {https://doi.org/10.1016/0003-4916(59)90003-X} {\bibfield  {journal}
  {\bibinfo  {journal} {Annals of Physics}\ }\textbf {\bibinfo {volume} {8}},\
  \bibinfo {pages} {343} (\bibinfo {year} {1959}{\natexlab{b}})}\BibitemShut
  {NoStop}%
\bibitem [{\citenamefont {Devreese}\ and\ \citenamefont
  {Alexandrov}(2009)}]{devreese2009}%
  \BibitemOpen
  \bibfield  {author} {\bibinfo {author} {\bibfnamefont {J.~T.}\ \bibnamefont
  {Devreese}}\ and\ \bibinfo {author} {\bibfnamefont {A.~S.}\ \bibnamefont
  {Alexandrov}},\ }\bibfield  {title} {\bibinfo {title} {Fr{\"o}hlich polaron
  and bipolaron: Recent developments},\ }\href
  {https://doi.org/10.1088/0034-4885/72/6/066501} {\bibfield  {journal}
  {\bibinfo  {journal} {Reports on Progress in Physics}\ }\textbf {\bibinfo
  {volume} {72}},\ \bibinfo {pages} {066501} (\bibinfo {year}
  {2009})}\BibitemShut {NoStop}%
\bibitem [{\citenamefont {Sio}\ \emph {et~al.}(2019{\natexlab{a}})\citenamefont
  {Sio}, \citenamefont {Verdi}, \citenamefont {Ponc{\'e}},\ and\ \citenamefont
  {Giustino}}]{sio2019}%
  \BibitemOpen
  \bibfield  {author} {\bibinfo {author} {\bibfnamefont {W.~H.}\ \bibnamefont
  {Sio}}, \bibinfo {author} {\bibfnamefont {C.}~\bibnamefont {Verdi}}, \bibinfo
  {author} {\bibfnamefont {S.}~\bibnamefont {Ponc{\'e}}},\ and\ \bibinfo
  {author} {\bibfnamefont {F.}~\bibnamefont {Giustino}},\ }\bibfield  {title}
  {\bibinfo {title} {Polarons from {{First Principles}}, without
  {{Supercells}}},\ }\href {https://doi.org/10.1103/PhysRevLett.122.246403}
  {\bibfield  {journal} {\bibinfo  {journal} {Physical Review Letters}\
  }\textbf {\bibinfo {volume} {122}},\ \bibinfo {pages} {246403} (\bibinfo
  {year} {2019}{\natexlab{a}})}\BibitemShut {NoStop}%
\bibitem [{\citenamefont {Sio}\ \emph {et~al.}(2019{\natexlab{b}})\citenamefont
  {Sio}, \citenamefont {Verdi}, \citenamefont {Ponc{\'e}},\ and\ \citenamefont
  {Giustino}}]{sio2019a}%
  \BibitemOpen
  \bibfield  {author} {\bibinfo {author} {\bibfnamefont {W.~H.}\ \bibnamefont
  {Sio}}, \bibinfo {author} {\bibfnamefont {C.}~\bibnamefont {Verdi}}, \bibinfo
  {author} {\bibfnamefont {S.}~\bibnamefont {Ponc{\'e}}},\ and\ \bibinfo
  {author} {\bibfnamefont {F.}~\bibnamefont {Giustino}},\ }\bibfield  {title}
  {\bibinfo {title} {Ab initio theory of polarons: {{Formalism}} and
  applications},\ }\href {https://doi.org/10.1103/PhysRevB.99.235139}
  {\bibfield  {journal} {\bibinfo  {journal} {Physical Review B}\ }\textbf
  {\bibinfo {volume} {99}},\ \bibinfo {pages} {235139} (\bibinfo {year}
  {2019}{\natexlab{b}})}\BibitemShut {NoStop}%
\bibitem [{\citenamefont {{Lafuente-Bartolome}}\ \emph
  {et~al.}(2022{\natexlab{a}})\citenamefont {{Lafuente-Bartolome}},
  \citenamefont {Lian}, \citenamefont {Sio}, \citenamefont {Gurtubay},
  \citenamefont {Eiguren},\ and\ \citenamefont
  {Giustino}}]{lafuente-bartolome2022}%
  \BibitemOpen
  \bibfield  {author} {\bibinfo {author} {\bibfnamefont {J.}~\bibnamefont
  {{Lafuente-Bartolome}}}, \bibinfo {author} {\bibfnamefont {C.}~\bibnamefont
  {Lian}}, \bibinfo {author} {\bibfnamefont {W.~H.}\ \bibnamefont {Sio}},
  \bibinfo {author} {\bibfnamefont {I.~G.}\ \bibnamefont {Gurtubay}}, \bibinfo
  {author} {\bibfnamefont {A.}~\bibnamefont {Eiguren}},\ and\ \bibinfo {author}
  {\bibfnamefont {F.}~\bibnamefont {Giustino}},\ }\bibfield  {title} {\bibinfo
  {title} {{\emph{Ab Initio}} self-consistent many-body theory of polarons at
  all couplings},\ }\href {https://doi.org/10.1103/PhysRevB.106.075119}
  {\bibfield  {journal} {\bibinfo  {journal} {Physical Review B}\ }\textbf
  {\bibinfo {volume} {106}},\ \bibinfo {pages} {075119} (\bibinfo {year}
  {2022}{\natexlab{a}})}\BibitemShut {NoStop}%
\bibitem [{\citenamefont {{Lafuente-Bartolome}}\ \emph
  {et~al.}(2022{\natexlab{b}})\citenamefont {{Lafuente-Bartolome}},
  \citenamefont {Lian}, \citenamefont {Sio}, \citenamefont {Gurtubay},
  \citenamefont {Eiguren},\ and\ \citenamefont
  {Giustino}}]{lafuente-bartolome2022a}%
  \BibitemOpen
  \bibfield  {author} {\bibinfo {author} {\bibfnamefont {J.}~\bibnamefont
  {{Lafuente-Bartolome}}}, \bibinfo {author} {\bibfnamefont {C.}~\bibnamefont
  {Lian}}, \bibinfo {author} {\bibfnamefont {W.~H.}\ \bibnamefont {Sio}},
  \bibinfo {author} {\bibfnamefont {I.~G.}\ \bibnamefont {Gurtubay}}, \bibinfo
  {author} {\bibfnamefont {A.}~\bibnamefont {Eiguren}},\ and\ \bibinfo {author}
  {\bibfnamefont {F.}~\bibnamefont {Giustino}},\ }\bibfield  {title} {\bibinfo
  {title} {Unified {{Approach}} to {{Polarons}} and {{Phonon-Induced Band
  Structure Renormalization}}},\ }\href
  {https://doi.org/10.1103/PhysRevLett.129.076402} {\bibfield  {journal}
  {\bibinfo  {journal} {Physical Review Letters}\ }\textbf {\bibinfo {volume}
  {129}},\ \bibinfo {pages} {076402} (\bibinfo {year}
  {2022}{\natexlab{b}})}\BibitemShut {NoStop}%
\bibitem [{\citenamefont {Feynman}(1955)}]{feynman1955}%
  \BibitemOpen
  \bibfield  {author} {\bibinfo {author} {\bibfnamefont {R.~P.}\ \bibnamefont
  {Feynman}},\ }\bibfield  {title} {\bibinfo {title} {Slow {{Electrons}} in a
  {{Polar Crystal}}},\ }\href {https://doi.org/10.1103/PhysRev.97.660}
  {\bibfield  {journal} {\bibinfo  {journal} {Physical Review}\ }\textbf
  {\bibinfo {volume} {97}},\ \bibinfo {pages} {660} (\bibinfo {year}
  {1955})}\BibitemShut {NoStop}%
\bibitem [{\citenamefont {Allen}\ and\ \citenamefont
  {Heine}(1976)}]{allen1976}%
  \BibitemOpen
  \bibfield  {author} {\bibinfo {author} {\bibfnamefont {P.~B.}\ \bibnamefont
  {Allen}}\ and\ \bibinfo {author} {\bibfnamefont {V.}~\bibnamefont {Heine}},\
  }\bibfield  {title} {\bibinfo {title} {Theory of the temperature dependence
  of electronic band structures},\ }\href
  {https://doi.org/10.1088/0022-3719/9/12/013} {\bibfield  {journal} {\bibinfo
  {journal} {Journal of Physics C: Solid State Physics}\ }\textbf {\bibinfo
  {volume} {9}},\ \bibinfo {pages} {2305} (\bibinfo {year} {1976})}\BibitemShut
  {NoStop}%
\bibitem [{\citenamefont {Giustino}(2017)}]{giustino2017}%
  \BibitemOpen
  \bibfield  {author} {\bibinfo {author} {\bibfnamefont {F.}~\bibnamefont
  {Giustino}},\ }\bibfield  {title} {\bibinfo {title} {Electron-phonon
  interactions from first principles},\ }\href
  {https://doi.org/10.1103/RevModPhys.89.015003} {\bibfield  {journal}
  {\bibinfo  {journal} {Reviews of Modern Physics}\ }\textbf {\bibinfo {volume}
  {89}},\ \bibinfo {pages} {015003} (\bibinfo {year} {2017})}\BibitemShut
  {NoStop}%
\bibitem [{\citenamefont {Eliashberg}(1960)}]{eliashberg1960}%
  \BibitemOpen
  \bibfield  {author} {\bibinfo {author} {\bibfnamefont {{\relax
  GM}.}~\bibnamefont {Eliashberg}},\ }\bibfield  {title} {\bibinfo {title}
  {Interactions between electrons and lattice vibrations in a superconductor},\
  }\href@noop {} {\bibfield  {journal} {\bibinfo  {journal} {Soviet
  Physics--JETP [translation of Zhurnal Eksperimentalnoi i Teoreticheskoi
  Fiziki]}\ }\textbf {\bibinfo {volume} {11}},\ \bibinfo {pages} {696}
  (\bibinfo {year} {1960})}\BibitemShut {NoStop}%
\bibitem [{\citenamefont {McMillan}(1968)}]{mcmillan1968}%
  \BibitemOpen
  \bibfield  {author} {\bibinfo {author} {\bibfnamefont {W.~L.}\ \bibnamefont
  {McMillan}},\ }\bibfield  {title} {\bibinfo {title} {Transition
  {{Temperature}} of {{Strong-Coupled Superconductors}}},\ }\href
  {https://doi.org/10.1103/PhysRev.167.331} {\bibfield  {journal} {\bibinfo
  {journal} {Physical Review}\ }\textbf {\bibinfo {volume} {167}},\ \bibinfo
  {pages} {331} (\bibinfo {year} {1968})}\BibitemShut {NoStop}%
\bibitem [{\citenamefont {Allen}\ and\ \citenamefont
  {Dynes}(1975)}]{allen1975}%
  \BibitemOpen
  \bibfield  {author} {\bibinfo {author} {\bibfnamefont {P.~B.}\ \bibnamefont
  {Allen}}\ and\ \bibinfo {author} {\bibfnamefont {R.~C.}\ \bibnamefont
  {Dynes}},\ }\bibfield  {title} {\bibinfo {title} {Transition temperature of
  strong-coupled superconductors reanalyzed},\ }\href
  {https://doi.org/10.1103/PhysRevB.12.905} {\bibfield  {journal} {\bibinfo
  {journal} {Physical Review B}\ }\textbf {\bibinfo {volume} {12}},\ \bibinfo
  {pages} {905} (\bibinfo {year} {1975})}\BibitemShut {NoStop}%
\bibitem [{\citenamefont {Marsiglio}\ and\ \citenamefont
  {Carbotte}(2008)}]{marsiglio2008}%
  \BibitemOpen
  \bibfield  {author} {\bibinfo {author} {\bibfnamefont {F.}~\bibnamefont
  {Marsiglio}}\ and\ \bibinfo {author} {\bibfnamefont {J.~P.}\ \bibnamefont
  {Carbotte}},\ }\bibfield  {title} {\bibinfo {title} {Electron-{{Phonon
  Superconductivity}}},\ }in\ \href
  {https://doi.org/10.1007/978-3-540-73253-2_3} {\emph {\bibinfo {booktitle}
  {Superconductivity: {{Conventional}} and {{Unconventional
  Superconductors}}}}},\ \bibinfo {editor} {edited by\ \bibinfo {editor}
  {\bibfnamefont {K.~H.}\ \bibnamefont {Bennemann}}\ and\ \bibinfo {editor}
  {\bibfnamefont {J.~B.}\ \bibnamefont {Ketterson}}}\ (\bibinfo  {publisher}
  {Springer},\ \bibinfo {address} {Berlin, Heidelberg},\ \bibinfo {year}
  {2008})\ pp.\ \bibinfo {pages} {73--162}\BibitemShut {NoStop}%
\bibitem [{\citenamefont {Feynman}\ \emph {et~al.}(1962)\citenamefont
  {Feynman}, \citenamefont {Hellwarth}, \citenamefont {Iddings},\ and\
  \citenamefont {Platzman}}]{feynman1962}%
  \BibitemOpen
  \bibfield  {author} {\bibinfo {author} {\bibfnamefont {R.~P.}\ \bibnamefont
  {Feynman}}, \bibinfo {author} {\bibfnamefont {R.~W.}\ \bibnamefont
  {Hellwarth}}, \bibinfo {author} {\bibfnamefont {C.~K.}\ \bibnamefont
  {Iddings}},\ and\ \bibinfo {author} {\bibfnamefont {P.~M.}\ \bibnamefont
  {Platzman}},\ }\bibfield  {title} {\bibinfo {title} {Mobility of {{Slow
  Electrons}} in a {{Polar Crystal}}},\ }\href
  {https://doi.org/10.1103/PhysRev.127.1004} {\bibfield  {journal} {\bibinfo
  {journal} {Physical Review}\ }\textbf {\bibinfo {volume} {127}},\ \bibinfo
  {pages} {1004} (\bibinfo {year} {1962})}\BibitemShut {NoStop}%
\bibitem [{\citenamefont {Kadanoff}(1963)}]{kadanoff1963}%
  \BibitemOpen
  \bibfield  {author} {\bibinfo {author} {\bibfnamefont {L.~P.}\ \bibnamefont
  {Kadanoff}},\ }\bibfield  {title} {\bibinfo {title} {Boltzmann {{Equation}}
  for {{Polarons}}},\ }\href {https://doi.org/10.1103/PhysRev.130.1364}
  {\bibfield  {journal} {\bibinfo  {journal} {Physical Review}\ }\textbf
  {\bibinfo {volume} {130}},\ \bibinfo {pages} {1364} (\bibinfo {year}
  {1963})}\BibitemShut {NoStop}%
\bibitem [{\citenamefont {Tempere}\ and\ \citenamefont
  {Devreese}(2001)}]{tempere2001}%
  \BibitemOpen
  \bibfield  {author} {\bibinfo {author} {\bibfnamefont {J.}~\bibnamefont
  {Tempere}}\ and\ \bibinfo {author} {\bibfnamefont {J.~T.}\ \bibnamefont
  {Devreese}},\ }\bibfield  {title} {\bibinfo {title} {Optical absorption of an
  interacting many-polaron gas},\ }\href
  {https://doi.org/10.1103/PhysRevB.64.104504} {\bibfield  {journal} {\bibinfo
  {journal} {Physical Review B}\ }\textbf {\bibinfo {volume} {64}},\ \bibinfo
  {pages} {104504} (\bibinfo {year} {2001})}\BibitemShut {NoStop}%
\bibitem [{\citenamefont {Ponc{\'e}}\ \emph {et~al.}(2020)\citenamefont
  {Ponc{\'e}}, \citenamefont {Li}, \citenamefont {Reichardt},\ and\
  \citenamefont {Giustino}}]{ponce2020}%
  \BibitemOpen
  \bibfield  {author} {\bibinfo {author} {\bibfnamefont {S.}~\bibnamefont
  {Ponc{\'e}}}, \bibinfo {author} {\bibfnamefont {W.}~\bibnamefont {Li}},
  \bibinfo {author} {\bibfnamefont {S.}~\bibnamefont {Reichardt}},\ and\
  \bibinfo {author} {\bibfnamefont {F.}~\bibnamefont {Giustino}},\ }\bibfield
  {title} {\bibinfo {title} {First-principles calculations of charge carrier
  mobility and conductivity in bulk semiconductors and two-dimensional
  materials},\ }\href {https://doi.org/10.1088/1361-6633/ab6a43} {\bibfield
  {journal} {\bibinfo  {journal} {Reports on Progress in Physics}\ }\textbf
  {\bibinfo {volume} {83}},\ \bibinfo {pages} {036501} (\bibinfo {year}
  {2020})}\BibitemShut {NoStop}%
\bibitem [{\citenamefont {Kresse}\ and\ \citenamefont
  {Hafner}(1993)}]{kresse1993}%
  \BibitemOpen
  \bibfield  {author} {\bibinfo {author} {\bibfnamefont {G.}~\bibnamefont
  {Kresse}}\ and\ \bibinfo {author} {\bibfnamefont {J.}~\bibnamefont
  {Hafner}},\ }\bibfield  {title} {\bibinfo {title} {Ab initio molecular
  dynamics for liquid metals},\ }\href
  {https://doi.org/10.1103/PhysRevB.47.558} {\bibfield  {journal} {\bibinfo
  {journal} {Physical Review B}\ }\textbf {\bibinfo {volume} {47}},\ \bibinfo
  {pages} {558} (\bibinfo {year} {1993})}\BibitemShut {NoStop}%
\bibitem [{\citenamefont {Kresse}\ and\ \citenamefont
  {Furthm{\"u}ller}(1996{\natexlab{a}})}]{kresse1996}%
  \BibitemOpen
  \bibfield  {author} {\bibinfo {author} {\bibfnamefont {G.}~\bibnamefont
  {Kresse}}\ and\ \bibinfo {author} {\bibfnamefont {J.}~\bibnamefont
  {Furthm{\"u}ller}},\ }\bibfield  {title} {\bibinfo {title} {Efficiency of
  ab-initio total energy calculations for metals and semiconductors using a
  plane-wave basis set},\ }\href {https://doi.org/10.1016/0927-0256(96)00008-0}
  {\bibfield  {journal} {\bibinfo  {journal} {Computational Materials Science}\
  }\textbf {\bibinfo {volume} {6}},\ \bibinfo {pages} {15} (\bibinfo {year}
  {1996}{\natexlab{a}})}\BibitemShut {NoStop}%
\bibitem [{\citenamefont {Kresse}\ and\ \citenamefont
  {Furthm{\"u}ller}(1996{\natexlab{b}})}]{kresse1996a}%
  \BibitemOpen
  \bibfield  {author} {\bibinfo {author} {\bibfnamefont {G.}~\bibnamefont
  {Kresse}}\ and\ \bibinfo {author} {\bibfnamefont {J.}~\bibnamefont
  {Furthm{\"u}ller}},\ }\bibfield  {title} {\bibinfo {title} {Software
  {{VASP}}, vienna (1999)},\ }\href@noop {} {\bibfield  {journal} {\bibinfo
  {journal} {Physical Review B}\ }\textbf {\bibinfo {volume} {54}},\ \bibinfo
  {pages} {169} (\bibinfo {year} {1996}{\natexlab{b}})}\BibitemShut {NoStop}%
\bibitem [{\citenamefont {Giannozzi}\ \emph {et~al.}(2009)\citenamefont
  {Giannozzi}, \citenamefont {Baroni}, \citenamefont {Bonini}, \citenamefont
  {Calandra}, \citenamefont {Car}, \citenamefont {Cavazzoni}, \citenamefont
  {Ceresoli}, \citenamefont {Chiarotti}, \citenamefont {Cococcioni},
  \citenamefont {Dabo}, \citenamefont {Corso}, \citenamefont {de~Gironcoli},
  \citenamefont {Fabris}, \citenamefont {Fratesi}, \citenamefont {Gebauer},
  \citenamefont {Gerstmann}, \citenamefont {Gougoussis}, \citenamefont
  {Kokalj}, \citenamefont {Lazzeri}, \citenamefont {{Martin-Samos}},
  \citenamefont {Marzari}, \citenamefont {Mauri}, \citenamefont {Mazzarello},
  \citenamefont {Paolini}, \citenamefont {Pasquarello}, \citenamefont
  {Paulatto}, \citenamefont {Sbraccia}, \citenamefont {Scandolo}, \citenamefont
  {Sclauzero}, \citenamefont {Seitsonen}, \citenamefont {Smogunov},
  \citenamefont {Umari},\ and\ \citenamefont {Wentzcovitch}}]{giannozzi2009}%
  \BibitemOpen
  \bibfield  {author} {\bibinfo {author} {\bibfnamefont {P.}~\bibnamefont
  {Giannozzi}}, \bibinfo {author} {\bibfnamefont {S.}~\bibnamefont {Baroni}},
  \bibinfo {author} {\bibfnamefont {N.}~\bibnamefont {Bonini}}, \bibinfo
  {author} {\bibfnamefont {M.}~\bibnamefont {Calandra}}, \bibinfo {author}
  {\bibfnamefont {R.}~\bibnamefont {Car}}, \bibinfo {author} {\bibfnamefont
  {C.}~\bibnamefont {Cavazzoni}}, \bibinfo {author} {\bibfnamefont
  {D.}~\bibnamefont {Ceresoli}}, \bibinfo {author} {\bibfnamefont {G.~L.}\
  \bibnamefont {Chiarotti}}, \bibinfo {author} {\bibfnamefont {M.}~\bibnamefont
  {Cococcioni}}, \bibinfo {author} {\bibfnamefont {I.}~\bibnamefont {Dabo}},
  \bibinfo {author} {\bibfnamefont {A.~D.}\ \bibnamefont {Corso}}, \bibinfo
  {author} {\bibfnamefont {S.}~\bibnamefont {de~Gironcoli}}, \bibinfo {author}
  {\bibfnamefont {S.}~\bibnamefont {Fabris}}, \bibinfo {author} {\bibfnamefont
  {G.}~\bibnamefont {Fratesi}}, \bibinfo {author} {\bibfnamefont
  {R.}~\bibnamefont {Gebauer}}, \bibinfo {author} {\bibfnamefont
  {U.}~\bibnamefont {Gerstmann}}, \bibinfo {author} {\bibfnamefont
  {C.}~\bibnamefont {Gougoussis}}, \bibinfo {author} {\bibfnamefont
  {A.}~\bibnamefont {Kokalj}}, \bibinfo {author} {\bibfnamefont
  {M.}~\bibnamefont {Lazzeri}}, \bibinfo {author} {\bibfnamefont
  {L.}~\bibnamefont {{Martin-Samos}}}, \bibinfo {author} {\bibfnamefont
  {N.}~\bibnamefont {Marzari}}, \bibinfo {author} {\bibfnamefont
  {F.}~\bibnamefont {Mauri}}, \bibinfo {author} {\bibfnamefont
  {R.}~\bibnamefont {Mazzarello}}, \bibinfo {author} {\bibfnamefont
  {S.}~\bibnamefont {Paolini}}, \bibinfo {author} {\bibfnamefont
  {A.}~\bibnamefont {Pasquarello}}, \bibinfo {author} {\bibfnamefont
  {L.}~\bibnamefont {Paulatto}}, \bibinfo {author} {\bibfnamefont
  {C.}~\bibnamefont {Sbraccia}}, \bibinfo {author} {\bibfnamefont
  {S.}~\bibnamefont {Scandolo}}, \bibinfo {author} {\bibfnamefont
  {G.}~\bibnamefont {Sclauzero}}, \bibinfo {author} {\bibfnamefont {A.~P.}\
  \bibnamefont {Seitsonen}}, \bibinfo {author} {\bibfnamefont {A.}~\bibnamefont
  {Smogunov}}, \bibinfo {author} {\bibfnamefont {P.}~\bibnamefont {Umari}},\
  and\ \bibinfo {author} {\bibfnamefont {R.~M.}\ \bibnamefont {Wentzcovitch}},\
  }\bibfield  {title} {\bibinfo {title} {{{QUANTUM ESPRESSO}}: A modular and
  open-source software project for quantum simulations of materials},\ }\href
  {https://doi.org/10.1088/0953-8984/21/39/395502} {\bibfield  {journal}
  {\bibinfo  {journal} {Journal of Physics: Condensed Matter}\ }\textbf
  {\bibinfo {volume} {21}},\ \bibinfo {pages} {395502} (\bibinfo {year}
  {2009})}\BibitemShut {NoStop}%
\bibitem [{\citenamefont {Giannozzi}\ \emph {et~al.}(2017)\citenamefont
  {Giannozzi}, \citenamefont {Andreussi}, \citenamefont {Brumme}, \citenamefont
  {Bunau}, \citenamefont {Nardelli}, \citenamefont {Calandra}, \citenamefont
  {Car}, \citenamefont {Cavazzoni}, \citenamefont {Ceresoli}, \citenamefont
  {Cococcioni}, \citenamefont {Colonna}, \citenamefont {Carnimeo},
  \citenamefont {Corso}, \citenamefont {de~Gironcoli}, \citenamefont {Delugas},
  \citenamefont {DiStasio}, \citenamefont {Ferretti}, \citenamefont {Floris},
  \citenamefont {Fratesi}, \citenamefont {Fugallo}, \citenamefont {Gebauer},
  \citenamefont {Gerstmann}, \citenamefont {Giustino}, \citenamefont {Gorni},
  \citenamefont {Jia}, \citenamefont {Kawamura}, \citenamefont {Ko},
  \citenamefont {Kokalj}, \citenamefont {K{\"u}{\c c}{\"u}kbenli},
  \citenamefont {Lazzeri}, \citenamefont {Marsili}, \citenamefont {Marzari},
  \citenamefont {Mauri}, \citenamefont {Nguyen}, \citenamefont {Nguyen},
  \citenamefont {{Otero-de-la-Roza}}, \citenamefont {Paulatto}, \citenamefont
  {Ponc{\'e}}, \citenamefont {Rocca}, \citenamefont {Sabatini}, \citenamefont
  {Santra}, \citenamefont {Schlipf}, \citenamefont {Seitsonen}, \citenamefont
  {Smogunov}, \citenamefont {Timrov}, \citenamefont {Thonhauser}, \citenamefont
  {Umari}, \citenamefont {Vast}, \citenamefont {Wu},\ and\ \citenamefont
  {Baroni}}]{giannozzi2017}%
  \BibitemOpen
  \bibfield  {author} {\bibinfo {author} {\bibfnamefont {P.}~\bibnamefont
  {Giannozzi}}, \bibinfo {author} {\bibfnamefont {O.}~\bibnamefont
  {Andreussi}}, \bibinfo {author} {\bibfnamefont {T.}~\bibnamefont {Brumme}},
  \bibinfo {author} {\bibfnamefont {O.}~\bibnamefont {Bunau}}, \bibinfo
  {author} {\bibfnamefont {M.~B.}\ \bibnamefont {Nardelli}}, \bibinfo {author}
  {\bibfnamefont {M.}~\bibnamefont {Calandra}}, \bibinfo {author}
  {\bibfnamefont {R.}~\bibnamefont {Car}}, \bibinfo {author} {\bibfnamefont
  {C.}~\bibnamefont {Cavazzoni}}, \bibinfo {author} {\bibfnamefont
  {D.}~\bibnamefont {Ceresoli}}, \bibinfo {author} {\bibfnamefont
  {M.}~\bibnamefont {Cococcioni}}, \bibinfo {author} {\bibfnamefont
  {N.}~\bibnamefont {Colonna}}, \bibinfo {author} {\bibfnamefont
  {I.}~\bibnamefont {Carnimeo}}, \bibinfo {author} {\bibfnamefont {A.~D.}\
  \bibnamefont {Corso}}, \bibinfo {author} {\bibfnamefont {S.}~\bibnamefont
  {de~Gironcoli}}, \bibinfo {author} {\bibfnamefont {P.}~\bibnamefont
  {Delugas}}, \bibinfo {author} {\bibfnamefont {R.~A.}\ \bibnamefont
  {DiStasio}}, \bibinfo {author} {\bibfnamefont {A.}~\bibnamefont {Ferretti}},
  \bibinfo {author} {\bibfnamefont {A.}~\bibnamefont {Floris}}, \bibinfo
  {author} {\bibfnamefont {G.}~\bibnamefont {Fratesi}}, \bibinfo {author}
  {\bibfnamefont {G.}~\bibnamefont {Fugallo}}, \bibinfo {author} {\bibfnamefont
  {R.}~\bibnamefont {Gebauer}}, \bibinfo {author} {\bibfnamefont
  {U.}~\bibnamefont {Gerstmann}}, \bibinfo {author} {\bibfnamefont
  {F.}~\bibnamefont {Giustino}}, \bibinfo {author} {\bibfnamefont
  {T.}~\bibnamefont {Gorni}}, \bibinfo {author} {\bibfnamefont
  {J.}~\bibnamefont {Jia}}, \bibinfo {author} {\bibfnamefont {M.}~\bibnamefont
  {Kawamura}}, \bibinfo {author} {\bibfnamefont {H.-Y.}\ \bibnamefont {Ko}},
  \bibinfo {author} {\bibfnamefont {A.}~\bibnamefont {Kokalj}}, \bibinfo
  {author} {\bibfnamefont {E.}~\bibnamefont {K{\"u}{\c c}{\"u}kbenli}},
  \bibinfo {author} {\bibfnamefont {M.}~\bibnamefont {Lazzeri}}, \bibinfo
  {author} {\bibfnamefont {M.}~\bibnamefont {Marsili}}, \bibinfo {author}
  {\bibfnamefont {N.}~\bibnamefont {Marzari}}, \bibinfo {author} {\bibfnamefont
  {F.}~\bibnamefont {Mauri}}, \bibinfo {author} {\bibfnamefont {N.~L.}\
  \bibnamefont {Nguyen}}, \bibinfo {author} {\bibfnamefont {H.-V.}\
  \bibnamefont {Nguyen}}, \bibinfo {author} {\bibfnamefont {A.}~\bibnamefont
  {{Otero-de-la-Roza}}}, \bibinfo {author} {\bibfnamefont {L.}~\bibnamefont
  {Paulatto}}, \bibinfo {author} {\bibfnamefont {S.}~\bibnamefont {Ponc{\'e}}},
  \bibinfo {author} {\bibfnamefont {D.}~\bibnamefont {Rocca}}, \bibinfo
  {author} {\bibfnamefont {R.}~\bibnamefont {Sabatini}}, \bibinfo {author}
  {\bibfnamefont {B.}~\bibnamefont {Santra}}, \bibinfo {author} {\bibfnamefont
  {M.}~\bibnamefont {Schlipf}}, \bibinfo {author} {\bibfnamefont {A.~P.}\
  \bibnamefont {Seitsonen}}, \bibinfo {author} {\bibfnamefont {A.}~\bibnamefont
  {Smogunov}}, \bibinfo {author} {\bibfnamefont {I.}~\bibnamefont {Timrov}},
  \bibinfo {author} {\bibfnamefont {T.}~\bibnamefont {Thonhauser}}, \bibinfo
  {author} {\bibfnamefont {P.}~\bibnamefont {Umari}}, \bibinfo {author}
  {\bibfnamefont {N.}~\bibnamefont {Vast}}, \bibinfo {author} {\bibfnamefont
  {X.}~\bibnamefont {Wu}},\ and\ \bibinfo {author} {\bibfnamefont
  {S.}~\bibnamefont {Baroni}},\ }\bibfield  {title} {\bibinfo {title} {Advanced
  capabilities for materials modelling with {{Quantum ESPRESSO}}},\ }\href
  {https://doi.org/10.1088/1361-648X/aa8f79} {\bibfield  {journal} {\bibinfo
  {journal} {Journal of Physics: Condensed Matter}\ }\textbf {\bibinfo {volume}
  {29}},\ \bibinfo {pages} {465901} (\bibinfo {year} {2017})}\BibitemShut
  {NoStop}%
\bibitem [{\citenamefont {Gonze}\ \emph {et~al.}(2020)\citenamefont {Gonze},
  \citenamefont {Amadon}, \citenamefont {Antonius}, \citenamefont {Arnardi},
  \citenamefont {Baguet}, \citenamefont {Beuken}, \citenamefont {Bieder},
  \citenamefont {Bottin}, \citenamefont {Bouchet}, \citenamefont {Bousquet},
  \citenamefont {Brouwer}, \citenamefont {Bruneval}, \citenamefont {Brunin},
  \citenamefont {Cavignac}, \citenamefont {Charraud}, \citenamefont {Chen},
  \citenamefont {C{\^o}t{\'e}}, \citenamefont {Cottenier}, \citenamefont
  {Denier}, \citenamefont {Geneste}, \citenamefont {Ghosez}, \citenamefont
  {Giantomassi}, \citenamefont {Gillet}, \citenamefont {Gingras}, \citenamefont
  {Hamann}, \citenamefont {Hautier}, \citenamefont {He}, \citenamefont
  {Helbig}, \citenamefont {Holzwarth}, \citenamefont {Jia}, \citenamefont
  {Jollet}, \citenamefont {{Lafargue-Dit-Hauret}}, \citenamefont {Lejaeghere},
  \citenamefont {Marques}, \citenamefont {Martin}, \citenamefont {Martins},
  \citenamefont {Miranda}, \citenamefont {Naccarato}, \citenamefont {Persson},
  \citenamefont {Petretto}, \citenamefont {Planes}, \citenamefont {Pouillon},
  \citenamefont {Prokhorenko}, \citenamefont {Ricci}, \citenamefont
  {Rignanese}, \citenamefont {Romero}, \citenamefont {Schmitt}, \citenamefont
  {Torrent}, \citenamefont {{van Setten}}, \citenamefont {Van~Troeye},
  \citenamefont {Verstraete}, \citenamefont {Z{\'e}rah},\ and\ \citenamefont
  {Zwanziger}}]{gonze2020}%
  \BibitemOpen
  \bibfield  {author} {\bibinfo {author} {\bibfnamefont {X.}~\bibnamefont
  {Gonze}}, \bibinfo {author} {\bibfnamefont {B.}~\bibnamefont {Amadon}},
  \bibinfo {author} {\bibfnamefont {G.}~\bibnamefont {Antonius}}, \bibinfo
  {author} {\bibfnamefont {F.}~\bibnamefont {Arnardi}}, \bibinfo {author}
  {\bibfnamefont {L.}~\bibnamefont {Baguet}}, \bibinfo {author} {\bibfnamefont
  {J.-M.}\ \bibnamefont {Beuken}}, \bibinfo {author} {\bibfnamefont
  {J.}~\bibnamefont {Bieder}}, \bibinfo {author} {\bibfnamefont
  {F.}~\bibnamefont {Bottin}}, \bibinfo {author} {\bibfnamefont
  {J.}~\bibnamefont {Bouchet}}, \bibinfo {author} {\bibfnamefont
  {E.}~\bibnamefont {Bousquet}}, \bibinfo {author} {\bibfnamefont
  {N.}~\bibnamefont {Brouwer}}, \bibinfo {author} {\bibfnamefont
  {F.}~\bibnamefont {Bruneval}}, \bibinfo {author} {\bibfnamefont
  {G.}~\bibnamefont {Brunin}}, \bibinfo {author} {\bibfnamefont
  {T.}~\bibnamefont {Cavignac}}, \bibinfo {author} {\bibfnamefont {J.-B.}\
  \bibnamefont {Charraud}}, \bibinfo {author} {\bibfnamefont {W.}~\bibnamefont
  {Chen}}, \bibinfo {author} {\bibfnamefont {M.}~\bibnamefont {C{\^o}t{\'e}}},
  \bibinfo {author} {\bibfnamefont {S.}~\bibnamefont {Cottenier}}, \bibinfo
  {author} {\bibfnamefont {J.}~\bibnamefont {Denier}}, \bibinfo {author}
  {\bibfnamefont {G.}~\bibnamefont {Geneste}}, \bibinfo {author} {\bibfnamefont
  {P.}~\bibnamefont {Ghosez}}, \bibinfo {author} {\bibfnamefont
  {M.}~\bibnamefont {Giantomassi}}, \bibinfo {author} {\bibfnamefont
  {Y.}~\bibnamefont {Gillet}}, \bibinfo {author} {\bibfnamefont
  {O.}~\bibnamefont {Gingras}}, \bibinfo {author} {\bibfnamefont {D.~R.}\
  \bibnamefont {Hamann}}, \bibinfo {author} {\bibfnamefont {G.}~\bibnamefont
  {Hautier}}, \bibinfo {author} {\bibfnamefont {X.}~\bibnamefont {He}},
  \bibinfo {author} {\bibfnamefont {N.}~\bibnamefont {Helbig}}, \bibinfo
  {author} {\bibfnamefont {N.}~\bibnamefont {Holzwarth}}, \bibinfo {author}
  {\bibfnamefont {Y.}~\bibnamefont {Jia}}, \bibinfo {author} {\bibfnamefont
  {F.}~\bibnamefont {Jollet}}, \bibinfo {author} {\bibfnamefont
  {W.}~\bibnamefont {{Lafargue-Dit-Hauret}}}, \bibinfo {author} {\bibfnamefont
  {K.}~\bibnamefont {Lejaeghere}}, \bibinfo {author} {\bibfnamefont {M.~A.~L.}\
  \bibnamefont {Marques}}, \bibinfo {author} {\bibfnamefont {A.}~\bibnamefont
  {Martin}}, \bibinfo {author} {\bibfnamefont {C.}~\bibnamefont {Martins}},
  \bibinfo {author} {\bibfnamefont {H.~P.~C.}\ \bibnamefont {Miranda}},
  \bibinfo {author} {\bibfnamefont {F.}~\bibnamefont {Naccarato}}, \bibinfo
  {author} {\bibfnamefont {K.}~\bibnamefont {Persson}}, \bibinfo {author}
  {\bibfnamefont {G.}~\bibnamefont {Petretto}}, \bibinfo {author}
  {\bibfnamefont {V.}~\bibnamefont {Planes}}, \bibinfo {author} {\bibfnamefont
  {Y.}~\bibnamefont {Pouillon}}, \bibinfo {author} {\bibfnamefont
  {S.}~\bibnamefont {Prokhorenko}}, \bibinfo {author} {\bibfnamefont
  {F.}~\bibnamefont {Ricci}}, \bibinfo {author} {\bibfnamefont {G.-M.}\
  \bibnamefont {Rignanese}}, \bibinfo {author} {\bibfnamefont {A.~H.}\
  \bibnamefont {Romero}}, \bibinfo {author} {\bibfnamefont {M.~M.}\
  \bibnamefont {Schmitt}}, \bibinfo {author} {\bibfnamefont {M.}~\bibnamefont
  {Torrent}}, \bibinfo {author} {\bibfnamefont {M.~J.}\ \bibnamefont {{van
  Setten}}}, \bibinfo {author} {\bibfnamefont {B.}~\bibnamefont {Van~Troeye}},
  \bibinfo {author} {\bibfnamefont {M.~J.}\ \bibnamefont {Verstraete}},
  \bibinfo {author} {\bibfnamefont {G.}~\bibnamefont {Z{\'e}rah}},\ and\
  \bibinfo {author} {\bibfnamefont {J.~W.}\ \bibnamefont {Zwanziger}},\
  }\bibfield  {title} {\bibinfo {title} {The {{Abinit}}~project: {{Impact}},
  environment and recent developments},\ }\href
  {https://doi.org/10.1016/j.cpc.2019.107042} {\bibfield  {journal} {\bibinfo
  {journal} {Computer Physics Communications}\ }\textbf {\bibinfo {volume}
  {248}},\ \bibinfo {pages} {107042} (\bibinfo {year} {2020})}\BibitemShut
  {NoStop}%
\bibitem [{\citenamefont {Romero}\ \emph {et~al.}(2020)\citenamefont {Romero},
  \citenamefont {Allan}, \citenamefont {Amadon}, \citenamefont {Antonius},
  \citenamefont {Applencourt}, \citenamefont {Baguet}, \citenamefont {Bieder},
  \citenamefont {Bottin}, \citenamefont {Bouchet}, \citenamefont {Bousquet},
  \citenamefont {Bruneval}, \citenamefont {Brunin}, \citenamefont {Caliste},
  \citenamefont {C{\^o}t{\'e}}, \citenamefont {Denier}, \citenamefont {Dreyer},
  \citenamefont {Ghosez}, \citenamefont {Giantomassi}, \citenamefont {Gillet},
  \citenamefont {Gingras}, \citenamefont {Hamann}, \citenamefont {Hautier},
  \citenamefont {Jollet}, \citenamefont {Jomard}, \citenamefont {Martin},
  \citenamefont {Miranda}, \citenamefont {Naccarato}, \citenamefont {Petretto},
  \citenamefont {Pike}, \citenamefont {Planes}, \citenamefont {Prokhorenko},
  \citenamefont {Rangel}, \citenamefont {Ricci}, \citenamefont {Rignanese},
  \citenamefont {Royo}, \citenamefont {Stengel}, \citenamefont {Torrent},
  \citenamefont {{van Setten}}, \citenamefont {Troeye}, \citenamefont
  {Verstraete}, \citenamefont {Wiktor}, \citenamefont {Zwanziger},\ and\
  \citenamefont {Gonze}}]{romero2020}%
  \BibitemOpen
  \bibfield  {author} {\bibinfo {author} {\bibfnamefont {A.~H.}\ \bibnamefont
  {Romero}}, \bibinfo {author} {\bibfnamefont {D.~C.}\ \bibnamefont {Allan}},
  \bibinfo {author} {\bibfnamefont {B.}~\bibnamefont {Amadon}}, \bibinfo
  {author} {\bibfnamefont {G.}~\bibnamefont {Antonius}}, \bibinfo {author}
  {\bibfnamefont {T.}~\bibnamefont {Applencourt}}, \bibinfo {author}
  {\bibfnamefont {L.}~\bibnamefont {Baguet}}, \bibinfo {author} {\bibfnamefont
  {J.}~\bibnamefont {Bieder}}, \bibinfo {author} {\bibfnamefont
  {F.}~\bibnamefont {Bottin}}, \bibinfo {author} {\bibfnamefont
  {J.}~\bibnamefont {Bouchet}}, \bibinfo {author} {\bibfnamefont
  {E.}~\bibnamefont {Bousquet}}, \bibinfo {author} {\bibfnamefont
  {F.}~\bibnamefont {Bruneval}}, \bibinfo {author} {\bibfnamefont
  {G.}~\bibnamefont {Brunin}}, \bibinfo {author} {\bibfnamefont
  {D.}~\bibnamefont {Caliste}}, \bibinfo {author} {\bibfnamefont
  {M.}~\bibnamefont {C{\^o}t{\'e}}}, \bibinfo {author} {\bibfnamefont
  {J.}~\bibnamefont {Denier}}, \bibinfo {author} {\bibfnamefont
  {C.}~\bibnamefont {Dreyer}}, \bibinfo {author} {\bibfnamefont
  {P.}~\bibnamefont {Ghosez}}, \bibinfo {author} {\bibfnamefont
  {M.}~\bibnamefont {Giantomassi}}, \bibinfo {author} {\bibfnamefont
  {Y.}~\bibnamefont {Gillet}}, \bibinfo {author} {\bibfnamefont
  {O.}~\bibnamefont {Gingras}}, \bibinfo {author} {\bibfnamefont {D.~R.}\
  \bibnamefont {Hamann}}, \bibinfo {author} {\bibfnamefont {G.}~\bibnamefont
  {Hautier}}, \bibinfo {author} {\bibfnamefont {F.}~\bibnamefont {Jollet}},
  \bibinfo {author} {\bibfnamefont {G.}~\bibnamefont {Jomard}}, \bibinfo
  {author} {\bibfnamefont {A.}~\bibnamefont {Martin}}, \bibinfo {author}
  {\bibfnamefont {H.~P.~C.}\ \bibnamefont {Miranda}}, \bibinfo {author}
  {\bibfnamefont {F.}~\bibnamefont {Naccarato}}, \bibinfo {author}
  {\bibfnamefont {G.}~\bibnamefont {Petretto}}, \bibinfo {author}
  {\bibfnamefont {N.~A.}\ \bibnamefont {Pike}}, \bibinfo {author}
  {\bibfnamefont {V.}~\bibnamefont {Planes}}, \bibinfo {author} {\bibfnamefont
  {S.}~\bibnamefont {Prokhorenko}}, \bibinfo {author} {\bibfnamefont
  {T.}~\bibnamefont {Rangel}}, \bibinfo {author} {\bibfnamefont
  {F.}~\bibnamefont {Ricci}}, \bibinfo {author} {\bibfnamefont {G.-M.}\
  \bibnamefont {Rignanese}}, \bibinfo {author} {\bibfnamefont {M.}~\bibnamefont
  {Royo}}, \bibinfo {author} {\bibfnamefont {M.}~\bibnamefont {Stengel}},
  \bibinfo {author} {\bibfnamefont {M.}~\bibnamefont {Torrent}}, \bibinfo
  {author} {\bibfnamefont {M.~J.}\ \bibnamefont {{van Setten}}}, \bibinfo
  {author} {\bibfnamefont {B.~V.}\ \bibnamefont {Troeye}}, \bibinfo {author}
  {\bibfnamefont {M.~J.}\ \bibnamefont {Verstraete}}, \bibinfo {author}
  {\bibfnamefont {J.}~\bibnamefont {Wiktor}}, \bibinfo {author} {\bibfnamefont
  {J.~W.}\ \bibnamefont {Zwanziger}},\ and\ \bibinfo {author} {\bibfnamefont
  {X.}~\bibnamefont {Gonze}},\ }\bibfield  {title} {\bibinfo {title}
  {{{ABINIT}}: {{Overview}}, and focus on selected capabilities},\ }\href@noop
  {} {\bibfield  {journal} {\bibinfo  {journal} {Journal of Chemical Physics}\
  }\textbf {\bibinfo {volume} {152}},\ \bibinfo {pages} {124102} (\bibinfo
  {year} {2020})}\BibitemShut {NoStop}%
\bibitem [{\citenamefont {Lee}\ \emph {et~al.}(2023)\citenamefont {Lee},
  \citenamefont {Ponc{\'e}}, \citenamefont {Bushick}, \citenamefont
  {Hajinazar}, \citenamefont {{Lafuente-Bartolome}}, \citenamefont {Leveillee},
  \citenamefont {Lian}, \citenamefont {Lihm}, \citenamefont {Macheda},
  \citenamefont {Mori}, \citenamefont {Paudyal}, \citenamefont {Sio},
  \citenamefont {Tiwari}, \citenamefont {Zacharias}, \citenamefont {Zhang},
  \citenamefont {Bonini}, \citenamefont {Kioupakis}, \citenamefont {Margine},\
  and\ \citenamefont {Giustino}}]{lee2023}%
  \BibitemOpen
  \bibfield  {author} {\bibinfo {author} {\bibfnamefont {H.}~\bibnamefont
  {Lee}}, \bibinfo {author} {\bibfnamefont {S.}~\bibnamefont {Ponc{\'e}}},
  \bibinfo {author} {\bibfnamefont {K.}~\bibnamefont {Bushick}}, \bibinfo
  {author} {\bibfnamefont {S.}~\bibnamefont {Hajinazar}}, \bibinfo {author}
  {\bibfnamefont {J.}~\bibnamefont {{Lafuente-Bartolome}}}, \bibinfo {author}
  {\bibfnamefont {J.}~\bibnamefont {Leveillee}}, \bibinfo {author}
  {\bibfnamefont {C.}~\bibnamefont {Lian}}, \bibinfo {author} {\bibfnamefont
  {J.-M.}\ \bibnamefont {Lihm}}, \bibinfo {author} {\bibfnamefont
  {F.}~\bibnamefont {Macheda}}, \bibinfo {author} {\bibfnamefont
  {H.}~\bibnamefont {Mori}}, \bibinfo {author} {\bibfnamefont {H.}~\bibnamefont
  {Paudyal}}, \bibinfo {author} {\bibfnamefont {W.~H.}\ \bibnamefont {Sio}},
  \bibinfo {author} {\bibfnamefont {S.}~\bibnamefont {Tiwari}}, \bibinfo
  {author} {\bibfnamefont {M.}~\bibnamefont {Zacharias}}, \bibinfo {author}
  {\bibfnamefont {X.}~\bibnamefont {Zhang}}, \bibinfo {author} {\bibfnamefont
  {N.}~\bibnamefont {Bonini}}, \bibinfo {author} {\bibfnamefont
  {E.}~\bibnamefont {Kioupakis}}, \bibinfo {author} {\bibfnamefont {E.~R.}\
  \bibnamefont {Margine}},\ and\ \bibinfo {author} {\bibfnamefont
  {F.}~\bibnamefont {Giustino}},\ }\bibfield  {title} {\bibinfo {title}
  {Electron--phonon physics from first principles using the {{EPW}} code},\
  }\href {https://doi.org/10.1038/s41524-023-01107-3} {\bibfield  {journal}
  {\bibinfo  {journal} {npj Computational Materials}\ }\textbf {\bibinfo
  {volume} {9}},\ \bibinfo {pages} {1} (\bibinfo {year} {2023})}\BibitemShut
  {NoStop}%
\bibitem [{\citenamefont {Zhou}\ \emph {et~al.}(2021)\citenamefont {Zhou},
  \citenamefont {Park}, \citenamefont {Lu}, \citenamefont {Maliyov},
  \citenamefont {Tong},\ and\ \citenamefont {Bernardi}}]{zhou2021}%
  \BibitemOpen
  \bibfield  {author} {\bibinfo {author} {\bibfnamefont {J.-J.}\ \bibnamefont
  {Zhou}}, \bibinfo {author} {\bibfnamefont {J.}~\bibnamefont {Park}}, \bibinfo
  {author} {\bibfnamefont {I.-T.}\ \bibnamefont {Lu}}, \bibinfo {author}
  {\bibfnamefont {I.}~\bibnamefont {Maliyov}}, \bibinfo {author} {\bibfnamefont
  {X.}~\bibnamefont {Tong}},\ and\ \bibinfo {author} {\bibfnamefont
  {M.}~\bibnamefont {Bernardi}},\ }\bibfield  {title} {\bibinfo {title}
  {Perturbo: {{A}} software package for {\emph{ab initio}} electron--phonon
  interactions, charge transport and ultrafast dynamics},\ }\href
  {https://doi.org/10.1016/j.cpc.2021.107970} {\bibfield  {journal} {\bibinfo
  {journal} {Computer Physics Communications}\ }\textbf {\bibinfo {volume}
  {264}},\ \bibinfo {pages} {107970} (\bibinfo {year} {2021})}\BibitemShut
  {NoStop}%
\bibitem [{\citenamefont {Verdi}\ and\ \citenamefont
  {Giustino}(2015)}]{verdi2015}%
  \BibitemOpen
  \bibfield  {author} {\bibinfo {author} {\bibfnamefont {C.}~\bibnamefont
  {Verdi}}\ and\ \bibinfo {author} {\bibfnamefont {F.}~\bibnamefont
  {Giustino}},\ }\bibfield  {title} {\bibinfo {title} {Fr{\"o}hlich
  {{Electron-Phonon Vertex}} from {{First Principles}}},\ }\href
  {https://doi.org/10.1103/PhysRevLett.115.176401} {\bibfield  {journal}
  {\bibinfo  {journal} {Physical Review Letters}\ }\textbf {\bibinfo {volume}
  {115}},\ \bibinfo {pages} {176401} (\bibinfo {year} {2015})}\BibitemShut
  {NoStop}%
\bibitem [{\citenamefont {Collignon}\ \emph {et~al.}(2019)\citenamefont
  {Collignon}, \citenamefont {Lin}, \citenamefont {Rischau}, \citenamefont
  {Fauqu{\'e}},\ and\ \citenamefont {Behnia}}]{collignon2019}%
  \BibitemOpen
  \bibfield  {author} {\bibinfo {author} {\bibfnamefont {C.}~\bibnamefont
  {Collignon}}, \bibinfo {author} {\bibfnamefont {X.}~\bibnamefont {Lin}},
  \bibinfo {author} {\bibfnamefont {C.~W.}\ \bibnamefont {Rischau}}, \bibinfo
  {author} {\bibfnamefont {B.}~\bibnamefont {Fauqu{\'e}}},\ and\ \bibinfo
  {author} {\bibfnamefont {K.}~\bibnamefont {Behnia}},\ }\bibfield  {title}
  {\bibinfo {title} {Metallicity and {{Superconductivity}} in {{Doped Strontium
  Titanate}}},\ }\href
  {https://doi.org/10.1146/annurev-conmatphys-031218-013144} {\bibfield
  {journal} {\bibinfo  {journal} {Annual Review of Condensed Matter Physics}\
  }\textbf {\bibinfo {volume} {10}},\ \bibinfo {pages} {25} (\bibinfo {year}
  {2019})}\BibitemShut {NoStop}%
\bibitem [{\citenamefont {Gastiasoro}\ \emph {et~al.}(2020)\citenamefont
  {Gastiasoro}, \citenamefont {Ruhman},\ and\ \citenamefont
  {Fernandes}}]{gastiasoro2020}%
  \BibitemOpen
  \bibfield  {author} {\bibinfo {author} {\bibfnamefont {M.~N.}\ \bibnamefont
  {Gastiasoro}}, \bibinfo {author} {\bibfnamefont {J.}~\bibnamefont {Ruhman}},\
  and\ \bibinfo {author} {\bibfnamefont {R.~M.}\ \bibnamefont {Fernandes}},\
  }\bibfield  {title} {\bibinfo {title} {Superconductivity in dilute
  {{SrTi${\mathrm{O}}_{3}$}} : {{A}} review},\ }\href
  {https://doi.org/10.1016/j.aop.2020.168107} {\bibfield  {journal} {\bibinfo
  {journal} {Annals of Physics}\ }\textbf {\bibinfo {volume} {417}},\ \bibinfo
  {pages} {168107} (\bibinfo {year} {2020})}\BibitemShut {NoStop}%
\bibitem [{\citenamefont {Gupta}\ \emph {et~al.}(2022)\citenamefont {Gupta},
  \citenamefont {Silotia}, \citenamefont {Kumari}, \citenamefont {Dumen},
  \citenamefont {Goyal}, \citenamefont {Tomar}, \citenamefont {Wadehra},
  \citenamefont {Ayyub},\ and\ \citenamefont {Chakraverty}}]{gupta2022}%
  \BibitemOpen
  \bibfield  {author} {\bibinfo {author} {\bibfnamefont {A.}~\bibnamefont
  {Gupta}}, \bibinfo {author} {\bibfnamefont {H.}~\bibnamefont {Silotia}},
  \bibinfo {author} {\bibfnamefont {A.}~\bibnamefont {Kumari}}, \bibinfo
  {author} {\bibfnamefont {M.}~\bibnamefont {Dumen}}, \bibinfo {author}
  {\bibfnamefont {S.}~\bibnamefont {Goyal}}, \bibinfo {author} {\bibfnamefont
  {R.}~\bibnamefont {Tomar}}, \bibinfo {author} {\bibfnamefont
  {N.}~\bibnamefont {Wadehra}}, \bibinfo {author} {\bibfnamefont
  {P.}~\bibnamefont {Ayyub}},\ and\ \bibinfo {author} {\bibfnamefont
  {S.}~\bibnamefont {Chakraverty}},\ }\bibfield  {title} {\bibinfo {title}
  {{{KTa${\mathrm{O}}_{3}$}}---{{The New Kid}} on the {{Spintronics Block}}},\
  }\href {https://doi.org/10.1002/adma.202106481} {\bibfield  {journal}
  {\bibinfo  {journal} {Advanced Materials}\ }\textbf {\bibinfo {volume}
  {34}},\ \bibinfo {pages} {2106481} (\bibinfo {year} {2022})}\BibitemShut
  {NoStop}%
\bibitem [{\citenamefont {Ranalli}\ \emph {et~al.}(2023)\citenamefont
  {Ranalli}, \citenamefont {Verdi}, \citenamefont {Monacelli}, \citenamefont
  {Kresse}, \citenamefont {Calandra},\ and\ \citenamefont
  {Franchini}}]{ranalli2023}%
  \BibitemOpen
  \bibfield  {author} {\bibinfo {author} {\bibfnamefont {L.}~\bibnamefont
  {Ranalli}}, \bibinfo {author} {\bibfnamefont {C.}~\bibnamefont {Verdi}},
  \bibinfo {author} {\bibfnamefont {L.}~\bibnamefont {Monacelli}}, \bibinfo
  {author} {\bibfnamefont {G.}~\bibnamefont {Kresse}}, \bibinfo {author}
  {\bibfnamefont {M.}~\bibnamefont {Calandra}},\ and\ \bibinfo {author}
  {\bibfnamefont {C.}~\bibnamefont {Franchini}},\ }\bibfield  {title} {\bibinfo
  {title} {Temperature-{{Dependent Anharmonic Phonons}} in {{Quantum
  Paraelectric}} {{KTa${\mathrm{O}}_{3}$}} by {{First Principles}} and
  {{Machine-Learned Force Fields}}},\ }\href
  {https://doi.org/10.1002/qute.202200131} {\bibfield  {journal} {\bibinfo
  {journal} {Advanced Quantum Technologies}\ }\textbf {\bibinfo {volume} {6}},\
  \bibinfo {pages} {2200131} (\bibinfo {year} {2023})}\BibitemShut {NoStop}%
\bibitem [{\citenamefont {Verdi}\ \emph {et~al.}(2023)\citenamefont {Verdi},
  \citenamefont {Ranalli}, \citenamefont {Franchini},\ and\ \citenamefont
  {Kresse}}]{verdi2023}%
  \BibitemOpen
  \bibfield  {author} {\bibinfo {author} {\bibfnamefont {C.}~\bibnamefont
  {Verdi}}, \bibinfo {author} {\bibfnamefont {L.}~\bibnamefont {Ranalli}},
  \bibinfo {author} {\bibfnamefont {C.}~\bibnamefont {Franchini}},\ and\
  \bibinfo {author} {\bibfnamefont {G.}~\bibnamefont {Kresse}},\ }\bibfield
  {title} {\bibinfo {title} {Quantum paraelectricity and structural phase
  transitions in strontium titanate beyond density functional theory},\ }\href
  {https://doi.org/10.1103/PhysRevMaterials.7.L030801} {\bibfield  {journal}
  {\bibinfo  {journal} {Physical Review Materials}\ }\textbf {\bibinfo {volume}
  {7}},\ \bibinfo {pages} {L030801} (\bibinfo {year} {2023})}\BibitemShut
  {NoStop}%
\bibitem [{\citenamefont {Saidi}\ \emph {et~al.}(2016)\citenamefont {Saidi},
  \citenamefont {Ponc{\'e}},\ and\ \citenamefont {Monserrat}}]{saidi2016}%
  \BibitemOpen
  \bibfield  {author} {\bibinfo {author} {\bibfnamefont {W.~A.}\ \bibnamefont
  {Saidi}}, \bibinfo {author} {\bibfnamefont {S.}~\bibnamefont {Ponc{\'e}}},\
  and\ \bibinfo {author} {\bibfnamefont {B.}~\bibnamefont {Monserrat}},\
  }\bibfield  {title} {\bibinfo {title} {Temperature {{Dependence}} of the
  {{Energy Levels}} of {{Methylammonium Lead Iodide Perovskite}} from
  {{First-Principles}}},\ }\href {https://doi.org/10.1021/acs.jpclett.6b02560}
  {\bibfield  {journal} {\bibinfo  {journal} {The Journal of Physical Chemistry
  Letters}\ }\textbf {\bibinfo {volume} {7}},\ \bibinfo {pages} {5247}
  (\bibinfo {year} {2016})}\BibitemShut {NoStop}%
\bibitem [{\citenamefont {Jena}\ \emph {et~al.}(2019)\citenamefont {Jena},
  \citenamefont {Kulkarni},\ and\ \citenamefont {Miyasaka}}]{jena2019}%
  \BibitemOpen
  \bibfield  {author} {\bibinfo {author} {\bibfnamefont {A.~K.}\ \bibnamefont
  {Jena}}, \bibinfo {author} {\bibfnamefont {A.}~\bibnamefont {Kulkarni}},\
  and\ \bibinfo {author} {\bibfnamefont {T.}~\bibnamefont {Miyasaka}},\
  }\bibfield  {title} {\bibinfo {title} {Halide {{Perovskite Photovoltaics}}:
  {{Background}}, {{Status}}, and {{Future Prospects}}},\ }\href
  {https://doi.org/10.1021/acs.chemrev.8b00539} {\bibfield  {journal} {\bibinfo
   {journal} {Chemical Reviews}\ }\textbf {\bibinfo {volume} {119}},\ \bibinfo
  {pages} {3036} (\bibinfo {year} {2019})}\BibitemShut {NoStop}%
\bibitem [{\citenamefont {Schilcher}\ \emph {et~al.}(2021)\citenamefont
  {Schilcher}, \citenamefont {Robinson}, \citenamefont {Abramovitch},
  \citenamefont {Tan}, \citenamefont {Rappe}, \citenamefont {Reichman},\ and\
  \citenamefont {Egger}}]{schilcher2021}%
  \BibitemOpen
  \bibfield  {author} {\bibinfo {author} {\bibfnamefont {M.~J.}\ \bibnamefont
  {Schilcher}}, \bibinfo {author} {\bibfnamefont {P.~J.}\ \bibnamefont
  {Robinson}}, \bibinfo {author} {\bibfnamefont {D.~J.}\ \bibnamefont
  {Abramovitch}}, \bibinfo {author} {\bibfnamefont {L.~Z.}\ \bibnamefont
  {Tan}}, \bibinfo {author} {\bibfnamefont {A.~M.}\ \bibnamefont {Rappe}},
  \bibinfo {author} {\bibfnamefont {D.~R.}\ \bibnamefont {Reichman}},\ and\
  \bibinfo {author} {\bibfnamefont {D.~A.}\ \bibnamefont {Egger}},\ }\bibfield
  {title} {\bibinfo {title} {The {{Significance}} of {{Polarons}} and {{Dynamic
  Disorder}} in {{Halide Perovskites}}},\ }\href
  {https://doi.org/10.1021/acsenergylett.1c00506} {\bibfield  {journal}
  {\bibinfo  {journal} {ACS Energy Letters}\ }\textbf {\bibinfo {volume} {6}},\
  \bibinfo {pages} {2162} (\bibinfo {year} {2021})}\BibitemShut {NoStop}%
\bibitem [{\citenamefont {Yamada}\ and\ \citenamefont
  {Kanemitsu}(2022)}]{yamada2022}%
  \BibitemOpen
  \bibfield  {author} {\bibinfo {author} {\bibfnamefont {Y.}~\bibnamefont
  {Yamada}}\ and\ \bibinfo {author} {\bibfnamefont {Y.}~\bibnamefont
  {Kanemitsu}},\ }\bibfield  {title} {\bibinfo {title} {Electron-phonon
  interactions in halide perovskites},\ }\href
  {https://doi.org/10.1038/s41427-022-00394-4} {\bibfield  {journal} {\bibinfo
  {journal} {NPG Asia Materials}\ }\textbf {\bibinfo {volume} {14}},\ \bibinfo
  {pages} {48} (\bibinfo {year} {2022})}\BibitemShut {NoStop}%
\bibitem [{\citenamefont {Errea}\ \emph {et~al.}(2014)\citenamefont {Errea},
  \citenamefont {Calandra},\ and\ \citenamefont {Mauri}}]{errea2014}%
  \BibitemOpen
  \bibfield  {author} {\bibinfo {author} {\bibfnamefont {I.}~\bibnamefont
  {Errea}}, \bibinfo {author} {\bibfnamefont {M.}~\bibnamefont {Calandra}},\
  and\ \bibinfo {author} {\bibfnamefont {F.}~\bibnamefont {Mauri}},\ }\bibfield
   {title} {\bibinfo {title} {Anharmonic free energies and phonon dispersions
  from the stochastic self-consistent harmonic approximation: {{Application}}
  to platinum and palladium hydrides},\ }\href
  {https://doi.org/10.1103/PhysRevB.89.064302} {\bibfield  {journal} {\bibinfo
  {journal} {Physical Review B}\ }\textbf {\bibinfo {volume} {89}},\ \bibinfo
  {pages} {064302} (\bibinfo {year} {2014})}\BibitemShut {NoStop}%
\bibitem [{\citenamefont {Drozdov}\ \emph {et~al.}(2015)\citenamefont
  {Drozdov}, \citenamefont {Eremets}, \citenamefont {Troyan}, \citenamefont
  {Ksenofontov},\ and\ \citenamefont {Shylin}}]{drozdov2015}%
  \BibitemOpen
  \bibfield  {author} {\bibinfo {author} {\bibfnamefont {A.~P.}\ \bibnamefont
  {Drozdov}}, \bibinfo {author} {\bibfnamefont {M.~I.}\ \bibnamefont
  {Eremets}}, \bibinfo {author} {\bibfnamefont {I.~A.}\ \bibnamefont {Troyan}},
  \bibinfo {author} {\bibfnamefont {V.}~\bibnamefont {Ksenofontov}},\ and\
  \bibinfo {author} {\bibfnamefont {S.~I.}\ \bibnamefont {Shylin}},\ }\bibfield
   {title} {\bibinfo {title} {Conventional superconductivity at 203 kelvin at
  high pressures in the sulfur hydride system},\ }\href
  {https://doi.org/10.1038/nature14964} {\bibfield  {journal} {\bibinfo
  {journal} {Nature}\ }\textbf {\bibinfo {volume} {525}},\ \bibinfo {pages}
  {73} (\bibinfo {year} {2015})}\BibitemShut {NoStop}%
\bibitem [{\citenamefont {Errea}\ \emph {et~al.}(2015)\citenamefont {Errea},
  \citenamefont {Calandra}, \citenamefont {Pickard}, \citenamefont {Nelson},
  \citenamefont {Needs}, \citenamefont {Li}, \citenamefont {Liu}, \citenamefont
  {Zhang}, \citenamefont {Ma},\ and\ \citenamefont {Mauri}}]{errea2015}%
  \BibitemOpen
  \bibfield  {author} {\bibinfo {author} {\bibfnamefont {I.}~\bibnamefont
  {Errea}}, \bibinfo {author} {\bibfnamefont {M.}~\bibnamefont {Calandra}},
  \bibinfo {author} {\bibfnamefont {C.~J.}\ \bibnamefont {Pickard}}, \bibinfo
  {author} {\bibfnamefont {J.}~\bibnamefont {Nelson}}, \bibinfo {author}
  {\bibfnamefont {R.~J.}\ \bibnamefont {Needs}}, \bibinfo {author}
  {\bibfnamefont {Y.}~\bibnamefont {Li}}, \bibinfo {author} {\bibfnamefont
  {H.}~\bibnamefont {Liu}}, \bibinfo {author} {\bibfnamefont {Y.}~\bibnamefont
  {Zhang}}, \bibinfo {author} {\bibfnamefont {Y.}~\bibnamefont {Ma}},\ and\
  \bibinfo {author} {\bibfnamefont {F.}~\bibnamefont {Mauri}},\ }\bibfield
  {title} {\bibinfo {title} {High-{{Pressure Hydrogen Sulfide}} from {{First
  Principles}}: {{A Strongly Anharmonic Phonon-Mediated Superconductor}}},\
  }\href {https://doi.org/10.1103/PhysRevLett.114.157004} {\bibfield  {journal}
  {\bibinfo  {journal} {Physical Review Letters}\ }\textbf {\bibinfo {volume}
  {114}},\ \bibinfo {pages} {157004} (\bibinfo {year} {2015})}\BibitemShut
  {NoStop}%
\bibitem [{\citenamefont {Somayazulu}\ \emph {et~al.}(2019)\citenamefont
  {Somayazulu}, \citenamefont {Ahart}, \citenamefont {Mishra}, \citenamefont
  {Geballe}, \citenamefont {Baldini}, \citenamefont {Meng}, \citenamefont
  {Struzhkin},\ and\ \citenamefont {Hemley}}]{somayazulu2019}%
  \BibitemOpen
  \bibfield  {author} {\bibinfo {author} {\bibfnamefont {M.}~\bibnamefont
  {Somayazulu}}, \bibinfo {author} {\bibfnamefont {M.}~\bibnamefont {Ahart}},
  \bibinfo {author} {\bibfnamefont {A.~K.}\ \bibnamefont {Mishra}}, \bibinfo
  {author} {\bibfnamefont {Z.~M.}\ \bibnamefont {Geballe}}, \bibinfo {author}
  {\bibfnamefont {M.}~\bibnamefont {Baldini}}, \bibinfo {author} {\bibfnamefont
  {Y.}~\bibnamefont {Meng}}, \bibinfo {author} {\bibfnamefont {V.~V.}\
  \bibnamefont {Struzhkin}},\ and\ \bibinfo {author} {\bibfnamefont {R.~J.}\
  \bibnamefont {Hemley}},\ }\bibfield  {title} {\bibinfo {title} {Evidence for
  {{Superconductivity}} above 260 {{K}} in {{Lanthanum Superhydride}} at
  {{Megabar Pressures}}},\ }\href
  {https://doi.org/10.1103/PhysRevLett.122.027001} {\bibfield  {journal}
  {\bibinfo  {journal} {Physical Review Letters}\ }\textbf {\bibinfo {volume}
  {122}},\ \bibinfo {pages} {027001} (\bibinfo {year} {2019})}\BibitemShut
  {NoStop}%
\bibitem [{\citenamefont {Errea}\ \emph {et~al.}(2020)\citenamefont {Errea},
  \citenamefont {Belli}, \citenamefont {Monacelli}, \citenamefont {Sanna},
  \citenamefont {Koretsune}, \citenamefont {Tadano}, \citenamefont {Bianco},
  \citenamefont {Calandra}, \citenamefont {Arita}, \citenamefont {Mauri},\ and\
  \citenamefont {{Flores-Livas}}}]{errea2020}%
  \BibitemOpen
  \bibfield  {author} {\bibinfo {author} {\bibfnamefont {I.}~\bibnamefont
  {Errea}}, \bibinfo {author} {\bibfnamefont {F.}~\bibnamefont {Belli}},
  \bibinfo {author} {\bibfnamefont {L.}~\bibnamefont {Monacelli}}, \bibinfo
  {author} {\bibfnamefont {A.}~\bibnamefont {Sanna}}, \bibinfo {author}
  {\bibfnamefont {T.}~\bibnamefont {Koretsune}}, \bibinfo {author}
  {\bibfnamefont {T.}~\bibnamefont {Tadano}}, \bibinfo {author} {\bibfnamefont
  {R.}~\bibnamefont {Bianco}}, \bibinfo {author} {\bibfnamefont
  {M.}~\bibnamefont {Calandra}}, \bibinfo {author} {\bibfnamefont
  {R.}~\bibnamefont {Arita}}, \bibinfo {author} {\bibfnamefont
  {F.}~\bibnamefont {Mauri}},\ and\ \bibinfo {author} {\bibfnamefont {J.~A.}\
  \bibnamefont {{Flores-Livas}}},\ }\bibfield  {title} {\bibinfo {title}
  {Quantum crystal structure in the 250-kelvin superconducting lanthanum
  hydride},\ }\href {https://doi.org/10.1038/s41586-020-1955-z} {\bibfield
  {journal} {\bibinfo  {journal} {Nature}\ }\textbf {\bibinfo {volume} {578}},\
  \bibinfo {pages} {66} (\bibinfo {year} {2020})}\BibitemShut {NoStop}%
\bibitem [{\citenamefont {Hirsch}\ and\ \citenamefont
  {Marsiglio}(2021)}]{hirsch2021}%
  \BibitemOpen
  \bibfield  {author} {\bibinfo {author} {\bibfnamefont {J.~E.}\ \bibnamefont
  {Hirsch}}\ and\ \bibinfo {author} {\bibfnamefont {F.}~\bibnamefont
  {Marsiglio}},\ }\bibfield  {title} {\bibinfo {title} {Nonstandard
  superconductivity or no superconductivity in hydrides under high pressure},\
  }\href {https://doi.org/10.1103/PhysRevB.103.134505} {\bibfield  {journal}
  {\bibinfo  {journal} {Physical Review B}\ }\textbf {\bibinfo {volume}
  {103}},\ \bibinfo {pages} {134505} (\bibinfo {year} {2021})}\BibitemShut
  {NoStop}%
\bibitem [{\citenamefont {Shipley}\ \emph {et~al.}(2021)\citenamefont
  {Shipley}, \citenamefont {Hutcheon}, \citenamefont {Needs},\ and\
  \citenamefont {Pickard}}]{shipley2021}%
  \BibitemOpen
  \bibfield  {author} {\bibinfo {author} {\bibfnamefont {A.~M.}\ \bibnamefont
  {Shipley}}, \bibinfo {author} {\bibfnamefont {M.~J.}\ \bibnamefont
  {Hutcheon}}, \bibinfo {author} {\bibfnamefont {R.~J.}\ \bibnamefont
  {Needs}},\ and\ \bibinfo {author} {\bibfnamefont {C.~J.}\ \bibnamefont
  {Pickard}},\ }\bibfield  {title} {\bibinfo {title} {High-throughput discovery
  of high-temperature conventional superconductors},\ }\href
  {https://doi.org/10.1103/PhysRevB.104.054501} {\bibfield  {journal} {\bibinfo
   {journal} {Physical Review B}\ }\textbf {\bibinfo {volume} {104}},\ \bibinfo
  {pages} {054501} (\bibinfo {year} {2021})}\BibitemShut {NoStop}%
\bibitem [{\citenamefont {Troyan}\ \emph {et~al.}(2021)\citenamefont {Troyan},
  \citenamefont {Semenok}, \citenamefont {Kvashnin}, \citenamefont {Sadakov},
  \citenamefont {Sobolevskiy}, \citenamefont {Pudalov}, \citenamefont
  {Ivanova}, \citenamefont {Prakapenka}, \citenamefont {Greenberg},
  \citenamefont {Gavriliuk}, \citenamefont {Lyubutin}, \citenamefont
  {Struzhkin}, \citenamefont {Bergara}, \citenamefont {Errea}, \citenamefont
  {Bianco}, \citenamefont {Calandra}, \citenamefont {Mauri}, \citenamefont
  {Monacelli}, \citenamefont {Akashi},\ and\ \citenamefont
  {Oganov}}]{troyan2021}%
  \BibitemOpen
  \bibfield  {author} {\bibinfo {author} {\bibfnamefont {I.~A.}\ \bibnamefont
  {Troyan}}, \bibinfo {author} {\bibfnamefont {D.~V.}\ \bibnamefont {Semenok}},
  \bibinfo {author} {\bibfnamefont {A.~G.}\ \bibnamefont {Kvashnin}}, \bibinfo
  {author} {\bibfnamefont {A.~V.}\ \bibnamefont {Sadakov}}, \bibinfo {author}
  {\bibfnamefont {O.~A.}\ \bibnamefont {Sobolevskiy}}, \bibinfo {author}
  {\bibfnamefont {V.~M.}\ \bibnamefont {Pudalov}}, \bibinfo {author}
  {\bibfnamefont {A.~G.}\ \bibnamefont {Ivanova}}, \bibinfo {author}
  {\bibfnamefont {V.~B.}\ \bibnamefont {Prakapenka}}, \bibinfo {author}
  {\bibfnamefont {E.}~\bibnamefont {Greenberg}}, \bibinfo {author}
  {\bibfnamefont {A.~G.}\ \bibnamefont {Gavriliuk}}, \bibinfo {author}
  {\bibfnamefont {I.~S.}\ \bibnamefont {Lyubutin}}, \bibinfo {author}
  {\bibfnamefont {V.~V.}\ \bibnamefont {Struzhkin}}, \bibinfo {author}
  {\bibfnamefont {A.}~\bibnamefont {Bergara}}, \bibinfo {author} {\bibfnamefont
  {I.}~\bibnamefont {Errea}}, \bibinfo {author} {\bibfnamefont
  {R.}~\bibnamefont {Bianco}}, \bibinfo {author} {\bibfnamefont
  {M.}~\bibnamefont {Calandra}}, \bibinfo {author} {\bibfnamefont
  {F.}~\bibnamefont {Mauri}}, \bibinfo {author} {\bibfnamefont
  {L.}~\bibnamefont {Monacelli}}, \bibinfo {author} {\bibfnamefont
  {R.}~\bibnamefont {Akashi}},\ and\ \bibinfo {author} {\bibfnamefont {A.~R.}\
  \bibnamefont {Oganov}},\ }\bibfield  {title} {\bibinfo {title} {Anomalous
  {{High}}-{{Temperature Superconductivity}} in {{YH}} {\textsubscript{6}}},\
  }\href {https://doi.org/10.1002/adma.202006832} {\bibfield  {journal}
  {\bibinfo  {journal} {Advanced Materials}\ }\textbf {\bibinfo {volume}
  {33}},\ \bibinfo {pages} {2006832} (\bibinfo {year} {2021})}\BibitemShut
  {NoStop}%
\bibitem [{\citenamefont {Zhang}\ \emph {et~al.}(2022)\citenamefont {Zhang},
  \citenamefont {Zhao},\ and\ \citenamefont {Yang}}]{zhang2022}%
  \BibitemOpen
  \bibfield  {author} {\bibinfo {author} {\bibfnamefont {X.}~\bibnamefont
  {Zhang}}, \bibinfo {author} {\bibfnamefont {Y.}~\bibnamefont {Zhao}},\ and\
  \bibinfo {author} {\bibfnamefont {G.}~\bibnamefont {Yang}},\ }\bibfield
  {title} {\bibinfo {title} {Superconducting ternary hydrides under high
  pressure},\ }\href {https://doi.org/10.1002/wcms.1582} {\bibfield  {journal}
  {\bibinfo  {journal} {WIREs Computational Molecular Science}\ }\textbf
  {\bibinfo {volume} {12}},\ \bibinfo {pages} {e1582} (\bibinfo {year}
  {2022})}\BibitemShut {NoStop}%
\bibitem [{\citenamefont {Deinzer}\ \emph {et~al.}(2003)\citenamefont
  {Deinzer}, \citenamefont {Birner},\ and\ \citenamefont
  {Strauch}}]{deinzer2003}%
  \BibitemOpen
  \bibfield  {author} {\bibinfo {author} {\bibfnamefont {G.}~\bibnamefont
  {Deinzer}}, \bibinfo {author} {\bibfnamefont {G.}~\bibnamefont {Birner}},\
  and\ \bibinfo {author} {\bibfnamefont {D.}~\bibnamefont {Strauch}},\
  }\bibfield  {title} {\bibinfo {title} {{\emph{Ab Initio}} calculation of the
  linewidth of various phonon modes in germanium and silicon},\ }\href
  {https://doi.org/10.1103/PhysRevB.67.144304} {\bibfield  {journal} {\bibinfo
  {journal} {Physical Review B}\ }\textbf {\bibinfo {volume} {67}},\ \bibinfo
  {pages} {144304} (\bibinfo {year} {2003})}\BibitemShut {NoStop}%
\bibitem [{\citenamefont {Ward}\ \emph {et~al.}(2009)\citenamefont {Ward},
  \citenamefont {Broido}, \citenamefont {Stewart},\ and\ \citenamefont
  {Deinzer}}]{ward2009}%
  \BibitemOpen
  \bibfield  {author} {\bibinfo {author} {\bibfnamefont {A.}~\bibnamefont
  {Ward}}, \bibinfo {author} {\bibfnamefont {D.~A.}\ \bibnamefont {Broido}},
  \bibinfo {author} {\bibfnamefont {D.~A.}\ \bibnamefont {Stewart}},\ and\
  \bibinfo {author} {\bibfnamefont {G.}~\bibnamefont {Deinzer}},\ }\bibfield
  {title} {\bibinfo {title} {{\emph{Ab Initio}} theory of the lattice thermal
  conductivity in diamond},\ }\href
  {https://doi.org/10.1103/PhysRevB.80.125203} {\bibfield  {journal} {\bibinfo
  {journal} {Physical Review B}\ }\textbf {\bibinfo {volume} {80}},\ \bibinfo
  {pages} {125203} (\bibinfo {year} {2009})}\BibitemShut {NoStop}%
\bibitem [{\citenamefont {Togo}\ \emph {et~al.}(2023)\citenamefont {Togo},
  \citenamefont {Chaput}, \citenamefont {Tadano},\ and\ \citenamefont
  {Tanaka}}]{togo2023}%
  \BibitemOpen
  \bibfield  {author} {\bibinfo {author} {\bibfnamefont {A.}~\bibnamefont
  {Togo}}, \bibinfo {author} {\bibfnamefont {L.}~\bibnamefont {Chaput}},
  \bibinfo {author} {\bibfnamefont {T.}~\bibnamefont {Tadano}},\ and\ \bibinfo
  {author} {\bibfnamefont {I.}~\bibnamefont {Tanaka}},\ }\bibfield  {title}
  {\bibinfo {title} {Implementation strategies in phonopy and phono3py},\
  }\href {https://doi.org/10.1088/1361-648X/acd831} {\bibfield  {journal}
  {\bibinfo  {journal} {Journal of Physics: Condensed Matter}\ }\textbf
  {\bibinfo {volume} {35}},\ \bibinfo {pages} {353001} (\bibinfo {year}
  {2023})}\BibitemShut {NoStop}%
\bibitem [{\citenamefont {Togo}(2023)}]{togo2023a}%
  \BibitemOpen
  \bibfield  {author} {\bibinfo {author} {\bibfnamefont {A.}~\bibnamefont
  {Togo}},\ }\bibfield  {title} {\bibinfo {title} {First-principles {{Phonon
  Calculations}} with {{Phonopy}} and {{Phono3py}}},\ }\href
  {https://doi.org/10.7566/JPSJ.92.012001} {\bibfield  {journal} {\bibinfo
  {journal} {Journal of the Physical Society of Japan}\ }\textbf {\bibinfo
  {volume} {92}},\ \bibinfo {pages} {012001} (\bibinfo {year}
  {2023})}\BibitemShut {NoStop}%
\bibitem [{\citenamefont {Monacelli}\ \emph {et~al.}(2021)\citenamefont
  {Monacelli}, \citenamefont {Bianco}, \citenamefont {Cherubini}, \citenamefont
  {Calandra}, \citenamefont {Errea},\ and\ \citenamefont
  {Mauri}}]{monacelli2021}%
  \BibitemOpen
  \bibfield  {author} {\bibinfo {author} {\bibfnamefont {L.}~\bibnamefont
  {Monacelli}}, \bibinfo {author} {\bibfnamefont {R.}~\bibnamefont {Bianco}},
  \bibinfo {author} {\bibfnamefont {M.}~\bibnamefont {Cherubini}}, \bibinfo
  {author} {\bibfnamefont {M.}~\bibnamefont {Calandra}}, \bibinfo {author}
  {\bibfnamefont {I.}~\bibnamefont {Errea}},\ and\ \bibinfo {author}
  {\bibfnamefont {F.}~\bibnamefont {Mauri}},\ }\bibfield  {title} {\bibinfo
  {title} {The stochastic self-consistent harmonic approximation: Calculating
  vibrational properties of materials with full quantum and anharmonic
  effects},\ }\href {https://doi.org/10.1088/1361-648X/ac066b} {\bibfield
  {journal} {\bibinfo  {journal} {Journal of Physics: Condensed Matter}\
  }\textbf {\bibinfo {volume} {33}},\ \bibinfo {pages} {363001} (\bibinfo
  {year} {2021})}\BibitemShut {NoStop}%
\bibitem [{\citenamefont {Riseborough}(1984)}]{riseborough1984}%
  \BibitemOpen
  \bibfield  {author} {\bibinfo {author} {\bibfnamefont {P.~S.}\ \bibnamefont
  {Riseborough}},\ }\bibfield  {title} {\bibinfo {title} {The small polaron
  with nonlinear electron-phonon interactions},\ }\href
  {https://doi.org/10.1016/0003-4916(84)90183-0} {\bibfield  {journal}
  {\bibinfo  {journal} {Annals of Physics}\ }\textbf {\bibinfo {volume}
  {153}},\ \bibinfo {pages} {1} (\bibinfo {year} {1984})}\BibitemShut {NoStop}%
\bibitem [{\citenamefont {Adolphs}\ and\ \citenamefont
  {Berciu}(2013)}]{adolphs2013}%
  \BibitemOpen
  \bibfield  {author} {\bibinfo {author} {\bibfnamefont {C.~P.~J.}\
  \bibnamefont {Adolphs}}\ and\ \bibinfo {author} {\bibfnamefont
  {M.}~\bibnamefont {Berciu}},\ }\bibfield  {title} {\bibinfo {title} {Going
  beyond the linear approximation in describing electron-phonon coupling:
  {{Relevance}} for the {{Holstein}} model},\ }\href
  {https://doi.org/10.1209/0295-5075/102/47003} {\bibfield  {journal} {\bibinfo
   {journal} {EPL (Europhysics Letters)}\ }\textbf {\bibinfo {volume} {102}},\
  \bibinfo {pages} {47003} (\bibinfo {year} {2013})}\BibitemShut {NoStop}%
\bibitem [{\citenamefont {Adolphs}\ and\ \citenamefont
  {Berciu}(2014)}]{adolphs2014}%
  \BibitemOpen
  \bibfield  {author} {\bibinfo {author} {\bibfnamefont {C.~P.~J.}\
  \bibnamefont {Adolphs}}\ and\ \bibinfo {author} {\bibfnamefont
  {M.}~\bibnamefont {Berciu}},\ }\bibfield  {title} {\bibinfo {title}
  {Single-polaron properties for double-well electron-phonon coupling},\ }\href
  {https://doi.org/10.1103/PhysRevB.89.035122} {\bibfield  {journal} {\bibinfo
  {journal} {Physical Review B}\ }\textbf {\bibinfo {volume} {89}},\ \bibinfo
  {pages} {035122} (\bibinfo {year} {2014})}\BibitemShut {NoStop}%
\bibitem [{\citenamefont {Li}\ and\ \citenamefont {Johnston}(2015)}]{li2015}%
  \BibitemOpen
  \bibfield  {author} {\bibinfo {author} {\bibfnamefont {S.}~\bibnamefont
  {Li}}\ and\ \bibinfo {author} {\bibfnamefont {S.}~\bibnamefont {Johnston}},\
  }\bibfield  {title} {\bibinfo {title} {The effects of non-linear
  electron-phonon interactions on superconductivity and charge-density-wave
  correlations},\ }\href {https://doi.org/10.1209/0295-5075/109/27007}
  {\bibfield  {journal} {\bibinfo  {journal} {EPL (Europhysics Letters)}\
  }\textbf {\bibinfo {volume} {109}},\ \bibinfo {pages} {27007} (\bibinfo
  {year} {2015})}\BibitemShut {NoStop}%
\bibitem [{\citenamefont {Li}\ \emph {et~al.}(2015)\citenamefont {Li},
  \citenamefont {Nowadnick},\ and\ \citenamefont {Johnston}}]{li2015a}%
  \BibitemOpen
  \bibfield  {author} {\bibinfo {author} {\bibfnamefont {S.}~\bibnamefont
  {Li}}, \bibinfo {author} {\bibfnamefont {E.~A.}\ \bibnamefont {Nowadnick}},\
  and\ \bibinfo {author} {\bibfnamefont {S.}~\bibnamefont {Johnston}},\
  }\bibfield  {title} {\bibinfo {title} {Quasiparticle properties of the
  nonlinear {{Holstein}} model at finite doping and temperature},\ }\href
  {https://doi.org/10.1103/PhysRevB.92.064301} {\bibfield  {journal} {\bibinfo
  {journal} {Physical Review B}\ }\textbf {\bibinfo {volume} {92}},\ \bibinfo
  {pages} {064301} (\bibinfo {year} {2015})}\BibitemShut {NoStop}%
\bibitem [{\citenamefont {Kennes}\ \emph {et~al.}(2017)\citenamefont {Kennes},
  \citenamefont {Wilner}, \citenamefont {Reichman},\ and\ \citenamefont
  {Millis}}]{kennes2017}%
  \BibitemOpen
  \bibfield  {author} {\bibinfo {author} {\bibfnamefont {D.~M.}\ \bibnamefont
  {Kennes}}, \bibinfo {author} {\bibfnamefont {E.~Y.}\ \bibnamefont {Wilner}},
  \bibinfo {author} {\bibfnamefont {D.~R.}\ \bibnamefont {Reichman}},\ and\
  \bibinfo {author} {\bibfnamefont {A.~J.}\ \bibnamefont {Millis}},\ }\bibfield
   {title} {\bibinfo {title} {Transient superconductivity from electronic
  squeezing of optically pumped phonons},\ }\href
  {https://doi.org/10.1038/nphys4024} {\bibfield  {journal} {\bibinfo
  {journal} {Nature Physics}\ }\textbf {\bibinfo {volume} {13}},\ \bibinfo
  {pages} {479} (\bibinfo {year} {2017})}\BibitemShut {NoStop}%
\bibitem [{\citenamefont {Sentef}(2017)}]{sentef2017}%
  \BibitemOpen
  \bibfield  {author} {\bibinfo {author} {\bibfnamefont {M.~A.}\ \bibnamefont
  {Sentef}},\ }\bibfield  {title} {\bibinfo {title} {Light-enhanced
  electron-phonon coupling from nonlinear electron-phonon coupling},\ }\href
  {https://doi.org/10.1103/PhysRevB.95.205111} {\bibfield  {journal} {\bibinfo
  {journal} {Physical Review B}\ }\textbf {\bibinfo {volume} {95}},\ \bibinfo
  {pages} {205111} (\bibinfo {year} {2017})}\BibitemShut {NoStop}%
\bibitem [{\citenamefont {Dee}\ \emph {et~al.}(2020)\citenamefont {Dee},
  \citenamefont {Coulter}, \citenamefont {Kleiner},\ and\ \citenamefont
  {Johnston}}]{dee2020}%
  \BibitemOpen
  \bibfield  {author} {\bibinfo {author} {\bibfnamefont {P.~M.}\ \bibnamefont
  {Dee}}, \bibinfo {author} {\bibfnamefont {J.}~\bibnamefont {Coulter}},
  \bibinfo {author} {\bibfnamefont {K.~G.}\ \bibnamefont {Kleiner}},\ and\
  \bibinfo {author} {\bibfnamefont {S.}~\bibnamefont {Johnston}},\ }\bibfield
  {title} {\bibinfo {title} {Relative importance of nonlinear electron-phonon
  coupling and vertex corrections in the {{Holstein}} model},\ }\href
  {https://doi.org/10.1038/s42005-020-00413-2} {\bibfield  {journal} {\bibinfo
  {journal} {Communications Physics}\ }\textbf {\bibinfo {volume} {3}},\
  \bibinfo {pages} {145} (\bibinfo {year} {2020})}\BibitemShut {NoStop}%
\bibitem [{\citenamefont {Grandi}\ \emph {et~al.}(2021)\citenamefont {Grandi},
  \citenamefont {Li},\ and\ \citenamefont {Eckstein}}]{grandi2021}%
  \BibitemOpen
  \bibfield  {author} {\bibinfo {author} {\bibfnamefont {F.}~\bibnamefont
  {Grandi}}, \bibinfo {author} {\bibfnamefont {J.}~\bibnamefont {Li}},\ and\
  \bibinfo {author} {\bibfnamefont {M.}~\bibnamefont {Eckstein}},\ }\bibfield
  {title} {\bibinfo {title} {Ultrafast {{Mott}} transition driven by nonlinear
  electron-phonon interaction},\ }\href
  {https://doi.org/10.1103/PhysRevB.103.L041110} {\bibfield  {journal}
  {\bibinfo  {journal} {Physical Review B}\ }\textbf {\bibinfo {volume}
  {103}},\ \bibinfo {pages} {L041110} (\bibinfo {year} {2021})}\BibitemShut
  {NoStop}%
\bibitem [{\citenamefont {Sous}\ \emph {et~al.}(2021)\citenamefont {Sous},
  \citenamefont {Kloss}, \citenamefont {Kennes}, \citenamefont {Reichman},\
  and\ \citenamefont {Millis}}]{sous2021}%
  \BibitemOpen
  \bibfield  {author} {\bibinfo {author} {\bibfnamefont {J.}~\bibnamefont
  {Sous}}, \bibinfo {author} {\bibfnamefont {B.}~\bibnamefont {Kloss}},
  \bibinfo {author} {\bibfnamefont {D.~M.}\ \bibnamefont {Kennes}}, \bibinfo
  {author} {\bibfnamefont {D.~R.}\ \bibnamefont {Reichman}},\ and\ \bibinfo
  {author} {\bibfnamefont {A.~J.}\ \bibnamefont {Millis}},\ }\bibfield  {title}
  {\bibinfo {title} {Phonon-induced disorder in dynamics of optically pumped
  metals from nonlinear electron-phonon coupling},\ }\href
  {https://doi.org/10.1038/s41467-021-26030-3} {\bibfield  {journal} {\bibinfo
  {journal} {Nature Communications}\ }\textbf {\bibinfo {volume} {12}},\
  \bibinfo {pages} {5803} (\bibinfo {year} {2021})}\BibitemShut {NoStop}%
\bibitem [{\citenamefont {Prokof'ev}\ and\ \citenamefont
  {Svistunov}(2022)}]{prokofev2022}%
  \BibitemOpen
  \bibfield  {author} {\bibinfo {author} {\bibfnamefont {N.~V.}\ \bibnamefont
  {Prokof'ev}}\ and\ \bibinfo {author} {\bibfnamefont {B.~V.}\ \bibnamefont
  {Svistunov}},\ }\bibfield  {title} {\bibinfo {title} {Phonon modulated
  hopping polarons: x -representation technique},\ }\href
  {https://doi.org/10.1103/PhysRevB.106.L041117} {\bibfield  {journal}
  {\bibinfo  {journal} {Physical Review B}\ }\textbf {\bibinfo {volume}
  {106}},\ \bibinfo {pages} {L041117} (\bibinfo {year} {2022})}\BibitemShut
  {NoStop}%
\bibitem [{\citenamefont {Ragni}\ \emph {et~al.}(2023)\citenamefont {Ragni},
  \citenamefont {Hahn}, \citenamefont {Zhang}, \citenamefont {Prokof'ev},
  \citenamefont {Kuklov}, \citenamefont {Klimin}, \citenamefont {Houtput},
  \citenamefont {Svistunov}, \citenamefont {Tempere}, \citenamefont {Nagaosa},
  \citenamefont {Franchini},\ and\ \citenamefont {Mishchenko}}]{ragni2023}%
  \BibitemOpen
  \bibfield  {author} {\bibinfo {author} {\bibfnamefont {S.}~\bibnamefont
  {Ragni}}, \bibinfo {author} {\bibfnamefont {T.}~\bibnamefont {Hahn}},
  \bibinfo {author} {\bibfnamefont {Z.}~\bibnamefont {Zhang}}, \bibinfo
  {author} {\bibfnamefont {N.}~\bibnamefont {Prokof'ev}}, \bibinfo {author}
  {\bibfnamefont {A.}~\bibnamefont {Kuklov}}, \bibinfo {author} {\bibfnamefont
  {S.}~\bibnamefont {Klimin}}, \bibinfo {author} {\bibfnamefont
  {M.}~\bibnamefont {Houtput}}, \bibinfo {author} {\bibfnamefont
  {B.}~\bibnamefont {Svistunov}}, \bibinfo {author} {\bibfnamefont
  {J.}~\bibnamefont {Tempere}}, \bibinfo {author} {\bibfnamefont
  {N.}~\bibnamefont {Nagaosa}}, \bibinfo {author} {\bibfnamefont
  {C.}~\bibnamefont {Franchini}},\ and\ \bibinfo {author} {\bibfnamefont
  {A.~S.}\ \bibnamefont {Mishchenko}},\ }\bibfield  {title} {\bibinfo {title}
  {Polaron with quadratic electron-phonon interaction},\ }\href
  {https://doi.org/10.1103/PhysRevB.107.L121109} {\bibfield  {journal}
  {\bibinfo  {journal} {Physical Review B}\ }\textbf {\bibinfo {volume}
  {107}},\ \bibinfo {pages} {L121109} (\bibinfo {year} {2023})}\BibitemShut
  {NoStop}%
\bibitem [{\citenamefont {Zhang}\ \emph {et~al.}(2023)\citenamefont {Zhang},
  \citenamefont {Kuklov}, \citenamefont {Prokof'ev},\ and\ \citenamefont
  {Svistunov}}]{zhang2023}%
  \BibitemOpen
  \bibfield  {author} {\bibinfo {author} {\bibfnamefont {Z.}~\bibnamefont
  {Zhang}}, \bibinfo {author} {\bibfnamefont {A.}~\bibnamefont {Kuklov}},
  \bibinfo {author} {\bibfnamefont {N.}~\bibnamefont {Prokof'ev}},\ and\
  \bibinfo {author} {\bibfnamefont {B.}~\bibnamefont {Svistunov}},\ }\bibfield
  {title} {\bibinfo {title} {Soliton states from quadratic electron-phonon
  interaction},\ }\href {https://doi.org/10.1103/PhysRevB.108.245127}
  {\bibfield  {journal} {\bibinfo  {journal} {Physical Review B}\ }\textbf
  {\bibinfo {volume} {108}},\ \bibinfo {pages} {245127} (\bibinfo {year}
  {2023})}\BibitemShut {NoStop}%
\bibitem [{\citenamefont {Han}\ \emph {et~al.}(2024)\citenamefont {Han},
  \citenamefont {Kivelson},\ and\ \citenamefont {Volkov}}]{han2024}%
  \BibitemOpen
  \bibfield  {author} {\bibinfo {author} {\bibfnamefont {Z.}~\bibnamefont
  {Han}}, \bibinfo {author} {\bibfnamefont {S.~A.}\ \bibnamefont {Kivelson}},\
  and\ \bibinfo {author} {\bibfnamefont {P.~A.}\ \bibnamefont {Volkov}},\
  }\bibfield  {title} {\bibinfo {title} {Quantum {{Bipolaron
  Superconductivity}} from {{Quadratic Electron-Phonon Coupling}}},\ }\href
  {https://doi.org/10.1103/PhysRevLett.132.226001} {\bibfield  {journal}
  {\bibinfo  {journal} {Physical Review Letters}\ }\textbf {\bibinfo {volume}
  {132}},\ \bibinfo {pages} {226001} (\bibinfo {year} {2024})}\BibitemShut
  {NoStop}%
\bibitem [{\citenamefont {Kova{\v c}}\ \emph {et~al.}(2024)\citenamefont
  {Kova{\v c}}, \citenamefont {Gole{\v z}}, \citenamefont {Mierzejewski},\ and\
  \citenamefont {Bon{\v c}a}}]{kovac2024}%
  \BibitemOpen
  \bibfield  {author} {\bibinfo {author} {\bibfnamefont {K.}~\bibnamefont
  {Kova{\v c}}}, \bibinfo {author} {\bibfnamefont {D.}~\bibnamefont {Gole{\v
  z}}}, \bibinfo {author} {\bibfnamefont {M.}~\bibnamefont {Mierzejewski}},\
  and\ \bibinfo {author} {\bibfnamefont {J.}~\bibnamefont {Bon{\v c}a}},\
  }\bibfield  {title} {\bibinfo {title} {Optical {{Manipulation}} of
  {{Bipolarons}} in a {{System}} with {{Nonlinear Electron-Phonon Coupling}}},\
  }\href {https://doi.org/10.1103/PhysRevLett.132.106001} {\bibfield  {journal}
  {\bibinfo  {journal} {Physical Review Letters}\ }\textbf {\bibinfo {volume}
  {132}},\ \bibinfo {pages} {106001} (\bibinfo {year} {2024})}\BibitemShut
  {NoStop}%
\bibitem [{\citenamefont {Klimin}\ \emph {et~al.}(2024)\citenamefont {Klimin},
  \citenamefont {Tempere}, \citenamefont {Houtput}, \citenamefont {Ragni},
  \citenamefont {Hahn}, \citenamefont {Franchini},\ and\ \citenamefont
  {Mishchenko}}]{klimin2024}%
  \BibitemOpen
  \bibfield  {author} {\bibinfo {author} {\bibfnamefont {S.~N.}\ \bibnamefont
  {Klimin}}, \bibinfo {author} {\bibfnamefont {J.}~\bibnamefont {Tempere}},
  \bibinfo {author} {\bibfnamefont {M.}~\bibnamefont {Houtput}}, \bibinfo
  {author} {\bibfnamefont {S.}~\bibnamefont {Ragni}}, \bibinfo {author}
  {\bibfnamefont {T.}~\bibnamefont {Hahn}}, \bibinfo {author} {\bibfnamefont
  {C.}~\bibnamefont {Franchini}},\ and\ \bibinfo {author} {\bibfnamefont
  {A.~S.}\ \bibnamefont {Mishchenko}},\ }\bibfield  {title} {\bibinfo {title}
  {Analytic method for quadratic polarons in nonparabolic bands},\ }\href
  {https://doi.org/10.1103/PhysRevB.110.075107} {\bibfield  {journal} {\bibinfo
   {journal} {Physical Review B}\ }\textbf {\bibinfo {volume} {110}},\ \bibinfo
  {pages} {075107} (\bibinfo {year} {2024})}\BibitemShut {NoStop}%
\bibitem [{\citenamefont {Ngai}(1974)}]{ngai1974}%
  \BibitemOpen
  \bibfield  {author} {\bibinfo {author} {\bibfnamefont {K.~L.}\ \bibnamefont
  {Ngai}},\ }\bibfield  {title} {\bibinfo {title} {Two-{{Phonon Deformation
  Potential}} and {{Superconductivity}} in {{Degenerate Semiconductors}}},\
  }\href {https://doi.org/10.1103/PhysRevLett.32.215} {\bibfield  {journal}
  {\bibinfo  {journal} {Physical Review Letters}\ }\textbf {\bibinfo {volume}
  {32}},\ \bibinfo {pages} {215} (\bibinfo {year} {1974})}\BibitemShut
  {NoStop}%
\bibitem [{\citenamefont {Epifanov}\ \emph {et~al.}(1981)\citenamefont
  {Epifanov}, \citenamefont {Levanyuk},\ and\ \citenamefont
  {Levanyuk}}]{epifanov1981}%
  \BibitemOpen
  \bibfield  {author} {\bibinfo {author} {\bibfnamefont {{\relax Yu}.~N.}\
  \bibnamefont {Epifanov}}, \bibinfo {author} {\bibfnamefont {A.~P.}\
  \bibnamefont {Levanyuk}},\ and\ \bibinfo {author} {\bibfnamefont {G.~M.}\
  \bibnamefont {Levanyuk}},\ }\bibfield  {title} {\bibinfo {title} {Interaction
  of carriers with to-phonons and electrical conductivity of ferroelectrics},\
  }\href {https://doi.org/10.1080/00150198108017687} {\bibfield  {journal}
  {\bibinfo  {journal} {Ferroelectrics}\ }\textbf {\bibinfo {volume} {35}},\
  \bibinfo {pages} {199} (\bibinfo {year} {1981})}\BibitemShut {NoStop}%
\bibitem [{\citenamefont {Van Der~Marel}\ \emph {et~al.}(2019)\citenamefont
  {Van Der~Marel}, \citenamefont {Barantani},\ and\ \citenamefont
  {Rischau}}]{vandermarel2019}%
  \BibitemOpen
  \bibfield  {author} {\bibinfo {author} {\bibfnamefont {D.}~\bibnamefont {Van
  Der~Marel}}, \bibinfo {author} {\bibfnamefont {F.}~\bibnamefont
  {Barantani}},\ and\ \bibinfo {author} {\bibfnamefont {C.~W.}\ \bibnamefont
  {Rischau}},\ }\bibfield  {title} {\bibinfo {title} {Possible mechanism for
  superconductivity in doped {{SrTiO}} 3},\ }\href
  {https://doi.org/10.1103/PhysRevResearch.1.013003} {\bibfield  {journal}
  {\bibinfo  {journal} {Physical Review Research}\ }\textbf {\bibinfo {volume}
  {1}},\ \bibinfo {pages} {013003} (\bibinfo {year} {2019})}\BibitemShut
  {NoStop}%
\bibitem [{\citenamefont {Kiselov}\ and\ \citenamefont
  {Feigel'man}(2021)}]{kiselov2021}%
  \BibitemOpen
  \bibfield  {author} {\bibinfo {author} {\bibfnamefont {D.~E.}\ \bibnamefont
  {Kiselov}}\ and\ \bibinfo {author} {\bibfnamefont {M.~V.}\ \bibnamefont
  {Feigel'man}},\ }\bibfield  {title} {\bibinfo {title} {Theory of
  superconductivity due to {{Ngai}}'s mechanism in lightly doped
  {{SrTi${\mathrm{O}}_{3}$}}},\ }\href
  {https://doi.org/10.1103/PhysRevB.104.L220506} {\bibfield  {journal}
  {\bibinfo  {journal} {Physical Review B}\ }\textbf {\bibinfo {volume}
  {104}},\ \bibinfo {pages} {L220506} (\bibinfo {year} {2021})}\BibitemShut
  {NoStop}%
\bibitem [{\citenamefont {Kumar}\ \emph {et~al.}(2021)\citenamefont {Kumar},
  \citenamefont {Yudson},\ and\ \citenamefont {Maslov}}]{kumar2021}%
  \BibitemOpen
  \bibfield  {author} {\bibinfo {author} {\bibfnamefont {A.}~\bibnamefont
  {Kumar}}, \bibinfo {author} {\bibfnamefont {V.~I.}\ \bibnamefont {Yudson}},\
  and\ \bibinfo {author} {\bibfnamefont {D.~L.}\ \bibnamefont {Maslov}},\
  }\bibfield  {title} {\bibinfo {title} {Quasiparticle and {{Nonquasiparticle
  Transport}} in {{Doped Quantum Paraelectrics}}},\ }\href
  {https://doi.org/10.1103/PhysRevLett.126.076601} {\bibfield  {journal}
  {\bibinfo  {journal} {Physical Review Letters}\ }\textbf {\bibinfo {volume}
  {126}},\ \bibinfo {pages} {076601} (\bibinfo {year} {2021})}\BibitemShut
  {NoStop}%
\bibitem [{\citenamefont {Nazaryan}\ and\ \citenamefont
  {Feigel'man}(2021)}]{nazaryan2021}%
  \BibitemOpen
  \bibfield  {author} {\bibinfo {author} {\bibfnamefont {{\relax Kh}.~G.}\
  \bibnamefont {Nazaryan}}\ and\ \bibinfo {author} {\bibfnamefont {M.~V.}\
  \bibnamefont {Feigel'man}},\ }\bibfield  {title} {\bibinfo {title}
  {Conductivity and thermoelectric coefficients of doped
  {{SrTi${\mathrm{O}}_{3}$}} at high temperatures},\ }\href
  {https://doi.org/10.1103/PhysRevB.104.115201} {\bibfield  {journal} {\bibinfo
   {journal} {Physical Review B}\ }\textbf {\bibinfo {volume} {104}},\ \bibinfo
  {pages} {115201} (\bibinfo {year} {2021})}\BibitemShut {NoStop}%
\bibitem [{\citenamefont {Zacharias}\ and\ \citenamefont
  {Giustino}(2016)}]{zacharias2016}%
  \BibitemOpen
  \bibfield  {author} {\bibinfo {author} {\bibfnamefont {M.}~\bibnamefont
  {Zacharias}}\ and\ \bibinfo {author} {\bibfnamefont {F.}~\bibnamefont
  {Giustino}},\ }\bibfield  {title} {\bibinfo {title} {One-shot calculation of
  temperature-dependent optical spectra and phonon-induced band-gap
  renormalization},\ }\href {https://doi.org/10.1103/PhysRevB.94.075125}
  {\bibfield  {journal} {\bibinfo  {journal} {Physical Review B}\ }\textbf
  {\bibinfo {volume} {94}},\ \bibinfo {pages} {075125} (\bibinfo {year}
  {2016})}\BibitemShut {NoStop}%
\bibitem [{\citenamefont {Monserrat}(2018)}]{monserrat2018}%
  \BibitemOpen
  \bibfield  {author} {\bibinfo {author} {\bibfnamefont {B.}~\bibnamefont
  {Monserrat}},\ }\bibfield  {title} {\bibinfo {title} {Electron--phonon
  coupling from finite differences},\ }\href
  {https://doi.org/10.1088/1361-648X/aaa737} {\bibfield  {journal} {\bibinfo
  {journal} {Journal of Physics: Condensed Matter}\ }\textbf {\bibinfo {volume}
  {30}},\ \bibinfo {pages} {083001} (\bibinfo {year} {2018})}\BibitemShut
  {NoStop}%
\bibitem [{\citenamefont {Zacharias}\ \emph {et~al.}(2020)\citenamefont
  {Zacharias}, \citenamefont {Scheffler},\ and\ \citenamefont
  {Carbogno}}]{zacharias2020}%
  \BibitemOpen
  \bibfield  {author} {\bibinfo {author} {\bibfnamefont {M.}~\bibnamefont
  {Zacharias}}, \bibinfo {author} {\bibfnamefont {M.}~\bibnamefont
  {Scheffler}},\ and\ \bibinfo {author} {\bibfnamefont {C.}~\bibnamefont
  {Carbogno}},\ }\bibfield  {title} {\bibinfo {title} {Fully anharmonic
  nonperturbative theory of vibronically renormalized electronic band
  structures},\ }\href {https://doi.org/10.1103/PhysRevB.102.045126} {\bibfield
   {journal} {\bibinfo  {journal} {Physical Review B}\ }\textbf {\bibinfo
  {volume} {102}},\ \bibinfo {pages} {045126} (\bibinfo {year}
  {2020})}\BibitemShut {NoStop}%
\bibitem [{\citenamefont {Kundu}\ \emph {et~al.}(2021)\citenamefont {Kundu},
  \citenamefont {Govoni}, \citenamefont {Yang}, \citenamefont {Ceriotti},
  \citenamefont {Gygi},\ and\ \citenamefont {Galli}}]{kundu2021}%
  \BibitemOpen
  \bibfield  {author} {\bibinfo {author} {\bibfnamefont {A.}~\bibnamefont
  {Kundu}}, \bibinfo {author} {\bibfnamefont {M.}~\bibnamefont {Govoni}},
  \bibinfo {author} {\bibfnamefont {H.}~\bibnamefont {Yang}}, \bibinfo {author}
  {\bibfnamefont {M.}~\bibnamefont {Ceriotti}}, \bibinfo {author}
  {\bibfnamefont {F.}~\bibnamefont {Gygi}},\ and\ \bibinfo {author}
  {\bibfnamefont {G.}~\bibnamefont {Galli}},\ }\bibfield  {title} {\bibinfo
  {title} {Quantum vibronic effects on the electronic properties of solid and
  molecular carbon},\ }\href
  {https://doi.org/10.1103/PhysRevMaterials.5.L070801} {\bibfield  {journal}
  {\bibinfo  {journal} {Physical Review Materials}\ }\textbf {\bibinfo {volume}
  {5}},\ \bibinfo {pages} {L070801} (\bibinfo {year} {2021})}\BibitemShut
  {NoStop}%
\bibitem [{\citenamefont {Yildirim}\ \emph {et~al.}(2001)\citenamefont
  {Yildirim}, \citenamefont {G{\"u}lseren}, \citenamefont {Lynn}, \citenamefont
  {Brown}, \citenamefont {Udovic}, \citenamefont {Huang}, \citenamefont
  {Rogado}, \citenamefont {Regan}, \citenamefont {Hayward}, \citenamefont
  {Slusky}, \citenamefont {He}, \citenamefont {Haas}, \citenamefont {Khalifah},
  \citenamefont {Inumaru},\ and\ \citenamefont {Cava}}]{yildirim2001}%
  \BibitemOpen
  \bibfield  {author} {\bibinfo {author} {\bibfnamefont {T.}~\bibnamefont
  {Yildirim}}, \bibinfo {author} {\bibfnamefont {O.}~\bibnamefont
  {G{\"u}lseren}}, \bibinfo {author} {\bibfnamefont {J.~W.}\ \bibnamefont
  {Lynn}}, \bibinfo {author} {\bibfnamefont {C.~M.}\ \bibnamefont {Brown}},
  \bibinfo {author} {\bibfnamefont {T.~J.}\ \bibnamefont {Udovic}}, \bibinfo
  {author} {\bibfnamefont {Q.}~\bibnamefont {Huang}}, \bibinfo {author}
  {\bibfnamefont {N.}~\bibnamefont {Rogado}}, \bibinfo {author} {\bibfnamefont
  {K.~A.}\ \bibnamefont {Regan}}, \bibinfo {author} {\bibfnamefont {M.~A.}\
  \bibnamefont {Hayward}}, \bibinfo {author} {\bibfnamefont {J.~S.}\
  \bibnamefont {Slusky}}, \bibinfo {author} {\bibfnamefont {T.}~\bibnamefont
  {He}}, \bibinfo {author} {\bibfnamefont {M.~K.}\ \bibnamefont {Haas}},
  \bibinfo {author} {\bibfnamefont {P.}~\bibnamefont {Khalifah}}, \bibinfo
  {author} {\bibfnamefont {K.}~\bibnamefont {Inumaru}},\ and\ \bibinfo {author}
  {\bibfnamefont {R.~J.}\ \bibnamefont {Cava}},\ }\bibfield  {title} {\bibinfo
  {title} {Giant {{Anharmonicity}} and {{Nonlinear Electron-Phonon Coupling}}
  in {{Mg${\mathrm{B}}_{2}$}} : {{A Combined First-Principles Calculation}} and
  {{Neutron Scattering Study}}},\ }\href
  {https://doi.org/10.1103/PhysRevLett.87.037001} {\bibfield  {journal}
  {\bibinfo  {journal} {Physical Review Letters}\ }\textbf {\bibinfo {volume}
  {87}},\ \bibinfo {pages} {037001} (\bibinfo {year} {2001})}\BibitemShut
  {NoStop}%
\bibitem [{\citenamefont {Liu}\ \emph {et~al.}(2001)\citenamefont {Liu},
  \citenamefont {Mazin},\ and\ \citenamefont {Kortus}}]{liu2001}%
  \BibitemOpen
  \bibfield  {author} {\bibinfo {author} {\bibfnamefont {A.~Y.}\ \bibnamefont
  {Liu}}, \bibinfo {author} {\bibfnamefont {I.~I.}\ \bibnamefont {Mazin}},\
  and\ \bibinfo {author} {\bibfnamefont {J.}~\bibnamefont {Kortus}},\
  }\bibfield  {title} {\bibinfo {title} {Beyond {{Eliashberg
  Superconductivity}} in {{Mg${\mathrm{B}}_{2}$}}: {{Anharmonicity}},
  {{Two-Phonon Scattering}}, and {{Multiple Gaps}}},\ }\href
  {https://doi.org/10.1103/PhysRevLett.87.087005} {\bibfield  {journal}
  {\bibinfo  {journal} {Physical Review Letters}\ }\textbf {\bibinfo {volume}
  {87}},\ \bibinfo {pages} {087005} (\bibinfo {year} {2001})}\BibitemShut
  {NoStop}%
\bibitem [{\citenamefont {Bianco}\ and\ \citenamefont
  {Errea}(2023)}]{bianco2023}%
  \BibitemOpen
  \bibfield  {author} {\bibinfo {author} {\bibfnamefont {R.}~\bibnamefont
  {Bianco}}\ and\ \bibinfo {author} {\bibfnamefont {I.}~\bibnamefont {Errea}},\
  }\href {https://doi.org/10.48550/arXiv.2303.02621} {\bibinfo {title}
  {Non-perturbative theory of the electron-phonon coupling and its
  first-principles implementation}} (\bibinfo {year} {2023}),\ \Eprint
  {https://arxiv.org/abs/2303.02621} {arXiv:2303.02621 [cond-mat]} \BibitemShut
  {NoStop}%
\bibitem [{\citenamefont {Giustino}\ \emph {et~al.}(2010)\citenamefont
  {Giustino}, \citenamefont {Louie},\ and\ \citenamefont
  {Cohen}}]{giustino2010}%
  \BibitemOpen
  \bibfield  {author} {\bibinfo {author} {\bibfnamefont {F.}~\bibnamefont
  {Giustino}}, \bibinfo {author} {\bibfnamefont {S.~G.}\ \bibnamefont
  {Louie}},\ and\ \bibinfo {author} {\bibfnamefont {M.~L.}\ \bibnamefont
  {Cohen}},\ }\bibfield  {title} {\bibinfo {title} {Electron-{{Phonon
  Renormalization}} of the {{Direct Band Gap}} of {{Diamond}}},\ }\href
  {https://doi.org/10.1103/PhysRevLett.105.265501} {\bibfield  {journal}
  {\bibinfo  {journal} {Physical Review Letters}\ }\textbf {\bibinfo {volume}
  {105}},\ \bibinfo {pages} {265501} (\bibinfo {year} {2010})}\BibitemShut
  {NoStop}%
\bibitem [{\citenamefont {Landau}\ \emph {et~al.}(2013)\citenamefont {Landau},
  \citenamefont {Bell}, \citenamefont {Kearsley}, \citenamefont {Pitaevskii},
  \citenamefont {Lifshitz},\ and\ \citenamefont {Sykes}}]{landau2013}%
  \BibitemOpen
  \bibfield  {author} {\bibinfo {author} {\bibfnamefont {L.~D.}\ \bibnamefont
  {Landau}}, \bibinfo {author} {\bibfnamefont {J.~S.}\ \bibnamefont {Bell}},
  \bibinfo {author} {\bibfnamefont {M.~J.}\ \bibnamefont {Kearsley}}, \bibinfo
  {author} {\bibfnamefont {L.~P.}\ \bibnamefont {Pitaevskii}}, \bibinfo
  {author} {\bibfnamefont {E.~M.}\ \bibnamefont {Lifshitz}},\ and\ \bibinfo
  {author} {\bibfnamefont {J.~B.}\ \bibnamefont {Sykes}},\ }\href@noop {}
  {\emph {\bibinfo {title} {Electrodynamics of {{Continuous Media}}}}}\
  (\bibinfo  {publisher} {Elsevier},\ \bibinfo {year} {2013})\BibitemShut
  {NoStop}%
\bibitem [{\citenamefont {Nunes}\ and\ \citenamefont
  {Gonze}(2001)}]{nunes2001}%
  \BibitemOpen
  \bibfield  {author} {\bibinfo {author} {\bibfnamefont {R.~W.}\ \bibnamefont
  {Nunes}}\ and\ \bibinfo {author} {\bibfnamefont {X.}~\bibnamefont {Gonze}},\
  }\bibfield  {title} {\bibinfo {title} {Berry-phase treatment of the
  homogeneous electric field perturbation in insulators},\ }\href
  {https://doi.org/10.1103/PhysRevB.63.155107} {\bibfield  {journal} {\bibinfo
  {journal} {Physical Review B}\ }\textbf {\bibinfo {volume} {63}},\ \bibinfo
  {pages} {155107} (\bibinfo {year} {2001})}\BibitemShut {NoStop}%
\bibitem [{\citenamefont {Souza}\ \emph {et~al.}(2002)\citenamefont {Souza},
  \citenamefont {{\'I}{\~n}iguez},\ and\ \citenamefont
  {Vanderbilt}}]{souza2002}%
  \BibitemOpen
  \bibfield  {author} {\bibinfo {author} {\bibfnamefont {I.}~\bibnamefont
  {Souza}}, \bibinfo {author} {\bibfnamefont {J.}~\bibnamefont
  {{\'I}{\~n}iguez}},\ and\ \bibinfo {author} {\bibfnamefont {D.}~\bibnamefont
  {Vanderbilt}},\ }\bibfield  {title} {\bibinfo {title} {First-{{Principles
  Approach}} to {{Insulators}} in {{Finite Electric Fields}}},\ }\href
  {https://doi.org/10.1103/PhysRevLett.89.117602} {\bibfield  {journal}
  {\bibinfo  {journal} {Physical Review Letters}\ }\textbf {\bibinfo {volume}
  {89}},\ \bibinfo {pages} {117602} (\bibinfo {year} {2002})}\BibitemShut
  {NoStop}%
\bibitem [{\citenamefont {Gonze}\ and\ \citenamefont {Lee}(1997)}]{gonze1997}%
  \BibitemOpen
  \bibfield  {author} {\bibinfo {author} {\bibfnamefont {X.}~\bibnamefont
  {Gonze}}\ and\ \bibinfo {author} {\bibfnamefont {C.}~\bibnamefont {Lee}},\
  }\bibfield  {title} {\bibinfo {title} {Dynamical matrices, {{Born}} effective
  charges, dielectric permittivity tensors, and interatomic force constants
  from density-functional perturbation theory},\ }\href
  {https://doi.org/10.1103/PhysRevB.55.10355} {\bibfield  {journal} {\bibinfo
  {journal} {Physical Review B}\ }\textbf {\bibinfo {volume} {55}},\ \bibinfo
  {pages} {10355} (\bibinfo {year} {1997})}\BibitemShut {NoStop}%
\bibitem [{\citenamefont {Srivastava}(1990)}]{srivastava1990}%
  \BibitemOpen
  \bibfield  {author} {\bibinfo {author} {\bibfnamefont {G.~P.}\ \bibnamefont
  {Srivastava}},\ }\href@noop {} {\emph {\bibinfo {title} {The Physics of
  Phonons}}}\ (\bibinfo  {publisher} {A. Hilger},\ \bibinfo {address} {Bristol
  Philadelphia New York},\ \bibinfo {year} {1990})\BibitemShut {NoStop}%
\bibitem [{\citenamefont {Guster}\ \emph {et~al.}(2022)\citenamefont {Guster},
  \citenamefont {Melo}, \citenamefont {Martin}, \citenamefont
  {{Brousseau-Couture}}, \citenamefont {De~Abreu}, \citenamefont {Miglio},
  \citenamefont {Giantomassi}, \citenamefont {C{\^o}t{\'e}}, \citenamefont
  {Frost}, \citenamefont {Verstraete},\ and\ \citenamefont
  {Gonze}}]{guster2022}%
  \BibitemOpen
  \bibfield  {author} {\bibinfo {author} {\bibfnamefont {B.}~\bibnamefont
  {Guster}}, \bibinfo {author} {\bibfnamefont {P.}~\bibnamefont {Melo}},
  \bibinfo {author} {\bibfnamefont {B.~A.~A.}\ \bibnamefont {Martin}}, \bibinfo
  {author} {\bibfnamefont {V.}~\bibnamefont {{Brousseau-Couture}}}, \bibinfo
  {author} {\bibfnamefont {J.~C.}\ \bibnamefont {De~Abreu}}, \bibinfo {author}
  {\bibfnamefont {A.}~\bibnamefont {Miglio}}, \bibinfo {author} {\bibfnamefont
  {M.}~\bibnamefont {Giantomassi}}, \bibinfo {author} {\bibfnamefont
  {M.}~\bibnamefont {C{\^o}t{\'e}}}, \bibinfo {author} {\bibfnamefont {J.~M.}\
  \bibnamefont {Frost}}, \bibinfo {author} {\bibfnamefont {M.~J.}\ \bibnamefont
  {Verstraete}},\ and\ \bibinfo {author} {\bibfnamefont {X.}~\bibnamefont
  {Gonze}},\ }\bibfield  {title} {\bibinfo {title} {Erratum: {{Fr{\"o}hlich}}
  polaron effective mass and localization length in cubic materials:
  {{Degenerate}} and anisotropic electronic bands [{{Phys}}. {{Rev}}. {{B}}
  {\textbf{104}} , 235123 (2021)]},\ }\href
  {https://doi.org/10.1103/PhysRevB.105.119902} {\bibfield  {journal} {\bibinfo
   {journal} {Physical Review B}\ }\textbf {\bibinfo {volume} {105}},\ \bibinfo
  {pages} {119902} (\bibinfo {year} {2022})}\BibitemShut {NoStop}%
\bibitem [{\citenamefont {Mahan}(2000)}]{mahan2000}%
  \BibitemOpen
  \bibfield  {author} {\bibinfo {author} {\bibfnamefont {G.~D.}\ \bibnamefont
  {Mahan}},\ }\href@noop {} {\emph {\bibinfo {title} {Many Particle
  Physics}}},\ \bibinfo {edition} {3rd}\ ed.,\ Physics of Solids and Liquids\
  (\bibinfo  {publisher} {Kluwer academic/Plenum pub},\ \bibinfo {address} {New
  York},\ \bibinfo {year} {2000})\BibitemShut {NoStop}%
\bibitem [{\citenamefont {Maradudin}\ and\ \citenamefont
  {Vosko}(1968)}]{maradudin1968}%
  \BibitemOpen
  \bibfield  {author} {\bibinfo {author} {\bibfnamefont {A.~A.}\ \bibnamefont
  {Maradudin}}\ and\ \bibinfo {author} {\bibfnamefont {S.~H.}\ \bibnamefont
  {Vosko}},\ }\bibfield  {title} {\bibinfo {title} {Symmetry {{Properties}} of
  the {{Normal Vibrations}} of a {{Crystal}}},\ }\href
  {https://doi.org/10.1103/RevModPhys.40.1} {\bibfield  {journal} {\bibinfo
  {journal} {Reviews of Modern Physics}\ }\textbf {\bibinfo {volume} {40}},\
  \bibinfo {pages} {1} (\bibinfo {year} {1968})}\BibitemShut {NoStop}%
\bibitem [{\citenamefont {Miglio}\ \emph {et~al.}(2020)\citenamefont {Miglio},
  \citenamefont {{Brousseau-Couture}}, \citenamefont {Godbout}, \citenamefont
  {Antonius}, \citenamefont {Chan}, \citenamefont {Louie}, \citenamefont
  {C{\^o}t{\'e}}, \citenamefont {Giantomassi},\ and\ \citenamefont
  {Gonze}}]{miglio2020}%
  \BibitemOpen
  \bibfield  {author} {\bibinfo {author} {\bibfnamefont {A.}~\bibnamefont
  {Miglio}}, \bibinfo {author} {\bibfnamefont {V.}~\bibnamefont
  {{Brousseau-Couture}}}, \bibinfo {author} {\bibfnamefont {E.}~\bibnamefont
  {Godbout}}, \bibinfo {author} {\bibfnamefont {G.}~\bibnamefont {Antonius}},
  \bibinfo {author} {\bibfnamefont {Y.-H.}\ \bibnamefont {Chan}}, \bibinfo
  {author} {\bibfnamefont {S.~G.}\ \bibnamefont {Louie}}, \bibinfo {author}
  {\bibfnamefont {M.}~\bibnamefont {C{\^o}t{\'e}}}, \bibinfo {author}
  {\bibfnamefont {M.}~\bibnamefont {Giantomassi}},\ and\ \bibinfo {author}
  {\bibfnamefont {X.}~\bibnamefont {Gonze}},\ }\bibfield  {title} {\bibinfo
  {title} {Predominance of non-adiabatic effects in zero-point renormalization
  of the electronic band gap},\ }\href
  {https://doi.org/10.1038/s41524-020-00434-z} {\bibfield  {journal} {\bibinfo
  {journal} {npj Computational Materials}\ }\textbf {\bibinfo {volume} {6}},\
  \bibinfo {pages} {167} (\bibinfo {year} {2020})}\BibitemShut {NoStop}%
\bibitem [{\citenamefont {Houtput}\ and\ \citenamefont
  {Tempere}(2021)}]{houtput2021}%
  \BibitemOpen
  \bibfield  {author} {\bibinfo {author} {\bibfnamefont {M.}~\bibnamefont
  {Houtput}}\ and\ \bibinfo {author} {\bibfnamefont {J.}~\bibnamefont
  {Tempere}},\ }\bibfield  {title} {\bibinfo {title} {Beyond the {{Fr{\"o}hlich
  Hamiltonian}}: {{Path-integral}} treatment of large polarons in anharmonic
  solids},\ }\href {https://doi.org/10.1103/PhysRevB.103.184306} {\bibfield
  {journal} {\bibinfo  {journal} {Physical Review B}\ }\textbf {\bibinfo
  {volume} {103}},\ \bibinfo {pages} {184306} (\bibinfo {year}
  {2021})}\BibitemShut {NoStop}%
\bibitem [{\citenamefont {Brunin}\ \emph
  {et~al.}(2020{\natexlab{a}})\citenamefont {Brunin}, \citenamefont {Miranda},
  \citenamefont {Giantomassi}, \citenamefont {Royo}, \citenamefont {Stengel},
  \citenamefont {Verstraete}, \citenamefont {Gonze}, \citenamefont
  {Rignanese},\ and\ \citenamefont {Hautier}}]{brunin2020}%
  \BibitemOpen
  \bibfield  {author} {\bibinfo {author} {\bibfnamefont {G.}~\bibnamefont
  {Brunin}}, \bibinfo {author} {\bibfnamefont {H.~P.~C.}\ \bibnamefont
  {Miranda}}, \bibinfo {author} {\bibfnamefont {M.}~\bibnamefont
  {Giantomassi}}, \bibinfo {author} {\bibfnamefont {M.}~\bibnamefont {Royo}},
  \bibinfo {author} {\bibfnamefont {M.}~\bibnamefont {Stengel}}, \bibinfo
  {author} {\bibfnamefont {M.~J.}\ \bibnamefont {Verstraete}}, \bibinfo
  {author} {\bibfnamefont {X.}~\bibnamefont {Gonze}}, \bibinfo {author}
  {\bibfnamefont {G.-M.}\ \bibnamefont {Rignanese}},\ and\ \bibinfo {author}
  {\bibfnamefont {G.}~\bibnamefont {Hautier}},\ }\bibfield  {title} {\bibinfo
  {title} {Electron-{{Phonon}} beyond {{Fr{\"o}hlich}}: {{Dynamical
  Quadrupoles}} in {{Polar}} and {{Covalent Solids}}},\ }\href
  {https://doi.org/10.1103/PhysRevLett.125.136601} {\bibfield  {journal}
  {\bibinfo  {journal} {Physical Review Letters}\ }\textbf {\bibinfo {volume}
  {125}},\ \bibinfo {pages} {136601} (\bibinfo {year}
  {2020}{\natexlab{a}})}\BibitemShut {NoStop}%
\bibitem [{\citenamefont {Brunin}\ \emph
  {et~al.}(2020{\natexlab{b}})\citenamefont {Brunin}, \citenamefont {Miranda},
  \citenamefont {Giantomassi}, \citenamefont {Royo}, \citenamefont {Stengel},
  \citenamefont {Verstraete}, \citenamefont {Gonze}, \citenamefont
  {Rignanese},\ and\ \citenamefont {Hautier}}]{brunin2020a}%
  \BibitemOpen
  \bibfield  {author} {\bibinfo {author} {\bibfnamefont {G.}~\bibnamefont
  {Brunin}}, \bibinfo {author} {\bibfnamefont {H.~P.~C.}\ \bibnamefont
  {Miranda}}, \bibinfo {author} {\bibfnamefont {M.}~\bibnamefont
  {Giantomassi}}, \bibinfo {author} {\bibfnamefont {M.}~\bibnamefont {Royo}},
  \bibinfo {author} {\bibfnamefont {M.}~\bibnamefont {Stengel}}, \bibinfo
  {author} {\bibfnamefont {M.~J.}\ \bibnamefont {Verstraete}}, \bibinfo
  {author} {\bibfnamefont {X.}~\bibnamefont {Gonze}}, \bibinfo {author}
  {\bibfnamefont {G.-M.}\ \bibnamefont {Rignanese}},\ and\ \bibinfo {author}
  {\bibfnamefont {G.}~\bibnamefont {Hautier}},\ }\bibfield  {title} {\bibinfo
  {title} {Phonon-limited electron mobility in {{Si}}, {{GaAs}}, and {{GaP}}
  with exact treatment of dynamical quadrupoles},\ }\href
  {https://doi.org/10.1103/PhysRevB.102.094308} {\bibfield  {journal} {\bibinfo
   {journal} {Physical Review B}\ }\textbf {\bibinfo {volume} {102}},\ \bibinfo
  {pages} {094308} (\bibinfo {year} {2020}{\natexlab{b}})}\BibitemShut
  {NoStop}%
\bibitem [{\citenamefont {Park}\ \emph {et~al.}(2020)\citenamefont {Park},
  \citenamefont {Zhou}, \citenamefont {Jhalani}, \citenamefont {Dreyer},\ and\
  \citenamefont {Bernardi}}]{park2020}%
  \BibitemOpen
  \bibfield  {author} {\bibinfo {author} {\bibfnamefont {J.}~\bibnamefont
  {Park}}, \bibinfo {author} {\bibfnamefont {J.-J.}\ \bibnamefont {Zhou}},
  \bibinfo {author} {\bibfnamefont {V.~A.}\ \bibnamefont {Jhalani}}, \bibinfo
  {author} {\bibfnamefont {C.~E.}\ \bibnamefont {Dreyer}},\ and\ \bibinfo
  {author} {\bibfnamefont {M.}~\bibnamefont {Bernardi}},\ }\bibfield  {title}
  {\bibinfo {title} {Long-range quadrupole electron-phonon interaction from
  first principles},\ }\href {https://doi.org/10.1103/PhysRevB.102.125203}
  {\bibfield  {journal} {\bibinfo  {journal} {Physical Review B}\ }\textbf
  {\bibinfo {volume} {102}},\ \bibinfo {pages} {125203} (\bibinfo {year}
  {2020})}\BibitemShut {NoStop}%
\bibitem [{\citenamefont {Jhalani}\ \emph {et~al.}(2020)\citenamefont
  {Jhalani}, \citenamefont {Zhou}, \citenamefont {Park}, \citenamefont
  {Dreyer},\ and\ \citenamefont {Bernardi}}]{jhalani2020}%
  \BibitemOpen
  \bibfield  {author} {\bibinfo {author} {\bibfnamefont {V.~A.}\ \bibnamefont
  {Jhalani}}, \bibinfo {author} {\bibfnamefont {J.-J.}\ \bibnamefont {Zhou}},
  \bibinfo {author} {\bibfnamefont {J.}~\bibnamefont {Park}}, \bibinfo {author}
  {\bibfnamefont {C.~E.}\ \bibnamefont {Dreyer}},\ and\ \bibinfo {author}
  {\bibfnamefont {M.}~\bibnamefont {Bernardi}},\ }\bibfield  {title} {\bibinfo
  {title} {Piezoelectric {{Electron-Phonon Interaction}} from {{{\emph{Ab
  Initio}}}} {{Dynamical Quadrupoles}}: {{Impact}} on {{Charge Transport}} in
  {{Wurtzite GaN}}},\ }\href {https://doi.org/10.1103/PhysRevLett.125.136602}
  {\bibfield  {journal} {\bibinfo  {journal} {Physical Review Letters}\
  }\textbf {\bibinfo {volume} {125}},\ \bibinfo {pages} {136602} (\bibinfo
  {year} {2020})}\BibitemShut {NoStop}%
\bibitem [{\citenamefont {Smondyrev}(1986)}]{smondyrev1986}%
  \BibitemOpen
  \bibfield  {author} {\bibinfo {author} {\bibfnamefont {M.~A.}\ \bibnamefont
  {Smondyrev}},\ }\bibfield  {title} {\bibinfo {title} {Diagrams in the polaron
  model},\ }\href {https://doi.org/10.1007/BF01017794} {\bibfield  {journal}
  {\bibinfo  {journal} {Theoretical and Mathematical Physics}\ }\textbf
  {\bibinfo {volume} {68}},\ \bibinfo {pages} {653} (\bibinfo {year}
  {1986})}\BibitemShut {NoStop}%
\bibitem [{\citenamefont {Alexandrov}\ and\ \citenamefont
  {Devreese}(2010)}]{alexandrov2010}%
  \BibitemOpen
  \bibfield  {author} {\bibinfo {author} {\bibfnamefont {A.~S.}\ \bibnamefont
  {Alexandrov}}\ and\ \bibinfo {author} {\bibfnamefont {J.~T.}\ \bibnamefont
  {Devreese}},\ }\href {https://doi.org/10.1007/978-3-642-01896-1} {\emph
  {\bibinfo {title} {Advances in {{Polaron Physics}}}}},\ \bibinfo {series}
  {Springer {{Series}} in {{Solid-State Sciences}}}, Vol.\ \bibinfo {volume}
  {159}\ (\bibinfo  {publisher} {Springer Berlin Heidelberg},\ \bibinfo
  {address} {Berlin, Heidelberg},\ \bibinfo {year} {2010})\BibitemShut
  {NoStop}%
\bibitem [{\citenamefont {Lee}\ \emph {et~al.}(2020)\citenamefont {Lee},
  \citenamefont {Zhou}, \citenamefont {Chen},\ and\ \citenamefont
  {Bernardi}}]{lee2020}%
  \BibitemOpen
  \bibfield  {author} {\bibinfo {author} {\bibfnamefont {N.-E.}\ \bibnamefont
  {Lee}}, \bibinfo {author} {\bibfnamefont {J.-J.}\ \bibnamefont {Zhou}},
  \bibinfo {author} {\bibfnamefont {H.-Y.}\ \bibnamefont {Chen}},\ and\
  \bibinfo {author} {\bibfnamefont {M.}~\bibnamefont {Bernardi}},\ }\bibfield
  {title} {\bibinfo {title} {Ab initio electron-two-phonon scattering in
  {{GaAs}} from next-to-leading order perturbation theory},\ }\href
  {https://doi.org/10.1038/s41467-020-15339-0} {\bibfield  {journal} {\bibinfo
  {journal} {Nature Communications}\ }\textbf {\bibinfo {volume} {11}},\
  \bibinfo {pages} {1607} (\bibinfo {year} {2020})}\BibitemShut {NoStop}%
\bibitem [{\citenamefont {Goldberg}(1985)}]{goldberg1985}%
  \BibitemOpen
  \bibfield  {author} {\bibinfo {author} {\bibfnamefont {G.}~\bibnamefont
  {Goldberg}},\ }\bibfield  {title} {\bibinfo {title} {A rule for the
  combinatoric factors of {{Feynman}} diagrams},\ }\href
  {https://doi.org/10.1103/PhysRevD.32.3331} {\bibfield  {journal} {\bibinfo
  {journal} {Physical Review D}\ }\textbf {\bibinfo {volume} {32}},\ \bibinfo
  {pages} {3331} (\bibinfo {year} {1985})}\BibitemShut {NoStop}%
\bibitem [{\citenamefont {Kittel}(2018)}]{kittel2018}%
  \BibitemOpen
  \bibfield  {author} {\bibinfo {author} {\bibfnamefont {C.}~\bibnamefont
  {Kittel}},\ }\href@noop {} {\emph {\bibinfo {title} {Introduction to Solid
  State Physics}}},\ \bibinfo {edition} {global edition, [9th edition]}\ ed.\
  (\bibinfo  {publisher} {Wiley},\ \bibinfo {address} {Hoboken, NJ},\ \bibinfo
  {year} {2018})\BibitemShut {NoStop}%
\bibitem [{\citenamefont {Guster}\ \emph {et~al.}(2021)\citenamefont {Guster},
  \citenamefont {Melo}, \citenamefont {Martin}, \citenamefont
  {{Brousseau-Couture}}, \citenamefont {De~Abreu}, \citenamefont {Miglio},
  \citenamefont {Giantomassi}, \citenamefont {C{\^o}t{\'e}}, \citenamefont
  {Frost}, \citenamefont {Verstraete},\ and\ \citenamefont
  {Gonze}}]{guster2021}%
  \BibitemOpen
  \bibfield  {author} {\bibinfo {author} {\bibfnamefont {B.}~\bibnamefont
  {Guster}}, \bibinfo {author} {\bibfnamefont {P.}~\bibnamefont {Melo}},
  \bibinfo {author} {\bibfnamefont {B.~A.~A.}\ \bibnamefont {Martin}}, \bibinfo
  {author} {\bibfnamefont {V.}~\bibnamefont {{Brousseau-Couture}}}, \bibinfo
  {author} {\bibfnamefont {J.~C.}\ \bibnamefont {De~Abreu}}, \bibinfo {author}
  {\bibfnamefont {A.}~\bibnamefont {Miglio}}, \bibinfo {author} {\bibfnamefont
  {M.}~\bibnamefont {Giantomassi}}, \bibinfo {author} {\bibfnamefont
  {M.}~\bibnamefont {C{\^o}t{\'e}}}, \bibinfo {author} {\bibfnamefont {J.~M.}\
  \bibnamefont {Frost}}, \bibinfo {author} {\bibfnamefont {M.~J.}\ \bibnamefont
  {Verstraete}},\ and\ \bibinfo {author} {\bibfnamefont {X.}~\bibnamefont
  {Gonze}},\ }\bibfield  {title} {\bibinfo {title} {Fr{\"o}hlich polaron
  effective mass and localization length in cubic materials: {{Degenerate}} and
  anisotropic electronic bands},\ }\href
  {https://doi.org/10.1103/PhysRevB.104.235123} {\bibfield  {journal} {\bibinfo
   {journal} {Physical Review B}\ }\textbf {\bibinfo {volume} {104}},\ \bibinfo
  {pages} {235123} (\bibinfo {year} {2021})}\BibitemShut {NoStop}%
\bibitem [{\citenamefont {Ranalli}\ \emph {et~al.}(2024)\citenamefont
  {Ranalli}, \citenamefont {Verdi}, \citenamefont {Zacharias}, \citenamefont
  {Even}, \citenamefont {Giustino},\ and\ \citenamefont
  {Franchini}}]{ranalli2024}%
  \BibitemOpen
  \bibfield  {author} {\bibinfo {author} {\bibfnamefont {L.}~\bibnamefont
  {Ranalli}}, \bibinfo {author} {\bibfnamefont {C.}~\bibnamefont {Verdi}},
  \bibinfo {author} {\bibfnamefont {M.}~\bibnamefont {Zacharias}}, \bibinfo
  {author} {\bibfnamefont {J.}~\bibnamefont {Even}}, \bibinfo {author}
  {\bibfnamefont {F.}~\bibnamefont {Giustino}},\ and\ \bibinfo {author}
  {\bibfnamefont {C.}~\bibnamefont {Franchini}},\ }\bibfield  {title} {\bibinfo
  {title} {Electron mobilities in {{SrTi${\mathrm{O}}_{3}$}} and
  {{KTa${\mathrm{O}}_{3}$}}: {{Role}} of phonon anharmonicity, mass
  renormalization, and disorder},\ }\href
  {https://doi.org/10.1103/PhysRevMaterials.8.104603} {\bibfield  {journal}
  {\bibinfo  {journal} {Physical Review Materials}\ }\textbf {\bibinfo {volume}
  {8}},\ \bibinfo {pages} {104603} (\bibinfo {year} {2024})}\BibitemShut
  {NoStop}%
\bibitem [{\citenamefont {{King-Smith}}\ and\ \citenamefont
  {Vanderbilt}(1993)}]{king-smith1993}%
  \BibitemOpen
  \bibfield  {author} {\bibinfo {author} {\bibfnamefont {R.~D.}\ \bibnamefont
  {{King-Smith}}}\ and\ \bibinfo {author} {\bibfnamefont {D.}~\bibnamefont
  {Vanderbilt}},\ }\bibfield  {title} {\bibinfo {title} {Theory of polarization
  of crystalline solids},\ }\href {https://doi.org/10.1103/PhysRevB.47.1651}
  {\bibfield  {journal} {\bibinfo  {journal} {Physical Review B}\ }\textbf
  {\bibinfo {volume} {47}},\ \bibinfo {pages} {1651} (\bibinfo {year}
  {1993})}\BibitemShut {NoStop}%
\bibitem [{\citenamefont {Veithen}\ \emph {et~al.}(2005)\citenamefont
  {Veithen}, \citenamefont {Gonze},\ and\ \citenamefont
  {Ghosez}}]{veithen2005}%
  \BibitemOpen
  \bibfield  {author} {\bibinfo {author} {\bibfnamefont {M.}~\bibnamefont
  {Veithen}}, \bibinfo {author} {\bibfnamefont {X.}~\bibnamefont {Gonze}},\
  and\ \bibinfo {author} {\bibfnamefont {{\relax Ph}.}~\bibnamefont {Ghosez}},\
  }\bibfield  {title} {\bibinfo {title} {Nonlinear optical susceptibilities,
  {{Raman}} efficiencies, and electro-optic tensors from first-principles
  density functional perturbation theory},\ }\href
  {https://doi.org/10.1103/PhysRevB.71.125107} {\bibfield  {journal} {\bibinfo
  {journal} {Physical Review B}\ }\textbf {\bibinfo {volume} {71}},\ \bibinfo
  {pages} {125107} (\bibinfo {year} {2005})}\BibitemShut {NoStop}%
\bibitem [{\citenamefont {Perdew}\ \emph {et~al.}(2008)\citenamefont {Perdew},
  \citenamefont {Ruzsinszky}, \citenamefont {Csonka}, \citenamefont {Vydrov},
  \citenamefont {Scuseria}, \citenamefont {Constantin}, \citenamefont {Zhou},\
  and\ \citenamefont {Burke}}]{perdew2008}%
  \BibitemOpen
  \bibfield  {author} {\bibinfo {author} {\bibfnamefont {J.~P.}\ \bibnamefont
  {Perdew}}, \bibinfo {author} {\bibfnamefont {A.}~\bibnamefont {Ruzsinszky}},
  \bibinfo {author} {\bibfnamefont {G.~I.}\ \bibnamefont {Csonka}}, \bibinfo
  {author} {\bibfnamefont {O.~A.}\ \bibnamefont {Vydrov}}, \bibinfo {author}
  {\bibfnamefont {G.~E.}\ \bibnamefont {Scuseria}}, \bibinfo {author}
  {\bibfnamefont {L.~A.}\ \bibnamefont {Constantin}}, \bibinfo {author}
  {\bibfnamefont {X.}~\bibnamefont {Zhou}},\ and\ \bibinfo {author}
  {\bibfnamefont {K.}~\bibnamefont {Burke}},\ }\bibfield  {title} {\bibinfo
  {title} {Restoring the {{Density-Gradient Expansion}} for {{Exchange}} in
  {{Solids}} and {{Surfaces}}},\ }\href
  {https://doi.org/10.1103/PhysRevLett.100.136406} {\bibfield  {journal}
  {\bibinfo  {journal} {Physical Review Letters}\ }\textbf {\bibinfo {volume}
  {100}},\ \bibinfo {pages} {136406} (\bibinfo {year} {2008})}\BibitemShut
  {NoStop}%
\bibitem [{\citenamefont {King}\ \emph {et~al.}(2012)\citenamefont {King},
  \citenamefont {He}, \citenamefont {Eknapakul}, \citenamefont {Buaphet},
  \citenamefont {Mo}, \citenamefont {Kaneko}, \citenamefont {Harashima},
  \citenamefont {Hikita}, \citenamefont {Bahramy}, \citenamefont {Bell},
  \citenamefont {Hussain}, \citenamefont {Tokura}, \citenamefont {Shen},
  \citenamefont {Hwang}, \citenamefont {Baumberger},\ and\ \citenamefont
  {Meevasana}}]{king2012}%
  \BibitemOpen
  \bibfield  {author} {\bibinfo {author} {\bibfnamefont {P.~D.~C.}\
  \bibnamefont {King}}, \bibinfo {author} {\bibfnamefont {R.~H.}\ \bibnamefont
  {He}}, \bibinfo {author} {\bibfnamefont {T.}~\bibnamefont {Eknapakul}},
  \bibinfo {author} {\bibfnamefont {P.}~\bibnamefont {Buaphet}}, \bibinfo
  {author} {\bibfnamefont {S.-K.}\ \bibnamefont {Mo}}, \bibinfo {author}
  {\bibfnamefont {Y.}~\bibnamefont {Kaneko}}, \bibinfo {author} {\bibfnamefont
  {S.}~\bibnamefont {Harashima}}, \bibinfo {author} {\bibfnamefont
  {Y.}~\bibnamefont {Hikita}}, \bibinfo {author} {\bibfnamefont {M.~S.}\
  \bibnamefont {Bahramy}}, \bibinfo {author} {\bibfnamefont {C.}~\bibnamefont
  {Bell}}, \bibinfo {author} {\bibfnamefont {Z.}~\bibnamefont {Hussain}},
  \bibinfo {author} {\bibfnamefont {Y.}~\bibnamefont {Tokura}}, \bibinfo
  {author} {\bibfnamefont {Z.-X.}\ \bibnamefont {Shen}}, \bibinfo {author}
  {\bibfnamefont {H.~Y.}\ \bibnamefont {Hwang}}, \bibinfo {author}
  {\bibfnamefont {F.}~\bibnamefont {Baumberger}},\ and\ \bibinfo {author}
  {\bibfnamefont {W.}~\bibnamefont {Meevasana}},\ }\bibfield  {title} {\bibinfo
  {title} {Subband {{Structure}} of a {{Two-Dimensional Electron Gas Formed}}
  at the {{Polar Surface}} of the {{Strong Spin-Orbit Perovskite}}
  {{KTa${\mathrm{O}}_{3}$}}},\ }\href
  {https://doi.org/10.1103/PhysRevLett.108.117602} {\bibfield  {journal}
  {\bibinfo  {journal} {Physical Review Letters}\ }\textbf {\bibinfo {volume}
  {108}},\ \bibinfo {pages} {117602} (\bibinfo {year} {2012})}\BibitemShut
  {NoStop}%
\bibitem [{\citenamefont {Esswein}\ and\ \citenamefont
  {Spaldin}(2023)}]{esswein2023}%
  \BibitemOpen
  \bibfield  {author} {\bibinfo {author} {\bibfnamefont {T.}~\bibnamefont
  {Esswein}}\ and\ \bibinfo {author} {\bibfnamefont {N.~A.}\ \bibnamefont
  {Spaldin}},\ }\bibfield  {title} {\bibinfo {title} {First-principles
  calculation of electron-phonon coupling in doped {{KTa${\mathrm{O}}_{3}$}}},\
  }\href {https://doi.org/10.12688/openreseurope.16312.1} {\bibfield  {journal}
  {\bibinfo  {journal} {Open Research Europe}\ }\textbf {\bibinfo {volume}
  {3}},\ \bibinfo {pages} {177} (\bibinfo {year} {2023})}\BibitemShut {NoStop}%
\bibitem [{\citenamefont {Dressler}\ \emph {et~al.}(1987)\citenamefont
  {Dressler}, \citenamefont {Griebner},\ and\ \citenamefont
  {Kittner}}]{dressler1987}%
  \BibitemOpen
  \bibfield  {author} {\bibinfo {author} {\bibfnamefont {L.}~\bibnamefont
  {Dressler}}, \bibinfo {author} {\bibfnamefont {U.}~\bibnamefont {Griebner}},\
  and\ \bibinfo {author} {\bibfnamefont {R.}~\bibnamefont {Kittner}},\
  }\bibfield  {title} {\bibinfo {title} {Precision measurement of lattice
  parameters in {{LiF}} monocrystals},\ }\href
  {https://doi.org/10.1002/crat.2170221116} {\bibfield  {journal} {\bibinfo
  {journal} {Crystal Research and Technology}\ }\textbf {\bibinfo {volume}
  {22}},\ \bibinfo {pages} {1431} (\bibinfo {year} {1987})}\BibitemShut
  {NoStop}%
\bibitem [{\citenamefont {Dolling}\ \emph {et~al.}(1968)\citenamefont
  {Dolling}, \citenamefont {Smith}, \citenamefont {Nicklow}, \citenamefont
  {Vijayaraghavan},\ and\ \citenamefont {Wilkinson}}]{dolling1968}%
  \BibitemOpen
  \bibfield  {author} {\bibinfo {author} {\bibfnamefont {G.}~\bibnamefont
  {Dolling}}, \bibinfo {author} {\bibfnamefont {H.~G.}\ \bibnamefont {Smith}},
  \bibinfo {author} {\bibfnamefont {R.~M.}\ \bibnamefont {Nicklow}}, \bibinfo
  {author} {\bibfnamefont {P.~R.}\ \bibnamefont {Vijayaraghavan}},\ and\
  \bibinfo {author} {\bibfnamefont {M.~K.}\ \bibnamefont {Wilkinson}},\
  }\bibfield  {title} {\bibinfo {title} {Lattice {{Dynamics}} of {{Lithium
  Fluoride}}},\ }\href {https://doi.org/10.1103/PhysRev.168.970} {\bibfield
  {journal} {\bibinfo  {journal} {Physical Review}\ }\textbf {\bibinfo {volume}
  {168}},\ \bibinfo {pages} {970} (\bibinfo {year} {1968})}\BibitemShut
  {NoStop}%
\bibitem [{\citenamefont {Andeen}\ \emph {et~al.}(1970)\citenamefont {Andeen},
  \citenamefont {Fontanella},\ and\ \citenamefont {Schuele}}]{andeen1970}%
  \BibitemOpen
  \bibfield  {author} {\bibinfo {author} {\bibfnamefont {C.}~\bibnamefont
  {Andeen}}, \bibinfo {author} {\bibfnamefont {J.}~\bibnamefont {Fontanella}},\
  and\ \bibinfo {author} {\bibfnamefont {D.}~\bibnamefont {Schuele}},\
  }\bibfield  {title} {\bibinfo {title} {Low-{{Frequency Dielectric Constant}}
  of {{LiF}}, {{NaF}}, {{NaCl}}, {{NaBr}}, {{KCl}}, and {{KBr}} by the
  {{Method}} of {{Substitution}}},\ }\href
  {https://doi.org/10.1103/PhysRevB.2.5068} {\bibfield  {journal} {\bibinfo
  {journal} {Physical Review B}\ }\textbf {\bibinfo {volume} {2}},\ \bibinfo
  {pages} {5068} (\bibinfo {year} {1970})}\BibitemShut {NoStop}%
\bibitem [{\citenamefont {Levin}\ and\ \citenamefont
  {Offenbacher}(1960)}]{levin1960}%
  \BibitemOpen
  \bibfield  {author} {\bibinfo {author} {\bibfnamefont {E.~R.}\ \bibnamefont
  {Levin}}\ and\ \bibinfo {author} {\bibfnamefont {E.~L.}\ \bibnamefont
  {Offenbacher}},\ }\bibfield  {title} {\bibinfo {title} {Theory of
  {{Dielectric Constants}} of {{LiF}}},\ }\href
  {https://doi.org/10.1103/PhysRev.118.1142} {\bibfield  {journal} {\bibinfo
  {journal} {Physical Review}\ }\textbf {\bibinfo {volume} {118}},\ \bibinfo
  {pages} {1142} (\bibinfo {year} {1960})}\BibitemShut {NoStop}%
\bibitem [{\citenamefont {Wemple}(1965)}]{wemple1965}%
  \BibitemOpen
  \bibfield  {author} {\bibinfo {author} {\bibfnamefont {S.~H.}\ \bibnamefont
  {Wemple}},\ }\bibfield  {title} {\bibinfo {title} {Some {{Transport
  Properties}} of {{Oxygen-Deficient Single-Crystal Potassium Tantalate}}
  {{KTa${\mathrm{O}}_{3}$}}},\ }\href
  {https://doi.org/10.1103/PhysRev.137.A1575} {\bibfield  {journal} {\bibinfo
  {journal} {Physical Review}\ }\textbf {\bibinfo {volume} {137}},\ \bibinfo
  {pages} {A1575} (\bibinfo {year} {1965})}\BibitemShut {NoStop}%
\bibitem [{\citenamefont {Senhouse}\ \emph {et~al.}(1965)\citenamefont
  {Senhouse}, \citenamefont {Smith},\ and\ \citenamefont
  {DePaolis}}]{senhouse1965}%
  \BibitemOpen
  \bibfield  {author} {\bibinfo {author} {\bibfnamefont {L.~S.}\ \bibnamefont
  {Senhouse}}, \bibinfo {author} {\bibfnamefont {G.~E.}\ \bibnamefont
  {Smith}},\ and\ \bibinfo {author} {\bibfnamefont {M.~V.}\ \bibnamefont
  {DePaolis}},\ }\bibfield  {title} {\bibinfo {title} {Cyclotron {{Resonance}}
  in {{Potassium Tantalate}}},\ }\href
  {https://doi.org/10.1103/PhysRevLett.15.776} {\bibfield  {journal} {\bibinfo
  {journal} {Physical Review Letters}\ }\textbf {\bibinfo {volume} {15}},\
  \bibinfo {pages} {776} (\bibinfo {year} {1965})}\BibitemShut {NoStop}%
\bibitem [{\citenamefont {Mattheiss}(1972)}]{mattheiss1972}%
  \BibitemOpen
  \bibfield  {author} {\bibinfo {author} {\bibfnamefont {L.~F.}\ \bibnamefont
  {Mattheiss}},\ }\bibfield  {title} {\bibinfo {title} {Energy bands for
  {{KNi${\mathrm{F}}_{3}$}}, {{SrTi${\mathrm{O}}_{3}$}},
  {{KMo${\mathrm{O}}_{3}$}}, and {{KTa${\mathrm{O}}_{3}$}}},\ }\href
  {https://doi.org/10.1103/PhysRevB.6.4718} {\bibfield  {journal} {\bibinfo
  {journal} {Physical Review B}\ }\textbf {\bibinfo {volume} {6}},\ \bibinfo
  {pages} {4718} (\bibinfo {year} {1972})}\BibitemShut {NoStop}%
\bibitem [{\citenamefont {Perry}\ \emph {et~al.}(1989)\citenamefont {Perry},
  \citenamefont {Currat}, \citenamefont {Buhay}, \citenamefont {Migoni},
  \citenamefont {Stirling},\ and\ \citenamefont {Axe}}]{perry1989}%
  \BibitemOpen
  \bibfield  {author} {\bibinfo {author} {\bibfnamefont {C.~H.}\ \bibnamefont
  {Perry}}, \bibinfo {author} {\bibfnamefont {R.}~\bibnamefont {Currat}},
  \bibinfo {author} {\bibfnamefont {H.}~\bibnamefont {Buhay}}, \bibinfo
  {author} {\bibfnamefont {R.~M.}\ \bibnamefont {Migoni}}, \bibinfo {author}
  {\bibfnamefont {W.~G.}\ \bibnamefont {Stirling}},\ and\ \bibinfo {author}
  {\bibfnamefont {J.~D.}\ \bibnamefont {Axe}},\ }\bibfield  {title} {\bibinfo
  {title} {Phonon dispersion and lattice dynamics of {{KTa${\mathrm{O}}_{3}$}}
  from 4 to 1220 {{K}}},\ }\href {https://doi.org/10.1103/PhysRevB.39.8666}
  {\bibfield  {journal} {\bibinfo  {journal} {Physical Review B}\ }\textbf
  {\bibinfo {volume} {39}},\ \bibinfo {pages} {8666} (\bibinfo {year}
  {1989})}\BibitemShut {NoStop}%
\bibitem [{\citenamefont {Fujii}\ and\ \citenamefont
  {Sakudo}(1976)}]{fujii1976}%
  \BibitemOpen
  \bibfield  {author} {\bibinfo {author} {\bibfnamefont {Y.}~\bibnamefont
  {Fujii}}\ and\ \bibinfo {author} {\bibfnamefont {T.}~\bibnamefont {Sakudo}},\
  }\bibfield  {title} {\bibinfo {title} {Dielectric and {{Optical Properties}}
  of {{KTa${\mathrm{O}}_{3}$}}},\ }\href {https://doi.org/10.1143/JPSJ.41.888}
  {\bibfield  {journal} {\bibinfo  {journal} {Journal of the Physical Society
  of Japan}\ }\textbf {\bibinfo {volume} {41}},\ \bibinfo {pages} {888}
  (\bibinfo {year} {1976})}\BibitemShut {NoStop}%
\bibitem [{\citenamefont {Setyawan}\ and\ \citenamefont
  {Curtarolo}(2010)}]{setyawan2010}%
  \BibitemOpen
  \bibfield  {author} {\bibinfo {author} {\bibfnamefont {W.}~\bibnamefont
  {Setyawan}}\ and\ \bibinfo {author} {\bibfnamefont {S.}~\bibnamefont
  {Curtarolo}},\ }\bibfield  {title} {\bibinfo {title} {High-throughput
  electronic band structure calculations: {{Challenges}} and tools},\ }\href
  {https://doi.org/10.1016/j.commatsci.2010.05.010} {\bibfield  {journal}
  {\bibinfo  {journal} {Computational Materials Science}\ }\textbf {\bibinfo
  {volume} {49}},\ \bibinfo {pages} {299} (\bibinfo {year} {2010})}\BibitemShut
  {NoStop}%
\bibitem [{\citenamefont {Refson}\ \emph {et~al.}(2006)\citenamefont {Refson},
  \citenamefont {Tulip},\ and\ \citenamefont {Clark}}]{refson2006}%
  \BibitemOpen
  \bibfield  {author} {\bibinfo {author} {\bibfnamefont {K.}~\bibnamefont
  {Refson}}, \bibinfo {author} {\bibfnamefont {P.~R.}\ \bibnamefont {Tulip}},\
  and\ \bibinfo {author} {\bibfnamefont {S.~J.}\ \bibnamefont {Clark}},\
  }\bibfield  {title} {\bibinfo {title} {Variational density-functional
  perturbation theory for dielectrics and lattice dynamics},\ }\href
  {https://doi.org/10.1103/PhysRevB.73.155114} {\bibfield  {journal} {\bibinfo
  {journal} {Physical Review B}\ }\textbf {\bibinfo {volume} {73}},\ \bibinfo
  {pages} {155114} (\bibinfo {year} {2006})}\BibitemShut {NoStop}%
\bibitem [{\citenamefont {Houtput}\ and\ \citenamefont
  {Tempere}(2022)}]{houtput2022}%
  \BibitemOpen
  \bibfield  {author} {\bibinfo {author} {\bibfnamefont {M.}~\bibnamefont
  {Houtput}}\ and\ \bibinfo {author} {\bibfnamefont {J.}~\bibnamefont
  {Tempere}},\ }\bibfield  {title} {\bibinfo {title} {Optical conductivity of
  an anharmonic large polaron gas at weak coupling},\ }\href
  {https://doi.org/10.1103/PhysRevB.106.214315} {\bibfield  {journal} {\bibinfo
   {journal} {Physical Review B}\ }\textbf {\bibinfo {volume} {106}},\ \bibinfo
  {pages} {214315} (\bibinfo {year} {2022})}\BibitemShut {NoStop}%
\bibitem [{\citenamefont {Brivio}\ \emph {et~al.}(2015)\citenamefont {Brivio},
  \citenamefont {Frost}, \citenamefont {Skelton}, \citenamefont {Jackson},
  \citenamefont {Weber}, \citenamefont {Weller}, \citenamefont {Go{\~n}i},
  \citenamefont {Leguy}, \citenamefont {Barnes},\ and\ \citenamefont
  {Walsh}}]{brivio2015}%
  \BibitemOpen
  \bibfield  {author} {\bibinfo {author} {\bibfnamefont {F.}~\bibnamefont
  {Brivio}}, \bibinfo {author} {\bibfnamefont {J.~M.}\ \bibnamefont {Frost}},
  \bibinfo {author} {\bibfnamefont {J.~M.}\ \bibnamefont {Skelton}}, \bibinfo
  {author} {\bibfnamefont {A.~J.}\ \bibnamefont {Jackson}}, \bibinfo {author}
  {\bibfnamefont {O.~J.}\ \bibnamefont {Weber}}, \bibinfo {author}
  {\bibfnamefont {M.~T.}\ \bibnamefont {Weller}}, \bibinfo {author}
  {\bibfnamefont {A.~R.}\ \bibnamefont {Go{\~n}i}}, \bibinfo {author}
  {\bibfnamefont {A.~M.~A.}\ \bibnamefont {Leguy}}, \bibinfo {author}
  {\bibfnamefont {P.~R.~F.}\ \bibnamefont {Barnes}},\ and\ \bibinfo {author}
  {\bibfnamefont {A.}~\bibnamefont {Walsh}},\ }\bibfield  {title} {\bibinfo
  {title} {Lattice dynamics and vibrational spectra of the orthorhombic,
  tetragonal, and cubic phases of methylammonium lead iodide},\ }\href
  {https://doi.org/10.1103/PhysRevB.92.144308} {\bibfield  {journal} {\bibinfo
  {journal} {Physical Review B}\ }\textbf {\bibinfo {volume} {92}},\ \bibinfo
  {pages} {144308} (\bibinfo {year} {2015})}\BibitemShut {NoStop}%
\bibitem [{\citenamefont {Gu}\ \emph {et~al.}(2021)\citenamefont {Gu},
  \citenamefont {Yin},\ and\ \citenamefont {Gong}}]{gu2021}%
  \BibitemOpen
  \bibfield  {author} {\bibinfo {author} {\bibfnamefont {H.-Y.}\ \bibnamefont
  {Gu}}, \bibinfo {author} {\bibfnamefont {W.-J.}\ \bibnamefont {Yin}},\ and\
  \bibinfo {author} {\bibfnamefont {X.-G.}\ \bibnamefont {Gong}},\ }\bibfield
  {title} {\bibinfo {title} {Significant phonon anharmonicity drives phase
  transitions in cspb${\mathrm{i}}_{3}$},\ }\href
  {https://doi.org/10.1063/5.0072367} {\bibfield  {journal} {\bibinfo
  {journal} {Applied Physics Letters}\ }\textbf {\bibinfo {volume} {119}},\
  \bibinfo {pages} {191101} (\bibinfo {year} {2021})}\BibitemShut {NoStop}%
\bibitem [{\citenamefont {Houtput}(2025)}]{1e2ph-spectral}%
  \BibitemOpen
  \bibfield  {author} {\bibinfo {author} {\bibfnamefont {M.}~\bibnamefont
  {Houtput}},\ }\href {https://github.com/MHoutput/1e2ph-spectral/} {\bibinfo
  {title} {{{https://github.com/MHoutput/1e2ph-spectral/}}}} (\bibinfo {year}
  {2025})\BibitemShut {NoStop}%
\end{thebibliography}%

\end{document}